\documentclass[11pt]{article}
\pdfoutput=1
\usepackage{amssymb,amsmath,times}
\usepackage[T1]{fontenc}
\usepackage{mathptmx}
\usepackage[dvipsnames]{xcolor}
\usepackage{graphicx}
\usepackage{fancyhdr}
\usepackage{multirow}
\usepackage[numbers,sort&compress]{natbib}
\usepackage{color}
\usepackage[nolist]{acronym}
\usepackage[plainpages=false,pdfpagelabels,colorlinks=true]{hyperref}
\usepackage[latin1]{inputenc}
\usepackage{grffile}
\usepackage{multirow}
\usepackage{adjustbox}
\usepackage{subfiles}
\usepackage[small,compact]{titlesec}
\usepackage[font=small]{caption}
\usepackage{multirow}
\usepackage{afterpage}
\usepackage{multicol}
\usepackage[normalem]{ulem}
\usepackage{wrapfig}
\usepackage{pdfpages}
\usepackage{bm}

\usepackage{cleveref}
\crefformat{footnote}{#2\footnotemark[#1]#3}

\usepackage{chngcntr}
\counterwithin{figure}{section}
\counterwithin{table}{section}

\usepackage{enumitem}
\setitemize{noitemsep,topsep=0pt,parsep=0pt,partopsep=0pt}
\setenumerate{noitemsep,topsep=0pt,parsep=0pt,partopsep=0pt}

\usepackage[font={small},margin=0em]{caption}

\usepackage[small,compact]{titlesec}
\titleformat{\section}
  {\normalfont\sffamily\Large\bfseries\color{MidnightBlue}\filcenter}
  {\thesection}{1em}{}
\titleformat{\subsection}
  {\normalfont\sffamily\large\bfseries\color{MidnightBlue}}
  {\thesubsection}{1em}{}
\titleformat{\subsubsection}
  {\normalfont\sffamily\bfseries\itshape\color{MidnightBlue}}
  {\thesubsubsection}{0.5em}{}
\titleformat{\paragraph}[runin]
  {\normalfont\sffamily\bfseries\color{MidnightBlue}}
  {\theparagraph}{}{$\bullet$\hskip 0.5em}
\titlespacing*{\subsection}{0pt}{6pt}{2pt}
\titlespacing*{\subsubsection}{0pt}{4pt}{2pt}

\fancypagestyle{cost}{\fancyhf{}\fancyfoot[C]{\thepage\\\textit{\sf\color{gray}The cost information contained in this document is of a budgetary and planning nature and is intended for informational purposes only.}}}

\newlength{\pagewidthA}
\newlength{\pageheightA}
\newlength{\pagewidthB}
\newlength{\pageheightB}
\newlength{\stockwidth}
\newlength{\stockheight}

\usepackage{geometry}
\newcommand{\generatePageLayouts}{%
  \newgeometry{layoutwidth=\pagewidthA,layoutheight=\pageheightA, left=1in,right=1in,top=1in,bottom=1in}
  \savegeometry{LayoutPageA}

  \newgeometry{layoutwidth=\pagewidthB,layoutheight=\pageheightB, left=1in,right=1in,top=.75in,bottom=1in}
  \savegeometry{LayoutPageB}
}

\newcommand{\switchToLayoutPageA}{%
  \pdfpagewidth=\pagewidthA \pdfpageheight=\pageheightA % for PDF output
  \paperwidth=\pagewidthA \paperheight=\pageheightA     % for TikZ
  \stockwidth=\pagewidthA \stockheight=\pageheightA % hyperref (memoir)?!
  \loadgeometry{LayoutPageA} % note; \loadgeometry may reset paperwidth/h!
}

\newcommand{\switchToLayoutPageB}{%
  \pdfpagewidth=\pagewidthB \pdfpageheight=\pageheightB % for PDF output
  \paperwidth=\pagewidthB \paperheight=\pageheightB     % for TikZ
  \stockwidth=\pagewidthB \stockheight=\pageheightB % hyperref (memoir)?!
  \loadgeometry{LayoutPageB} % note; \loadgeometry may reset paperwidth/h!
}

\newcommand{\captiontext}{\small \setlength{\baselineskip}{0.90\baselineskip}}

\def\bei{\begin{itemize}}
\def\eei{\end{itemize}}

\def\Neff{{N_{\rm eff}}}
\newcommand{\hi}{H{\sc i}~}
\newcommand{\HI}{H{\sc i}}

\def\planck{{\it Planck}}

\def\wmap{{\it WMAP}}
\def\cobe{{\it COBE}}

\newcommand{\suffix}{pdf} % for pdflatex
\newcommand{\pico}{PICO}
\newcommand{\arcmin}{\ensuremath{'}}
\newcommand{\degree}{\ensuremath{^\circ}}

\def\microkamin{$\mu{\mbox{K}}\cdot \mbox{arcmin}$}

\def\ruo2{RuO$_{2}$}

\def\mathrelfun#1#2{\lower3.6pt\vbox{\baselineskip0pt\lineskip.9pt
  \ialign{$\mathsurround=0pt#1\hfil##\hfil$\crcr#2\crcr\sim\crcr}}}
\def\simlt{\mathrel{\mathpalette\mathrelfun <}}
\def\simgt{\mathrel{\mathpalette\mathrelfun >}}

\long\def\comment#1{}

\newbox\tablebox    \newdimen\tablewidth
\def\leaderfil{\leaders\hbox to 5pt{\hss.\hss}\hfil}

\def\tablenote#1 #2\par{\begingroup \parindent=0.8em
    \abovedisplayshortskip=0pt\belowdisplayshortskip=0pt
    \noindent
    $$\hss\vbox{\hsize\tablewidth \hangindent=\parindent \hangafter=1 \noindent
    \hbox to \parindent{$^#1$\hss}\strut#2\strut\par}\hss$$
    \endgroup}
\def\doubleline{\vskip 3pt\hrule \vskip 1.5pt \hrule \vskip 5pt}
\newcommand{\wisk}[1]{{\ifmmode{#1}\else{$#1$}\fi}}

\newbox\tablebox    \newdimen\tablewidth
\def\leaderfil{\leaders\hbox to 5pt{\hss.\hss}\hfil}
\def\endPlancktable{\tablewidth=\columnwidth 
    $$\hss\copy\tablebox\hss$$
    \vskip-\lastskip\vskip -2pt}

\def\tablenote#1 #2\par{\begingroup \parindent=0.8em
    \abovedisplayshortskip=0pt\belowdisplayshortskip=0pt
    \noindent
    $$\hss\vbox{\hsize\tablewidth \hangindent=\parindent \hangafter=1 \noindent
    \hbox to \parindent{$^#1$\hss}\strut#2\strut\par}\hss$$
    \endgroup}
\def\doubleline{\vskip 3pt\hrule \vskip 1.5pt \hrule \vskip 5pt}

\begin{document}

% generate page layouts first based on layoutwidth as page size;
  % don't switch actual page sizes yet:
  \generatePageLayouts{}

\setlength{\baselineskip}{0.96\baselineskip} %% measured, 5.0 lines/inch.  Can go to 0.96\baselineskip
\setlength{\parskip}{1.\parskip}
 \switchToLayoutPageA{}

%\tableofcontents

\pagenumbering{gobble}

%% Karl Young adding cover page as full page pdf.

%\includepdf{images/PICO_Cover_v6_final.pdf} %Full resolution cover
\includepdf{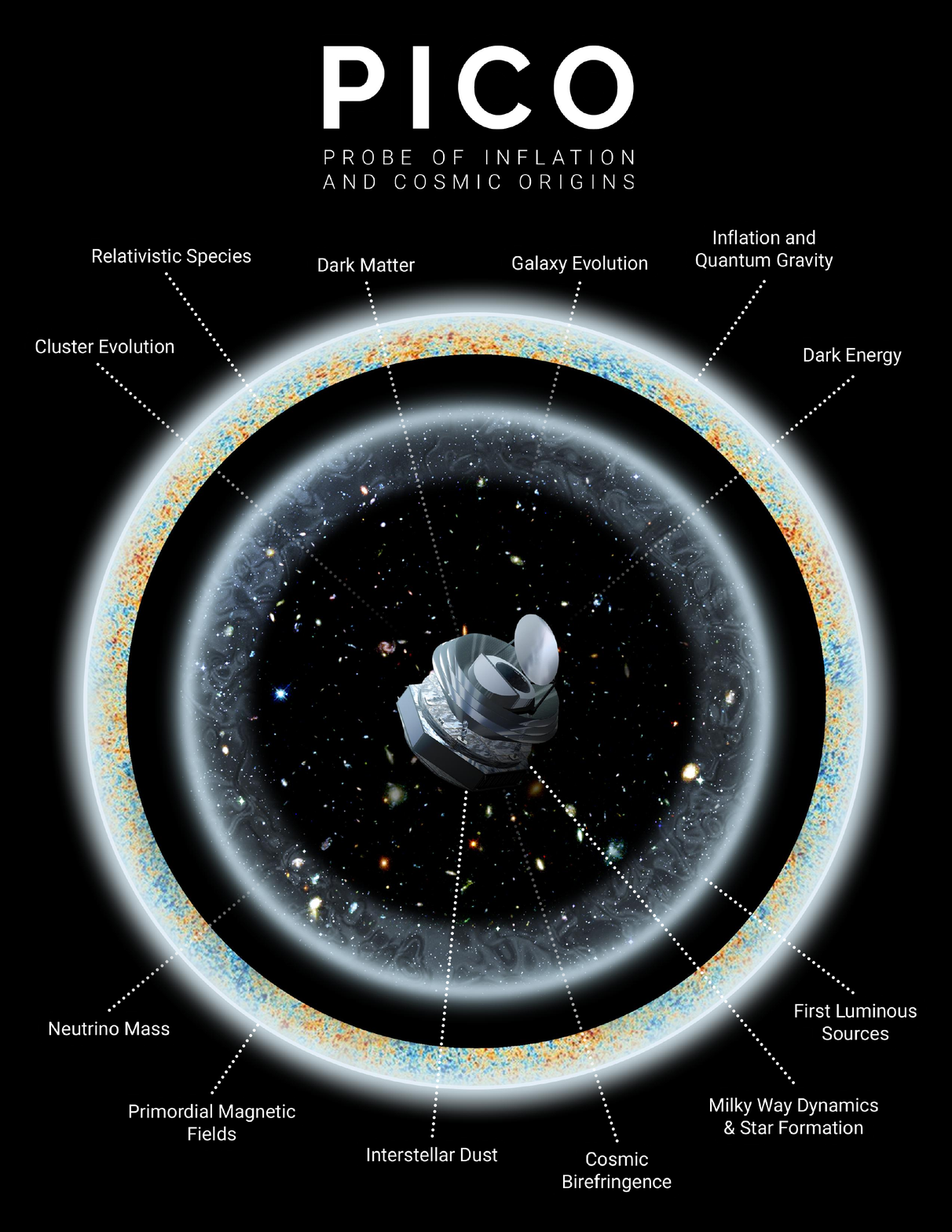} %Low resolution cover for faster pdf.

%%

%\LARGE{ \centerline{\bf{The Probe of Inflation and Cosmic Origins}}}
%\vspace{1.5in}
%%\Large{ \centerline{A Space Mission Study Report}}
%%\Large{ \centerline{December, 2018 }}
%%\vspace{0.5in}
%%\parindent = 0pt
%%\large{Principal Investigator:} \\
%%\large{Steering Committee:} \\
%%\large{Executive Committee:} \\
%%\large{Contributors:} \\
%%\large{Endorsers:} \\
%
%\normalsize
%
%\vspace{0.35in}
%\begin{figure}[!thb]
%%\begin{center}
%\hspace{-0.25in}
%\includegraphics[width=7.0in]{images/PICO_science goals and probes.png}
%%\end{center}
%%\hspace{-0.15in}
%%\caption{PICO science goals and probes}
%\label{fig:PICOsci_probes}
%\end{figure}

\parindent = 15pt

\newpage
\newpage
%\twocolumn

%test figure in main file
%
\pagenumbering{roman}
\setcounter{page}{1}

% Author list on seperate pages

%
% author list table.
% 4 columns, authors + affiliations
%
\LARGE{ \centerline{\bf{PICO: Probe of Inflation and Cosmic Origins}}}
%
%\vskip -16pt
\Large{ \centerline{Report from a Probe-Scale Mission Study}}
\Large{ \centerline{March, 2019 }}
\vspace{6pt}

\parindent = 0pt
%\normalsize{
\small{
Principal Investigator: Shaul Hanany (hanany@umn.edu)\,$^{1}$ } \\
\vspace{-6pt} \\
%\normalsize{
\small{
Steering Committee: Charles Bennett\,$^{2}$, Scott Dodelson\,$^{3}$, Lyman Page\,$^{4}$ } \\
\vspace{-6pt} \\
%\normalsize{
\small{
\raggedright
Executive Committee:
James~Bartlett\,$^{5,6}$,
Nick~Battaglia\,$^{7}$,
Jamie~Bock\,$^{8,6}$,
Julian~Borrill\,$^{9,10}$,
David~Chuss\,$^{11}$,
Brendan~P.~Crill\,$^{6}$,
Jacques~Delabrouille\,$^{5,12}$,
Mark~Devlin\,$^{13}$,
Laura~Fissel\,$^{14}$,
Raphael~Flauger\,$^{15}$,
Dan~Green\,$^{16}$,
J.~Colin~Hill\,$^{17,18}$,
Johannes~Hubmayr\,$^{19}$,
William~Jones\,$^{4}$,
Lloyd~Knox\,$^{20}$,
Al~Kogut\,$^{21}$,
Charles~Lawrence\,$^{6}$,
Jeff~McMahon\,$^{22}$,
Tim~Pearson\,$^{8}$,
Clem~Pryke\,$^{1}$,
Marcel~Schmittfull\,$^{18}$,
Amy~Trangsrud\,$^{6}$,
Alexander~van~Engelen\,$^{23}$
\\
}
\label{authorlist}

\vspace{9pt}

\Large  {\centerline {Authors}}
%\vskip 0pt
%
\footnotesize {
% change to left justified
\begin{multicols}{4}
%Marcelo Alvarez $^{9}$                \\
Marcelo Alvarez $^{9,16}$                 \\
Emmanuel Artis $^{12}$                  \\
%Peter Ashton $^{9}$                   \\
Peter Ashton $^{24,16,9}$                    \\
%Peter Ashton $^{24}$                   \\
Jonathan Aumont $^{25}$                 \\
Ragnhild Aurlien $^{26}$                \\
Ranajoy Banerji $^{26}$                 \\
R. Belen Barreiro $^{27}$               \\
James G. Bartlett $^{5,6}$               \\
%James G. Bartlett $^{6}$              \\
Soumen Basak $^{28}$                    \\
Nick Battaglia $^{7}$                  \\
Jamie Bock $^{8,6}$                      \\
%Jamie Bock   !!Caltech then JPL!! $^{6}$   \\
Kimberly K. Boddy $^{2}$               \\
Matteo Bonato $^{29}$                   \\
Julian Borrill $^{9,10}$                  \\
%Julian Borrill $^{10}$                 \\
Fran\c{c}ois Bouchet $^{30}$            \\
Fran\c{c}ois Boulanger $^{31}$          \\
Blakesley Burkhart $^{32}$              \\
Jens Chluba $^{33}$                     \\
David Chuss $^{11}$                     \\
Susan E. Clark $^{18,34}$                  \\
Joelle Cooperrider $^{6}$              \\
Brendan P. Crill $^{6}$                \\
Gianfranco De Zotti $^{35}$             \\
Jacques Delabrouille $^{5,12}$            \\
%Jacques Delabrouille $^{12}$           \\
%Mark Devlin $^{13}$   \\
Eleonora Di Valentino $^{36}$           \\
Joy Didier $^{37}$                      \\
Olivier Dor\'e $^{6,8}$                  \\
%Olivier Dor\'e    !!! JPL then Caltech!!! $^{8}$   \\
Hans K. Eriksen $^{26}$                 \\
Josquin Errard $^{5}$                  \\
Tom Essinger-Hileman $^{21}$            \\
Stephen Feeney $^{17}$                  \\
Jeffrey Filippini $^{38}$               \\
Laura Fissel $^{14}$                    \\
Raphael Flauger $^{15}$                 \\
Unni Fuskeland $^{26}$                  \\
Vera Gluscevic $^{39}$                  \\
Krzysztof M. Gorski $^{6}$             \\
Dan Green $^{16}$                       \\
Shaul Hanany $^{1}$                    \\
Brandon Hensley $^{4}$                 \\
Diego Herranz $^{27}$                   \\
J. Colin Hill $^{17,18}$                   \\
%J. Colin Hill $^{18}$                  \\
Eric Hivon $^{30}$                      \\
Ren\'{e}e  Hlo\v{z}ek $^{40}$           \\
Johannes Hubmayr $^{19}$                \\
Bradley R. Johnson $^{41}$              \\
William Jones $^{4}$                   \\
Terry Jones $^{1}$                     \\
Lloyd Knox $^{20}$                      \\
Al Kogut $^{21}$                        \\
Marcos L\'{o}pez-Caniego $^{42}$        \\
Charles Lawrence $^{6}$                \\
Alex Lazarian $^{43}$                   \\
Zack Li $^{4}$                         \\
Mathew Madhavacheril $^{4}$            \\
%Jeff McMahon $^{22}$   \\
Jean-Baptiste Melin $^{12}$             \\
Joel Meyers $^{44}$                     \\
Calum Murray $^{5}$                    \\
Mattia Negrello $^{45}$                 \\
Giles Novak $^{46}$                     \\
Roger O'Brient $^{6,8}$                  \\
%Roger O'Brient      !!! JPL then Caltech!!! $^{8}$   \\
Christopher Paine $^{6}$               \\
Tim Pearson $^{8}$                     \\
Levon Pogosian $^{47}$                  \\
Clem Pryke $^{1}$                      \\
%Giuseppe Puglisi $^{48}$               \\
Giuseppe Puglisi $^{48,49}$                \\
Mathieu Remazeilles $^{33}$             \\
Graca Rocha $^{6,8}$                     \\
%Graca Rocha $^{8}$                    \\
Marcel Schmittfull $^{18}$              \\
Douglas Scott $^{50}$                   \\
Peter Shirron $^{21}$                   \\
Ian Stephens $^{51}$                    \\
Brian Sutin $^{6}$                     \\
Maurizio Tomasi $^{52}$                 \\
Amy Trangsrud $^{6}$                   \\
Alexander van Engelen $^{23}$           \\
Flavien Vansyngel $^{53}$               \\
Ingunn K. Wehus $^{26}$                 \\
Qi Wen $^{1}$                          \\
Siyao Xu $^{43}$                        \\
Karl Young $^{1}$                      \\
Andrea Zonca $^{54}$
\end{multicols}
}

\Large { \centerline {Endorsers}}
%\vskip 0pt
%
\footnotesize {
%\begin{minipage}{1.2\textwidth}
\begin{multicols}{4}
Maximilian Abitbol              \\
Zeeshan Ahmed                   \\
David Alonso                    \\
Mustafa A. Amin                 \\
Adam Anderson                   \\
James Annis                     \\
Jason Austermann                \\
Carlo Baccigalupi               \\
Darcy Barron                    \\
Ritoban Basu Thakur             \\
Elia Battistelli                \\
Daniel Baumann                  \\
Karim Benabed                   \\
Bradford Benson                 \\
Paolo de Bernardis              \\
Marco Bersanelli                \\
Federico Bianchini              \\
Daniel Bilbao-Ahedo             \\
Colin Bischoff                  \\
Sebastian Bocquet               \\
J. Richard Bond                 \\
Jeff Booth                      \\
Sean Bryan                      \\
Carlo Burigana                  \\
Giovanni Cabass                 \\
Robert Caldwell                 \\
John Carlstrom                  \\
Xingang Chen                    \\
Francis-Yan Cyr-Racine          \\
Paolo de Bernardis              \\
Tijmen de Haan                  \\
C. Darren Dowell                \\
Cora Dvorkin                    \\
Chang Feng                      \\
Ivan Soares Ferreira            \\
Aurelien Fraisse                \\
Andrei V. Frolov                \\
Nicholas Galitzki               \\
Silvia Galli                    \\
Ken Ganga                       \\
Tuhin Ghosh                     \\
Sunil Golwala                   \\
Riccardo Gualtieri              \\
Jon E. Gudmundsson              \\
Nikhel Gupta                    \\
Nils Halverson                  \\
Kyle Helson                     \\
Sophie Henrot-Versill\'e        \\
Thiem Hoang                     \\
Kevin M. Huffenberger           \\
Kent Irwin                      \\
Marc Kamionkowski               \\
Reijo Keskitalo                 \\
Rishi Khatri                    \\
Chang-Goo Kim                   \\
Theodore Kisner                 \\
Arthur Kosowsky                 \\
Ely Kovetz                      \\
Kerstin Kunze                   \\
Guilaine Lagache                \\
Daniel Lenz                     \\
Fran\c{c}ois Levrier            \\
Marilena Loverde                \\
Philip Lubin                    \\
Juan Macias-Perez               \\
Nazzareno Mandolesi             \\
Enrique Mart\'{i}nez-Gonz\'{a}lez   \\
Carlos Martins                  \\
Silvia Masi                     \\
Tomotake Matsumura              \\
Darragh McCarthy                \\
P. Daniel Meerburg              \\
Alessandro Melchiorri           \\
Marius Millea                   \\
Amber Miller                    \\
Joseph Mohr                     \\
Lorenzo Moncelsi                \\
Pavel Motloch                   \\
Tony Mroczkowski                \\
Suvodip Mukherjee               \\
Johanna Nagy                    \\
Pavel Naselsky                  \\
Federico Nati                   \\
Paolo Natoli                    \\
Michael Niemack                 \\
Elena Orlando                   \\
Bruce Partridge                 \\
Marco Peloso                    \\
Francesco Piacentini            \\
Michel Piat                     \\
Giampaolo Pisano                \\
Nicolas Ponthieu                \\
Giuseppe Puglisi                \\
Benjamin Racine                 \\
Christian Reichardt             \\
Christophe Ringeval             \\
Karwan Rostem                   \\
Anirban Roy                     \\
Jose-Alberto Rubino-Martin      \\
Matarrese Sabino                \\
Maria Salatino                  \\
Benjamin Saliwanchik            \\
Neelima Sehgal                  \\
Sarah Shandera                  \\
Erik Shirokoff                  \\
An\v{z}e Slosar                 \\
Tarun Souradeep                 \\
Suzanne Staggs                  \\
George Stein                    \\
Radek Stompor                   \\
Rashid Sunyaev                  \\
Aritoki Suzuki                  \\
Eric Switzer                    \\
Andrea Tartari                  \\
Grant Teply                     \\
Peter Timbie                    \\
Matthieu Tristram               \\
Caterina Umilt\`{a}             \\
Rien van de Weygaert            \\
Vincent Vennin                  \\
Licia Verde                     \\
Patricio Vielva                 \\
Abigail Vieregg                 \\
Jan Vrtilek                     \\
Benjamin Wallisch               \\
Benjamin Wandelt                \\
Gensheng Wang                   \\
Scott Watson                    \\
Edward J. Wollack               \\
Zhilei Xu                       \\
Siavash Yasini
%Martin White    \\
\end{multicols}
%\end{minipage}
}

%\newpage
\Large  {\centerline {Affiliations}}
\vspace{-6pt}
\begin{multicols}{2}
\raggedright
\scriptsize {
%\footnotesize {
1. University of Minnesota - Twin Cities.  \\
2. Johns Hopkins University.  \\
3. Carnegie Melon University.  \\
4. Princeton University.  \\
5. APC, Univ Paris Diderot, CNRS/IN2P3, CEA/lrfu, Obs de Paris, Sorbonne Paris Cit\'e, France.  \\
6. Jet Propulsion Laboratory, California Institute of Technology.  \\
7. Cornell University.  \\
8. California Institute of Technology.  \\
9. Lawrence Berkeley National Laboratory.  \\
10. Space Sciences Laboratory, University of California, Berkeley.  \\
11. Villanova University.  \\
12. IRFU, CEA, Universit\'e Paris-Saclay, France.  \\
13. University of Pennsylvania.  \\
14. National Radio Astronomy Observatory.  \\
15. University of California, San Diego.  \\
16. University of California, Berkeley.  \\
17. Center for Computational Astrophysics, Flatiron Institute.  \\
18. Institute for Advanced Study, Princeton.  \\
19. National Institute of Standards and Technology.  \\
20. University of California, Davis.  \\
21. NASA Goddard Space Flight Center.  \\
22. University of Michigan.  \\
23. Canadian Institute for Theoretical Astrophysics, University of Toronto, Canada.  \\
24. Kavli Institute for the Physics and Mathematics of the Universe (WPI).  \\
25. IRAP, Universit\'e de Toulouse, France.  \\
26. University of Oslo, Norway.  \\
27. Instituto de F\'isica de Cantabria (CSIC-Universidad de Cantabria), Spain.  \\
28. School of Physics, Indian Institute of Science Education and Research Thiruvananthapuram, India.  \\
29. INAF-Istituto di Radioastronomia and Italian ALMA Regional Centre, Italy.  \\
30. Institut d'Astrophysique de Paris, CNRS and Sorbonne Universit\'e, France.  \\
31. Ecole Normale Superieure, Paris, France.  \\
32. Rutgers University.  \\
33. JBCA, University of Manchester.  \\
34. Hubble Fellow          \\
35. INAF-Osservatorio Astronomico di Padova, Italy.  \\
36. University of Manchester.  \\
37. University of Southern California.  \\
38. University of Illinois, Urbana-Champaign.  \\
39. University of Florida.  \\
40. Department of Astronomy \& Astrophysics and Dunlap Institute, University of Toronto, Canada.  \\
41. Columbia University.  \\
42. European Space Astronomy Centre.  \\
43. University of Wisconsin - Madison.  \\
44. Southern Methodist University.  \\
45. Cardiff University School of Physics and Astronomy.  \\
46. Northwestern University.  \\
47. Simon Fraser University.  \\
48. Stanford University.  \\
49. Kavli Institute for Particle Astrophysics and Cosmology.  \\
50. University of British Columbia, Canada.  \\
51. Harvard-Smithsonian Center for Astrophysics.  \\
52. Universit\`a degli studi di Milano.  \\
53. Institut d'Astrophysique Spatiale, CNRS, Univ. Paris-Sud, Universit\'e Paris-Saclay, France.  \\
54. San Diego Supercomputer Center, University of California, San Diego.
}
\end{multicols}

{
\begin{centering}
\small
{***}\\
\smallskip
{This research was funded by a NASA grant NNX17AK52G to the University of Minnesota / Twin Cities, by the Jet Propulsion Laboratory, California Institute of Technology, under a contract with the National Aeronautics and Space Administration, and by Lockheed Martin Corporation. \\ Substantial contributions to the development of PICO were volunteered by scientists at many institutions world-wide.  They are very gratefully acknowledged.} \\
\medskip

{The information presented about the PICO mission concept is pre-decisional and is provided for planning and discussion purposes only.}\\
\medskip

{The cost information contained in this document is of a budgetary and planning nature and is intended for informational purposes only.  It does not constitute a commitment on the part of JPL and/or Caltech.}\\
%\bigskip

%{\copyright\ 2018. All rights reserved.}\\
\end{centering}
}
\newpage
{
             % KY: making table of contents links black instead of red.
\hypersetup{linkcolor=black}
\tableofcontents
}
\newpage
\pagenumbering{arabic}
\setcounter{page}{1}
\setcounter{figure}{0}

\section{Executive Summary} % (2 pg, Hanany)}

%% Beginning of Martin

The \ac{CMB} comes to us from the furthest reaches of the observable Universe, and its photons experience all of cosmic history.  Created when the Universe was a hotter, simpler place, CMB photons carry information about fundamental physics, the constituents of the cosmos, and the theory of gravity.  On their journey they feel the impact of the gravitational potentials formed by the cosmic web of superclusters, clusters, and galaxies.  They interact with the ionized gas in the inter- and circum-galactic media, gas that eventually fuels star and galaxy formation.  Superposed upon the CMB is the emission from multiple extragalactic sources and from our Galaxy.  All of this leaves an imprint that sensitive measurements can disentangle so that CMB studies impact every aspect of cosmology and many areas of astrophysics.

%The \ac{CMB} comes to us from the furthest reaches of the observable Universe, and its photons experience all of cosmic history.  Created when the Universe was a hotter, simpler place, CMB photons probe fundamental physics, provide exquisite measurements of the constituents of the cosmos, and test relativity.  On their journey they feel the impact of the gravitational potentials formed by the cosmic web of superclusters, clusters, and galaxies.  They interact with the ionized gas in the inter- and circum-galactic media, gas that eventually fuels star and galaxy formation.  Superposed upon the CMB is the emission from multiple extragalactic sources and from our Galaxy.  All of this leaves an imprint that sensitive measurements can disentangle so that CMB studies impact every aspect of cosmology and many areas of astrophysics.

\begin{wrapfigure}{R}{0.31\textwidth}  % r is right aligned, l is left. Capital letters allow figure to float on page.
\vspace{-5pt} % if move up and reduce to 12 lines only (add [12] before {R}) saves 1 line.
\includegraphics[width=0.31\textwidth]{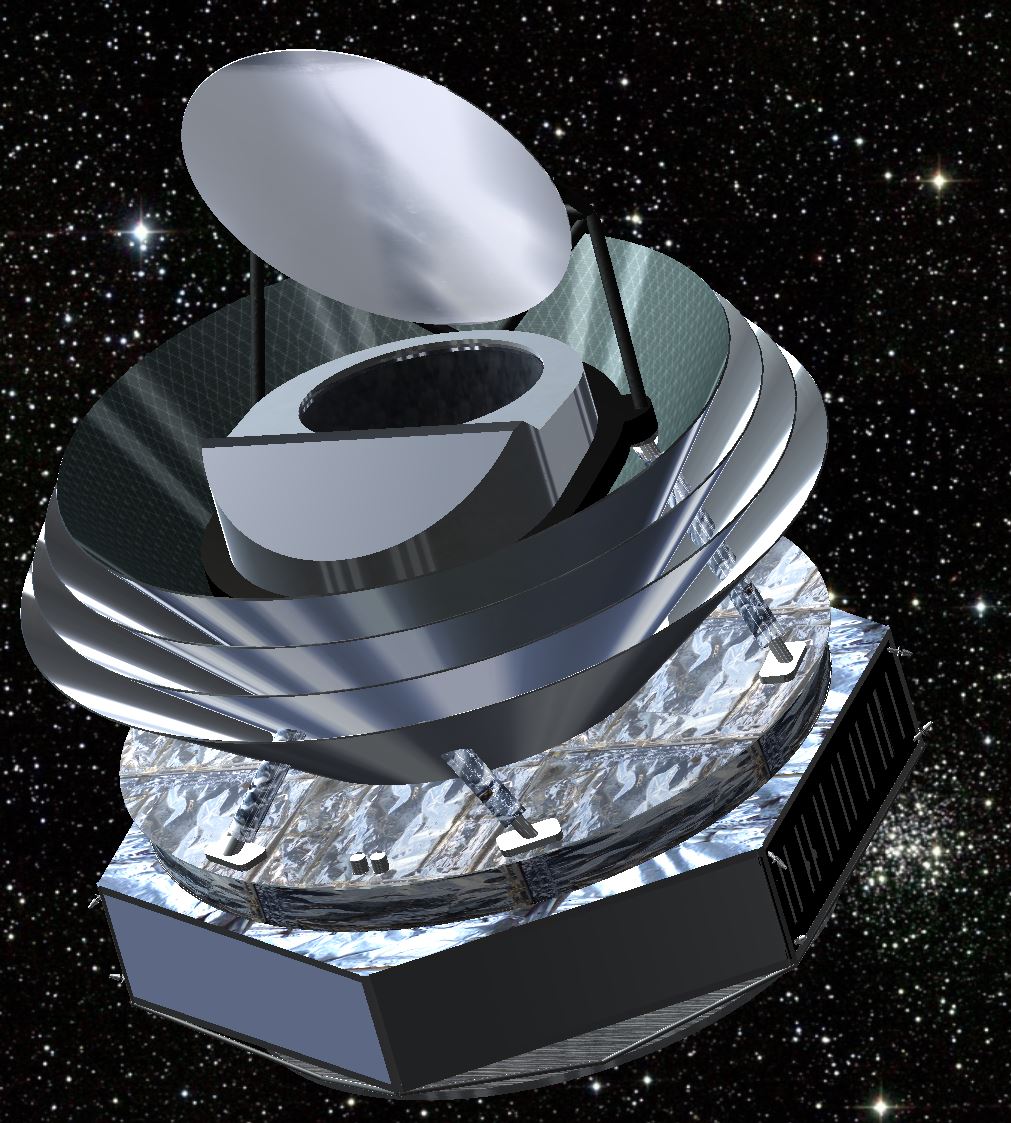}
\vspace{-0.25in}
\caption{\captiontext The PICO spacecraft 
\label{fig:pico_rendered} }
\end{wrapfigure}

Building upon a legacy of successful measurements, the next decade holds tremendous potential for new, exciting \ac{CMB} discoveries, to be delivered by the Probe of Inflation and Cosmic Origins (PICO, Fig.~\ref{fig:pico_rendered}). PICO is an imaging polarimeter that will scan the sky for 5 years in 21 frequency bands spread between 21 and 799~GHz; see Tables~\ref{tab:specs} and \ref{tab:spec_bands}. It will produce %ten independent 
full-sky surveys of intensity and polarization with a final combined-map noise level equivalent to 3300 \planck\ missions for the baseline required specifications, and according to our current best-estimate would perform as 6400 \planck\ missions.  

With these capabilities, unmatched by any other existing or proposed platform, PICO will address seven science objectives (SOs), which are listed in Table~\ref{tab:STM}. Delivering this set was the basis for selecting PICO's design and for setting instrument requirements. But, as described in this report, PICO's science reach is broader than the baseline set. 

PICO could determine the energy scale of inflation and give a first, direct probe of quantum gravity (SO1). %, \S~\ref{sec:fundamentalsci}). 
The mission will attempt to detect the signal that arises from gravitational waves sourced by inflation and parameterized by the tensor-to-scalar ratio $r$ at a level of $r =5\times10^{-4} \, (5\sigma)$. This level is 100 times lower than current upper limits, and more than 10 times lower than limits forecast by funded future experiments.  If the signal is not detected, PICO will constrain broad classes of inflationary models, exclude at $5 \sigma$ models for which the characteristic scale in the potential is the Planck scale, and distinguish between reheating scenarios at $3\sigma$ (SO1 and SO2). The combination of data from PICO and LSST could rule out all models of slow-roll single-field inflation, marking a watershed in studies of inflation. 
%If the signal is not detected, PICO will constrain broad classes of inflationary models and exclude at $5 \sigma$ models for which the characteristic scale in the potential is the Planck scale (SO1 and SO2). The combination of data from PICO and LSST can constrain features in the inflationary potential, the field content during inflation and could rule out all models of slow-roll single-field inflation, marking a watershed in studies of inflation. \comor{reword?}

The mission will have a deep impact on particle physics by measuring the minimum expected sum of the neutrino masses with $4\sigma$ confidence, rising to $7\sigma$ if the sum is near 0.1~eV (SO3). Reaching the $4\sigma$ level can only be achieved with an instrument that can measure the polarization of the CMB on the largest angular scales, a measurement best done from space, which gives access to the full sky, and with a broad band of frequencies to remove foreground contaminants.  
Cluster counts provided by PICO in combination with followup redshift measurements, and PICO's map of the projected gravitational potentials along the line of sight in combination with the LSST gold sample of galaxies, will give two additional independent and equally competitive constraints on the sum of neutrino masses. 

The measurements will either detect or strongly constrain deviations from the standard model of particle physics by counting the number of light particle species $N_{\rm eff}$ in the early universe.  The constraint of $\Delta N_{\rm eff} < 0.06 \, (2\sigma)$ will move the allowed decoupling temperature of a hypothetical new vector particle to temperatures that are 400 times higher than currently determined by \planck\ (SO4). The data will constrain generic models of dark matter, and enable a search for primordial magnetic fields with sufficient sensitivity: to rule them out as the sole source for the largest observed galactic magnetic fields; and to improve by a factor of 300 constraints on polarization rotation arising from early Universe fields that lead to cosmic birefringence, and thus constrain string theory-motivated axions. 

PICO will elucidate the processes affecting the evolution of cosmic structures. It will measure the optical depth to reionization $\tau$ with an error $\sigma(\tau) = 0.002$, limited only by the small number of spatial modes available in the largest angular scale CMB polarization (SO5). The measurement will be used to constrain models of the formation of the first luminous sources, and is a key input to all astrophysical attempts to improve the determination of the sum of neutrino masses. The data will give a map of the projected gravitational potential due to all structures with a \ac{SNR} 14 times higher than \planck , and a catalog of 150,000 clusters extending to their earliest formation redshift. Each of these datasets will be used in combination with other data -- from LSST and from future optical and infrared surveys -- to independently constrain the evolution of the amplitude of linear fluctuations $\sigma_{8}(z)$, with sub-percent accuracy.  

Cross-correlating PICO's map of the thermal Sunyaev--Zeldovich effect with LSST's gold sample of galaxies, a correlation that is forecast to have a \ac{SNR} exceeding 1000, will give precise tracing of the evolution of thermal pressure with $z$. This will be used to place constraints on models of baryonic feedback, which is the most uncertain ingredient in models of galaxy formation. 

%Galactic emissions which act as foregrounds and are stronger than the CMB polarized intensity will be separated using PICOs 21 bands spread over a broad frequency bandwidth. 

$\Lambda$CDM provides a good fit to most current data with only six parameters. But the model leaves fundamental questions open. Premier among them is the unknown content of the majority of the Universe. PICO data will reduce the allowed volume of uncertainty in a 12-dimensional $ \Lambda$CDM parameter space by a factor of nearly a billion relative to current \planck\ constraints. Such exquisite scrutiny of the prevailing paradigm will either give strong validation or require yet-to-be discovered revisions.

PICO's maps of the Milky Way will be used to resolve long-standing questions about our own Galaxy. Galactic interstellar dust grains are a link between atoms and molecules and planetary objects, yet their composition and their role in Galactic chemistry is still under debate. Galactic magnetic fields are known to play a key role in the dynamics of gas in the Galaxy, and in determining the efficiency of star formation, but their quantitative contribution relative to turbulence is yet to be determined. With the mission's Galactic dust polarization maps we will constrain dust properties, including composition, temperature, and emissivities (SO6), and we will make maps of the Galactic magnetic field. These detailed 1\arcmin\ resolution maps will be used to quantify the relative roles of gas turbulence and magnetic fields in the dynamics of the Galaxy and in the observed low star-formation efficiency (SO7). 

PICO will give deep, full-sky legacy maps with which astrophysicists will constrain the early phases of galaxy evolution by discovering 4500 strongly lensed dusty galaxies with $z$ up to 5; investigate the early phases of cluster evolution by discovering 50,000 proto-clusters out to $z\sim4.5$; perform a census of cold dust in 30,000 low $z$ galaxies; make cosmic infrared background maps of the anisotropies due to dusty star-forming galaxies; map magnetic fields in 70 nearby galaxies; and, with a 3,000-fold increase relative to \planck\ in the number of independent measurements of magnetic field in our own Galaxy, study how magnetic fields are generated through a combination of turbulence and large-scale gas motion. This rich harvest will be contained in maps of both intensity and polarization at 21 frequency bands, each much more sensitive than \planck 's nine frequency maps in intensity and seven in polarization. At 30, 155, and 385~GHz PICO's noise is 17, 40, and 100 times lower than \planck 's at 30, 143, and 353~GHz, respectively. Six of the PICO maps in bands between 321 and 800~GHz, which are key for the high $z$ science, are not accessible to ground-based instruments; \planck\ did not have polarization information above 353~GHz and PICO's highest resolution is five times finer than \planck 's. Only a space mission like PICO will provide such full-sky legacy maps.

%Five of the PICO bands between 385 and 800~GHz are key for the high $z$ science; \planck\ did not have polarization information above 353~GHz and PICO's highest resolution is five times finer than \planck 's. Only a space mission like PICO will provide such full-sky legacy maps. 

%PICO will give full-sky maps of intensity and polarization at 21 frequency bands, each much more sensitive than \planck 's nine frequency maps in intensity and seven in polarization. At 30, 155, and 385~GHz PICO's noise is 17, 40, and 100 times lower than \planck 's at 30, 143, and 353~GHz, respectively. Five PICO bands will have polarization information at frequencies between 385 and 800~GHz that \planck\ did not have, and PICO's highest resolution is five times finer than \planck 's. Only PICO will provide such full-sky legacy maps. With the six maps at frequencies not accessible to ground-based experiments we will: constrain the early phases of galaxy evolution by discovering 4500 strongly lensed dusty galaxies with $z$ up to 5; investigate the early phases of cluster evolution by discovering 50,000 proto-clusters out to $z\sim4.5$; perform a census of cold dust in 30,000 low $z$ galaxies; make cosmic infrared background maps of the anisotropies due to dusty star-forming galaxies; map magnetic fields in 70 nearby galaxies; and, with 3,000-fold increase relative to \planck\ in the number of independent measurements of magnetic field in our own Galaxy, study how magnetic fields are generated through a combination of turbulence and large-scale gas motion. 

With its broad frequency coverage, PICO is better equipped than any other current or planned instrument to separate the detected signals into their original sources of emission.  This capability is most important for unveiling the faintest of signals, the telltale signature of inflation, which is already known to be dominated by Galactic foregrounds. Our simulations indicate that PICO's combination of low noise and multitude of bands is sufficient to separate the inflationary signal from the foregrounds at the required level. But there are uncertainties in the modeling of Galactic foregrounds. To reduce these uncertainties and gain further confidence we recommend support for: (1) modeling, simulation, and algorithm development for effective foreground separation; and (2) improved Galactic emission measurements with sub-orbital experiments. 

%With its broad frequency coverage PICO is better equipped than any other current or planned instrument to separate the detected signals to their original sources of emission.  This capability is most important for the faintest of signals, the telltale of inflation, which is already known to be dominated by Galactic foregrounds. Our simulations indicate that PICO's combination of low noise and multitude of bands is sufficient to separate the inflationary signal from the foregrounds at the required level. But current uncertainties on the parameters characterizing Galactic foregrounds are large and we recommend support for (1) modeling, simulation, and algorithm development for effective foreground separation, and (2) improved Galactic emission measurements with sub-orbital experiments. 
%\comor{combine recommendations}?

% [TJP] ------------------------
% [TJP] The Science Traceability Matrix goes here. The table is in stm.tex, but the following LaTeX invocation makes it appear on a double-wide page
\afterpage{%
  % switch to LayoutPageB (includes switching page size)
  \switchToLayoutPageB{}
    %
% sensitivity table
% This is two tables in minipage environment
%
\begin{table}[tb]
\hskip3cm
\begin{minipage}[t]{0.3\textwidth}
\caption{\textbf{Mission Parameters}\label{tab:specs}}
\begingroup
%\openup 5pt
\newdimen\tblskip \tblskip=5pt
\nointerlineskip
\vskip -5mm
\footnotesize %\footnotesize
\setbox\tablebox=\vbox{
    \newdimen\digitwidth
    \setbox0=\hbox{\rm 0}
    \digitwidth=\wd0
    \catcode`*=\active
    \def*{\kern\digitwidth}
    \newdimen\signwidth
    \setbox0=\hbox{+}
    \signwidth=\wd0
    \catcode`!=\active
    \def!{\kern\signwidth}
\halign{%
\hbox to 1.8in{#\leaderfil}\tabskip=0.6em plus 0.6em&
#\hfil\tabskip=0pt\cr
\noalign{\doubleline}
\multispan2Combined polarization map depth (rms noise in $1\times1$ arcmin$^{2}$ pixel):\hfil\cr
\quad Baseline&0.87 $\mu$K$_{\rm CMB}$ arcmin equivalent to 3300 \textit{Planck} missions\cr
\quad CBE$^{a}$&0.61 $\mu$K$_{\rm CMB}$ arcmin equivalent to 6400 \textit{Planck} missions\cr
Survey duration / start & 5\,yrs / 2029 \cr
Orbit type & Sun-Earth L2 \cr
Launch mass & 2147\,kg \cr
Total power &1320\,W \cr
Data rate & 6.1\,Tbits/day \cr
Cost&\$\,958M\cr
%Launch&2029\cr
\noalign{\vskip 5pt\hrule\vskip 3pt}
\noalign{$^{a}$ CBE = Current best estimate.} % footnote to table
} % close halign
} % close vbox
\endPlancktable
\endgroup
%\end{table}
\end{minipage}
\hskip 3cm
\begin{minipage}[t]{0.5\textwidth}
\caption{\textbf{Frequency Bands, Resolution, and Noise Level}\label{tab:spec_bands}}
\begingroup
%\openup 5pt
\newdimen\tblskip \tblskip=5pt
\nointerlineskip
\vskip -5mm
\footnotesize %\footnotesize
\setbox\tablebox=\vbox{
    \newdimen\digitwidth
    \setbox0=\hbox{\rm 0}
    \digitwidth=\wd0
    \catcode`*=\active
    \def*{\kern\digitwidth}
    \newdimen\signwidth
    \setbox0=\hbox{+}
    \signwidth=\wd0
    \catcode`!=\active
    \def!{\kern\signwidth}
\halign{
\hbox to 1.8in{#\leaderfil}\tabskip=0.6em plus 0.6em&
\hfil#\hfil&
\hfil#\hfil&
\hfil#\hfil&
\hfil#\hfil&
\hfil#\hfil&
\hfil#\hfil&
\hfil#\hfil&
\hfil#\hfil&
\hfil#\hfil&
\hfil#\hfil&
\hfil#\hfil&
\hfil#\hfil&
\hfil#\hfil&
\hfil#\hfil&
\hfil#\hfil&
\hfil#\hfil&
\hfil#\hfil&
\hfil#\hfil&
\hfil#\hfil&
\hfil#\hfil&
\hfil#\hfil\tabskip=0pt\cr
\noalign{\doubleline}
Frequency [GHz]&21&25&30&36&43&52&62&75&90&108&129&155&186&223&268&321&385&462&555&666&799\cr
FWHM [arcmin]&38.4&32.0&28.3&23.6&22.2&18.4&12.8&10.7&9.5&7.9&7.4&6.2&4.3&3.6&3.2&2.6&2.5&2.1&1.5&1.3&1.1\cr
\noalign{Polarization map depth:}
\quad Baseline  [$\mu$K$_{\rm CMB}$\,arcmin]&23.9&18.4&12.4&*7.9&*7.9&*5.7&*5.4&*4.2&*2.8&*2.3&*2.1&*1.8&*4.0&*4.5&*3.1&*4.2&4.5&9.1&45.8&*177&1050\cr
\quad CBE$^a$  [$\mu$K$_{\rm CMB}$\,arcmin]&16.9&13.0&*8.7&*5.6&*5.6&*4.0&*3.8&*3.0&*2.0&*1.6&*1.5&*1.3&*2.8&*3.2&*2.2&*3.0&3.2&6.4&32.4&*125&*740\cr
\quad Baseline  [ Jy/sr]&*8.3&10.9&11.8&12.9&19.5&23.8&45.4&58.3&59.3&77.3&96.0&119&433&604&433&578&429&551&1580&2080&2880\cr
\quad CBE$^a$  [Jy/sr]&*5.9&*7.7&*8.3&*9.2&13.8&16.8&32.1&41.3&41.8&53.5&69.3&*84&302&436&304&411&303&387&1120&1470&2040\cr
\noalign{\vskip 5pt\hrule\vskip 3pt}
} % close halign
} % close vbox
\endPlancktable
\endgroup
%\end{table}
\end{minipage}
\end{table}

\vspace{8mm}
%
%
%   -----------------------------------------------------------------------------------------------------------------
%
\begin{table}
\caption{\textbf{Science Traceability Matrix (STM) }}\label{tab:STM}
%\small
\footnotesize
\begin{tabular}{@{}lcccccccc@{}}
%\noalign{\vskip 2mm}
\hline
\noalign{\vskip 2mm}    
% Header line 1
%\rowcolor[HTML]{EFEFEF} 
\multicolumn{1}{c}{\multirow{2}{1in}{\centering \bf Science Goals (from NASA Science Plan)}}&
\multicolumn{1}{c}{\multirow{2}{2in}{\centering \bf Science Objectives}}& 
\multicolumn{3}{c}{\bf Scientific Measurement Requirements}&
\multicolumn{1}{c}{}&
\multicolumn{2}{c}{\bf Instrument (single instrument, single mode)}&
\multicolumn{1}{c}{\multirow{2}{1.75in}{\centering \bf Mission Functional Requirements}} 
\\
%Header line 2
\noalign{\vskip 2mm}    
\cline{3-5}\cline{7-8}
\noalign{\vskip 2mm}    
%\rowcolor[HTML]{EFEFEF} 
\multicolumn{1}{c}{} &
\multicolumn{1}{c}{} &
\multicolumn{1}{c}{Model Parameters} &
\multicolumn{1}{c}{Physical Parameters} & 
\multicolumn{1}{c}{Observables} &
\multicolumn{1}{c}{} &
\multicolumn{1}{c}{Functional Requirements} &
\multicolumn{1}{c}{Projected Performance} & 
\\
% Line SO1
\noalign{\vskip 2mm}    
\hline
\multicolumn{1}{l}{\multirow{2}{1in}{\vskip5pt \textbf{\textit{Explore how the Universe began: Inflation}}}}&
\multicolumn{1}{l}{\parbox[t]{2in}{\textbf{SO1}. Probe the physics of the big bang by detecting the energy scale at which inflation occurred if it is above $5\times10^{15}$\,GeV, or place an upper limit if it is below (\S\,\ref{sec:inflation}, Fig.~\ref{fig:clbb})}}&
\multicolumn{1}{l}{\parbox[t]{2in}{Tensor-to-scalar ratio $r$: \\ $\sigma(r) = 1\times10^{-4}$ at $r = 0$; \\ $r < 5 \times 10^{-4}$ at $5\sigma$ confidence level$^a$}} &
\multicolumn{1}{l}{\parbox[t]{2in}{CMB polarization $BB$ power spectrum for modes $2<\ell<300$ to cosmic-variance limit, and CMB lensing power spectrum for modes $2<\ell<1000$ to cosmic-variance limit}}&
\multicolumn{1}{l}{\parbox[t]{2in}{Linear polarization across $62 < \nu < 223$\,GHz over entire sky; foreground separation requires $21 < \nu < 799$\,GHz}}& 
\multicolumn{1}{c}{} &
\multicolumn{1}{l}{\multirow{5}{1.75in}{%
\vskip15pt
Frequency coverage: central frequencies $\nu_c$ from 21 to 799\,GHz
\vskip5pt
Frequency resolution: $\Delta\nu/\nu_c = 25\%$
\vskip5pt
Sensitivity: See Table~\ref{tab:bands}
Combined instrument noise:  $< 0.61\,\mu{\rm K}_{\rm CMB}\sqrt{\rm s}$
\vskip5pt
Angular resolution [for delensing and foreground separation]: ${\rm FWHM} =  6.2' \times ( 155\,{\rm GHz} / \nu_c )$
%\vskip5pt
%Effective aperture: 1.4~m
\vskip5pt
Sampling rate: $( 3 / {\rm Beam FWHM} ) \times ( 336' / {\rm s})$
}}& 
\multicolumn{1}{l}{\parbox[t]{1.5in}{}}& 
\multicolumn{1}{l}{\multirow{7}{1.75in}{%
\vskip10pt
Sun-Earth L2 orbit with Sun-Probe-Earth $< 15^\circ$ (\S\,\ref{sec:mission_design}) 
\vskip5pt
5 yr survey (\S\,\ref{sec:operations})
\vskip5pt
Full sky survey: Spin instrument at 1 rpm; boresight $69^\circ$ off spin axis;
spin axis $26^\circ$ off anti-Sun line, precessing $360^\circ$ / 10hr (\S\,\ref{sec:survey_design})
\vskip5pt
Pointing control: Spin axis $60'$ ($3\sigma$, radial); spin \@ $1 \pm 0.1$ rpm ($3\sigma$) (\S\,\ref{sec:attitude_determination})
\vskip5pt
Pointing stability: Drift of spin axis $< 1'$/1min ($3\sigma$, radial);
jitter $< 20''$/20 ms ($3\sigma$, radial) (\S\,\ref{sec:attitude_determination})
\vskip5pt
Pointing knowledge
(telescope boresight):
$10'' \, (3\sigma$, each axis) from spacecraft attitude;
$1'' \, (1\sigma$, total) final reconstructed (\S\,\ref{sec:attitude_determination})
\vskip5pt
Return and process instrument data:
1.5 Tbits/day (after 4$\times$ compression) (\S\,\ref{sec:ground_segment}, \ref{sec:spacecraft})
\vskip5pt
Thermally isolate instrument from solar radiation and from spacecraft bus (\S\, \ref{sec:radiative_cooling}, \ref{sec:spacecraft})
}}\\
% Line SO2
\noalign{\vskip 1mm}
\cline{2-5}
\noalign{\vskip 1mm}
\multicolumn{1}{l}{}&
\multicolumn{1}{l}{\parbox[t]{2in}{\textbf{SO2}. Probe the physics of the big bang by excluding classes of potentials as the driving force of inflation (\S\,\ref{sec:inflation}, Fig.~\ref{fig:nsr})}}&
\multicolumn{1}{l}{\parbox[t]{2in}{Spectral index ($n_s$) and its derivative ($n_{\rm run}$): $\sigma(n_s) < 0.0015$; $\sigma(n_{\rm run}) < 0.002$}}&
\multicolumn{1}{l}{\parbox[t]{2in}{CMB polarization $BB$ power spectrum for modes $2<\ell<1000$ to cosmic-variance limit}}&
\multicolumn{1}{l}{\multirow{3}{2in}{
\vskip15pt
Intensity and linear polarization across $62 < \nu < 223$\,GHz over the entire sky; foreground separation encompassed by SO1}}& 
\multicolumn{1}{c}{} &
\multicolumn{1}{l}{\parbox[t]{1.75in}{}}& 
\multicolumn{1}{l}{\multirow{7}{1.5in}{%
Frequency coverage: See Tables~\ref{tab:spec_bands} and~\ref{tab:bands}.
\vskip 2pt 
21 bands with $\nu_c$ from 21 to 799\,GHz
\vskip5pt
Frequency resolution: $\Delta\nu/\nu_c = 25\%$
\vskip5pt
Sensitivity: See Table~\ref{tab:bands} %\ref{tab:sensitivity}.
\vskip2pt
Combined instrument noise: $0.43\,\mu{\rm K}_{\rm CMB}\sqrt{\rm s}$
\vskip5pt
Angular resolution: See Table~\ref{tab:spec_bands} and ~\ref{tab:bands}.
\vskip2pt
${\rm FWHM} = 6.2' \times (155\,{\rm GHz} / \nu_c )$;
$1.1'$ for $\nu_c = 799\,$GHz
\vskip5pt
Sampling rate: See Table~\ref{tab:focal_plane} %\ref{Sampling}.
$( 3 / {\rm Beam FWHM}) \times ( 336' / {\rm s})$ 
}}&
\multicolumn{1}{l}{\parbox[t]{1in}{}}\\
% Line SO3
\noalign{\vskip 1mm}
\cline{1-4}
\noalign{\vskip 1mm}
\multicolumn{1}{l}{\multirow{2}{1in}{\vskip5pt \textbf{\textit{Discover how the Universe works: neutrino mass and $N_{\rm eff}$}}}}&
\multicolumn{1}{l}{\parbox[t]{2in}{\textbf{SO3}. Determine the sum of neutrino masses. (\S\,\ref{sec:relics_neutrinos}, Fig.~\ref{fig:DM_baryons})}}&
\multicolumn{1}{l}{\parbox[t]{2in}{Sum of neutrino masses ($\Sigma m_\nu$): $\sigma(\Sigma m_\nu) = 14$\,meV with DESI or Euclid$^b$; independently  $\sigma(\Sigma m_\nu) = 14$\,meV using cluster counts$^c$ }}& %; \comred{$\Sigma m_\nu < ??$\,meV alone}}}&
\multicolumn{1}{l}{\parbox[t]{2in}{CMB polarization power spectra for modes $2<\ell<4000$; CMB intensity maps (to identify clusters using the Compton-$y$ signal)}}&
\multicolumn{1}{l}{\parbox[t]{2in}{}}& 
\multicolumn{1}{c}{} &
\multicolumn{1}{l}{\parbox[t]{1.75in}{}}& 
\multicolumn{1}{l}{\parbox[t]{1.5in}{}}& 
\multicolumn{1}{l}{\parbox[t]{1in}{}}
\\
% Line SO4
\noalign{\vskip 1mm}
\cline{2-4}
\noalign{\vskip 1mm}
&
\multicolumn{1}{l}{\parbox[t]{2in}{\textbf{SO4}. Tightly constrain the thermalized fundamental particle content of the early Universe (\S\,\ref{sec:relics_neutrinos}, Fig.~\ref{fig:Neff_future})}}&
\multicolumn{1}{l}{\parbox[t]{2in}{Number of light relic particle species $N_{\rm eff}$: $ \Delta N_{\rm eff}<0.06\,\, (95\%)$ }}&
\multicolumn{1}{l}{\parbox[t]{2in}{CMB temperature and polarization auto and cross power spectra $2<\ell<4000$}}&
\multicolumn{1}{l}{\parbox[t]{2in}{}}& 
\multicolumn{1}{c}{} &
\multicolumn{1}{l}{\parbox[t]{1.75in}{}}& 
\multicolumn{1}{l}{\parbox[t]{1.5in}{}}& 
\multicolumn{1}{l}{\parbox[t]{1in}{}}
\\
% Line SO5
\noalign{\vskip 1mm}
\cline{1-5}
\noalign{\vskip 1mm}
\multicolumn{1}{l}{\multirow{1}{1in}{\textbf{\textit{Explore how the Universe evolved: reionization}}}}&
\multicolumn{1}{l}{\parbox[t]{2in}{\textbf{SO5}. Distinguish between models that describe the formation of the earliest luminous sources in the Universe (\S\,\ref{sec:extragalacticsci}, Fig.~\ref{fig:ReionizationPICO})}}&
\multicolumn{1}{l}{\parbox[t]{2in}{Optical depth to reionization ($\tau$): $\sigma(\tau) < 0.002$}}&
\multicolumn{1}{l}{\parbox[t]{2in}{CMB polarization $EE$ power spectrum for modes $2<\ell<40$ to cosmic-variance limit}}&
\multicolumn{1}{l}{\parbox[t]{2in}{Linear polarization across $62 < \nu < 223$\,GHz over entire sky; foreground separation encompassed by SO1}}& 
\multicolumn{1}{c}{} &
\multicolumn{1}{l}{\parbox[t]{1.75in}{}}& 
\multicolumn{1}{l}{\parbox[t]{1.5in}{}}& 
\multicolumn{1}{l}{\parbox[t]{1in}{}}
\\
% Line SO6
\noalign{\vskip 1mm}
\cline{1-7}
\noalign{\vskip 1mm}
\multicolumn{1}{l}{\multirow{2}{1in}{{\vskip5pt \textbf{\textit{Explore how the Universe evolved: Galactic structure and dynamics}}}}}&
\multicolumn{1}{l}{\parbox[t]{2in}{\textbf{SO6}. Test models of the composition of Galactic interstellar dust (\S\,\ref{sec:test_composition_models})}}&
\multicolumn{1}{l}{\parbox[t]{2in}{Intrinsic polarization fractions of Galactic dust components to accuracy better than 3\% when averaged over $10'$ pixels }}&
\multicolumn{1}{l}{\parbox[t]{2in}{Spectral energy distribution of interstellar dust polarized emission between 108 and 799\,GHz}}&
\multicolumn{1}{l}{\parbox[t]{2in}{Intensity and linear polarization maps in 12 frequency bands between 108 and 799\,GHz}}& 
\multicolumn{1}{c}{} &
\multicolumn{1}{l}{\parbox[t]{1.75in}{ Encompassed by SO1--5}
}& 
\multicolumn{1}{l}{\parbox[t]{1.5in}{}}& 
\multicolumn{1}{l}{\parbox[t]{1in}{}}
\\
% Line SO7
\noalign{\vskip 1mm}
\cline{2-7}
\noalign{\vskip 1mm}
\multicolumn{1}{l}{}&
\multicolumn{1}{l}{\parbox[t]{2in}{\textbf{SO7}. Determine if magnetic fields are the dominant cause of low Galactic star-formation efficiency (\S\,\ref{sec:magnetic_fields})}}&
\multicolumn{1}{l}{\parbox[t]{2in}{Ratio of cloud mass to maximum mass that can be supported by magnetic field (``Mass to flux ratio'' $\mu$); %\comred{$\sigma(\mu) < ??$};
ratio of gas turbulent energy to magnetic energy (quantified through the Alfv\'{e}n Mach number $\mathcal{M}_A$) on scales 0.05--100\,pc  }}&%\comred{$\sigma(\mathcal{M}_A) < ??$}}}&
\multicolumn{1}{l}{\parbox[t]{2in}{Turbulence power spectrum on scales 0.05--100\,pc; magnetic field strength ($B$) as a function of spatial scale and density; hydrogen column density; gas velocity dispersion$^d$
}}&
\multicolumn{1}{l}{\parbox[t]{2in}{Intensity and linear polarization with $<1$\,pc resolution for thousands of molecular clouds and with $< 0.05$\,pc for the 10 nearest molecular clouds; maps of polarization with 1' resolution over the entire sky}}& 
\multicolumn{1}{c}{} &
\multicolumn{1}{l}{\parbox[t]{1.75in}{
Encompassed by SO1--5, except:
\vskip4pt
Angular resolution: $\le 1.1'$ (at highest frequency)
\vskip4pt
Sensitivity at 799\,GHz: 27.4\, kJy/sr
}}& 
\multicolumn{1}{l}{\parbox[t]{1.5in}}& 
\multicolumn{1}{l}{\parbox[t]{1in}{}}
\\
\noalign{\vskip 1mm}
\hline
\noalign{\vskip 1mm}
\end{tabular}
{\footnotesize
$^a$ The values predicted include delensing and foreground subtraction; see \S~\ref{sec:inflation}. \\
$^b$ Using $\tau$ and the power spectrum of the reconstructed lensing map (\S~\ref{sec:gravitationallensing}), both from PICO's measurements, and baryon acoustic oscillation data from DESI or Euclid. \\
$^c$ The constraint using clusters requires redshifts by future optical and IR surveys. \\
$^d$ Hydrogen column density and gas velocity dispersion will be provided by 21-cm surveys including HI4PI and GALFA-HI. 
}
\end{table}

   \clearpage
% start with LayoutPageA (includes switching page size)
\switchToLayoutPageA{}
% \input stm2.tex  stm2 moved to Legacy science section. (Karl)
% \clearpage
}
% [TJP] end --------------------

PICO's large multiplicity of independent maps and sky surveys, and its stable thermal environment will give control of systematic uncertainties unmatched by any other platform. Similar to its successful predecessors, \wmap\ and \planck , PICO will conduct observations from L2, a location proven to give thermal stability.  It will execute ten redundant,  full-sky surveys, each complete within 6 months. The scan pattern on the sky, which is optimized for control of polarimetric systematic uncertainties, ensures that the measured $I,\, Q$, and $U$ Stokes parameters can be reconstructed by each of the 12,996 polarization-sensitive detectors. 

%Similar to its successful predecessors, WMAP and \planck , PICO will conduct observations from L2, a location that ensures a stable thermal environment.  It will execute ten redundant,  full-sky surveys, each complete within 6 months. The sky scan pattern, which is optimized for control of polarimetric systematic uncertainties, ensures that the measured $I,\, Q$, and $U$ Stokes parameters can be reconstructed by each of the 12,996 polarization-sensitive detectors. The large multiplicity of independent maps and sky surveys, and the stable environment will together give control of systematic uncertainties unmatched by any other platform.

The mission has a single instrument that surveys the sky with a repetitive pattern.  The telescope is a 1.4~m entrance-aperture, two-reflector system, with passively cooled primary, and 4.5~K actively cooled aperture-stop and secondary. The 0.1~K cooled focal plane is based on three-color pixels coupling the incident radiation to transition-edge-sensor bolometers that are read out using a time-domain multiplexed system. All of these technologies are either already in use by sub-orbital experiments, or are simple extensions to higher or lower frequency bands. To ensure full readiness for mission initiation, we recommend continued support for technology development and maturation in the laboratory and by sub-orbital experiments. 

The science PICO will deliver addresses some of the most fundamental quests of human knowledge. Its science advances will enrich many areas of astrophysics, and will form the basis for the cosmological paradigm of the 2030s and beyond.  Many of these advances can only be achieved by a space-based mission. The design of PICO is informed by science breakthroughs made by \planck\ and sub-orbital experiments over the last decade. Further breakthroughs require a scale-up that is most optimally achieved by PICO.  There is a long heritage of space and sub-orbital measurements in these frequency bands and the PICO implementation is a conservative extension of past successes. The mission relies on today's technologies; no new fundamental developments are required. PICO is the only single-platform instrument with the combination of sensitivity, angular resolution, frequency bands, and control of systematic effects that can deliver the compelling, timely, and broad science. We recommend a start for the mission in the next decade.

\section{Science}
\label{sec:science}

\subsection{Introduction} % (1.5 pgs)}

%{\it NASA suggested table of contents says Science Intro or Landscape section should include: State of the Art in the Field ; Compelling Outstanding Questions; Needed Capabilities for Progress. }

%\vskip 10pt

The Probe of Inflation and Cosmic Origins (PICO) has seven \ac{SOs}. They derive from three strategic goals that are well-articulated by NASA's science plan~\citep{latest_nasa_science_plan,latest_nasa_strategic_plan}: to explore how the Universe began; to discover how the Universe works; and to explore how the Universe evolved. The \ac{SOs}, which include probing the inflationary epoch after the big bang, constraining the properties of fundamental particles and potentially detecting new ones, probing the structure and evolution of the Universe, and understanding the structure of our own Galaxy, require measurements in and around frequency bands in which the cosmic microwave background is most intense. The \ac{SOs} and the measurement requirements derived from them are given in Table~\ref{tab:STM} and define the PICO `baseline design.' 
This report focuses on the primary \ac{SOs} listed in the Table but also describes the much broader set of science deliverables that the mission design enables.
%This design gives rise to a mission that delivers a much broader set of science deliverables.  This report describes the broad array of science deliverables, but focuses primarily on the primary \ac{SOs} listed in the Table. 

The PICO mission consists of a single instrument: an imaging polarimeter that surveys the entire sky at 21 frequency bands spread between 21 and 799~GHz.  The telescope has an aperture of 1.4~m, giving diffraction-limited resolution between 38\arcmin\ and 1\arcmin . The instrument incorporates a 0.1~K cooled focal plane that hosts 12,996 \ac{TES} bolometric detectors. The baseline design contains a margin of 40\% in detector noise (\S\,\ref{sec:focal_plane}). We include throughout this report performance estimates that are based also on our current estimate for the actual performance. Those are labeled `\ac{CBE}' (\S\,\ref{sec:focal_plane}). Table~\ref{tab:specs} gives key mission parameters and Table~\ref{tab:spec_bands} gives the frequency bands, resolution, and both baseline and \ac{CBE} noise levels. Experience with past space missions, most recently with \planck , shows that pre-mission calculated detector performance is in fact achieved in space~\citep{planck1101.2038,planck1101.2039,Jarosik}.

Mission operations throughout the 5-year duration of the survey are simple and are optimized for polarimetric measurements as each sky pixel is scanned along multiple orientations. The spacecraft spins around its symmetry axis at 1~rpm and the symmetry axis precesses around the anti-Sun direction with a period of 10~hours. With this repetitive scan pattern, the entire sky is scanned every 6~months, giving ten independent full-sky maps of the intensity and polarization Stokes parameters $T$, $Q$, and $U$.  
%Each sky pixel is scanned along multiple orientations and therefore independent full-sky maps of all Stokes parameters can be reconstructed from the data of {\it each} of the detectors. \comor{give more explicit contrast to Planck and detector differencing in systematics?}
%This is in contrast to e.g. \planck ,  which relied on differencing 

% \comor{words about 1/f?}. 

Some of the PICO polarization science goals are more appropriately described in terms of $E$ and $B$ polarization maps rather than $Q$ and $U$~\cite{seljak97,kamionkowski97a, zaldarriaga97b,kamionkowski97b}. This is because sources of polarization signatures that are scalar in nature, such as primordial density perturbations, can only produce $E$-mode polarization. Sources that are tensor in nature, such as gravitational waves, can produce both $E$- and $B$-mode polarization. The angular power spectra of $E$ and $B$ maps will be denoted as `$EE$' and `$BB$'.

This report assumes that PICO's Phase~A will start in 2023. The science outcomes are expected to break new ground, and to be complementary to data sets available at the end of 2020s and the beginning of the following decade. Where appropriate we highlight complementarities with funded projects that are in implementation phases, such as LSST, Euclid, and WFIRST. We also include performance comparisons to funded CMB projects that are in implementation and for which final design specifications and projections exist in the literature. Such next-generation US-based sub-orbital CMB experiments are collectively denoted as `Stage-3 (S3)'~\citep{advancedact,spt3g,so,class_overview,biceparray,spider,piper}. 

%Therefore we are including performance comparisons to funded projects that are in implementation and for which final design specifications and projections exist in the literature. Such next-generation US-based CMB experiments are collectively denoted as Stage-3 (S3)~\citep{advancedact,spt3g,so,class_overview,biceparray,spider,piper}. 

This section describes PICO's science objectives, places them in the context of current knowledge, and provides performance forecasts (\S~\ref{sec:fundamentalsci}--\ref{sec:legacy}). It gives our estimates of: the efficacy of separating the detected radiation into the several astrophysical sources of emission~(\S~\ref{sec:signal_separation}); an assessment of anticipated systematic uncertainties~(\S~\ref{sec:systematics}); a discussion of PICO's complementarity with sub-orbital measurements~(\S~\ref{sec:complementarity}); and the measurement requirements that derive from the combination of these topics~(\S~\ref{sec:requirements}).  

\subsection{Fundamental Physics} %(6 pgs)
\label{sec:fundamentalsci}
%\vspace{-0.05in}

\subsubsection{Gravitational Waves and Inflation}
\label{sec:inflation}

%\paragraph{Targets}
\noindent$\bullet$ {\bf Targets} \hspace{0.1in} 

%According to Einstein's theory of general relativity, the density perturbations responsible for the observed CMB anisotropy must have been created long before the \ac{CMB} was released, and even before the Universe became filled with a hot and dense plasma of fundamental particles. Understanding the mechanism generating these perturbations, which evolved to fill the Universe with structures, is one of the most important open questions in cosmology. In addition to density perturbations, this mechanism may have also produced gravitational waves that would have left a $B$-mode polarization signature in the CMB~\cite{seljak97,kamionkowski97a}.  Any detection of primordial $B$-mode polarization by PICO will constitute evidence for gravitational waves from the same primordial period that created the density perturbations and will open a new window onto this early epoch. Because the dynamics of gravitational waves is essentially unaffected by the plasma, they would be a pristine relic from the earliest moments of our Universe, and their properties would shed light on the mechanism that created the primordial perturbations. 

According to Einstein's theory of general relativity, the density perturbations responsible for the observed CMB anisotropy must have been created long before the \ac{CMB} was released, and even before the Universe became filled with a hot and dense plasma of fundamental particles. Understanding the mechanism generating these perturbations, which evolved to fill the Universe with structures, is one of the most important open questions in cosmology. In addition to density perturbations, this mechanism may have also produced gravitational waves that would have left a $B$-mode polarization signature in the CMB~\cite{seljak97,kamionkowski97a}.  Any detection of primordial $B$-mode polarization by PICO will constitute evidence for gravitational waves from the same primordial period that created the density perturbations. Finding the signals will open a new window onto the earliest moments of our Universe, and studying their properties would shed light on the mechanism that created the primordial perturbations. 

Inflation, a period of nearly exponential expansion of the early Universe~\cite{Guth:1980zm,Linde:1981mu,Albrecht:1982wi,Starobinsky:1980te}, is the leading paradigm explaining the origin of the primordial density perturbations~\cite{Mukhanov:1981xt,Guth:1982ec,Hawking:1982cz,Starobinsky:1982ee,Bardeen:1983qw}. It predicts a nearly scale-invariant spectrum of primordial gravitational waves originating from quantum fluctuations~\cite{Starobinsky:1979ty}. Measurements of the \ac{CMB} are the only foreseeable way to detect these gravitational waves.
%\sout{In this sense, a detection of primordial $B$-modes would be the first observation of a phenomenon associated with quantum gravity~\cite{Krauss:2013pha}.} \comor{moved} 

\begin{figure}[!thb]
\centering
\vspace{-0.05in}
\hspace{-0.15in}
\includegraphics[width=3in]{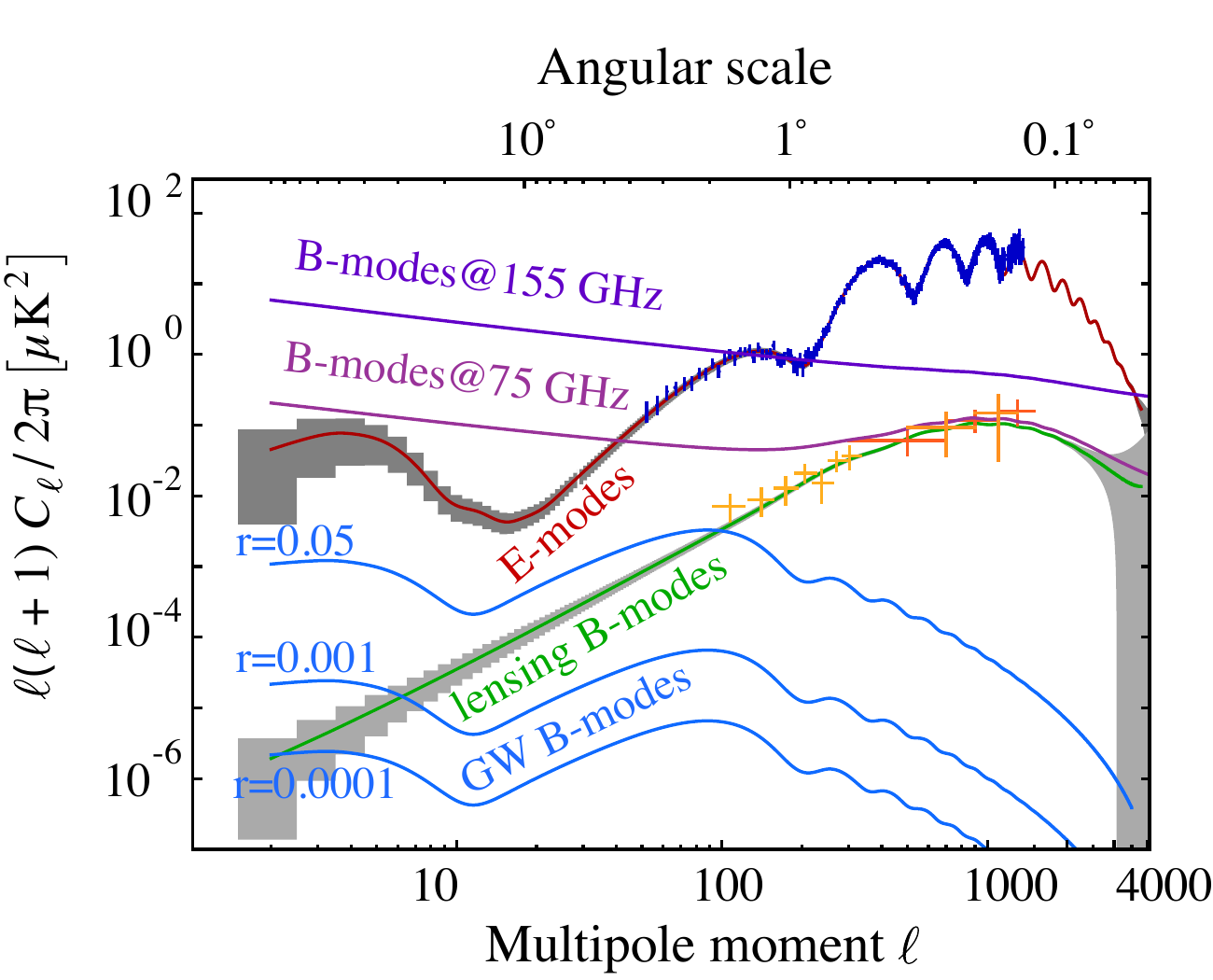}
\includegraphics[width=3in]{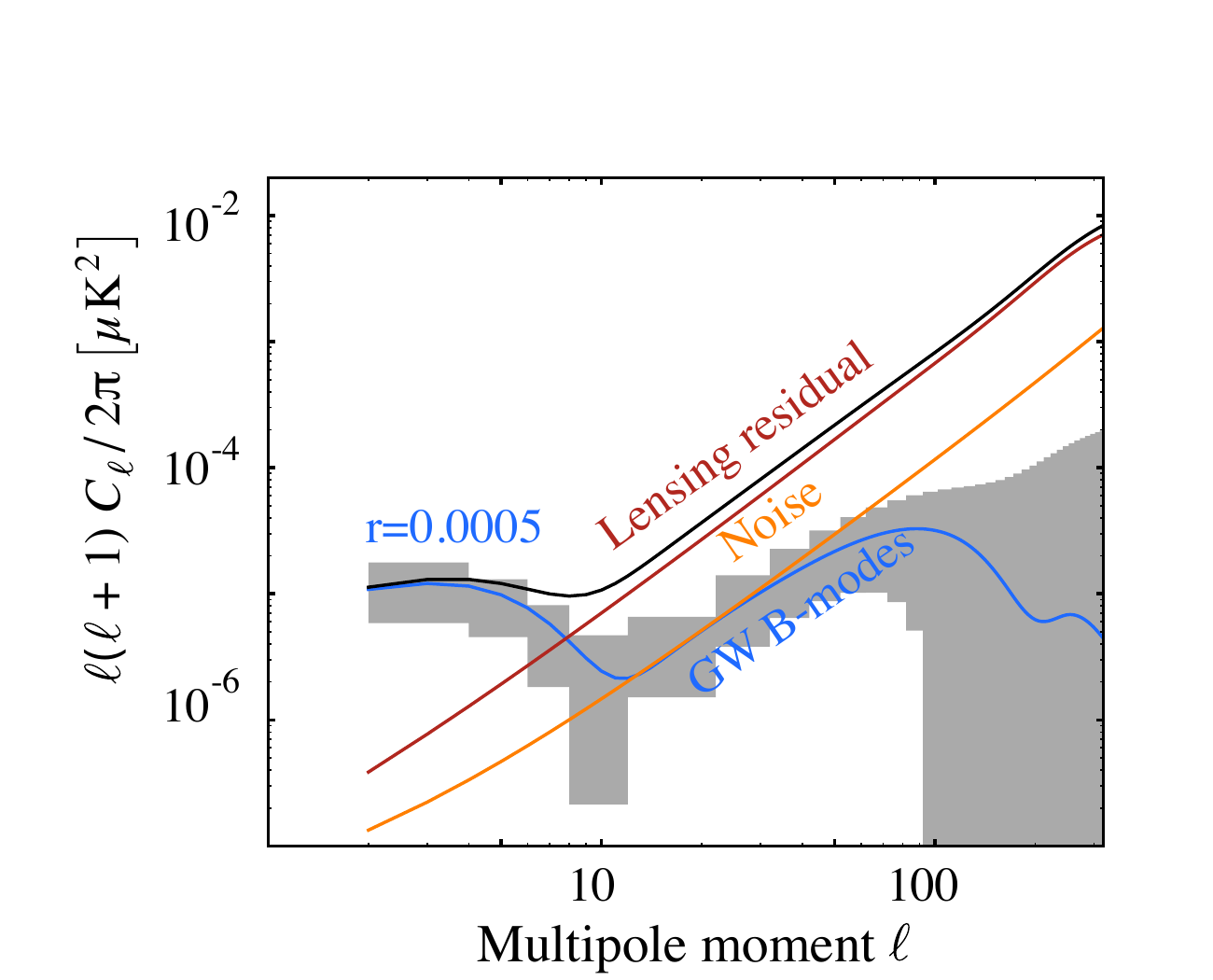}
\vspace{-0.1in}
\caption{\captiontext With PICO's baseline configuration we will measure the $EE$ (left, red) and lensing $BB$ (green) angular power spectra with high precision (grey). PICO's goal is to detect $r= 5\times 10^{-4}\, (5\sigma)$ (right, grey). This forecast includes PICO's 80\% delensing (red) and foreground separation. The baseline noise level (right, orange) allows detection of even lower levels; we expect foreground separation to limit performance.  As an example we show the total $BB$ spectra on the cleanest $60\%$ of the sky at 75 and 155~GHz (left, purple). The foregrounds largely dominate the cosmological signals. Also shown are measurements of lensing from current experiments (left, orange)~\citep{PB_BB, keisler2015, actpol_lensing_BB, Array:2015xqh}, \planck 's $EE$ measurements (left, dark blue)~\citep{Planck2018_I}, and the $BB$ spectrum produced by an inflationary gravity wave (GW) signal with different values of $r$ (cyan). 
\label{fig:clbb} }
\vspace{-0.05in}
\end{figure}

The strength of the signal, quantified by the tensor-to-scalar ratio $r$, is a direct measure of the expansion rate of the Universe during inflation. Together with the Friedmann equation, it reveals one of the most important characteristics of inflation: its energy scale.\footnote{In some models of inflation the one-to-one correspondence between $r$ and the energy scale of inflation does not hold because there are additional sources of gravitational waves~\cite{Namba:2015gja}. However, in these models the signal is highly non-Gaussian and could be distinguished from models without such sources, for which the signal is Gaussian.}  A detection of $r$  ``would be a watershed discovery'', a quote from the 2010 decadal panel report~\citep{blandford2010}. The combination of data from \planck\ and the BICEP/Keck Array give the strongest constraint to date, $r<0.06\,\, (95\%)$~\citep{bicep_keck2018}. Next decade S3 efforts strive to reach $\sigma(r)=2 \times10^{-3}$~\citep{SOscience, biceparray}.

PICO will detect primordial gravitational waves if inflation occurred at an energy scale of at least $5\times 10^{15}\,\rm{GeV}$, or equivalently $r= 5\times 10^{-4} \, (5\sigma)$ (SO1 in Table~\ref{tab:STM} and Fig.~\ref{fig:clbb}).  A detection will have profound implications for fundamental physics. It will provide evidence for a new energy scale tantalizingly close to the energy scale associated with grand unified theories, probe physics at energies far beyond the reach of terrestrial colliders, and be the first observation of a phenomenon associated with quantum gravity~\cite{Krauss:2013pha}.

%PICO's goal is to detect primordial gravitational waves if inflation occurred at an energy scale of at least $5\times 10^{15}\,\rm{GeV}$, or equivalently $r= 5\times 10^{-4} \, (5\sigma)$ (SO1 in Table~\ref{tab:STM} and Fig.~\ref{fig:clbb}).  A detection will have profound implications for fundamental physics. It will provide evidence for a new energy scale tantalizingly close to the energy scale associated with grand unified theories, probe physics at energies far beyond the reach of terrestrial colliders, and be the first observation of a phenomenon associated with quantum gravity~\cite{Krauss:2013pha}.

\begin{figure}[!thb]
\parbox{4.5in}{\centerline{
\includegraphics[width=4.5in]{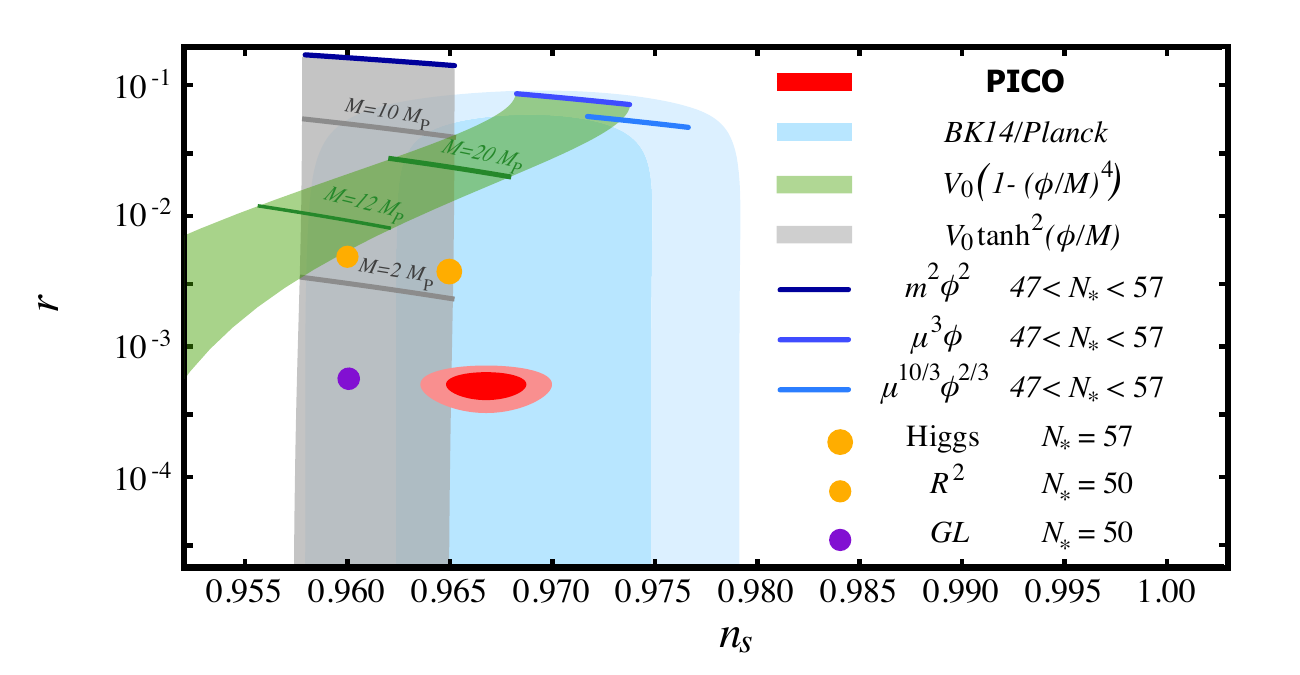} } }
\parbox{1.8in}{
\caption{\captiontext  Current $1\sigma$ and 2$\sigma$ limits on $r$ and $n_{\rm s}$ (cyan) and forecasted constraints for a fiducial model with $r = 0.0005$ for PICO, together with predictions for selected models of inflation. Characteristic super-Planckian scales in the potentials are marked with darker lines. GL is the Goncharev-Linde model (see text). }
\label{fig:nsr}}
\vspace{-0.1in}
\end{figure}

There are only two classes of slow-roll inflation in agreement with current data that naturally explain the observed value of the spectral index of primordial fluctuations $n_{\rm s}$~\cite{Aghanim:2018eyx}. The first class is characterized by potentials of the form $V(\phi)\propto\phi^p$. This class includes many of the simplest models of inflation, some of which have already been strongly disfavored by existing observations. Select models in this class are shown as blue lines in Fig.~\ref{fig:nsr}. When the constraints on $n_{\rm s}$ tighten by about a factor of two with the central value unchanged, and the upper limit on $r$ improves by an order of magnitude, this class would be ruled out. 

The second class is characterized by potentials that approach a constant as a function of field value, either like a power law or exponentially. Two representative examples in this class are shown as the green and gray bands in Fig.~\ref{fig:nsr}. This class also includes $R^2$ inflation, which predicts a tensor-to-scalar ratio of $r\sim 0.004$. All models in this class, with a characteristic scale in the potential that is larger than the Planck scale, predict a tensor-to-scalar ratio of $r\gtrsim 0.001$.  PICO will exclude these models with high confidence ($>5\sigma$), and is the only proposed mission for the next decade to reach an exclusion at more than $2\sigma$. While many microphysical models in this second class possess a characteristic scale that is super-Planckian, some have a somewhat smaller scale. One example is the Goncharov-Linde model, which predicts a tensor-to-scalar ratio of $r\sim 4\times 10^{-4}$~\cite{Goncharov:1983mw}, still within $4\sigma$ detection by PICO~(Fig.~\ref{fig:nsr}). There are models with much smaller values that are out of reach.  

%The second class is characterized by potentials that approach a constant as a function of field value, either like a power law or exponentially. Two representative examples in this class are shown as the green and gray bands in Fig.~\ref{fig:nsr}. This class also includes $R^2$ inflation, which predicts a tensor-to-scalar ratio of $r\sim 0.004$. All models in this class, with a characteristic scale in the potential that is larger than the Planck scale, predict a tensor-to-scalar ratio of $r\gtrsim 0.001$. PICO will definitively exclude these models, or will detect gravitational waves; either result will be achieved with more than $5\sigma$ confidence.  While many microphysical models in this second class possess a characteristic scale that is super-Planckian, some have a somewhat smaller scale. One example is the Goncharov-Linde model, which predicts a tensor-to-scalar ratio of $r\sim 4\times 10^{-4}$~\cite{Goncharov:1983mw}, still within $4\sigma$ detection by PICO~(Fig.~\ref{fig:nsr}). There are models with smaller values that are out of reach.  

Distinguishing between models with sub- and super-Planckian characteristic scales would provide much needed guidance to discriminate between classes of ideas for the physics of the earliest moments of our Universe. And this much is clear: PICO will either detect gravitational waves, or, if its required threshold is passed without a detection, most textbook models of inflation will be ruled out and the data would force a significant change in our understanding of the primordial universe.

%Distinguishing between models with sub- and super-Planckian characteristic scales would provide much needed guidance to discriminate between classes of ideas for the physics of the earliest moments of our Universe.

%While many microphysical models in this second class possess a characteristic scale that is super-Planckian, some have a somewhat smaller scale. One example is the Goncharov-Linde model, which predicts a tensor-to-scalar ratio of $r\sim 4\times 10^{-4}$~\cite{Goncharov:1983mw}, still within detection reach by PICO~(Fig.~\ref{fig:nsr}). There are models with significantly smaller values that are out of reach.  

%Distinguishing between models with sub- and super-Planckian characteristic scales would provide much needed guidance to discriminate between classes of ideas for the physics of the earliest moments of our Universe.

%\paragraph{Observational Considerations}
\noindent$\bullet$ {\bf Observational Considerations} \hspace{0.1in} 
The $BB$ angular power spectrum measured by PICO will have contributions from Galactic sources of emission and `lensing' $B$-modes, created by gravitational lensing of $E$-modes as the CMB photons traverse the gravitational potentials throughout the Universe (Fig.~\ref{fig:clbb} and \S\,\ref{sec:gravitationallensing}). In case of an $r$ detection, there will be two additional features due to the inflationary signal. One is the `recombination peak' at $\ell = 80$ and the other is the `reionization peak' at multipoles of $\ell\lesssim 10$. PICO's strong constraints on $r$ derive from using all available $\ell$ modes.

The Galactic signals act as foregrounds, and uncertainty in the characterization of these foregrounds already limits our ability to constrain $r$. 
%PICO's goal for reaching $\sigma(r) = 1\times10^{-4}$ is driven by estimates of the efficacy of foreground separation, not noise.  
An analytic performance forecast accounting for PICO's statistical noise level and a foreground model that has polarized emission from two components of dust, synchrotron radiation, and correlations between synchrotron and dust emission, gives $\sigma(r) = 2\times10^{-5}$, five times lower than our baseline requirement. This margin allows for degradation in foreground removal through inclusion of physical effects known to exist but not captured in the analytic forecasts. These effects are included in map-based simulations, which indicate that PICO will achieve its requirement; see Section~\ref{sec:signal_separation}. 
%\comor{say more?}

When the tensor-to-scalar ratio $r \simeq 0.01$, the $BB$ lensing and inflation spectra are comparable in magnitude at the recombination peak $(\ell = 80)$. For lower levels of $r$, the lensing $B$-mode dominates, but the $B$-mode maps can be `delensed' if the polarization maps are measured with few-arcmin resolution and sufficient depth~\citep{2004PhRvD..69d3005S,2012JCAP...06..014S}. Forecasts for PICO show that at least 73\% of the lensing $B$-mode power can be removed for the baseline configuration, after accounting for conservative Galactic foreground separation. As much as 85\% will be removed for the CBE and for milder foreground contamination. For measuring the recombination peak, delensing is essential in order to reach PICO's limits on $r$, and this was a driver in choosing the resolution of the instrument. 
% PICO will be relying on its own data to conduct delensing, thus avoiding increased noise from the need to cross-calibrate experiments, identify common observing areas on the sky, not having frequency-band coverage at the appropriate resolution to remove foregrounds, or from other systematic uncertainties.

For the levels of $r$ targeted by PICO, the $BB$ reionization signal $(\ell < 10)$ has a somewhat higher level than the lensing spectrum, but the map-level foregrounds at this angular scale are at least two orders of magnitude brighter.  There are currently no $BB$ measurements at these scales, and no S3 experiments plan to measure $B$-modes that reach to $\sigma(r) < 0.006$ in the lowest multipoles~\citep{class,piper}. PICO's instrument temporal stability, absence of atmospheric noise, full-sky coverage, and unmatched capability to characterize and separate foregrounds make it the most suitable instrument to measure these lowest multipoles (\S\,\ref{sec:signal_separation}).

If an inflationary $B$-mode signal is detected, it is important to characterize its entire $\ell$ dependence in the predicted reionization and  recombination peaks, in order to confirm -- rather than assume -- its expected dependence on angular scale.  Furthermore, the PICO full-sky coverage will enable detection of the recombination peak in several independent patches of the sky, giving an important systematic cross-check. Only a space mission can provide these important benefits. 

\noindent$\bullet$ {\bf Scalar Spectral Index and Non-Gaussianity} \hspace{0.1in} Models of the early Universe differ not only in their predictions for $r$ and the scalar spectral index $n_{\rm s}$, but also for the scale-dependence of $n_{\rm s}$, a parameter commonly called ``the running of $n_{\rm s}$'' and labeled $n_{\rm run}$. PICO will improve $n_{\rm s}$ and $n_{\rm run}$ constraints by a factor of three relative to \planck\ to achieve $\sigma(n_{\rm s})=0.0015$ and $\sigma(n_{\rm run})=0.002$. For many models of inflation, how reheating occurred is unknown, and this translates to different predictions for $n_{\rm s}$ and $n_{\rm run}$~\citep{planck2018_inflation}. PICO's precision is sufficient to distinguish between different possible reheating scenarios at $>3\sigma$ (SO2). 

\begin{figure}[h]
\hspace{-0.in}
\parbox{3.0in}{{
\includegraphics[width=2.95in]{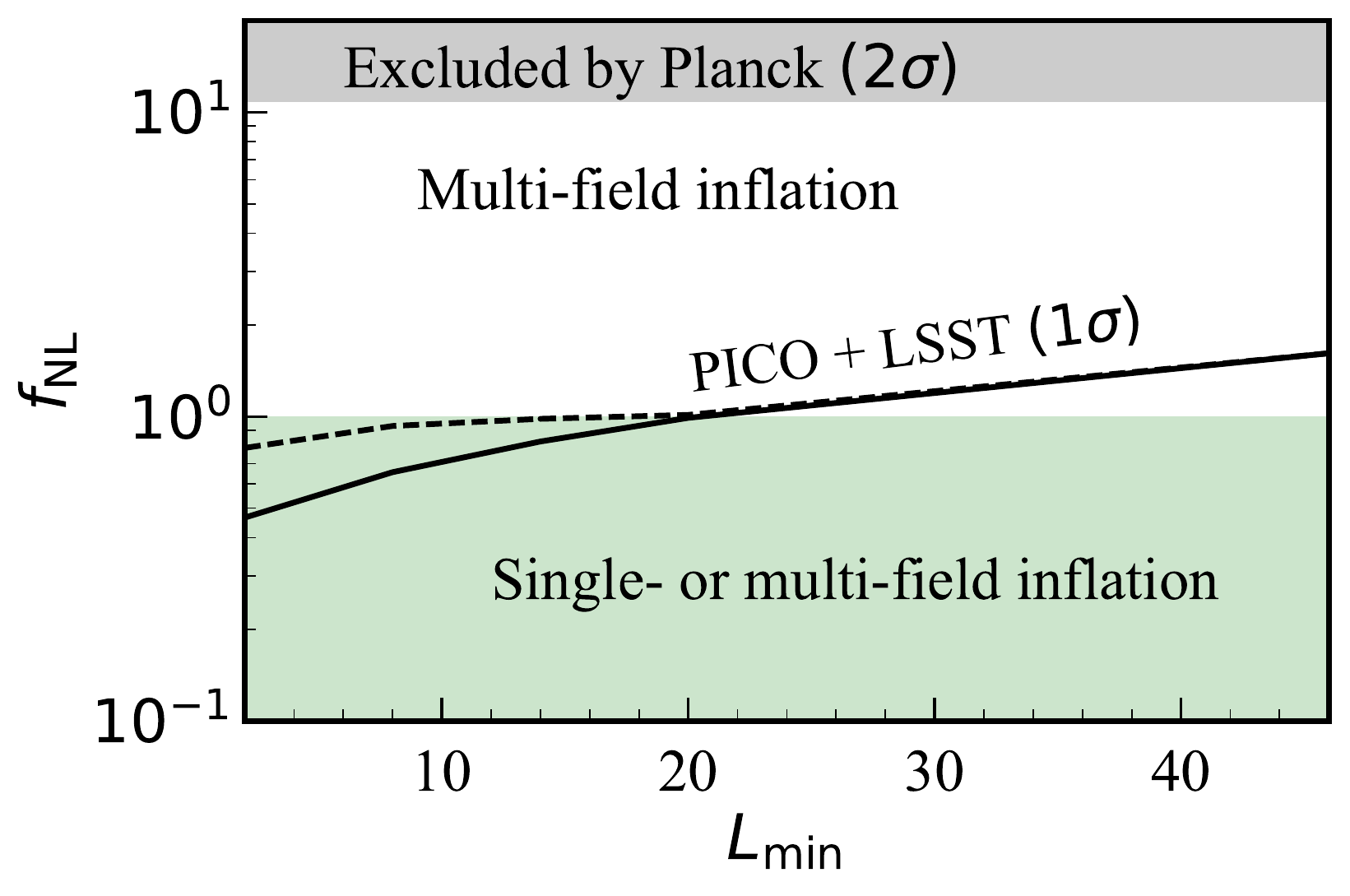} } }   % new version, LSST gold, 10year
\hspace{0.in}
\parbox{3.4in}{
\caption{\captiontext
Cross-correlating PICO's lensing potential map with LSST galaxies will allow detecting or excluding  $f^{\rm local}_{\rm NL}=2$ with $3\sigma$ evidence  if the data can be used at  angular scales $L \ge 8$ (solid black). A detection above $|f^{\rm local}_{\rm NL}| =1$ indicates that inflation is driven by multiple fields; single-field inflation has $|f^{\rm local}_{\rm NL}|<1$ (green region). The \planck\ constraint is $f^{\rm local}_{\rm NL} < 10.8\, (2\sigma)$. The cross-correlations will allow excluding or detecting $f^{\rm local}_{\rm NL}=2\, (2\sigma)$ if LSST data are used only for $L\ge 20$ (dash). Fig.~\ref{fig:sigma8} gives the assumptions used here for the LSST data. 
\label{fig:fnlconstraint}
} }
\vspace{-0.1in}
\end{figure}

The simplest models of inflation, in which there is a single inflaton field, predict primordial fluctuations that are very nearly Gaussian with $|f^{\rm local}_{\rm NL}| <1$, where $f^{\rm local}_{\rm NL}$ is a parameter quantifying the level of local non-Gaussianity~\citep{planck2015_17}. A detection of $|f^{\rm local}_{\rm NL}| >1$ points exclusively to models of inflation with multiple fields (Fig.~\ref{fig:fnlconstraint}). 
%, making $\sigma (f^{\rm local}_{\rm NL}) < 1$ a compelling target. 
\planck\ gives a constraint of $f^{\rm local}_{\rm NL} = 0.8 \pm 5 \, (1\sigma)$~\citep{planck2015_17}, and further measurements of the \ac{CMB} alone cannot improve on this constraint by more than a factor of 2--3. However, correlating large-scale structure tracers that have different clustering bias factors can enhance the signature of non-Gaussianity~\citep{2009PhRvL.102b1302S,2018PhRvD..97l3540S,2008PhRvD..77l3514D}. Fig.~\ref{fig:fnlconstraint} shows expected constraints from correlations between the PICO lensing potential maps (\S~\ref{sec:gravitationallensing}) and LSST galaxies. For $f^{\rm local}_{\rm NL}=2$, $3\sigma$ evidence will be reached if large angular scale ($L\ge 8$)\footnote{$L$ refers to multipoles in galaxy clustering fields and in CMB lensing~(\S~\ref{sec:gravitationallensing}), in contrast to the use of $\ell$  for the CMB itself. \label{foot:L}} auto- and cross-correlation spectra can be used. If LSST's auto-correlation can only be used on smaller angular scales $L\ge 20$, the $3\sigma$ evidence weakens to $2\sigma$. 
%Specifically, using data spanning angular scales with $L>6$ a compelling constraint of $|f_{NL}|<1\, (2\sigma)$ will be reached. 

%\comor{One can use correlations between large-scale structure tracers with different clustering bias factors and measure the relative difference between their clustering power spectra to effectively cancel cosmic variance~\citep{2009PhRvL.102b1302S,2018PhRvD..97l3540S}; this can constrain physics that affects the biasing of objects on large scales, such as primordial local non-Gaussianity~\citep{2008PhRvD..77l3514D}.  In Fig.~\ref{fig:fnlconstraint} we show the expected constraints for the CMB lensing field as reconstructed with PICO, in cross-correlation with  three years of the LSST survey. It can be seen that depending on the minimal multipole that can be used in the cross correlation, which is uncertain in both LSST and the PICO lensing map, the well-motivated theory target of $\sigma (f_\mathrm{NL}) \simeq 1$ \citep{2014arXiv1412.4671A} can be within reach. Values of $f_\mathrm{NL}$ at or above this level are a generic prediction of multi-field inflationary models.}

\subsubsection{Fundamental Particles: Light Relics, Dark Matter, and Neutrinos}
\label{sec:relics_neutrinos}

%%%%%%%%%%%%%%%%%%%%%%%%%
$\bullet$ {\bf Light Relics} \hspace{0.1in} In the inflationary paradigm, the Universe was reheated to temperatures of 
at least 10 MeV and perhaps as 
high as $10^{12}$ GeV.  At these high temperatures, even very weakly interacting or very massive particles, 
such as those arising in extensions of the Standard Model of particle physics, can be produced in large 
abundances~\cite{1979ARNPS..29..313S,Bolz:2000fu}.  As the Universe expands and cools,
the particles fall out of equilibrium, an event referred to as `decoupling,' and characterized by a decoupling temperature $T_{F}$.  The decoupling leaves observable signatures in the CMB power spectra. Through these effects the CMB is a sensitive probe of neutrino and other particles' properties.  

% sensitive probe of the fundamental particle content in the Universe
% large abundance, but not large enough to leave present day signatures? or they decay?
% don't like the words 'extensions of ...' suggests very unlikely things. 

One particularly compelling target is the effective number of light relic particle species $\Neff$. The canonical value with three neutrino families is $\Neff = 3.046$. Additional light particles contribute a change $\Delta \Neff$ that is a function only of the decoupling temperature and the spin of the particle~$g$. The magnitude of $\Delta\Neff$ is quite restricted, even for widely varying decoupling temperatures $T_{F}$. A range $ 0.027\,g \leq \ \Delta \Neff \leq 0.07\,g$ corresponds to a range in $T_{F}$ spanning decoupling during post-inflation reheating (0.027$g$) down to lower $T_{F}$ with decoupling occurring just prior to the QCD phase transition ($0.07g$).
%Additional light particles contribute a universal change to $\Neff$ that is a function only of the decoupling temperature and the effective degrees of freedom of the particle, $g$. Furthermore, the range of $\Delta\Neff$ is quite restricted, even for widely varying decoupling temperatures $T_{F}$ with the range $ 0.027\,g \leq \ \Delta \Neff \leq 0.07\,g$ corresponding to decoupling at higher temperatures during post-inflation reheating (0.027$g$) to lower temperatures shortly prior to the QCD phase transition ($0.07g$).

\begin{figure}[t!]
\begin{center}
\includegraphics[width=0.45\textwidth]{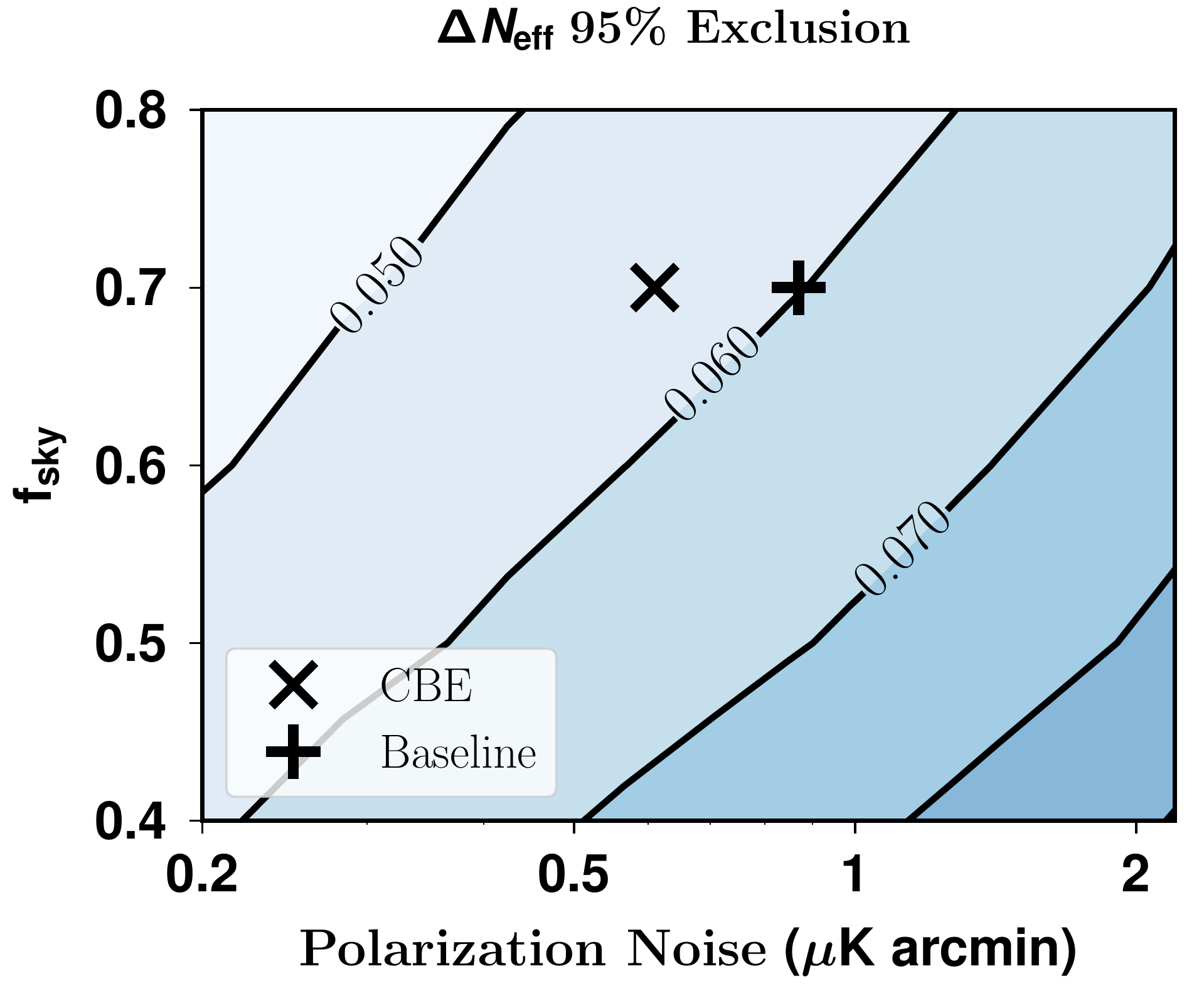}
\includegraphics[width=0.47\textwidth]{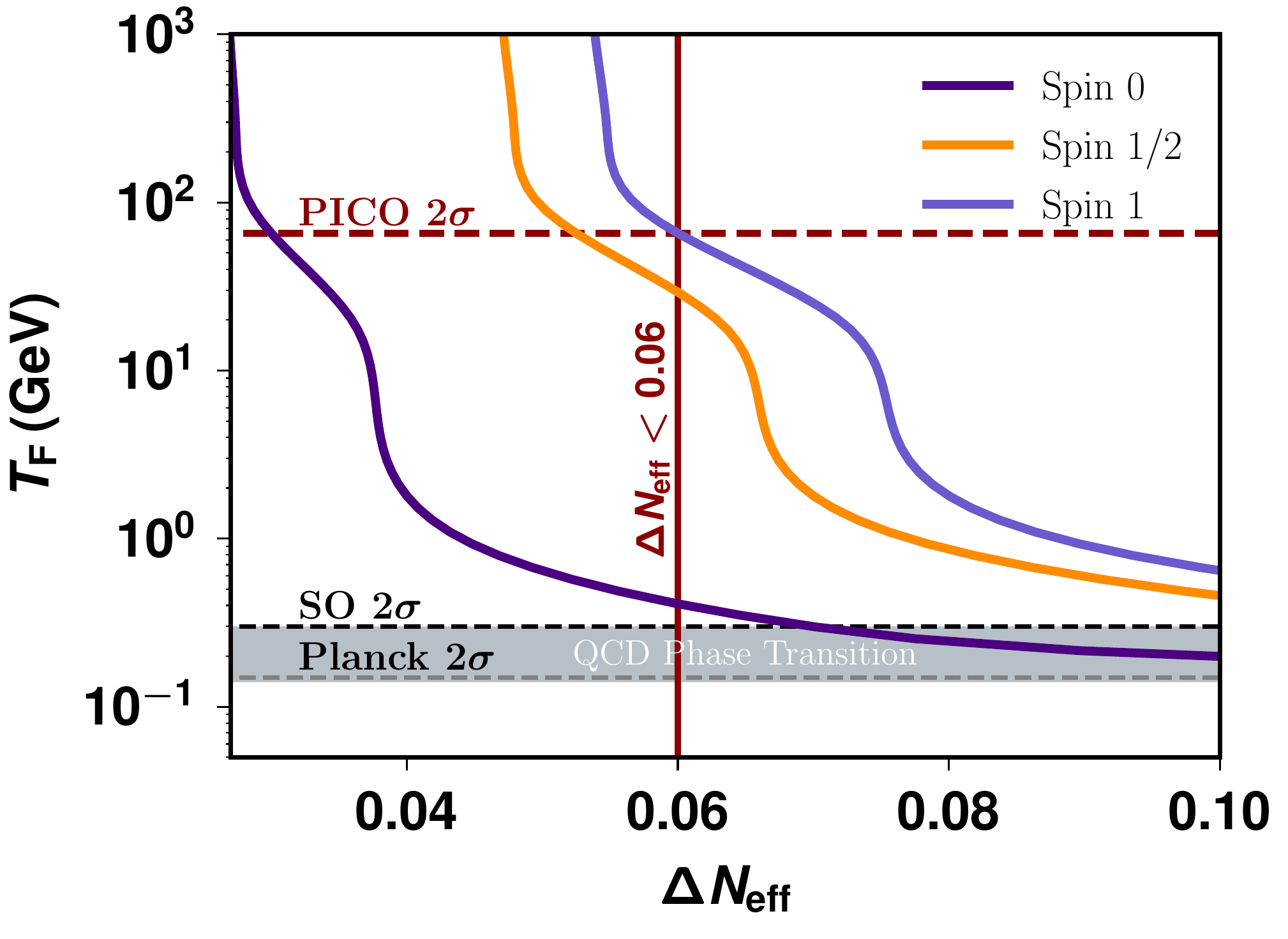}
\vspace{-0.15in}
\caption{ \captiontext PICO will achieve a constraint $\Delta \Neff < 0.06\, (95\%)$ (left, $2\sigma$ contours shown) in the baseline configuration (cross) using its cosmic-variance-limited measurement of $EE$ for $\ell \leq 2300$, and 21 frequency bands to utilize data over 70\% of the sky (5\arcmin\ resolution assumed). This constraint translates to moving up the lowest decoupling temperature $T_{F}$ for particles with spin 1, 1/2, and 0 by factors of 400, 200, and 6, respectively, relative to \planck\ (right, dashed black, only $T_{F}$ for vector particles is shown). We also show the projected vector particle limit for the Simons Observatory~\citep{SOscience}. }
\label{fig:Neff_future}  
\end{center}
\vspace{-0.2in}
\end{figure}

Information about $\Neff$ is gleaned primarily from the $TT,\, TE$, and $EE$ power spectra. For an experiment like PICO, which has sufficient resolution to reach a cosmic-variance-limited measurement\footnote{\label{CVL}A measurement is cosmic-variance-limited when the measurement uncertainty is dominated by the statistics of observing the finite number of spherical harmonic decomposition modes available in our Universe.} of $EE$ up to $\ell =2300$, the two additional most important parameters for improving constraints are the fraction of sky observed, $f_{\rm sky}$, and the noise (Fig.~\ref{fig:Neff_future}, left). The PICO baseline will use data from 70\% of the sky to constrain $\Delta \Neff < 0.06 \, (95\%)$ (S04).\footnote{The CMB $EE$ and the Galactic foregrounds $EE$ and $BB$ spectra are comparable in level (Fig.~\ref{fig:clbb}). With 21 frequency bands PICO should be able to separate signals at the mild levels necessary for $EE$ over 70\% of the sky~(\S~\ref{sec:signal_separation}).} This constraint, which is a factor of 4.7 improvement relative to \planck~($\Delta \Neff < 0.28$, 95\%) and will not be matched by any currently funded effort, opens up a new range of temperatures in which to detect the signature of light relic species. If no new species are detected, then the lowest temperature $T_{F}$ at which any %particle with spin 
vector particle (spin 1) could have fallen out of equilibrium will move up by a factor of 400 (Fig.~\ref{fig:Neff_future}, right). 

While our theoretical target for $\Neff$ is defined by particles that decoupled long before neutrinos did, there are a number of well-motivated scenarios in which the thermal evolution of the Standard Model is altered after the time of neutrino decoupling.  These scenarios will change the relationship between $\Neff$ as measured in the CMB and the value of $\Neff$ that affects the primordial abundance of the helium fraction $Y_p$ as inferred from big bang nucleosynthesis calculations. 
%While our theoretical target for $\Neff$ is defined by particles that decoupled long before neutrinos did, there are a number of well-motivated scenarios in which the thermal evolution of the Standard Model is altered after the time of neutrino decoupling.  These scenarios will change the relationship between $\Neff$ as measured in the CMB relative to the value of $\Neff$ that affects big bang nucleosynthesis calculations and the production of helium, which is quantified through the helium fraction $Y_p$.  
For example, the decay of a thermal relic into photons after nucleosynthesis would reduce $\Neff$ in the CMB but could leave $Y_p$ unaltered from its Standard Model value.  PICO will make a simultaneous measurement of $\Neff$ and $Y_p$ with $\sigma(\Neff) = 0.08$ and $\sigma(Y_p) =0.005$, giving a 2\% uncertainty on the value of $Y_p$. These uncertainties are equivalent to those available with other astrophysical measurements, but the systematic uncertainties are entirely different. Systematic uncertainties currently limit our knowledge of $Y_{p}$. 

%%%%%

\noindent$\bullet$ {\bf Dark Matter} \hspace{0.1in} Cosmological measurements have already confirmed the existence of one relic that lies beyond the Standard Model: dark matter. \ac{CMB} experiments are effective in constraining dark matter candidates in the lower mass range, which is not available for terrestrial direct detection experiments~\citep{Slatyer2009,Galli2009,Huetsi2009,Huetsi2011,Madhavacheril:2013cna,Green:2018pmd}. 

Interactions between dark matter and protons in the early Universe create a drag force between the two cosmological fluids, damping acoustic oscillations and suppressing power in density perturbations on small scales. As a result, the CMB temperature, polarization, and lensing power spectra are suppressed at high multipoles relative to a Universe without such drag forces.  This effect has been used to search for evidence of dark-matter--proton scattering over a range of masses, couplings, and interaction models~\citep{2002astro.ph..2496C, 2004PhRvD..70h3501S, Dvorkin:2013cea, 2018PhRvL.121h1301G,2018arXiv180108609B, 2018PhRvD..97j3530X, 2018arXiv180800001B, 2018PhRvD..98b3013S}, to test the possibility of an interacting dark-matter sub-component~\citep{2018arXiv180800001B}, and to provide consistency tests of dark matter in the context of the anomalous 21-cm signal reported by the EDGES collaboration~\citep{2018Natur.555...71B,2018Natur.555...67B,2018arXiv180800001B,2018arXiv180711482K}.
 
\begin{figure}[t]
\begin{center}
\includegraphics[width=0.50\textwidth]{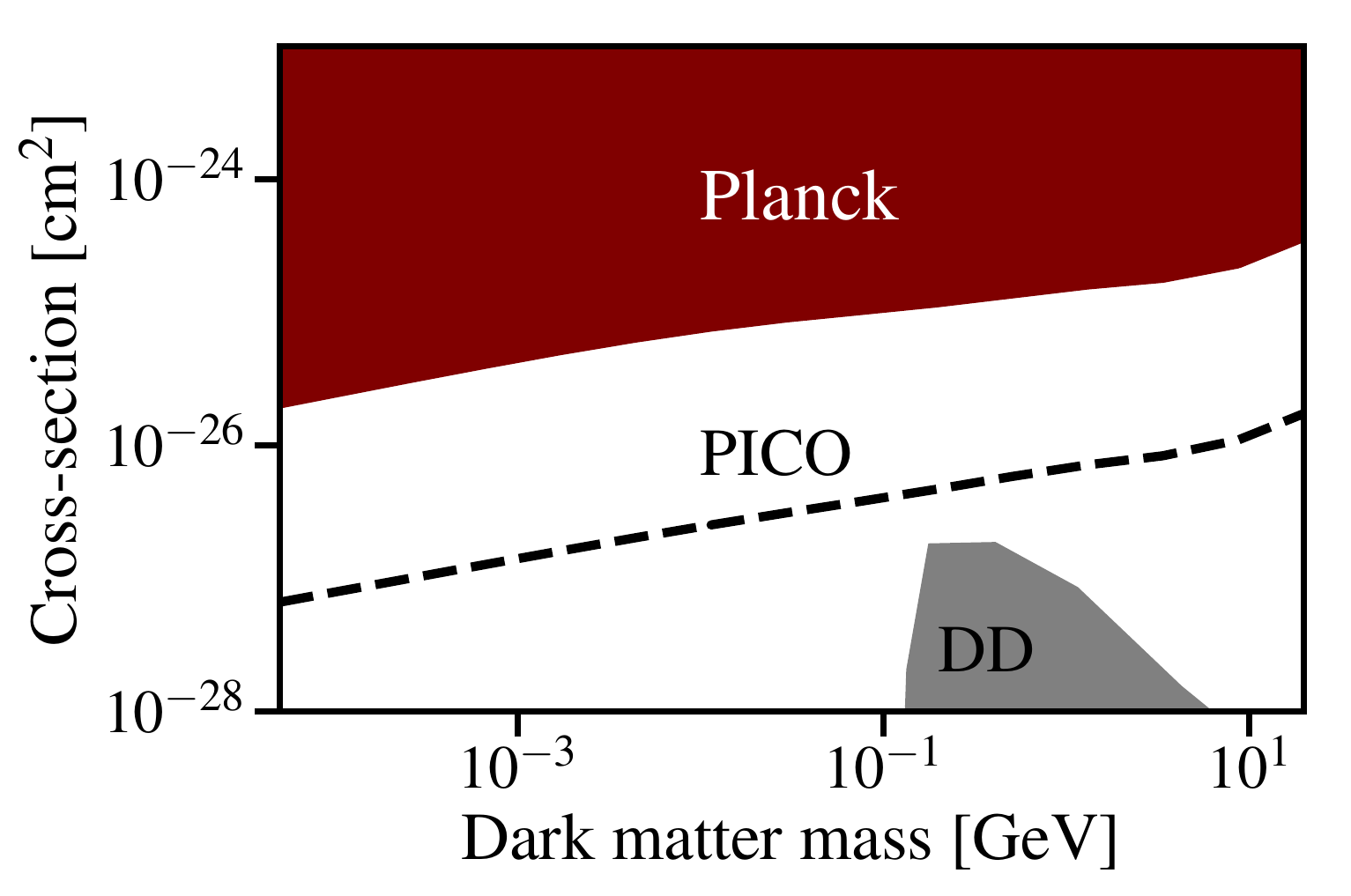}
\includegraphics[width=0.48\textwidth]{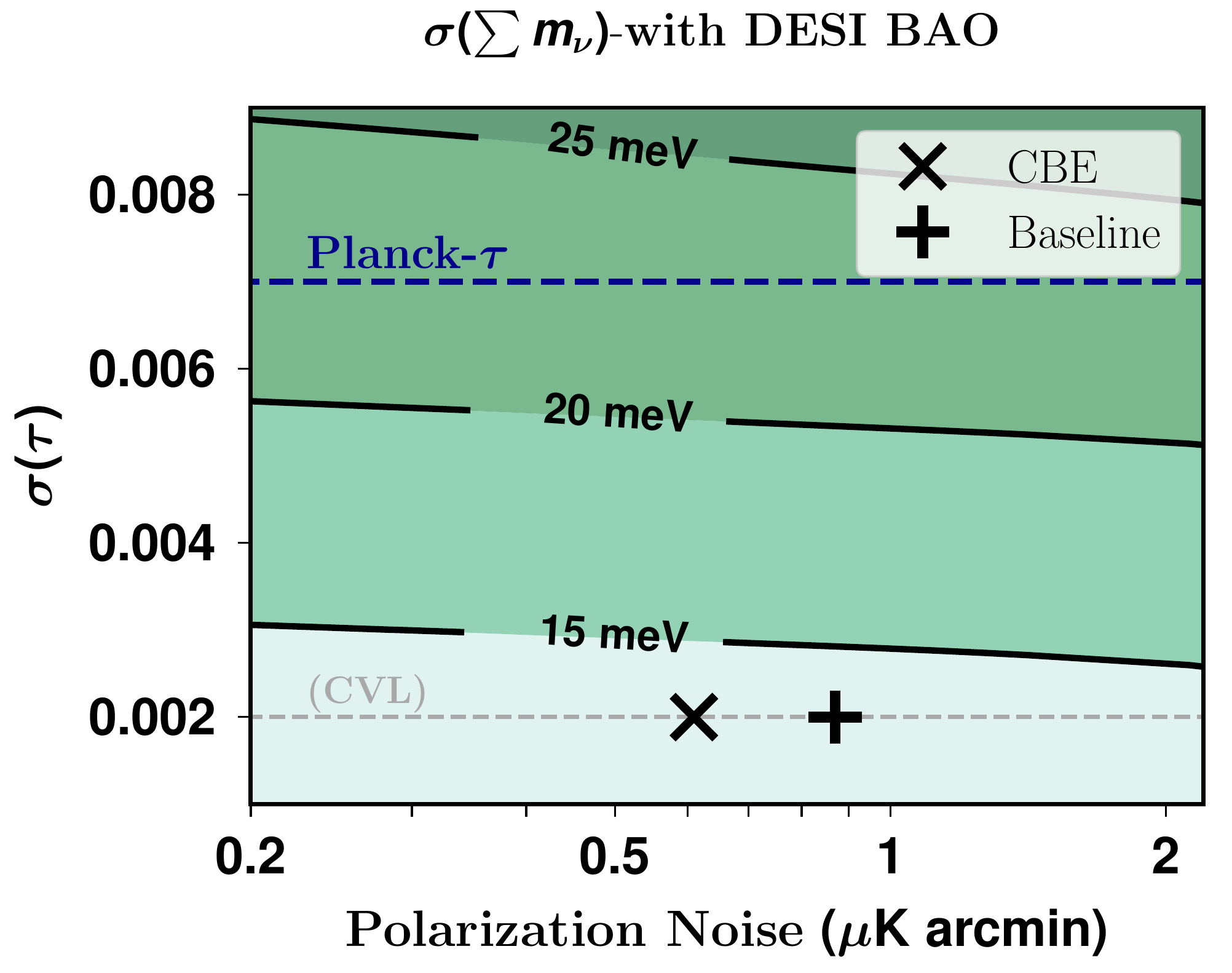}
\vspace{-0.15in}
\caption{\captiontext {\bf Left:} PICO will give a factor of 25 more stringent constraint on the spin-independent velocity-independent dark matter scattering cross-section (dashed) relative to current \planck\ 95\% confidence limit (red)~\citep{2018PhRvL.121h1301G}. Terrestrial direct detection experiments are expected to give complementary and stronger constraints, but only for the higher dark matter masses (grey)~\cite{2018PhRvD..97l3013K}. 
{\bf Right:} Using a cosmic-variance-limited (CVL) measurement of $\tau,\, \sigma(\tau)=0.002$, \ac{BAO} information from DESI, and separation of foregrounds over 70\% of the sky, PICO will reach $\sigma(\Sigma m_{\nu}) = 14$~meV (contours), giving at least a $4\sigma$ detection of the minimal expected sum of neutrino masses $\Sigma m_{\nu} = 58$~meV. 
\label{fig:DM_baryons} }
\end{center}
\vspace{-0.2in}
\end{figure}

PICO's constraining power comes primarily from making high \ac{SNR} maps of the lensing-induced deflections of polarized photons, which are discussed in Section~\ref{sec:extragalacticsci}.  For a spin-independent velocity-independent contact-interaction, chosen as our fiducial model, PICO will improve upon \planck 's dark matter cross-section constraints by a factor of 25 over a broad range of candidate masses that are largely unavailable for traditional direct detection experiments (Fig.~\ref{fig:DM_baryons}, right). 
%The constraints are complementary to those forthcoming from direct detection experiments, which are more sensitive at the high mass range.  
%\comor{need to decide what to do with this section.}
%which is largely unavailable to traditional direct detection experiments

The axion is another dark matter candidate that is well motivated by string theory~\citep{Arvanitaki_etal} and that is consistent with straightforward extensions of the Standard Model of particle physics~\citep{peccei,weinberg,wilczek}. For an axion mass in the intermediate range $10^{-30} < m_a< 10^{-26} $~eV, current measurements constrain its fraction to be $\le 2$\% $(1\sigma)$ of the total dark-matter density. If 2\% of the total dark content is made of axions, PICO's measurement of the $TT$, $TE$ and $EE$ spectra with additional constraints from the lensing reconstruction will detect this species at between $7$ and $13\sigma$, depending on the mass range. %, as shown in Figure~\ref{fig:axions}. 
This is an average improvement of a factor of 10 relative to \planck .
% and a factor of two relative to the combination of \planck\ and S3, respectively, profoundly shaping our understanding of the nature of dark matter and its small-scale clustering. 

\noindent$\bullet$ {\bf Neutrino Mass} \hspace{0.1in} \label{neutrino_fundamental} The origin and structure of the neutrino masses is one of the great outstanding  questions about the nature of the Standard Model particles.  
%Measurements of neutrinos in the lab have revealed much  about the mass differences and mixing angles.  
Cosmology offers a  measurement of the sum of the neutrino masses $\sum m_\nu$ through the gravitational influence of the non-relativistic  cosmic neutrinos.  The current measurement of $\Neff = 2.99 \pm 0.17$~\citep{Planck2018_VI} already confirms the existence of these neutrinos at $>10\sigma$ and their mass implies that they will contribute to the matter density at low redshifts.  The best current mass constraint arises from a combination of  \planck~and BOSS \ac{BAO} giving $\sum m_\nu < 0.12$ eV (95\%) \cite{Planck2018_VI}.

Cosmological measurements are primarily sensitive to the suppression of power on small scales after the neutrinos become non-relativistic, which can be measured via CMB lensing (\S~\ref{sec:gravitationallensing}), or weak lensing in galaxy surveys.  However, these measurements are limited by our knowledge of the amplitude of the primordial fluctuation power spectrum $A_s$ because they only constrain the combination $A_s e^{-2 \tau}$, where $\tau$ is the optical depth to reionization. Although many astrophysical surveys hope to detect $\sum m_\nu$, any detection of the minimum value expected from particle physics, $\sum m_\nu = 58$~meV, at more than $2 \sigma$, will require a better measurement of $\tau$.
% and thus do not provide a high-precision measurement of either $A_s$ or $\tau$ separately.  \comor{we mention 'galaxy survey', but don't complete the thread of thought. Do we mean to say that the galaxy survey will be limited by the same degeneracy?}

The strongest constraints on $\tau$ come from the $EE$ spectrum at $\ell < 10$, which requires measurements over the largest angular scales  and good separation of Galactic foreground sources of emission. The best current measurement with $\sigma({\tau}) = 0.007$ is from \planck~\cite{Planck2018_VI}. With this uncertainty in $\tau$ one is limited to  $\sigma(\sum m_\nu) \gtrsim 25$ meV, after including forthcoming \ac{BAO} information (Fig.~\ref{fig:DM_baryons}, right); no other survey or cosmological probe will improve this constraint, unless a more accurate measurement of $\tau$ is made. One of the S3 experiments is attempting to measure the lowest $\ell$s and improve upon the \planck\ precision by a factor of about two~\citep{class}. A space mission with its access to the entire sky and broad frequency coverage is the most suitable platform for the measurement~(\S~\ref{sec:signal_separation} and \S~\ref{sec:complementarity}). PICO will reach the cosmic-variance limit uncertainty on $\tau$, $\sigma(\tau) = 0.002$~(\S~\ref{sec:luminoussources}), and using its deep CMB lensing map (\S~\ref{sec:gravitationallensing}) will therefore reach $\sigma(\sum m_\nu) = 14$ meV when combined with measurements of \ac{BAO} from DESI or Euclid~\cite{Levi:2013gra}.
%(without \ac{BAO} data the constraint is $\sigma(\sum m_\nu) = 43$ meV).
This measurement will give a $4\sigma$ detection of the minimum sum (SO3). 

%Robustly detecting neutrino mass at  $> 3\sigma$ in any cosmological setting is only possible with an improved measurement of $\tau$, like the one achievable with PICO. This measurement will give $\sum m_\nu>0$ at greater than $4\sigma$ or would exclude the inverted hierarchy ($\sum m_\nu > 100$ meV) at 95\% confidence, depending on the central value of the measurement.  Lab-based measurement could determine the hierarchy before PICO, but only cosmology can measure $\sum m_\nu$.

%%%%%%%%%%%%%%%%%%%%%%%%%

\subsubsection{Fundamental Fields: Primordial Magnetic Fields and Cosmic Birefringence}

$\bullet$ {\bf Primordial Magnetic Fields} \hspace{0.1in} One of the long-standing puzzles in astrophysics is the origin of observed 1--10~$\mu$G galactic magnetic fields~\citep{Widrow:2002ud}. Producing such fields through a dynamo mechanism requires a primordial seed field~\citep{Widrow:2011hs}. Moreover, $\mu$G-strength fields have been observed in proto-galaxies that are too young to have gone through the number of revolutions necessary for the dynamo to work~\citep{Athreya:1998}. A primordial magnetic field (PMF), present at the time of galaxy formation, could provide the seed or even eliminate the need for the dynamo altogether. Specifically, a 0.1~nG field in the intergalactic plasma would be adiabatically compressed in the collapse to form a $\sim$1~$\mu$G galactic field~\citep{Grasso:2000wj}.
PMFs could have been generated in the aftermath of phase transitions in the early Universe~\citep{Vachaspati:1991nm}, during inflation~\cite{Turner:1987bw,Ratra:1991bn}, or at the end of inflation~\cite{DiazGil:2007dy}. A detection of PMFs with the CMB would be a major discovery because it would establish the magnetic field's primordial origin, signal new physics beyond standard models of particle physics and cosmology, and discriminate among different theories of the early Universe~\cite{Barnaby:2012tk,Long:2013tha,Durrer:2013pga}.

The current CMB bounds on PMF strength are $B_{\rm 1\,Mpc}<1.2$ nG at 95\% CL for the scale-invariant~PMF spectrum \cite{Planck2015PMF,Kunze2015,Chluba2015PMF,Zucca:2016iur}, based on measurements of the $TT$, $TE$, $EE$, and $BB$ spectra.\footnote{It is conventional to quote limits on the PMF strength smoothed over a $1$~Mpc region in comoving units,  i.e., rescaled to $z=0$: $\mathbf{B}_{\rm today} = a^2\mathbf{B}(a)$.} 
%In particular, PMF sourced vector modes contribute to the $BB$ power spectrum at high $\ell$ \cite{Lewis:2004ef}. 
The much more accurate measurement of $BB$ by PICO would only marginally improve the PMF bound because CMB spectra scale as $B^4_{\rm 1\,Mpc}$. However, Faraday rotation provides a signature that scales linearly with the strength of PMFs~\cite{Kosowsky:1996yc}. It converts CMB $E$ modes into $B$ modes, generating mode-coupling $EB$ and $TB$ correlations. So far this signature has been out of reach because prior experiments did not have sufficient sensitivity. Using Faraday rotation, PICO will probe PMFs as weak as 0.1~nG ($1\sigma$), a precision that already includes the effects of imperfect lensing subtraction, Galactic foregrounds~\cite{Oppermann:2011td,De:2013dra,Pogosian:2013dya}, and other systematic effects. With this precision, which is a factor of five stronger than achievable with S3 experiments, PICO can conclusively rule out the purely primordial (i.e., no-dynamo driven) origin of the largest galactic magnetic fields. \\
$\bullet$ {\bf Cosmic Birefringence} \hspace{0.1in}
A number of well-motivated extensions of the Standard Model involve (nearly) massless axion-like pseudo-scalar fields coupled to photons via the Chern-Simons interaction term~\citep{Freese:1990rb,Frieman:1995pm,Carroll:1998zi,Kaloper:2005aj}. These couplings also generically arise within quintessence models for dark energy~\citep{Carroll:1998zi}, chiral-gravity models~\citep{2008PhRvL.101n1101C}, and models that produce parity-violation during inflation~\cite{Gluscevic:2010vv}. Regardless of the source of the parity-violating coupling, its presence may cause cosmic birefringence -- a rotation of the polarization of an electromagnetic wave as it propagates across cosmological distances~\cite{Harari:1992ea,Carroll:1989vb,Carroll:1998zi}. Cosmic birefringence converts primordial $E$-modes into $B$-modes, producing $TB$ and $EB$ cross-correlations whose magnitude depends on the statistical properties of the rotation field in the sky~\cite{Kamionkowski:2008fp,Gluscevic:2009mm,Gluscevic:2012me}. Previous studies have constrained both a uniform rotation angle as well as anisotropic rotation described by a power spectrum \cite{Gluscevic:2012me}. The current bound on a uniform angle is 30\arcmin\ (68\%)~\cite{Planck2016_XLIX}, and the bound on the amplitude of a scale-invariant rotation angle spectrum, which could be caused by fluctuations in a light pseudo-scalar field present during inflation~\cite{Pospelov:2008gg}, is 0.11~deg$^2$ (95\%)~\citep{Array:2017rlf}). Using the combination of five bands in the 70--156~GHz range, PICO will reduce the 95\% CL bound on the uniform rotation angle by a factor of 300, to 0.1\arcmin.  
%, assuming that the beam can be calibrated to reach the main science goals without \emph{assuming} vanishing parity-odd spectra of $EB$ and $TB$ type. 
The 95\% CL bound on the amplitude of a scale-invariant rotation spectrum will be reduced by a factor of 275 to $4\times10^{-4}$~deg$^2$, giving important constraints on string-theory-motivated axions~\cite{Svrcek:2006yi,Pospelov:2008gg}.

%The simplest model for late-time acceleration of the Universe is with a slowly-evolving scalar field, also called quintessence~\cite{Carroll:1998zi}. Such a field generically couples to electromagnetism through a Chern Simons-like term, and causes linear polarization of photons propagating cosmological distances to rotate. This is known as cosmic birefringence~\cite{Carroll:1998zi}. The birefringence converts primordial $E$-modes into $B$-modes. It thus produces parity-violating $TB$ and $EB$ cross-correlations whose magnitude depends on the statistical properties of the rotation field in the sky~\cite{Kamionkowski:2008fp,Gluscevic:2009mm}. There are no theoretical predictions for the level of birefringence, but if observed, it would be evidence for physics beyond the standard model and a potential probe of dark-energy microphysics~\citep{Gluscevic:2009mm,Caldwell:2011,yadav2009}. 

%\vspace{-0.05in}

% ------------

\subsection{Cosmic Structure Formation and Evolution} % (4 pgs)
\label{sec:extragalacticsci}

%To cover: Galaxy Formation, Clusters, Reionization, point sources (probably moves to a new section called 'Legacy Science')
\subsubsection{The Formation of the First Luminous Sources} 
\label{sec:luminoussources}  

A few hundred million years after the Big Bang, the neutral hydrogen gas permeating the Universe was reionized by photons emitted by the first luminous sources to have formed.  The nature of these sources and the exact history of this epoch are key missing links in our understanding of structure formation (SO5).  
\begin{figure}
\hspace{-0.2in}
\parbox{3.1in}{\centerline {
\includegraphics[width=2.6in]{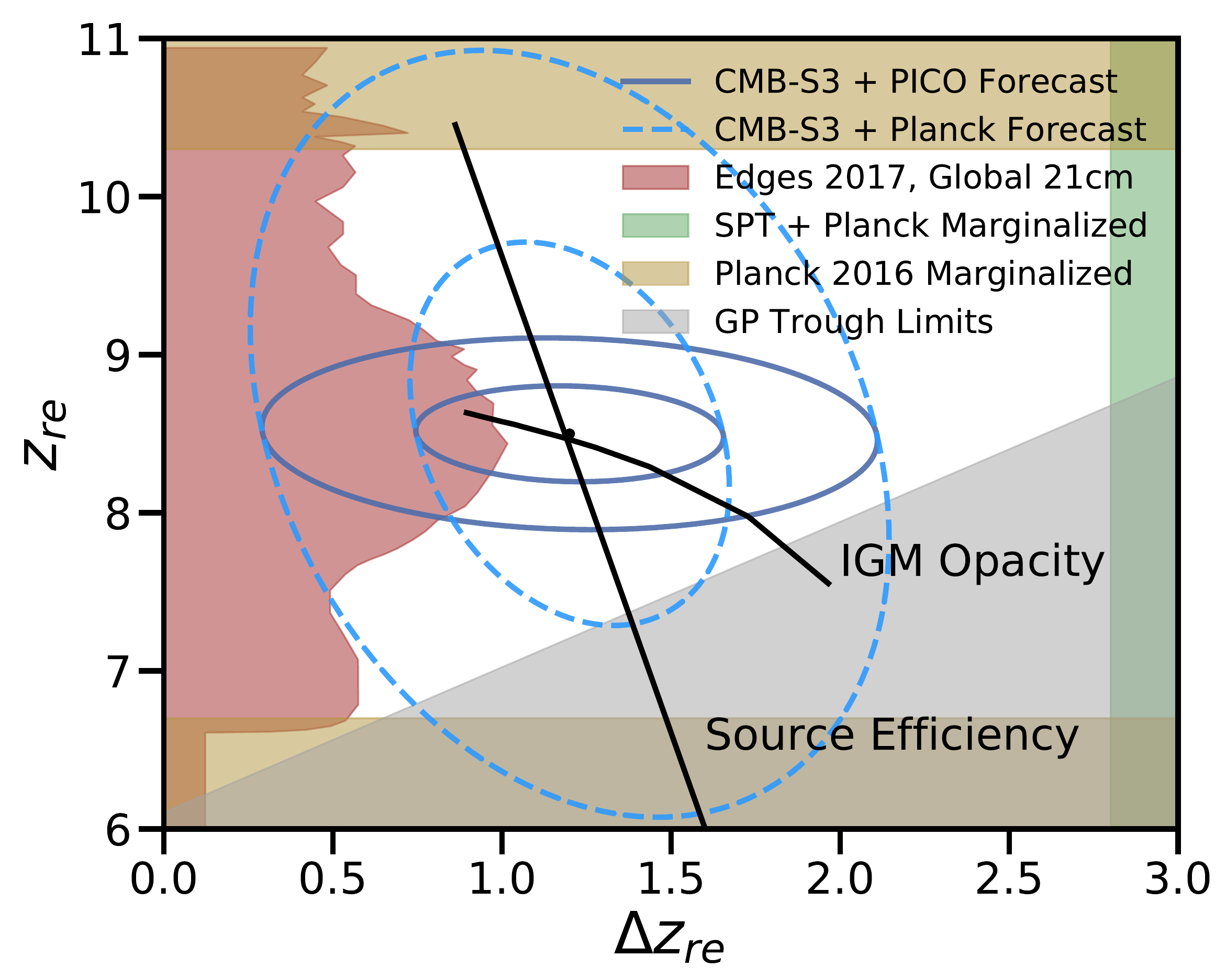} } }
\hspace{0.in}
\parbox{3.5in}{
\caption{\captiontext 
Contours of $1\sigma$ and $2\sigma$ constraints on the mean redshift and duration of reionization using PICO and CMB-S3 data (solid dark blue), and comparison with \planck\ and CMB-S3~(dash light blue). Source efficiency and IGM opacity (dark lines) are two physical parameters controlling the reionization process in current models. The PICO measurements, together with higher-resolution data of the kSZ effect, will significantly constrain the range of models allowed. Shaded regions show already existing constraints from EDGES, \planck\ + the South Pole Telescope,  \planck , and Gunn--Peterson (GP) trough measurements~\citep{Planck2018_VI,EDGES2017,Fan2006,Planck2016_reion}.  
\label{fig:ReionizationPICO} } }
\vspace{-0.15in}
\end{figure}

The reionization of the Universe imprints multiple signals in the temperature and polarization of the CMB.  In polarization, the most important signature is an enhancement in the $EE$ power spectrum at large angular scales $\ell \simlt 10$ (Fig.~\ref{fig:clbb}). This signal gives a direct measurement of the optical depth to the reionization epoch $\tau$ and thus to the mean redshift of reionization $z_{\rm re}$, with very little degeneracy with other cosmological parameters (Fig.~\ref{fig:ReionizationPICO}).\footnote{The mean redshift to reionization is defined to be the redshift when $50$\% of the cosmic volume was reionized.} \planck 's determination of the optical depth to reionization $\tau = 0.054 \pm 0.007\, (1\sigma) $ has indicated that reionization concluded by $z \sim 6$, but the measurement uncertainty leaves many unanswered questions including: were the ionizing sources primarily star-forming galaxies or more exotic sources such as supermassive black holes or annihilating dark matter? What was the mean free path of ionizing photons during this epoch?  What was the efficiency with which such photons were produced by ionizing sources?  
%What were the masses and environments of the dark matter halos that hosted the sources?  
Did the reionization epoch extend to $z \sim 15$--$20$, as has been claimed recently~\citep{Miranda2017}?
With ten independent maps of the entire sky, multiple frequency bands and ample sensitivity to remove foregrounds, PICO is uniquely suited to make the low-$\ell$ $EE$-spectrum measurements and reach cosmic-variance-limited precision with $\sigma(\tau)=0.002$, settling some of these questions and significantly constraining the others (SO5).

%With ten independent maps of the entire sky, multiple frequency bands and ample sensitivity to remove foregrounds, PICO is uniquely suited to make the low $\ell$ $EE$-spectrum measurements that are needed to elucidate the formation of the first luminous sources. It will reach the cosmic-variance-limited precision of $\sigma{\tau} = 0.002$. No such measurements have yet been demonstrated by sub-orbital instruments. 

Figure~\ref{fig:ReionizationPICO} presents forecasts for reionization constraints in the $z_{\rm re} - \Delta z_{\rm re}$ parameter space. These are obtained from PICO's measurement of $\tau$ in combination with S3 experiments' measurements of the ``patchy'' kinematic Sunyaev--Zeldovich (kSZ) effect, due to the peculiar velocities of free-electron bubbles around ionizing sources~\citep{Calabrese2014}.
The figure includes curves of constant efficiency of production of ionizing photons in the sources and of intergalactic-medium opacity, two parameters that quantify models of reionization. The curves shown are illustrative; families of models, that would be represented by parallel \lq source efficiency\rq~and \lq IGM Opacity\rq~lines, are allowed by current data. PICO's data will give simultaneous constraints on these physical parameters, yielding important information on the nature of the first luminous sources. For example, models in which the first sources are quasars rather than galaxies have significantly different IGM opacities and source efficiencies.

The process of reionization leaves specific non-Gaussian signatures in the CMB.  In particular, patchy reionization induces non-trivial 4-point functions in both temperature and polarization~\citep{SmithFerraro2017,DvorkinSmith2009}.  The temperature 4-point function can be used to separate reionization and late-time kSZ contributions.  Combinations of temperature and polarization data can be used to build quadratic estimators for reconstruction of the patchy $\tau$ field, analogous to CMB lensing reconstruction (\S\,\ref{sec:gravitationallensing}).  These estimators generally require high angular resolution, but also rely on foreground-cleaned CMB maps. Data from PICO's high-frequency bands -- which have better than 2~arcmin resolution and cover frequencies that are not optimal for observations from the ground -- will enable these estimators to be robustly applied to high-resolution ground-based CMB data, a strong example of ground-space complementarity.  %\comor{if pico is complementarity by {\it only} providing foreground maps at sufficiently high resolution, I think we should move this to the 'complementarity'. It is not a direct science goal or outcome.}
%% JCH: I checked the Smith-Ferraro 4-pt estimator, and PICO does not have sufficient resolution to do this (see Fig. 3 of https://arxiv.org/pdf/1803.07036.pdf )
%
%Thus, while PICO alone may not enable high \ac{SNR} reconstructions,

%With ten independent maps of the entire sky, multiple frequency bands and ample sensitivity to remove foregrounds, PICO is uniquely suited to make the low $\ell$ $EE$-spectrum measurements that are needed to elucidate the formation of the first luminous sources. No such measurements have yet been done from the ground. Of the currently operating and planned S3 experiments, one is targeting the lowest $\ell$ $EE$ spectrum~\citep{class_overview}. 

Decreasing the uncertainty on $\tau$ is important to break the degeneracy  between this parameter and the amplitude of the primordial power spectrum $A_{\rm s}$, a degeneracy in the CMB power spectra that hinders all cosmological observables of the growth of structure~(\S~\ref{sec:relics_neutrinos}). The degeneracy can only be broken through measurements of the low-$\ell$ $EE$ power spectrum. PICO's cosmic-variance-limited polarization measurements will thus improve constraints on the sum of neutrino masses, dark energy, and modified gravity coming from all low-$z$ growth measurements, including galaxy lensing, velocity-field measurements, redshift-space distortions, and galaxy surveys.

%%%%%%%%%%
\subsubsection{Probing the Evolution of Structures via Gravitational Lensing and Cluster Counts} 
\label{sec:gravitationallensing}

The particle content of the Universe, gravitational collapse, the effects of dark energy, and baryonic feedback processes that recycle energy determine the evolution of structures in the Universe. The amplitude of linear fluctuations as a function of redshift, parameterized by $\sigma_8(z)$, is thus a sensitive probe that embodies the effects of physical processes affecting growth. \ac{CMB} photons are affected by, and thus probe, $\sigma_{8}(z)$ as they traverse the entire Universe. PICO will tightly constrain $\sigma_8(z)$ through measurements of gravitational lensing and cluster counts. 

\begin{figure}[h]
\hspace{-0.2in}
\parbox{3.0in}{\centerline {
\includegraphics[width=2.2in]{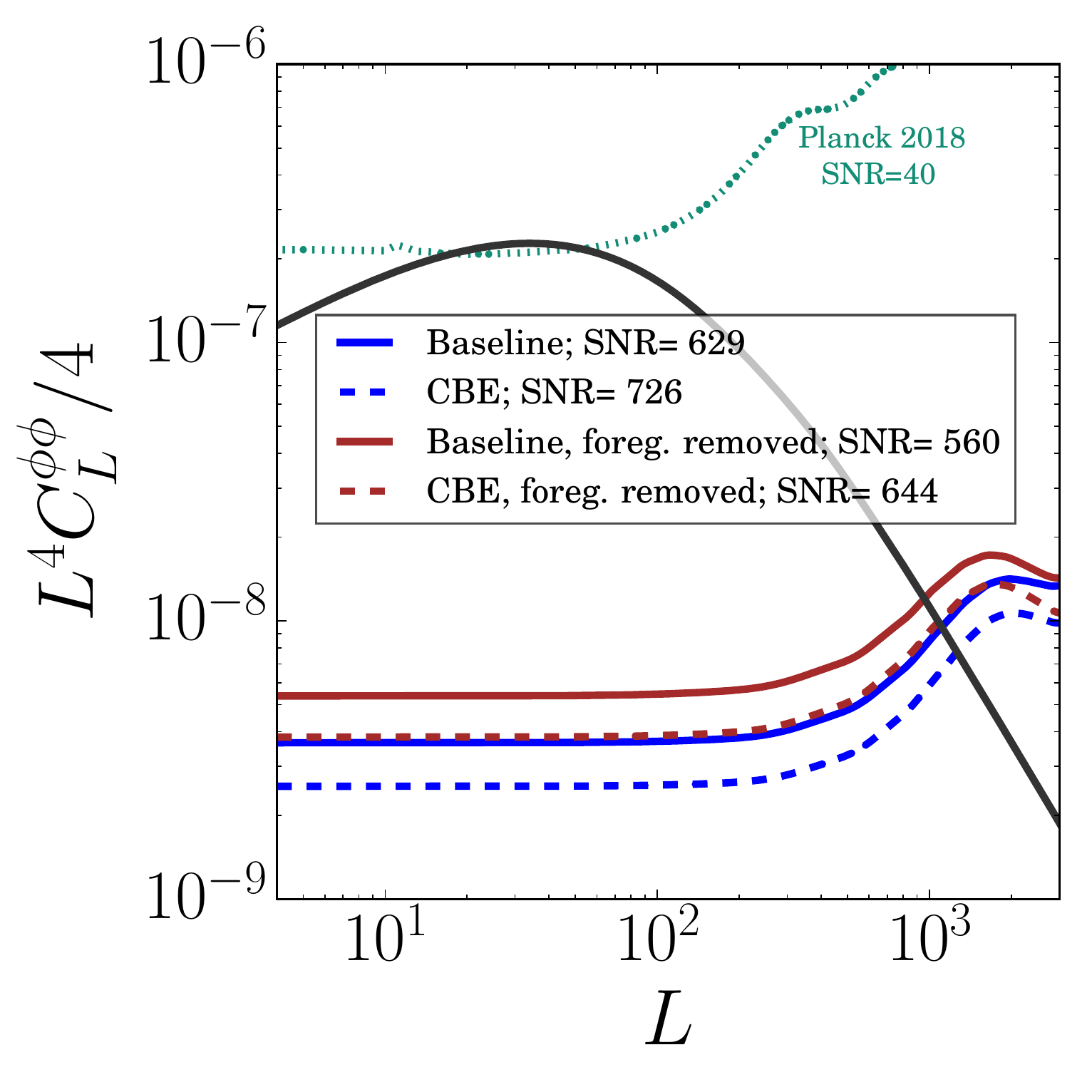} } }
\hspace{0.in}
\parbox{3.3in}{
\caption{\captiontext 
PICO will make a high \ac{SNR} full-sky map of the projected gravitational potential $\phi$ due to all matter between us and the last scattering surface at all angular scales $2 \leq L \lesssim 1000$~(Footnote~\ref{foot:L}) for which its noise (red and blue) is below the 
theoretically predicted power spectrum $C_{L}^{\phi \phi}$ (black). Noise predictions as a function of $L$ and anticipated \ac{SNR} values for the measurement of  $C_{L}^{\phi \phi}$ are given for the baseline (solid) and CBE (dashed) cases, both without (blue) and with (red) a process of foregrounds separation that degrades the \ac{SNR} by $\sim$10\,\%.  In all cases the \ac{SNR} is more than a factor 10 higher than currently available with \planck .
%The map of $\phi$ will be made using a map of $B$-mode arising from the gravitational lensing of \ac{CMB} photons by the large scale structure of the Universe. 
\label{fig:lensingNoisePICO} } }
\vspace{-0.1in}
\end{figure}

\noindent$\bullet$ {\bf Gravitational Lensing} \hspace{0.1in} \label{lensing} Matter between us and the last-scattering surface deflects the path of photons through gravitational lensing, imprinting the three-dimensional matter distribution across the volume of the Universe onto the CMB maps. The specific quantity being mapped by the data is the projected gravitational potential $\phi$ that is lensing the photons. From the lensing map, which receives contributions from all redshifts between us and the CMB, with the peak of the distribution at $z \sim 2$, we infer the angular power spectrum $C_{L}^{\phi \phi}$ (Fig.~\ref{fig:lensingNoisePICO}). Both the temperature and polarization maps of the CMB, and by extension the angular power spectra, are affected by lensing. 

\planck 's $\phi$ map had \ac{SNR} of $\sim$1 per $L$ mode over a narrow range of scales, $30 < L < 50$. PICO will make a true map, with \ac{SNR} $\gg1$ for each mode in the range $2 \leq L \lesssim 1000$.  While \planck\ had an \ac{SNR} of 40 integrated across the entire $C_{L}^{\phi \phi}$ power spectrum~\citep{2018arXiv180706210P}, PICO will give \ac{SNR} of 560 and 644 for the baseline and CBE configurations, respectively; both values already account for foreground separation (Fig.~\ref{fig:lensingNoisePICO}). 

PICO's $\phi$ map is a key ingredient in the delensing process that improves constraints on $r$~(\S~\ref{sec:inflation}) and in extracting neutrino mass constraints~(\S~\ref{sec:relics_neutrinos}). It will also be used to constrain the properties of quasars and other high-redshift astrophysical tracers of structure.  For example, cross-correlations with quasar samples from DESI will yield a precise determination of the quasar bias (and hence host halo mass) as a function of the quasar properties, such as (non-)obscuration.  Such studies are not possible with any other lensing techniques, due to their sensitivity to lower redshifts.

%\comor{Lensing will also be used to weigh dark matter halos hosting galaxies, quasars, and groups and clusters of galaxies.  In the context of quasar analyses described above, this halo mass information is in the ``one-halo'' term of the cross-correlation, while the bias information is in the ``two-halo'' term.  PICO will access both sets of information, allowing consistency tests and further tightening constraints.  Detailed forecasts for halo lensing are presented in the cluster subsection below.}

\noindent$\bullet$ {\bf \bm{$\sigma_{8}(z)$} from Gravitational Lensing} \hspace{0.1in} \label{sigma8_lensing}
Cross-correlations between the PICO lensing-potential map and wide-field samples of galaxies and quasars provide a powerful technique to measure the time dependence of the amplitude of matter fluctuations $\sigma_{8}(z)$ in tomographic redshift bins. This is achieved by overcoming the limitations of auto-correlations of these data sets.
The lensing $\phi$ map is sensitive to the projection of all matter back to the last-scattering surface, so it cannot resolve the time dependence of fluctuations, while galaxies and quasars trace matter in an unknown biased way so that the matter amplitude cannot be determined.
Cross-correlations of the two data sets, broken down to several tomographic redshift bins, will constrain how galaxies in each bin trace the dark matter, which will yield strong constraints on $\sigma_8(z)$ and thereby on structure growth and models of dark energy and modified gravity~\citep{2009PhRvL.102b1302S,2018PhRvD..97l3540S}.

\begin{figure}
\centering
\hspace{-0.15in}
\includegraphics[width=3in]{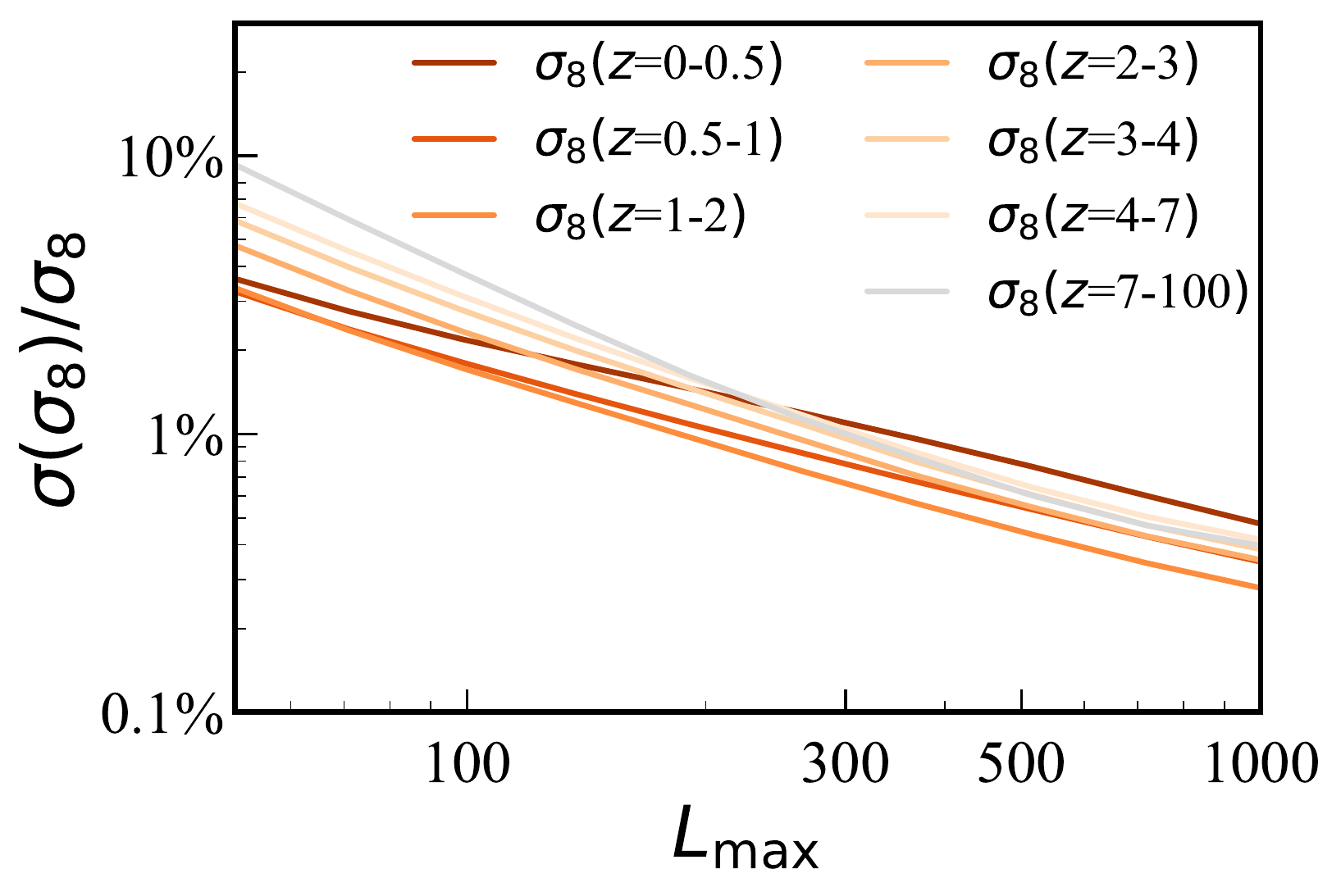}
\hspace{-0.1in}
\includegraphics[width=2.65in,trim= 0cm -0.25cm 0cm 0cm]{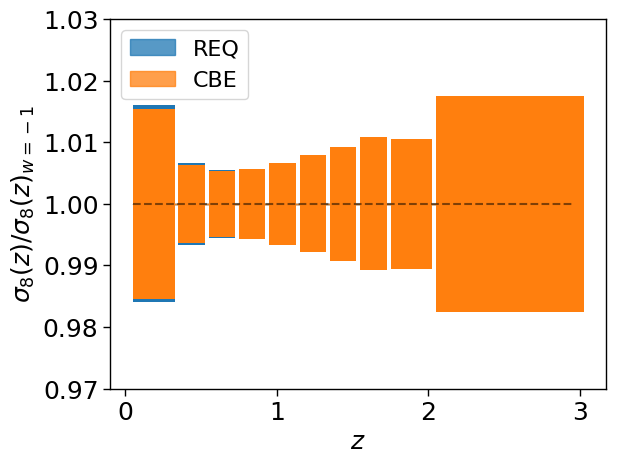}
\vspace{-0.14in}
\caption{\captiontext  
Sub-percent constraints on the evolution of $\sigma_{8}$ as a function of redshift will come from two independent PICO products: correlations between PICO's deep gravitational lensing map (Fig.~\ref{fig:lensingNoisePICO}) and LSST's gold sample of galaxies (left) and cluster counts (right). Fractional uncertainties in $\sigma_{8}$ relative to fiducial $\Lambda$CDM values are given as a function of the finest angular scale $L_{\rm max}$ of the correlation analysis for seven redshift bins (left).  The baseline and CBE configurations give essentially the same fractional errors of $\sigma_{8}(z)$ using cluster counts (right).  For LSST we assume: 10 years, 50\% sky fraction, 55 galaxies per ${\rm arcmin}^{2}$ at redshift $z<3$ with magnitude limit $i <25.3$~\citep{LSSTSciBook}, and dropout galaxies at $z>3$~\citep{dropouts} extrapolating recent Hyper Suprime-Cam observations~\cite{Schmittfull/Seljak,HSC1,HSC2}, with linear bias $b(z)=1+z$.
\label{fig:sigma8} }
\vspace{-0.16in}
\end{figure}

In the left panel of Fig.~\ref{fig:sigma8}  we show projected $1\sigma$ errors on $\sigma_8(z)$ when using cross-correlations with LSST's gold sample of galaxies~\citep{LSSTSciBook}.  Sub-percent accuracy is obtainable with PICO's resolution, which will give information extending to $L =1000$.\footnote{PICO's resolution is sufficient to give information for $L>1000$, but at these scales structures are non-linear and will not be used to constrain $\sigma_{8}(z)$.} This accuracy will be used to constrain dark energy or modified gravity, in the context of specific models, and to give a neutrino mass constraint that is independent from and competitive with that inferred from the CMB lensing auto-power spectrum (\S~\ref{neutrino_fundamental})~\citep{2018arXiv180902120Y} .

\noindent$\bullet$ {\bf Cluster Counts} \hspace{0.1in} \label{clusters}  
The distribution of galaxy clusters over redshift is one consequence of the evolution of structures and is thus a sensitive measure of $\sigma_{8}(z)$. The observational quantity of interest is $dN/(dz \,\, d m) $, the number of observed clusters per redshift and per mean mass interval, from which constraints on $\sigma_{8}(z)$ can be derived. Galaxy clusters found by PICO via the \ac{tSZ} effect (\S~\ref{sec:sz}) provide a catalog with a selection function that is simple to model and thus straightforward to use for cosmological inference. PICO's catalog will provide clusters with masses above $\sim2\times10^{14} M_\odot$ out to redshifts $z\sim3$. We forecast that PICO will find $\sim$150,000 galaxy clusters, assuming the cosmological parameters from \planck\  and using the 70\% of sky not obscured by the Milky Way.  Information provided by the high frequency bands will mitigate the potential reduction in detection efficiency due to dust emission by cluster members~\citep{melin_2018}, an advantage of space-based observations. Redshifts will be provided by future optical and infrared surveys, while cluster masses will be inferred by optical weak lensing for clusters with $z < 1.5$ and by PICO's own CMB halo lensing data at higher redshifts (see next paragraph).  This catalog will provide $\sigma_{8}$ with sub-percent precision for $0.5 < z < 2$ (Fig.~\ref{fig:sigma8}, right), and a neutrino mass constraint $\sigma(\sum m_{\nu}) = 14$~meV that is independent from the one coming from the CMB lensing measurements (SO3, \S~\ref{sec:relics_neutrinos}). A significant fraction of the PICO-detected clusters will also be detected by eROSITA, giving an exceptional catalog of multi-wavelength observations for detailed studies of cluster astrophysics.

\begin{figure}[h]
\vspace{-0.1in}
\hspace{-0.1in}
\parbox{3.1in}{\centerline {
\includegraphics[width=3.0in]{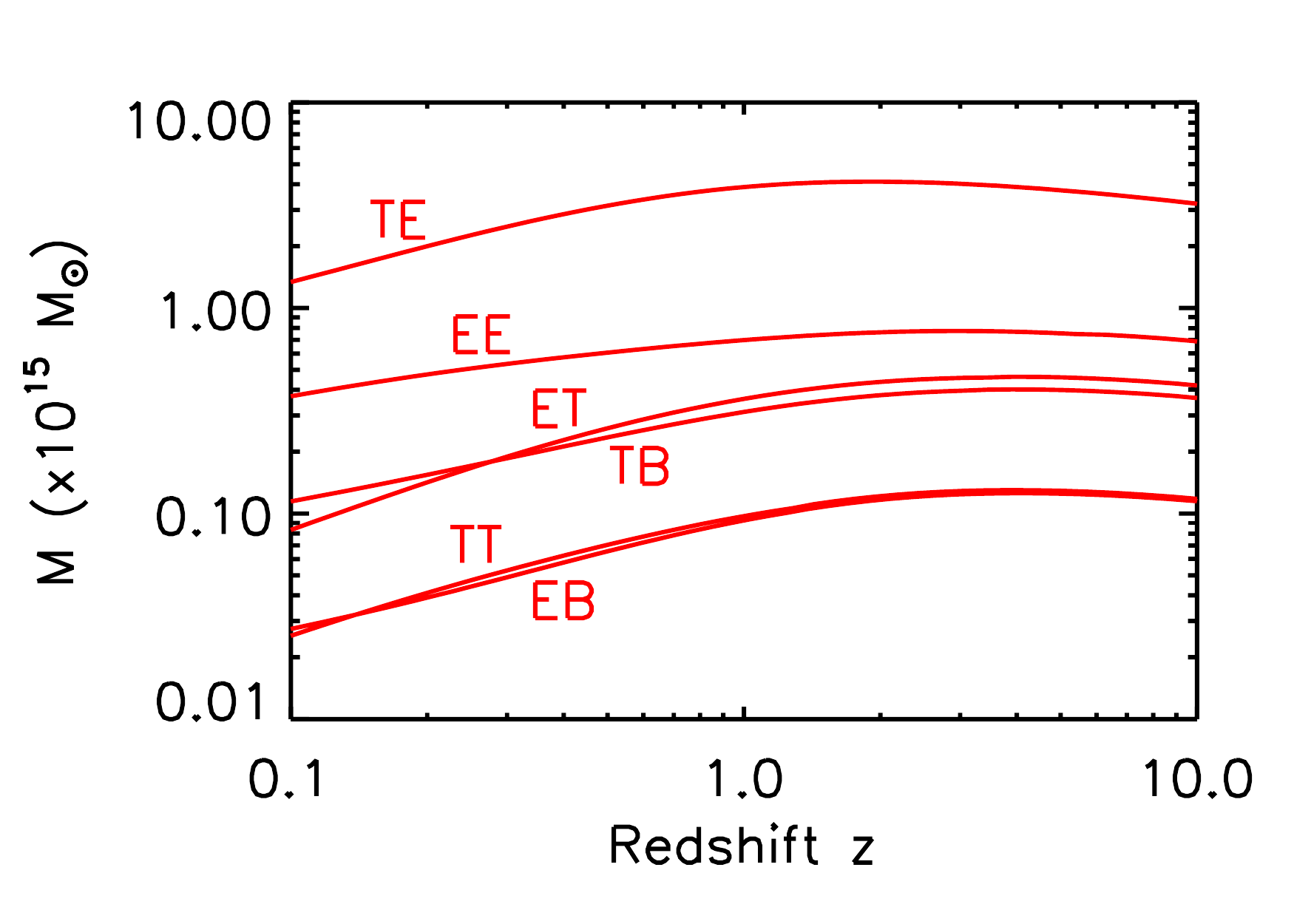} } }
\hspace{0.in}
\parbox{3.4in}{
\caption{\captiontext 
PICO will provide mass calibration for individual clusters and proto-clusters with mass as low as $10^{14}M_\odot$ at $z>2$ using `halo lensing'. Curves for different CMB signal correlations (red) give the $1\sigma$ sensitivity of an optimal mass filter~\citep{2015A&A...578A..21M} as a function of $z$.  The curves are flat at high redshift, demonstrating that the technique probes a broad range of redshifts. For PICO, the $EB$ and $TT$ estimators are equivalent, offering important cross-validation of measurements because the systematics are very different for temperature and polarization. 
\label{fig:HaloLensing} } }
\vspace{-0.2in}
\end{figure}

Calibrating the masses of clusters, i.e. determining $m(z)$, is the most uncertain step in inferring $\sigma_{8}$ and other cosmological parameters using cluster counts.  PICO will provide calibration using ``CMB halo lensing'', an approach that uses the small-scale effects of gravitational lensing due to dark matter halos around clusters and proto-clusters~\citep{2015ApJ...806..247B, 2015PhRvL.114o1302M, 2016A&A...594A..24P}. The technique is particularly effective for measuring halo masses out to high redshifts where gravitational lensing of background objects no longer works because there are no background sources. 
The approach is illustrated in Fig.~\ref{fig:HaloLensing}, which gives the $1\sigma$ uncertainty in a halo mass measurement as a function of the object's redshift. PICO will measure the mass of individual low-mass clusters ($\sim 10^{14}$\,M$_\odot$) over a wide redshift range, and by stacking will determine the mean mass of smaller halos, with masses of $\sim 10^{13}$\,M$_\odot$, which include those hosting individual galaxies. Because the vast majority of clusters have masses that are larger than $\sim 10^{14}$\,M$_\odot$, the PICO data will provide mass calibration for all objects of interest. The flattening at high redshift reflects the fact that the technique is sensitive over a broad range of redshifts. The high-frequency PICO data, for which the resolution matches that of ground-based instruments at lower frequencies, will play an essential role in cleaning foregrounds, particularly those derived from the temperature-based estimator, which is most contaminated by foregrounds. 

Beyond its role in calibrating masses for cluster counts, PICO's halo lensing measurements will also be a unique tool for measuring the relation between galaxies and their dark matter halos during the key epoch of cosmic star formation at $z\geq 2$, which is not reachable by other means.  This will provide valuable insight into the role of environment on galaxy formation during the rise to and fall from the peak of cosmic star formation at $z\sim 2$. 

%the mass sensitivity of PICO using a spatial filter optimized for extracting the mass of halos \citep{2015A&A...578A..21M}.  The curves give the $1\sigma$ noise in a mass measurement through the filter as a function of redshift. 

\subsubsection{Constraining Feedback Processes through the Sunyaev--Zeldovich Effect}
%Compton-$y$ map and tSZ auto-power spectrum} 
\label{sec:sz}
%\label{ymap}  

%\subsubsection{Constraining Structure Growth and Galaxy Formation via the Sunyaev-Zeldovich (SZ) Effects} 
% \label{sec:sz}

Not all CMB photons propagate through the Universe freely; about 6\% are Thomson-scattered by free electrons in the \ac{IGM} and \ac{ICM}.  These scattering events leave a measurable imprint on \ac{CMB} temperature fluctuations, which thereby contain a wealth of information about the growth of structures and the thermodynamic history of baryons. A fraction of these photons are responsible for the thermal and kinetic Sunyaev--Zeldovich effects (tSZ and kSZ)~\citep{zeldovich69,SZ1972}. The amplitudes of the tSZ and kSZ signals are proportional to the integrated electron pressure and momentum along the line of sight, respectively.  They thus contain information about the thermodynamic properties of the \ac{IGM} and \ac{ICM}, which are highly sensitive to astrophysical feedback. Feedback is the process of energy injection into the \ac{IGM} and \ac{ICM} from accreting supermassive black holes, supernovae, stellar winds, and other sources. Feedback processes are the most uncertain, yet crucial, ingredient in modern theories of galaxy formation; they are required in order to match observations of the stellar properties of galaxies, but the underlying details of the physical processes involved are still highly uncertain.

%%%%%%%%%%%%%%%%%%%%
%\subsubsection{Constraining Feedback Processes: Compton-$y$ map and tSZ auto-power spectrum} 
%\label{ymap}  

\begin{figure}[h]
\hspace{-0.1in}
\parbox{3.1in}{\centerline{
\includegraphics[width=3.0in]{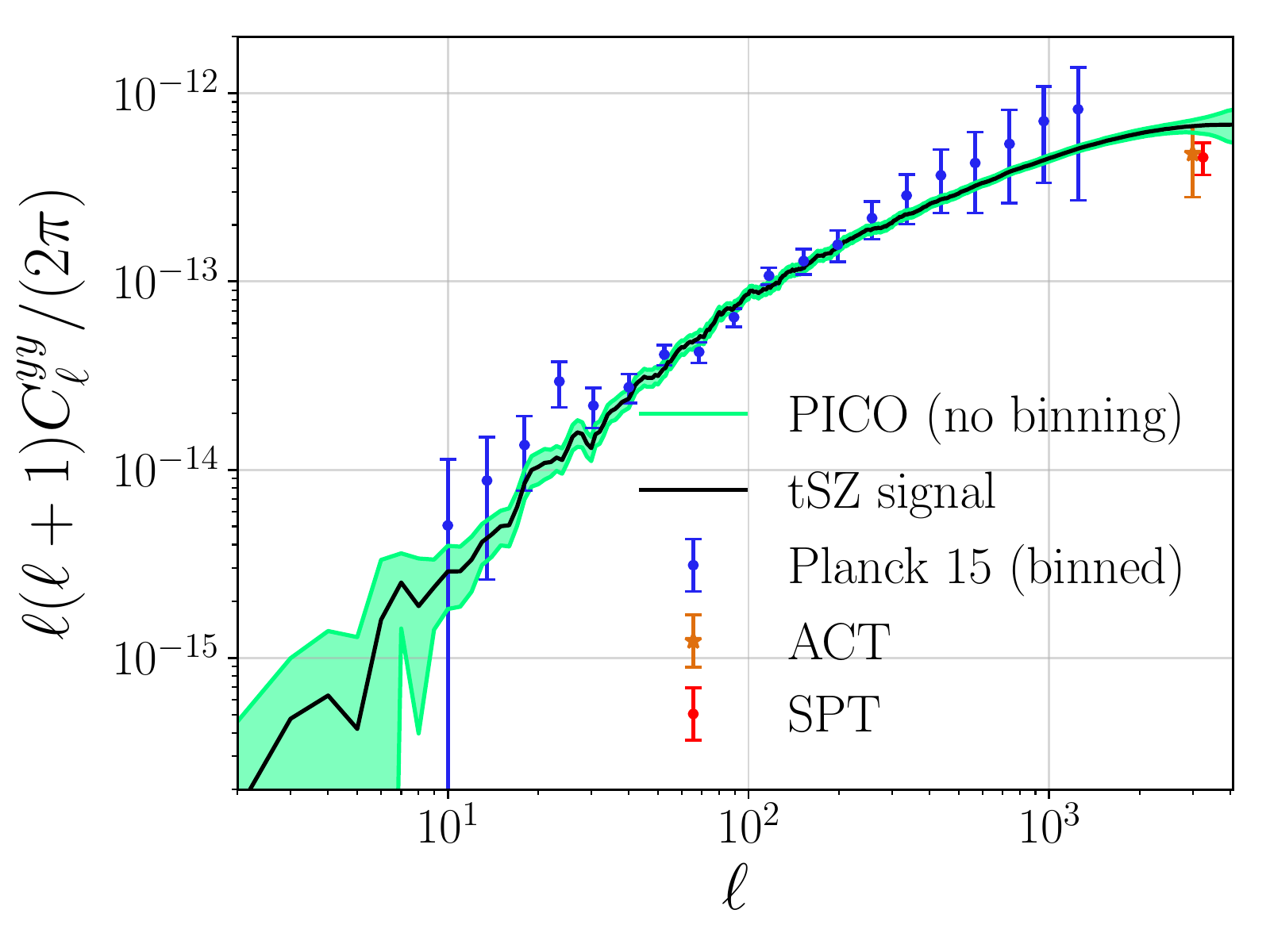} } }
\hspace{0.in}
\parbox{3.4in}{
\caption{\captiontext  
The PICO $y$-map will give a tSZ power spectrum with an \ac{SNR} of 1270 (green, $1\sigma$ per $\ell$ mode), which is nearly 100 times larger than from \planck\ (blue). Binning the data (not shown), as was done for \planck\, would further increase the \ac{SNR}.  We also show current measurements by the ground-based SPT and ACT~\citep{Sievers2013,George2015}. In these forecasts we reconstruct the Compton-$y$ field from maps that include Galactic foregrounds, CMB fluctuations, and PICO CBE noise using the needlet internal linear combination algorithm~\citep{Delabrouille2009}. The input maps use the \planck~sky model~\cite{delabrouille/etal:2013}.
%The black curve shows the simulated tSZ power spectrum signal.  The light green shaded region shows the error bars for PICO at each multipole, i.e., with no binning, as determined from NILC analysis of full-sky simulations.  The blue points show the current constraints from Planck, which have been averaged into broad multipole bins.  The orange and dark green points show the constraints from ACT and SPT, respectively, at a single multipole of $\ell=3000$.  The overall PICO $S/N = 1270$, nearly two orders of magnitude larger than current measurements.
\label{fig:PICO_tSZ_PS} } }
\vspace{-0.2in}
\end{figure}

Multifrequency \ac{CMB} data also allow the reconstruction of full-sky ``Compton-$y$ maps'' of the tSZ signal.  With low noise and broad frequency coverage, which is essential for separating out other signals, PICO will yield a definitive Compton-$y$ map over the full sky, with a total \ac{SNR} of 1270 for the CBE and $10$\% lower for the baseline configurations (Fig.~\ref{fig:PICO_tSZ_PS}). This is nearly two orders of magnitude higher \ac{SNR} than \planck , which already gave data with much higher \ac{SNR} than ground-based experiments. The tens of thousands of clusters forecast to be detected by PICO will be found in this $y$ map (\S~\ref{sec:gravitationallensing}).
%\comor{how does PICO compare to SO?.}

Strong constraints on models of astrophysical feedback will be obtained from the analysis of the PICO $y$-map, both from its auto-power spectrum and from cross-correlations with galaxy, group, cluster, and quasar samples. 
% Like the CMB-lensing map described above, the legacy value of the PICO $y$-map will be immense.  
As an example, we forecast the detection of cross-correlations between the PICO $y$-map and galaxy weak-lensing maps constructed from LSST and WFIRST data.  Considering the LSST gold weak-lensing sample, with a source density of 26 galaxies/arcmin${}^2$ covering 40\% of the sky, we forecast a detection of the tSZ--weak-lensing cross-correlation with \ac{SNR} = 3000.  Cross-correlations with the galaxies themselves will be measured at even higher \ac{SNR}.  At this immense significance, the signal can be broken down into dozens of tomographic redshift bins, precisely tracing the evolution of thermal pressure over cosmic time.  For PICO and WFIRST (assuming 45 galaxies/arcmin${}^2$ covering 5.3\% of the sky), we forecast \ac{SNR} = 1100 for the tSZ-weak lensing cross-correlation.  The WFIRST galaxy sample extends to higher redshift, and thus this high-\ac{SNR} measurement will allow the evolution of the thermal gas pressure to be probed to $z \approx 2$ (the peak of the cosmic star formation history) and beyond.  These measurements will revolutionize our understanding of galaxy formation and evolution by distinguishing between models of feedback energy injection at high significance.  Additional cross-correlations of the PICO $y$-map with quasar samples, filament catalogs, and other large-scale structure tracers will provide valuable information on baryonic physics that is complementary to inferences from the lensing cross-correlations described earlier.  
%Finally, beyond Compton-$y$, PICO's CMB halo lensing measurements will also be a unique tool for measuring the relation between galaxies and their dark matter halos during the key epochs of cosmic star formation at $z\geq 2$, not reachable by other means.  This will provide valuable insight into the role of environment on galaxy formation during the rise to and fall from the peak of cosmic star formation at $z\sim 2$. %\comor{is this already covered earlier in Marcel's lensing cross-correlations text?}

% ------------

\subsection{Testing $\Lambda$CDM} % (3 pgs)
\label{sec:testinglcdm}

The current cosmological model, as encoded by $\Lambda$CDM, provides a good fit to most current data. A host of cosmological observations including the CMB fit within the model that consists of only six parameters~\citep{Planck2018_I}. But the model is phenomenological and it leaves fundamental questions open. Premier among them is the unknown content of the majority of the Universe. Approximately 95\% of the Universe appears to be composed of dark matter and dark energy of unknown nature, both of which are necessary to explain observations at scales ranging from that of a galaxy to that of the Hubble volume. Yet, there are no detection of dark matter particles, and as for dark energy, it even lacks a compelling theoretical motivation.

In this context, tension between measurements of any $\Lambda$CDM parameter obtained by different probes compel additional stringent tests and investigation of alternatives to the prevailing paradigm. Examples of emerging tensions are: the $3.6\sigma$ discrepancy between the CMB- and local-Universe-anchored supernovae-based measurements of the Hubble constant~\citep{Aghanim:2018eyx,Riess2018}; the identification of lack of correlations at large angles in the $TT$ power spectrum that has an apparent probability of less than $10^{-3}$ of occurrence in standard $\Lambda$CDM~\citep{TT_correlations}; and the $\sim 2\sigma$ tension in measurements of the amplitude of late time perturbations $\sigma_{8}$ between the \planck\ CMB $TT$, $TE$, and $EE$ power spectra and those from cosmic shear surveys~\citep{Joudaki+2017, Abbott+2018, Hikage+2018, vanUitert+2018}.  A similar level  of tension for $\sigma_{8}$ ($\sim2\sigma$) arises when comparing \planck\ CMB spectra and cluster counts from \planck\ and other surveys~\citep{Planck, Bocquet+2018}. Such tensions, while perhaps only indicating the presence of systematic effects in the measurements, may in fact point toward new physics. One way to search for new physics is to better constrain the current measurements and the known extensions beyond the base six-parameter set. 

Given an experiment's baseline noise and angular resolution, and an input set of $N$ parameters, it is straightforward to calculate the uncertainty with which it will constrain the set~\citep{core_parameter}. A figure of merit (FOM) that quantifies the strength of the constraint is the volume of the uncertainty region in the $N$-dimensional parameter space. We use the same analytical approach and FOM that have also been used in other studies~\citep{core_parameter,Wang2008,pdg2018,Namikawa2010}.\footnote{The FOM is determined by the covariance of the Fisher information matrix, ${\rm FOM} = \left( \det \left[ \mbox{cov} ( p_i)  \right] \right)^{-1/2},\,\, i=1, ..., N$, where $p$ is the parameter set.} This FOM is defined such that a larger value linearly corresponds to {\it smaller} volume and thus to smaller parameter errors. 

\begin{figure}%[t,h]
\hspace{-0.2in}
\parbox{2.7in}{\centerline {
\includegraphics[width=3.0in]{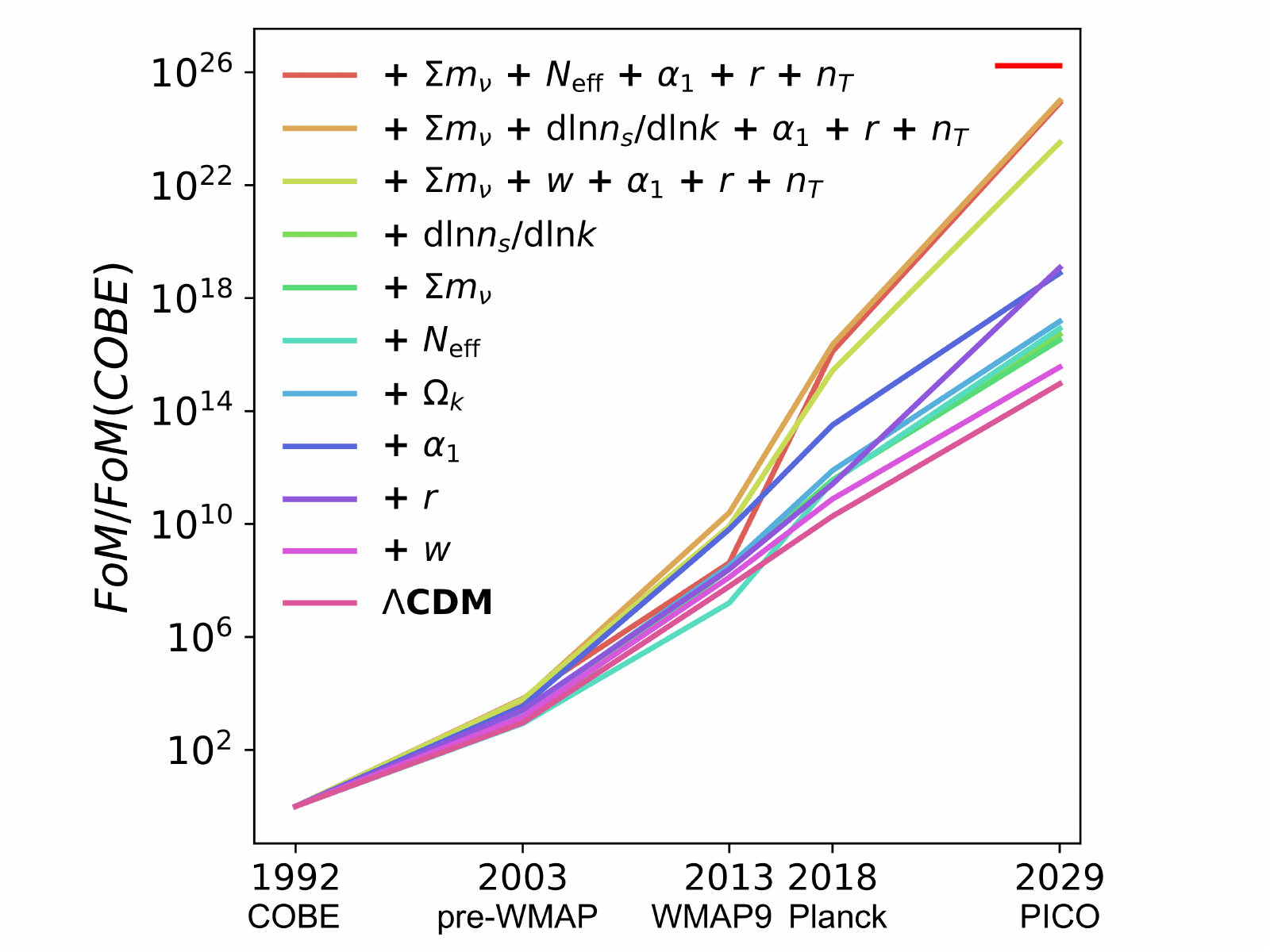} } }
\hspace{0.in}
\parbox{3.8in}{
\caption{\captiontext 
The increase in the FOM using data from CMB experiments since \cobe\ for the $\Lambda$CDM six-parameter model (dark purple) and when adding other cosmological parameters.  Increase in value represents increase in information content. PICO data will continue the average trend (blue line, $\Lambda$CDM + $\alpha_{1}$) of doubling the FOM every 10 months since 1992. For an 11-parameter set that includes $\Neff$ (red increasing line) PICO will improve the FOM by a factor of $0.5\times10^{9}$ relative to \planck , and will extract nearly the same information as that attainable by a mission with double the resolution and nine times lower baseline noise (top right red horizontal bar). The 11-parameter set includes: $w$-dark energy; $r$-the tensor to scalar ratio; $\alpha_{1}$-amplitude of correlated CDM isocurvature perturbations; $\Omega_{k}$-curvature; $\Neff$-effective number of light relics; $\sum m_{\nu}$-sum of neutrino masses; and $d\ln n_{s}/d\ln k$-running of the spectral index.   
\label{fig:fom} } }
\vspace{-0.13in}
\end{figure}

Fig.~\ref{fig:fom} shows the increase in the FOM since \cobe\ for the six-parameter $\Lambda$CDM model, as well as for additional cosmological parameters.\footnote{The six-parameter $\Lambda$CDM model includes: the baryon density; the dark matter density; the amplitude and spectral index of a power-law spectrum of initial perturbations; the angular scale of acoustic oscillations; and the optical depth to reionization.} The Figure only includes data from CMB experiments. The FOM for $\Lambda$CDM improved by a factor of 100 between {\it WMAP} and \planck , and will further improve by a factor of $10^{5}$ with PICO. For the 11-parameter set that includes $\Neff$ shown in the Figure PICO will improve upon \planck\ by a factor of $0.5\times10^{9}$. Having achieved this improvement, there would be only little information left to extract with this parameter set even by a mission with double the resolution and nearly ten times lower noise (Fig.~\ref{fig:fom}). Even stronger FOM improvements are obtained when a 12-parameter set is considered~\citep{picoweb_lcdm}, and when the PICO CMB data will be combined with data sets available in the next decade, including weak lensing, BAO, and cluster of galaxies. 

These improvements will test $\Lambda$CDM so stringently that it is hard to imagine it surviving such a scrutiny if it is not fundamentally correct. If tensions deepen to become discrepancies, it would be even more exciting if a new cosmological model emerged.

% ------------

\subsection{Galactic Structure and Star Formation} % (3 pgs)
\label{sec:galacticsci}

\begin{figure}[ht]
    \centering
    \includegraphics[width=6.5in]{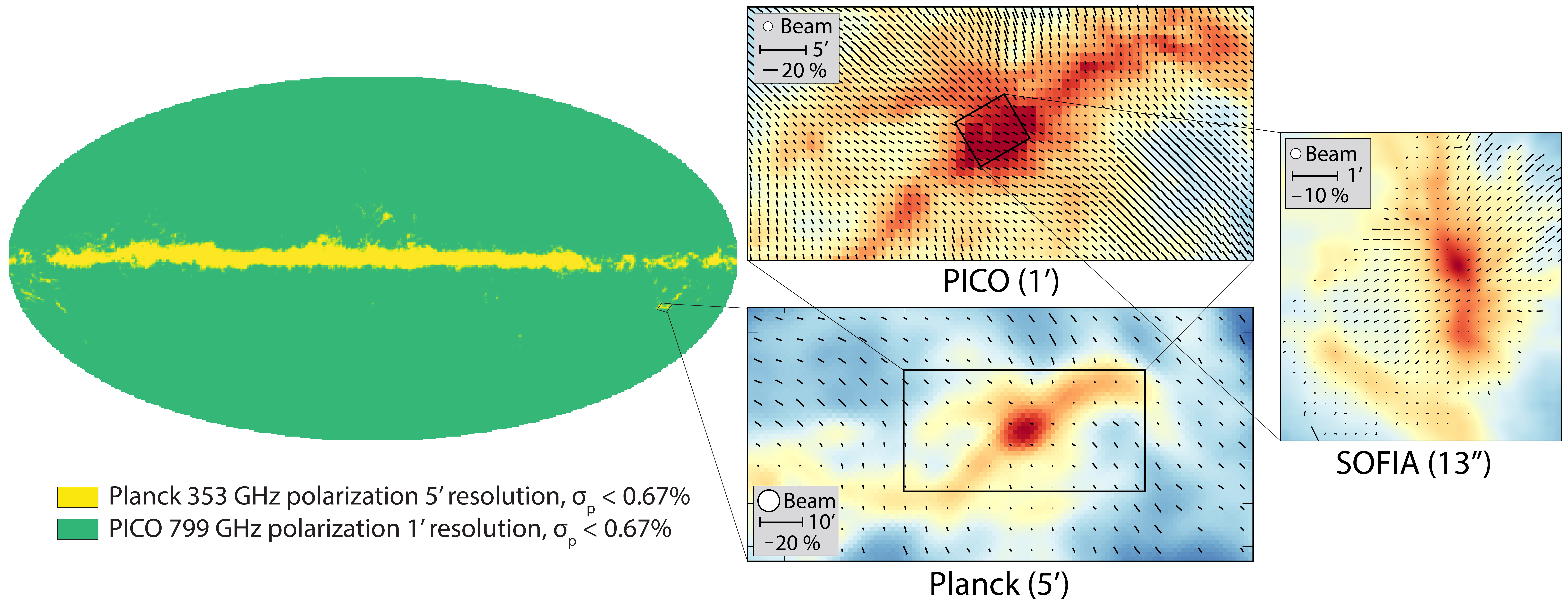}
\vspace{-0.25in}
    \caption{\captiontext  \planck 's 353~GHz polarization map gave a resolution of 5\arcmin~and sensitivity to polarization intensity of $\sigma_{p} < 0.67\%$ over a small portion of the sky (left, yellow).  At 799 GHz, the PICO baseline mission will give a polarization map of the {\it entire sky} and with 5 times higher resolution (left, green). In the middle panels, the \planck~map of the Orion region overlaid with vectors that are aligned with the inferred magnetic field (lower panel), and a simulated PICO observation (upper panel) illustrate the leap in information content (vector lengths are proportional to polarization fraction). With this map, and maps at other frequencies, PICO will characterize Galactic magnetized turbulence at scales spanning the diffuse ISM down to dense star-forming cores, which will be mapped with high-resolution polarimetry by instruments such as HAWC+/SOFIA~\citep{Chuss2018} (right panel) and ALMA~\citep{Bacciotti2018ApJ}.
    \label{fig:allsky} }
\vspace{-0.10in}
\end{figure}

\planck\ enabled an immense step forward in Galactic astrophysics~\citep{Planck2018:XII}. With seven full-sky polarization maps at frequencies between 30 and 353~GHz and a highest resolution of 5\arcmin, \planck\ provided entirely new and surprising data about the structure of the \ac{ISM}; the data have a lasting legacy for the foreseeable future. PICO will provide an even greater leap forward. It will produce 21 polarization maps of Galactic emission, and in the bands already probed by \planck\ they will be much deeper; for example, PICO's map at 321~GHz will be 105 times deeper than \planck's mean map depth at 353~GHz, and PICO's map at 30~GHz will be 17 times deeper than \planck's.  At 799~GHz PICO will have five times the resolution of \planck 's highest resolution map (Fig.~\ref{fig:allsky}). Such a data set can only be obtained from space. These data will complement a rich array of other polarization observations forthcoming in the next decade, including stellar polarization surveys to be combined with Gaia astrometry, and Faraday rotation measurements from  observations at radio wavelengths with the  Square Kilometer Array (SKA) and its precursors.

While the PICO data will likely provide many new insights and surprises, we focus here on two particularly important science objectives that are integral to NASA's science goal to explore how the Universe evolved; they relate to the structure and evolution of the Milky Way. These science objectives can only be achieved using the PICO dataset.

(1) {\em Test models of the composition of interstellar dust:} 
Less than $1\,\mu$m in size, dust grains are intermediate in the evolution from atoms and molecules to large solid bodies such as comets, asteroids, and planets. Encoded in the composition of dust are the pathways through which grains formed and grew. Dust grains also participate directly in interstellar chemistry, for example by catalyzing the formation of H$_2$ and organic molecules on their surfaces, in ways that depend upon their chemical makeup. Thus, the composition of dust grains is an essential aspect of the chemical evolution of interstellar matter, from the formation of complex molecules in space to the growth of planets. Through vastly improved spectral characterization of Galactic polarization, the PICO data will discriminate among models of Galactic dust composition to elucidate the chemical evolution of the Galaxy (SO6, \S~\ref{sec:test_composition_models}). The data will also guide the construction of methods for separating diffuse dust emission from cosmological signals of interest, particularly the inflationary signal. 
%\comor{can you please connect to 'galactic structure and evolution' (see example below)? also connect to foregrounds for B-mode if you think appropriate, or can leave this connection to later} \\

(2) {\em Determine how magnetic fields affect molecular cloud and star formation:}
Stars are formed through interactions between gravitational and magnetic fields, turbulence, and gas over more than four orders of magnitude of spatial scales, which span the diffuse ISM (kpc scale), molecular clouds (10~pc), and molecular cloud cores (0.1~pc). However, the role magnetic fields play in the large-scale structure of the diffuse \ac{ISM} and in the observed low star-formation efficiency has been elusive, owing to the dearth of data. By virtue of the strong dynamical coupling of dust and gas and the systematic alignment of dust grains with magnetic fields, PICO's dust polarization measurements will for the first time probe the large-scale Galactic magnetic field with enough resolution to trace the role of magnetic fields through the entirety of the star-formation process (SO7, \S~\ref{sec:magnetic_fields}).

\subsubsection{Test Models of the Composition of Interstellar Dust}
\label{sec:test_composition_models}

Strong extinction features at 9.7~$\mu$m and 18~$\mu$m indicate that much of interstellar dust is in the form of amorphous silicates, while features at 217.5~nm, 3.3~$\mu$m, and 3.4~$\mu$m attest to abundant hydrocarbons. It is unknown, however, whether the silicate and carbonaceous materials coexist on the same grains or whether grains of each composition grow through distinct, parallel pathways dictated by their surface chemistry. 
%If there are indeed multiple grain populations, this will induce additional challenges for modeling the emission from interstellar dust in both total intensity and polarization at levels relevant for B-mode science at all angular scales~\citep{Hensley2018}. 

%\comor{The last sentence implies that the issue of multiple grain populations is only important for CMB B-mode science. Is it? Is it not important on its own right, because it is important to know what dust is made of? This is the missing connection to the broader picture: why do we care about dust?}

Some data suggest that the populations are distinct. Spectropolarimetry of dust extinction reveals robust polarization in the 9.7\,$\mu$m silicate feature~\citep{Smith2000}, indicating that the silicate grains are aligned with the local magnetic field. In contrast, searches for polarization in the 3.4\,$\mu$m carbonaceous feature have yielded only upper limits, even along sightlines where silicate polarization is observed~\citep{Chiar2006,Mason2007}. These data are consistent with silicate and carbonaceous materials existing on separate grains with different alignment properties. 

%\comblue{However, the 3.4~$\mu$m feature arises only from aliphatic (chain-like) carbonaceous material, which may not trace all carbon-bearing grains. Additionally, available data are restricted to only a few highly-extincted sightlines that may not typify the entire diffuse \ac{ISM}.}

At odds with the spectropolarimetric evidence from dust extinction are current measurements of the polarization fraction of the far-infrared dust emission with \planck~\citep{Planck_Int_XXII} and BLASTPol~\citep{Ashton2018}. They show little to no frequency dependence, whereas substantial frequency dependence would be expected if two components with distinct polarization properties were contributing to the total emission. 
%\comblue{However, current uncertainties are relatively large and the BLASTPol data are from only high density sightlines that may not be representative of the diffuse \ac{ISM}. }

With excellent polarization sensitivity, even in diffuse regions, PICO will provide a definitive test of the two-component paradigm \citep{Meisner2015}. 
%\comblue{To assess PICO's ability to discriminate models quantitatively, we employed the analytic two component dust model of Meisner et al.~\cite{Meisner2015}, which provided a better fit to IRAS and \planck\ data than one component models. We ran 1000 simulations with different combinations of polarization fractions of the two components. We used PICO baseline noise levels, frequency bands at and above 107~GHz, and binned the simulated data in 7.9$^\prime$ pixels, matching the beam of the PICO 107~GHz band.} 
In this case, the PICO baseline mission will determine the intrinsic polarization fractions of each of the two components to a precision of 3\%. With this level of precision the data will validate or reject state-of-the-art dust models~\citep[e.g.][]{Draine2009,Guillet2018}, test for the presence of additional grain species with distinct polarization signatures, such as magnetic nanoparticles~\citep{Draine2013}, and will be used as a crucial input for the foreground separation necessary to extract cosmological $E$- and $B$-mode science. 

Anomalous microwave emission (AME) is a component of Galactic emission peaking in the 20--30~GHz range that has been tentatively identified with small, rapidly-spinning dust grains~\citep{dickinson/etal:2018}. As only upper limits have been placed on its polarization, its role as a foreground for cosmological $B$-mode science remains unclear; even small levels of polarization could prove challenging for determining $r$~(\S~\ref{sec:signal_separation}). PICO will finely sample the AME SED with its bands at 21, 25, 30, 36, and 43 GHz. Combined with ground-based maps at lower frequencies, for example C-BASS at 5~GHz~\citep{Dickinson2018a}, PICO will be used to efficiently separate the AME from synchrotron and free-free emission and either detect or place stringent upper limits on its polarization. Further, the enhanced frequency coverage will allow changes in the AME SED with interstellar environment to be characterized and thus elucidate its underlying physics.

\subsubsection{Determine How Magnetic Fields Affect Molecular Cloud and Star Formation}
\label{sec:magnetic_fields}

Stars form out of dense, gravitationally unstable regions within molecular gas clouds, which themselves form through the flow of diffuse, atomic-phase gas to denser regions. Magnetic fields play an important role throughout this process. 

On the largest scales, magnetized turbulence mediates the flow of the gaseous \ac{ISM} from the atomic to the denser, molecular phase. Recent observations suggest that the structure of the diffuse medium is highly anisotropic, and strongly coupled to the local magnetic field~\citep{Clark:2014, Clark:2015, Kalberla:2016, KalberlaKerp:2016}. As molecular gas clouds collapse to form stars, magnetic fields can slow the process of star formation by inhibiting movement of gas in the direction perpendicular to the field lines. Observations to date suggest that the outer envelopes of clouds can be supported against gravity by magnetic fields and turbulence, but in dense cores gravity tends to dominate, and so these dense structures can collapse to form stars \citep{Crutcher2010}.  The degree to which magnetic fields affect the formation of molecular clouds, as well as stars within these clouds, is poorly constrained, in large part due to the difficulty of making detailed maps of magnetic fields in the ISM.

%Stars form out of dense, gravitationally unstable regions within molecular gas clouds. The efficiency of this conversion from molecular gas to stars is very low, due to regulation from supersonic turbulent gas motions, magnetic fields, and feedback from young stars \citep{McKee2007}.  Magnetic fields may play an important role in slowing the process of star formation by inhibiting movement of gas in the direction perpendicular to the field lines.  Observations to date suggest that the outer envelopes of clouds can be supported against gravity by magnetic fields and turbulence, but in dense cores gravity tends to dominate, and so these dense structures can collapse to form stars \citep{Crutcher2010}.

%On larger scales, the formation of gravitationally unstable clouds is regulated by the flow of diffuse material into the molecular phase\comor{(?)}, a process that is mediated by magnetized turbulence in the low-density \ac{ISM}. Structure formation in the diffuse \ac{ISM} is poorly understood, but as a precursor to star formation it is crucial to understand what drives molecular cloud formation. Recent observations suggest that the structure of the diffuse medium is highly anisotropic, and strongly coupled to the local magnetic field~\citep{Clark:2014, Clark:2015, Kalberla:2016, KalberlaKerp:2016}. However, the degree to which magnetic fields affect the formation of molecular clouds, as well as stars within these clouds, is poorly constrained, in large part due to the difficulty of making detailed maps of magnetic fields in the interstellar medium.

\noindent$\bullet$ {\bf Formation of Magnetized Molecular Clouds from the Diffuse Interstellar Medium} \hspace{0.1in}
A comprehensive understanding of the magnetized diffuse \ac{ISM} is challenging because of its diverse composition, its sheer expanse, and the multi-scale nature of the physics that shapes it. To understand how matter and energy are exchanged between the diffuse and dense media, it is essential to measure the properties of the magnetic field over more than four orders of magnitude in column density. PICO is unique in its ability to provide the necessary data. \textit{Planck} achieved measurements of the diffuse sky at 60\arcmin\ resolution, resulting in $\sim$30,000 independent measurements of the magnetic field direction.  With 1.1\arcmin~resolution PICO will expand the number of independent polarization measurements to 86,000,000~(Fig.~\ref{fig:allsky}). The data will thus robustly characterize turbulent properties like the Alfv\'{e}n Mach number, $\mathcal{M}_A$, across a previously unexplored regime of parameter space. 

PICO's observations will complement recently completed high-dynamic-range neutral hydrogen surveys, such as \HI4PI \citep{HI4PI:2016} and GALFA-\hi \citep{Peek:2018}, as well as planned surveys of interstellar gas, most prominently with the SKA and its pathfinders. One of the open questions in diffuse structure formation is how gas flows within and between phases of the \ac{ISM}. A planned all-sky absorption line survey with the forthcoming SKA-1 will increase the number of measurements of the \ac{ISM} gas temperature by several orders of magnitude~\citep{McClure-Griffiths2015}. Quantitative comparisons of the \ac{ISM} temperature distribution from SKA-1 and estimates of the magnetic field strength and coherence length scale from PICO will elucidate the role of magnetized turbulence in the flow of matter in the \ac{ISM} from diffuse regions to regions of denser molecular gas.

\noindent$\bullet$ {\bf Formation of Stars within Magnetized Molecular Clouds} \hspace{0.1in}
The role of the magnetic field in star formation is quantified by the ratio of the energies stored in magnetic and gravitational fields, and the ratio of the energy stored in the magnetic field to that stored in gas turbulence. The first ratio is parameterized through a mass-to-flux ratio $\mu$, and the second through $\mathcal{M}_A$. 

With full-sky coverage and a resolution of 1.1\arcmin, PICO will map all the molecular clouds out to a distance of 3.4\,kpc with better than 1\,pc resolution.  Extrapolating from the Bolocam Galactic Plane Survey \citep[BGPS,][]{EllsworthBowers2015}, PICO is expected to make highly detailed magnetic field maps of over 2,000 molecular clouds, with $10^3$--$10^5$  independent polarization measurements per cloud. These are the {\it only foreseeable} measurements that will give $\mu$ and $\mathcal{M}_A$ over a statistically significant sample of molecular clouds. \planck , for example, mapped only ten nearby clouds to a similar level of detail~\citep{Planck:XXXV}. A large sample of clouds is crucial because: (1)~dust polarization observations are sensitive only to the magnetic field projected on the plane of the sky, and therefore polarization maps will look very different for molecular clouds observed at different viewing angles; and (2)~the relative importance of the magnetic field will likely be a function of cloud age and mass. By observing thousands of molecular clouds PICO will determine $\mu$ and  $\mathcal{M}_A$ for different sub-classes of cloud age and mass. 

\subsubsection{Galactic Legacy Science}
\label{sec:galactic_legacy}

PICO will also produce legacy datasets that will revolutionize our understanding of how magnetic fields influence physical processes ranging from planet formation to galaxy evolution.  For ten clouds closer than 500 pc, PICO will resolve magnetic fields on scales of 0.1~pc. This is the scale of dense cores and filaments for these clouds, and thus the observations will constrain how magnetic fields on these scales influence the formation of cloud cores.  Currently no experiment has the sensitivity and resolution to observe both the the large-scale (few parsec) and core-scale magnetic fields. By comparing the orientation of the core-scale magnetic fields with the orientation and sizes of proto-planetary disks, PICO will probe whether magnetic braking influences the growth of such disks~\citep{allen_2003,li_2014} and provide complementarity to higher angular resolution instruments such as ALMA and SOFIA~\citep{Bacciotti2018ApJ,Harper2018}~(Fig.~\ref{fig:allsky}).
%\comor{isn't ALMA or other instruments better suited for this?}

Key processes in the diffuse ISM, including heat transport, streaming of cosmic rays, and magnetic reconnection depend strongly on the level of the environment's magnetization~\citep{Lazarian:2006,Lazarian:2016,Lazarian_Vishniac:1999}.
PICO will give information about these processes with tens of millions of independent measurements of magnetic field orientation over the entire Galaxy. The measurements will also enable studies of the physical processes that generate magnetic fields through a combination of turbulence and large-scale gas motions~\citep{Xu_2018}.

Finally, PICO observations will create detailed magnetic field maps of about 70 nearby galaxies, with 100 or more measurements of magnetic field directions per galaxy. Currently, polarized dust emission has only been observed in M82 and NGC 253 using SOFIA~\citep{Jonesetal}. The PICO observations will determine how interaction between large-scale magnetic fields, turbulence, and feedback from previous generations of star formation affect galaxy evolution and star-formation efficiency.

% ------------

\subsection{Legacy Surveys} % (2 pgs)
\label{sec:legacy}

\definecolor{mygray}{gray}{0.6}

PICO was designed to respond to requirements posed by the seven \ac{SOs} listed in Table~\ref{tab:STM}. It will also generate a rich catalog of hundreds of thousands of new sources, consisting of proto-clusters, strongly lensed galaxies, and polarized radio and dusty galaxies. An abundance of information about galaxy and cluster evolution, dark matter, the physics of jets of active galactic nuclei, and magnetic fields of dusty galaxies will be stored in this catalog (Table~\ref{tab:STM2}). The catalog will be mined in future years through subsequent analysis and follow-up observations. 
 \begin{table}[h]
\caption{\textbf{Legacy Surveys } }\label{tab:STM2}
\footnotesize
\vspace{-0.1in}
\begin{tabular*}{\textwidth}{@{}l@{\extracolsep{\fill}}ll@{}}
\noalign{\vskip 2mm}
\hline
\noalign{\vskip 1mm}
\hline
\noalign{\vskip 2mm}    
% Header line 1
{\bf \hfil Catalog\hfil}&
{\bf \hfil Impact\hfil}&
{\bf \hfil Science\hfil}\\
% Line 1
\noalign{\vskip 2mm}    
\hline
\noalign{\vskip 1mm}    

\parbox[t]{0.8in}{Strongly\\ lensed galaxies}&
\parbox[t]{2.55in}{Discover 4500$^a$ strongly lensed and highly magnified dusty galaxies across redshift. 
\vspace{1mm}
{\color{mygray}\hrule}
\vspace{1mm}
Current knowledge: 13 sources confirmed in \planck\ data; a few hundred candidates in \textit{Herschel}, SPT and ACT data.}&
\parbox[t]{2.7in}{Gain information about the physics governing early, $z\simeq5$, galaxy evolution, taking advantage of magnification and extra resolution enabled by gravitational lensing;  learn about dark matter sub-structure in the lensing galaxies.}\\
% Line 3
\noalign{\vskip 1mm}    
\cline{1-3}
\noalign{\vskip 1mm}    

\parbox[t]{0.8in}{Proto-clusters}&
\parbox[t]{2.55in}{Discover 50,000$^{a}$ mm/sub-mm proto-clusters distributed over the sky out to $z\sim4.5$.  
\vspace{1mm}
{\color{mygray}\hrule}
\vspace{1mm}
Current knowledge: \planck\ + ACT/SPT data expected to yield a few tens.}&
\parbox[t]{2.7in}{Probe the earliest phases of cluster evolution, well beyond the reach of other instruments; test the formation history of the most massive virialized halos; investigate galaxy evolution in dense environments.}\\
% Line 2
\noalign{\vskip 1mm}    
\hline
\noalign{\vskip 1mm}    

\parbox[t]{0.8in}{Nearby galaxies}&
\parbox[t]{2.55in}{Detect 30,000 galaxies at $z\simlt 0.1$ at frequencies above 300~GHz.  
\vspace{1mm}
{\color{mygray}\hrule}
\vspace{1mm}
Current knowledge: 3400 (280) source candidates in the \planck\ 857 (353)~GHz  band. }&
\parbox[t]{2.7in}{Using frequencies that match cold ($15-25$~K) dust emission, give its spectral energy distribution as a function of galaxy properties to enable correlations with star-formation activity.} \\
\noalign{\vskip 1mm}
\hline 
\noalign{\vskip 1mm}

\parbox[t]{0.8in}{Polarized point\\ sources}&
\parbox[t]{2.55in}{Detect 2000$^{b}$ radio and several thousand dusty galaxies in polarization. 
\vspace{1mm}
{\color{mygray}\hrule}
\vspace{1mm}
Current knowledge:  about 200 radio sources up to 100~GHz; one polarization measurement of a dusty galaxy. }&
\parbox[t]{2.7in}{Study the physics of jets of extragalactic sources, close to their active nuclei; determine the large-scale structure of magnetic fields in dusty galaxies; determine the importance of polarized sources as a foreground for CMB polarization science.}\\
\noalign{\vskip 1mm}
\hline
\noalign{\vskip 1mm}

\parbox[t]{0.8in}{Cosmic infrared \\ background}&
\parbox[t]{2.55in}{Provide eight maps of the anisotropy from dusty star-forming galaxies for frequencies $\nu>200$~GHz, and with 1\arcmin\ resolution at 800~GHz.
\vspace{1mm}
{\color{mygray}\hrule}
\vspace{1mm}
Current knowledge:  Three \planck\ (higher noise) maps between 300 and 900~GHz with 5\arcmin\,~resolution. }&
\parbox[t]{2.7in}{Improve constraints on the parameters describing universal star-formation history. Construct a tracer of large-scale structure for CMB de-lensing. Cross-correlate with galaxy surveys and CMB lensing map.}\\
\noalign{\vskip 1mm}
\hline
\noalign{\vskip 1mm}

\end{tabular*}
{\footnotesize
$^a$ Confusion (not noise) limited\qquad
$^b$ Noise and confusion limited }
\end{table}

%Here we focus on two specific science deliverables that are enriched by PICO's unique capabilities. %catalog. \comor{Explain Why?}

%\comor{Kathy Romer says: Table 2 - I like the way this is broken down with ?Current knowledge? summaries. But it only talks about Planck. What about ACT, SPT, SO, balloons etc.? Gianfranco: SPT and ACT are already mentioned in connection to strongly lensed galaxies. So far there is not much from them, in the published literature, on source polarization. On proto-clusters there is the recent discovery of one at z=4.3 (Miller et al. 2018). Perhaps we could add that. }

\subsubsection{Early Phases of Galaxy Evolution}

%\comor{Kathy Romer: Section 2.3.1 - page 22 - I noted ?why do we care about lensed high-z galaxies? in the margin. So maybe you need to stress the motivation for this section more? Gianfranco: It is already said that strong lensing provides a unique possibility to look into the structure and kinematics of high-z dusty star-forming galaxies, i.e. to get crucial information on how they form and evolve. I don?t know what to say more.}

PICO's catalog of high-$z$ strongly-lensed galaxies will provide answers to major open issues in galaxy formation and evolution. What are the main physical mechanisms shaping the properties of galaxies~\citep{SilkMamon2012, SomervilleDave2015}: in situ processes, interactions, mergers, or cold flows  from the intergalactic medium? And how do feedback processes work? To settle these issues we need direct information on the structure and dynamics of high-$z$ galaxies. But these are compact, with typical sizes of 1--2~kpc~\cite{Fujimoto2018}), corresponding to angular sizes of 0.1--$0.2''$ at $z\simeq 2$--3. Thus they are hardly resolved, even by ALMA or by HST. If they {\it are} resolved, high enough \ac{SNR}s per resolution element are only achieved for the brightest galaxies, which are probably not representative of the general population.

Strong gravitational lensing provides a solution to these problems. Since lensing conserves the surface brightness, the effective angular size is stretched on average by a factor of $\mu^{1/2}$, where $\mu$ is the gravitational magnification, thus substantially increasing the resolving power. A spectacular example is ALMA observations of the \planck-discovered, strongly lensed galaxy PLCK\_G244.8\-+54.9 at $z \simeq 3.0$  with $\mu \simeq 30$~\citep{Canameras2017ALMA}. ALMA observations with a $0.1''$ resolution reached an astounding spatial resolution of 60~pc, substantially smaller than the size of Milky Way giant molecular clouds. CO spectroscopy of this object, measuring the kinematics of the molecular gas, gave an uncertainty of 40--50~km\,s$^{-1}$. Such precision allows a high \ac{SNR} detection of the predicted $\sim$1000~km\,s$^{-1}$ outflows capable of sweeping the galaxy clear of gas that would otherwise be available for star formation~\citep{KingPounds2015}. In this specific case, there were no clear indications that mergers or cold flows shaped the galaxy, but similar spectroscopy of another strongly lensed galaxy at $z=5.3$ detected a fast (800 km\,s$^{-1}$) molecular outflow due to feedback~\citep{spilker2018}.

%The outflow carries mass at a rate close to the star-formation rate, and can thus remove a large fraction of the gas that would otherwise be available for star formation.  
% Ca\~{n}ameras et al.~\citep{Canameras2017ALMA} have obtained CO spectroscopy for this object, measuring the kinematics of the molecular gas with an uncertainty of 40--50~km/s. This spectral resolution makes possible a direct investigation of massive outflows driven by AGN feedback at high $z$. Using this technique \citet{Spilker2018} detected a fast (800 km/s) molecular outflow due to feedback in a strongly lensed galaxy at $z=5.3$. They found that the outflow carries mass at a rate close to the star formation rate, and can thus remove a large fraction of the gas that would otherwise be available for star formation.

\begin{figure*}[t]
%\vskip-3cm
\begin{center}
\includegraphics[width=0.416\columnwidth, trim={0 0 0 0cm}, clip]{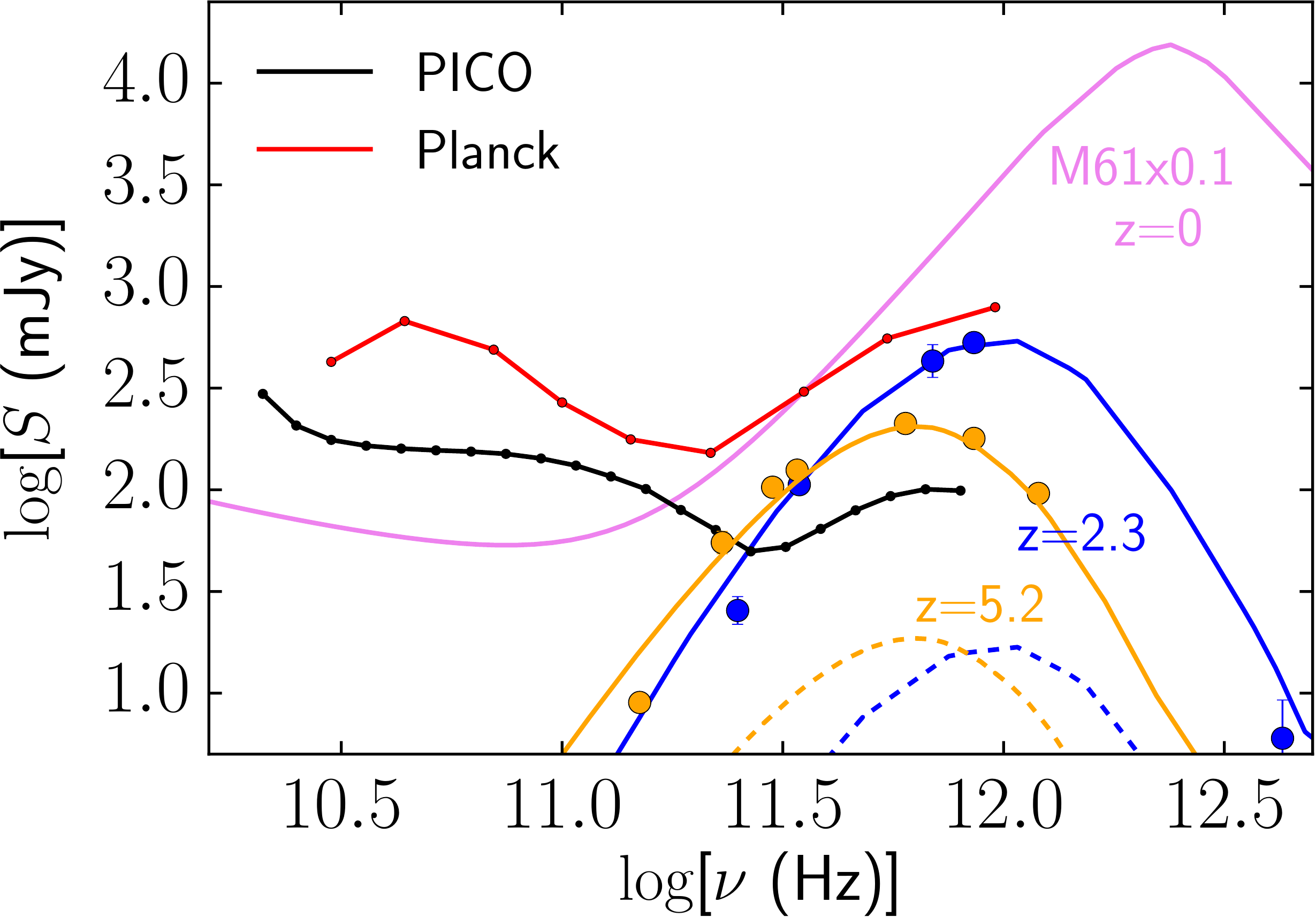}
\hspace{0.75cm}
\includegraphics[width=0.4\columnwidth, trim={0 0 0 0cm}, clip]{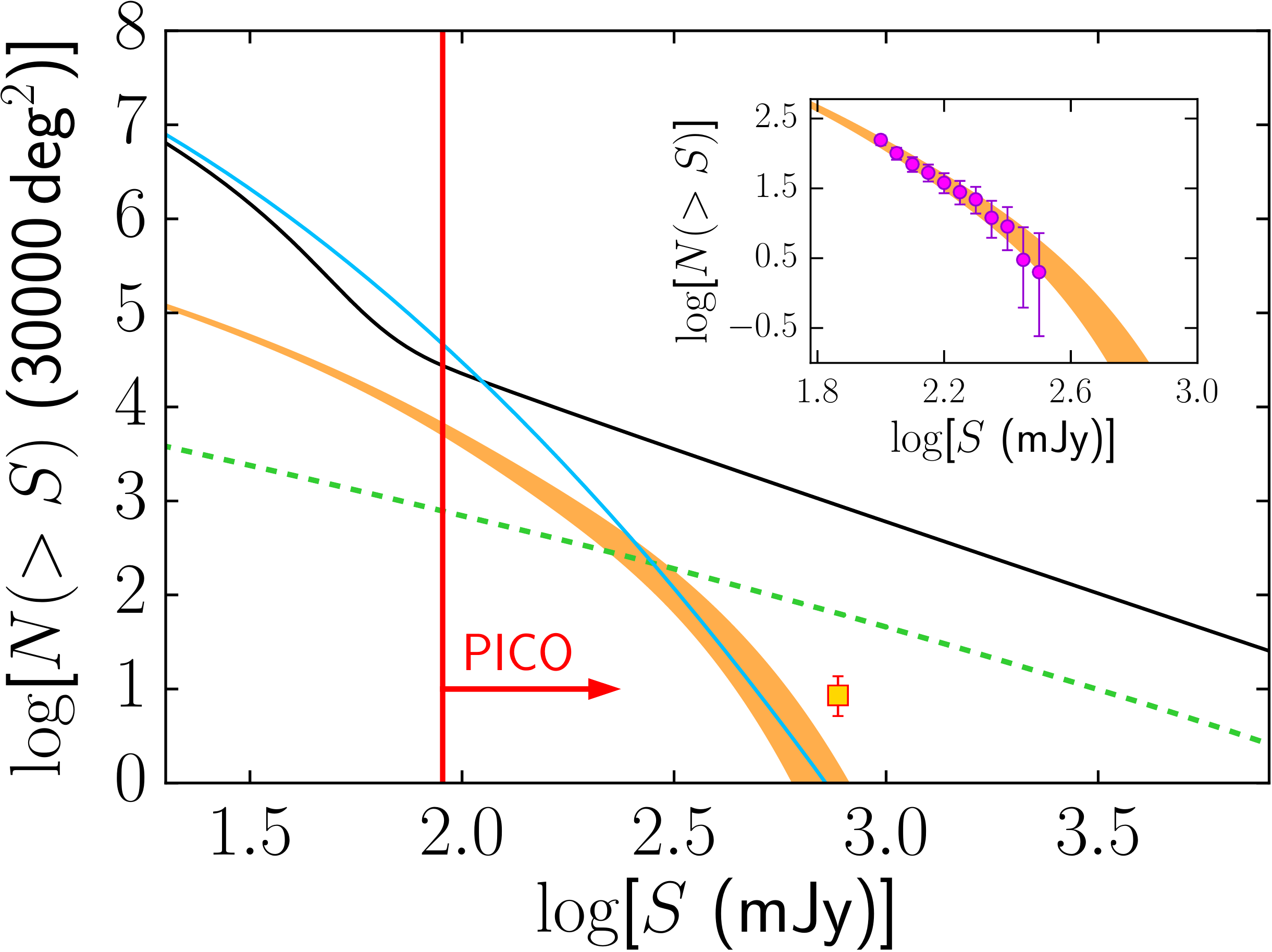}
\vskip-0.3cm
\caption{ \captiontext {\bf Left:} PICO will detect thousands of new strongly lensed galaxies near the peak of their spectral energy distributions (SEDs), such as SMM\,J2133$-$0102 (blue)  at $z=2.3$~\cite{Swinbank2010} and HLS\,J$091828.6{+}514223$ (orange) at $z=5.2$~\cite{Combes2012}. The dashed lines are the SEDs before magnification by lensing. PICO's higher resolution gives point-source detection limits (black line) that are up to 10 times fainter than \planck 's 90\% completeness limits (red line~\cite{PCCS2}). High-frequency measurements ($\nu>300$~GHz) of 30,000 low-$z$ galaxies, like M61 (magenta, SED was scaled down by a factor of ten), will give a census of their cold dust.  {\bf Right:} Integral counts of unlensed (black) and strongly lensed, high-$z$ (orange) star-forming galaxies for 70\% of the sky away from the Galactic plane at 600~GHz based on fits of \textit{Herschel} counts over 1000 deg$^2$ (inset~\citep{Negrello2017lensed}). The PICO detection region (right of vertical red line) will yield a factor of 1000 increase in strongly lensed galaxies relative to \planck~(yellow square), as well as about 50,000 proto-clusters (blue) and 2,000 radio sources (green)~\citep{Negrello2017protocl}.}
%Also shown are predicted radio source counts (green). }
\label{fig:SED3}
\end{center}
\vspace{-0.25in}
\end{figure*}

PICO will detect thousands of early forming galaxies whose flux densities are boosted by large factors due to strong lensing~(Fig.~\ref{fig:SED3}, right). Currently there are reports of just a few other high-$z$ galaxies that are spatially resolved thanks to gravitational lensing, albeit with less extreme magnifications~\citep{Dye2018, Lamarche2018, Sharda2018}. PICO's catalog will be transformative as it will probe the \ac{SED} of the lensed galaxies at their peaks. Two examples of known sources are shown in the left panel of Figure~\ref{fig:SED3}. While nearly all ground-based instruments observe at frequencies up to $\nu = 10^{11.45}$~Hz, PICO's data will extend to the peak of the \ac{SED}, up to $\nu = 10^{11.9}$~Hz.

An extrapolation of the \textit{Herschel} counts to the 70\% non-Galactic sky gives a detection of 4,500 strongly-lensed galaxies with a redshift distribution peaking at $2\simlt z \simlt 3$~\cite{Negrello2017lensed}, but extending up to $z> 5$ (Fig.~\ref{fig:SED3}, left panel).
%\footnote{Extrapolating from achieved performance by the South Pole Telescope~\cite{??}, we estimate that S3 experiments will detect $\sim$1,600 such sources, and only in the RJ part of the spectrum.} 
If objects like the $z=5.2$ strongly lensed galaxy HLS\,J091828.6+514223 exist at higher redshifts, they will be detectable by PICO out to $z>10$. At the 600~GHz detection limit, about 25\% of all detected extragalactic sources will be strongly lensed; for comparison, at optical/near-IR and radio wavelengths, where intensive searches have been carried out for many years, the yield is only about 0.1\%, more than two orders of magnitude lower~\cite{Treu2010}. To add to the extraordinary sub-mm lensing bonanza, the selection of PICO-detected strongly lensed galaxies will be easy because of their unique sub-mm colors (Fig.~\ref{fig:SED3}, left), resulting in a selection efficiency close to 100\% \citep{Negrello2010}. The survey will find the brightest objects over the entire sky, maximizing the efficiency of selecting sources for follow-up observations. 
%\comor{need to compare to SO, extrapolated from SPT}

The intensive high spectral and spatial resolution follow-up campaign of this large sample will enable a leap forward in our understanding of the processes driving early galaxy evolution and open up other exciting prospects, both on the astrophysical and cosmological sides \cite[e.g.,][]{Treu2010}.

\subsubsection{Early Phases of Cluster Evolution}

PICO will open a new window for the investigation of early phases of cluster evolution, when their member galaxies were actively star forming (and dusty), but the hot \ac{IGM} was not necessarily in place. In this phase, traditional approaches to cluster detection (X-ray and SZ surveys, and searches for galaxy red sequences) work only for the more evolved clusters, which do include a hot \ac{IGM}; indeed these methods have yielded only a handful of confirmed proto-clusters at $z\simgt 1.5$ \cite{Overzier2016}.\footnote{More high-$z$ proto-clusters have been found by targeting the environment of tracers of very massive halos, such as radio-galaxies, QSOs, and sub-mm galaxies. These searches are, however, obviously biased.} \planck~has demonstrated the power of low-resolution surveys for the study of large-scale structure~\cite{Planck2016high_z}, but its resolution was too poor to detect individual proto-clusters \cite{Negrello2017protocl}.  Studies of the high-$z$ two-point correlation function \cite{Chen2016, Negrello2017protocl} and \textit{Herschel} images of the few sub-mm bright protoclusters detected so far, at $z \le 4$ \cite{Ivison2013, Wang2016, Oteo2018}, all of which will be detected by PICO, indicate sizes of $\simeq 1'$ for the proto-cluster cores, nicely matching the PICO FWHM at the highest frequencies.

PICO will detect 50,000 proto-clusters as peaks in the high-frequency maps, which are not available for ground-based instruments (Table~\ref{tab:STM2}; blue line in the right-hand panel of Fig.~\ref{fig:SED3}).
%%% FULL SKY MAPS ARE NOT AVAILABLE FOR GROUND-BASED INSTRUMENTS %%%
%\footnote{Extrapolating from achieved performance by the South Pole Telescope~\cite{??}, we estimate that S3 experiments will detect few hundred sources.} 
The redshift distribution will extend out to $z\sim4.5$. This catalog will be augmented by 150,000 evolved clusters, detected by the SZ effect. This will constitute a breakthrough in the observational validation of the formation history of the most massive dark-matter halos, traced by clusters, representing a crucial test of models for structure formation. Follow-up observations will characterize the properties of member galaxies, probing galaxy evolution in dense environments and shedding light on the complex physical processes driving it.

\subsubsection{Additional Products of PICO Surveys}

PICO will yield a complete census of cold ($15$--$25$~K) dust, available to sustain star formation in the nearby Universe, by detecting tens of thousands of galaxies mostly at $z\simlt 0.1$; the \ac{SED} of M61 is a typical example (Fig.~\ref{fig:SED3}, left). With a statistical population, and information only available using data at frequencies above 300~GHz, we will investigate the spectral energy distribution of the dust as a function of galaxy properties, such as morphology and stellar mass. 

PICO will increase by an order of magnitude the number of blazars selected at sub-mm wavelengths and will determine the SEDs of many hundreds of them up to 800\,GHz and up to $z> 5$. Blazar searches are the most effective way to sample the most massive black holes at high $z$ because of the Doppler boosting of their flux densities. PICO's surveys of the largely unexplored mm/sub-mm spectral region will also offer the possibility to discover new transient sources or events, such as blazar outbursts~\cite{Metzger2015}.

PICO will make a leap forward in the determination of the polarization properties of both radio sources and dusty galaxies over a frequency range where ground-based surveys are impractical or impossible.
It will find  1,200 radio sources and 350 dusty galaxies above a flux density limit of 4~mJy at 320~GHz, and 500 radio sources and 15,000 dusty galaxies above 6~mJy at 800~GHz.
% At 320 (800)~GHz it will find 1,200~(500) radio sources and 350~(15,000) dusty galaxies above a flux limit of 4~(6)~mJy.  
These data will give information on the structure and ordering of large-scale magnetic fields in  dusty galaxies. In the case of radio sources, emission at higher frequencies comes from regions closer to the central engine, providing information on the innermost regions of the jets, close to the active nucleus. 

The anisotropy of the \ac{CIB}, produced by dusty star-forming galaxies over a wide redshift range $0 < z \lesssim 5$, is an excellent probe of the history of star formation across time. The \planck\ collaboration derived values for parameters describing the rate of star formation out to $z\sim4$~\cite{2014A&A...571A..30P,2014A&A...571A..18P,madau2014}. PICO's lower noise and twice the number of frequency bands will give an order of magnitude improvement on the statistical errors for these parameters~\cite{Wu:2016hej}. Similar improvement will be achieved in constraining $M_{\mathrm{eff}}$, the galaxy halo mass that is most efficient in producing star-formation activity. PICO's increased sensitivity to Galactic dust polarization will enhance the separation of signals coming from the largely unpolarized \ac{CIB} and polarized Galactic dust; an effective separation of signals currently limits making reliable, legacy-quality \ac{CIB} maps. 
By providing a nearly full-sky map of matter fluctuations traced by dusty star-forming galaxies, such a set of maps could be used for delensing the CMB~\cite{Sherwin/Schmittfull}, for measuring local primordial non-Gaussianity from \ac{CIB} auto-correlations~\cite{tucci}, or for cross-correlations with CMB lensing maps and with galaxy surveys~\cite{Schmittfull/Seljak}.

% ------------

\subsection{Signal Separation}%  (4 pages)}
\label{sec:signal_separation}

%Enumerate the various signals in polarization. Use the frequency band + signals figure. The challenge is to dig out the faintest of all signals, the one due to $r$. This sets the tone for the entire 'signal decomposition' or 'component separation'.  Removing the galactic signal to unmask $r \lesssim 0.001$ is a challenge for all future experiments searching for $r$ at that level, and is a strong advantage of a space platform. The physics of galactic signals suggests complexities in their combined emission properties; the level of this complexity is not known. } 

\subsubsection{The Signal Separation Challenge}
\label{sec:separation_challenge}

\begin{figure}[ht]
\includegraphics[width=0.49\textwidth]{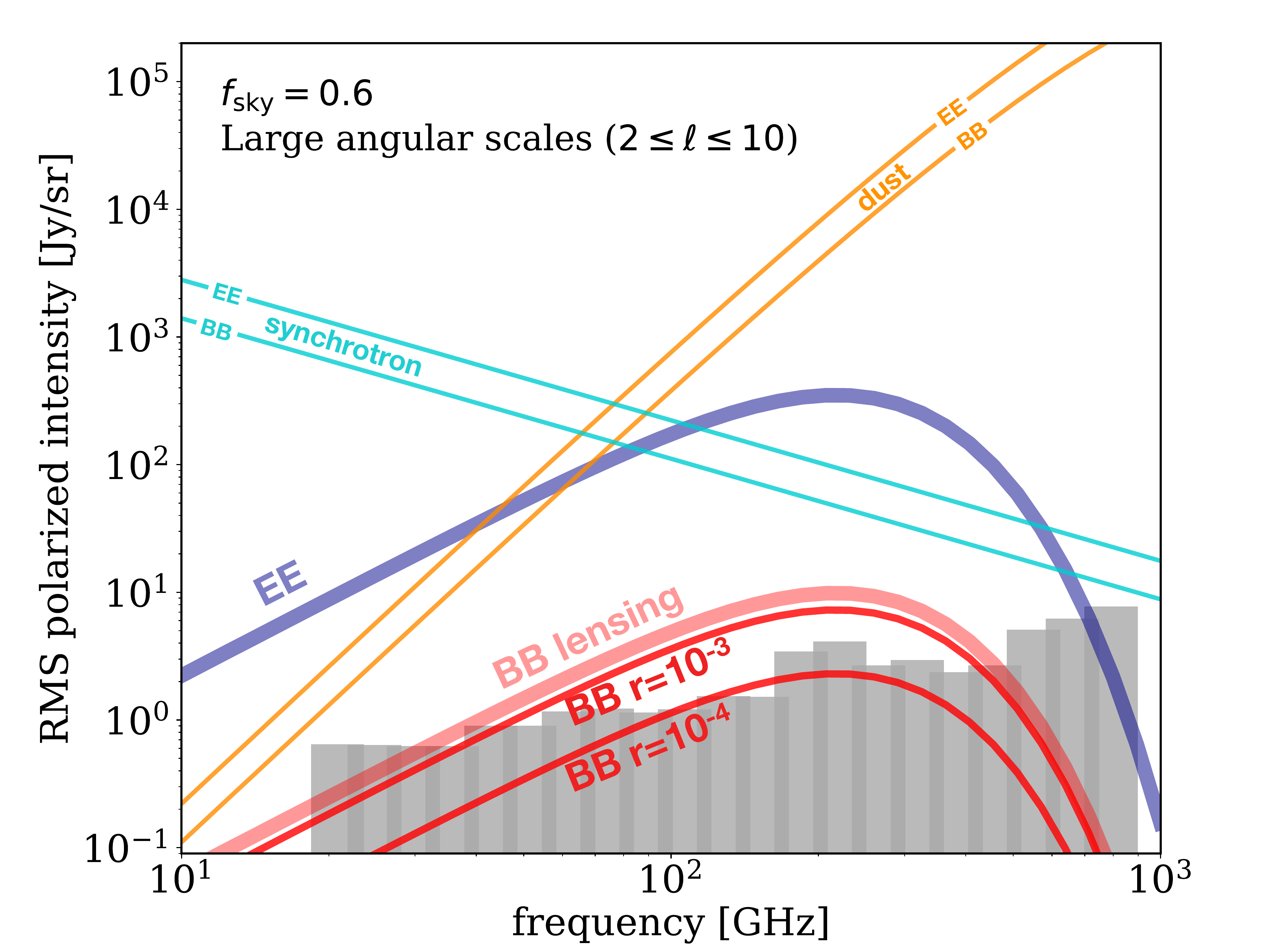}
\includegraphics[width=0.49\textwidth]{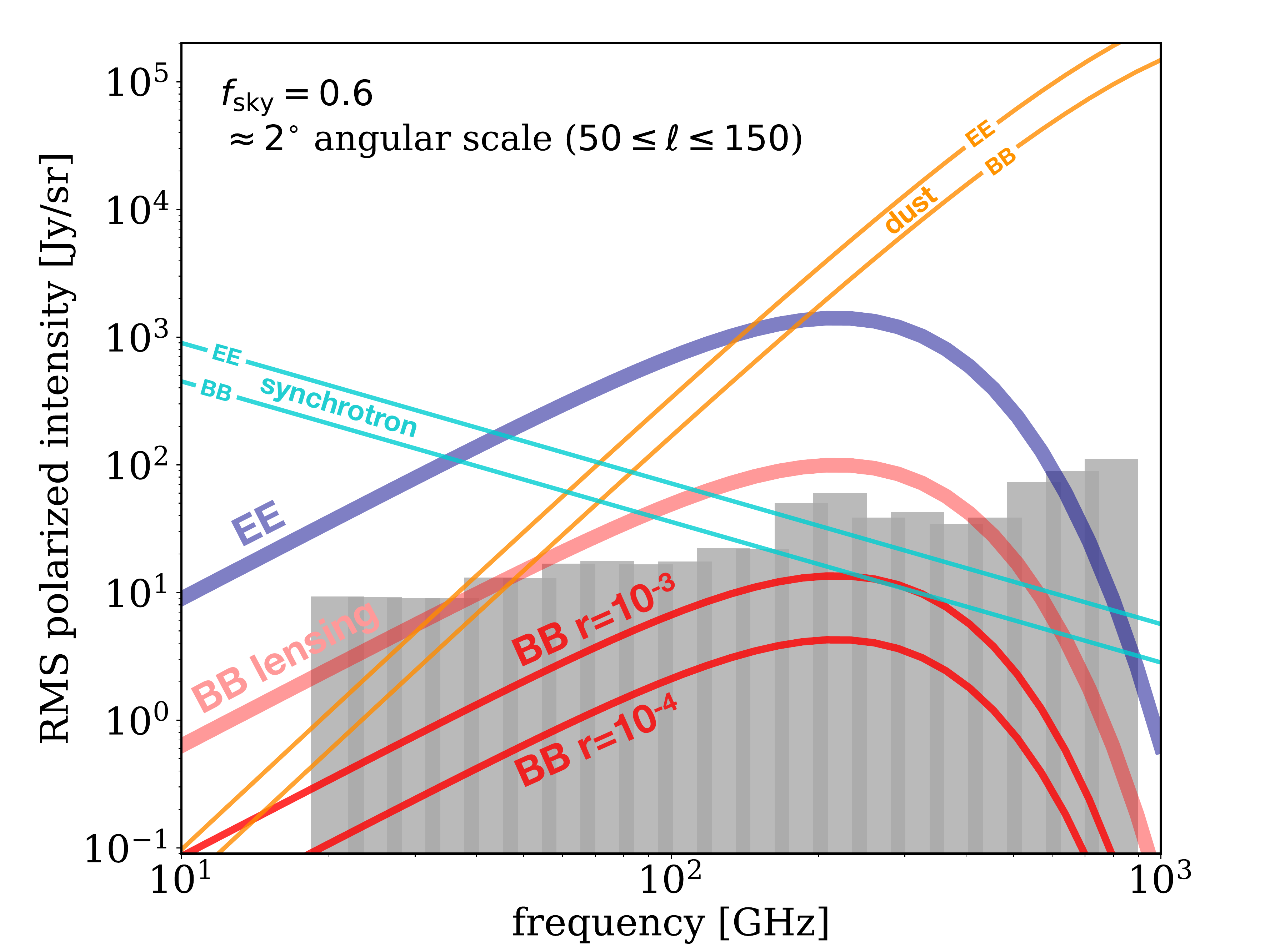}
\vspace{-0.1in}
\caption{\captiontext
Polarization $BB$ spectra of Galactic synchrotron and dust, compared to CMB polarization $EE$ and $BB$ spectra of different origins for two values of $r$ and for two ranges of angular scales: large-scale, $\ell \leq 10$, corresponding to the reionization peak (left panel); and intermediate scales $50 \leq \ell \leq 150$, corresponding to the recombination peak (right panel). 
Data from \planck\ indicate that for Galactic emission the level of the $E$-mode is approximately twice that of $B$~\cite{planckEB}.
The PICO baseline noise (grey bands) is low compared to the Galactic emission components, and thus they will be measured with high \ac{SNR} in many frequency bands.
\label{fig:pico-channels-and-fg} }
\vspace{-0.05in}
\end{figure}

In the PICO frequency range there are Galactic and extragalactic sources of emission. Galactic emissions are due to free-free, synchrotron, and dust, which arise respectively from photon emission in free electron-proton scattering, free electrons spiraling around Galactic magnetic field lines, and from $\sim$20~K elongated interstellar dust grains partially aligned with the local magnetic field. Free-free emission is expected to have negligible polarization. The emission from synchrotron and dust are linearly polarized, and has both $E$ and $B$ components (Fig.~\ref{fig:pico-channels-and-fg}).  Extragalactic sources of emission include the CMB, which has both $E$ and $B$ modes, and point sources %various types 
whose polarization level and type are not well constrained. The task of `separating the signal to its components' (sometimes shortened to `component separation') is to decompose the detected signal to its constituent sources. The required precision of signal separation is determined by the requirement to detect or set an upper limit on the inflationary $B$-mode, which is the faintest among PICO's targeted signals. In that context, the terms `foreground separation' and `foreground cleaning' are used as equivalents to `signal separation'. 

Galactic emission dominates the sky's polarized intensity on large angular scales ($\ell \lesssim 10$), it dominates the cosmological $B$-modes signals for $\ell \lesssim 150$ for all allowed levels of $r$, and it is expected to be significant even at $\ell \simeq 1000$, posing challenge for reconstructing the $B$-mode signal from lensing. This is illustrated in Figs.~\ref{fig:clbb} and~\ref{fig:pico-channels-and-fg}, which show Galactic emission power spectra calculated for the cleanest -- that is, the least Galactic-emission-contaminated -- 60\% of the sky. But even in small patches of the sky, far from the Galactic plane and with the least foreground contamination, Galactic emission levels are substantial relative to an inflationary signal of $r \sim 0.01$ and overwhelm it for $r \lesssim 0.001$~\cite{planckEB}. Separating the cosmological and Galactic emission signals is one of two primary challenges facing any next-decade experiment attempting to reach these levels of constraints on $r$ (the second is control of systematic uncertainties).

Foreground separation is challenging because the spatial power spectra and frequency spectra of the foregrounds are not known to sufficient accuracy anywhere across the sky.
%Galactic emissions are neither precisely characterized spatially, nor do they have , or were known to have simple, fittable spectral emission laws they could be separated. But neither is true. 
To a first approximation, the spectrum of synchrotron emission is a power law $I_{\rm sync} \propto \nu^{\alpha},$ with $\alpha \simeq -1$.  The spectrum of dust emission is $I_{\rm dust} \propto \nu^{\beta} B_\nu(T_{\rm dust}),$ where $\beta \simeq 1.6$, $T_{\rm dust} \simeq 20$\,K, and $B_\nu(T)$ is the Planck function; this is referred to as `modified blackbody emission'. If those models exactly reflected the properties of emitting sources, then in principle an experiment that had six frequency bands could determine the three emission parameters, as well as the three amplitudes for the dust, synchrotron, and CMB components. However, recent observations have shown that neither emission law is universal, that spectral parameters are not necessarily the same for intensity and polarization and that they vary across the sky \cite{SPASS_2018_variation,fuskeland2014_wmap_variation,planck_2013_xi}, and thus that the analytic forms and parameter values given above are only approximately valid for averages across the sky~\citep{chluba2017foregrounds}. Also, while both emission laws are well-motivated phenomenological descriptions, the fundamental physics of emissions from grains of different materials, sizes, and temperatures, and of electrons spiraling around magnetic fields, implies that these laws are expected to be neither exact, nor universal. 

At the low levels of $r$ targeted by PICO and by other next-decade experiments, even small inaccuracies in foreground modeling and characterization lead to biases and false detections. Several publications have demonstrated that fitting complicated dust temperature profiles using a simple one- or two-temperature model will bias the fitted CMB signal at levels $\delta r \approx 10^{-3}$, which is significant compared to PICO's goal~\citep{fantaye2011,armitage-caplan2012,kogut_fixsen2016,remazeilles/etal:2016,stompor2016}. 

%We can no longer impose specific models upon the data; rather, the data collected should provide information to constrain Galactic emissions with sufficient accuracy. 

% and other next decade will dramatically improve sensitivity to inflationary B-modes. The improved sensitivity requires concurrent improvements in foreground separation.  Simple foreground models, suitable for the current generation of CMB measurements, will fail at the higher PICO sensitivity.  For example, the \planck~modified blackbody model assumes that interstellar dust emits at a single temperature, which is clearly an approximation to the more complicated emission along lines of sight spanning hundreds of parsecs. Several publications have demonstrated that fitting complicated temperature profiles using a simple one- or two-temperature model will bias the fitted CMB signal at levels $\delta r \lesssim 10^{-3}$, large compared to the PICO goal~\citep{fantaye2011,armitage-caplan2012,kogut_fixsen2016,remazeilles/etal:2016,stompor2016}.

Further complicating the foreground-separation challenge is the fact that additional polarized foregrounds may exist.  
%`Anomalous microwave emission' (AME) has been observed at mm wavelengths, spatially correlated with thermal dust emission, but with intensity peaking at frequencies near 30 GHz.
Anomalous microwave emission (AME), dust-correlated emission peaking in intensity near 30~GHz, is an important low-frequency foreground in total intensity. 
It has been tentatively attributed to small, rapidly-spinning dust grains~\citep{dickinson2018}. Current $1\sigma$ upper limits on AME polarization are at the level of 1\%~\citep{dickinson2018}. If it is 1\% 
%Very few measurements of AME polarization exist, and there are only loose constraints on its fractional polarization; it is less than 3\% (2$\sigma$) at 18~GHz in one 0.5\% region of the sky~\citep{genova_santos:2015}.  If AME is 1\% 
polarized, left uncorrected it would give rise to a bias of $\delta r \simeq 5\times10^{-4}$~\citep{remazeilles2016}.  Astrophysical emission from CO lines at mm wavelengths is expected to be 0.1--1\,\% polarized~\citep{greeves1999, puglisi2017}.  Extragalactic radio sources show a median polarization of 2\%~\citep{Bonavera2018, puglisi2018_polsource, trombetti2018_fracpol}, and there is significant uncertainty about the polarization of dusty galaxies emitting in the PICO wavebands. Initial quantitative estimates show that ignoring radio sources and dusty galaxies may each lead to a bias $\delta r > 3\times10^{-3}$~\citep{toffolatti2012,Bonavera2018,remazeilles2018}  %at $\ell=80$ 
at low and high frequencies, respectively, and ignoring the CO\,$J=1\rightarrow0$ line could lead to a bias $\delta r > 2\times10^{-3}$~\citep{puglisi2017} at 115~GHz. % and the same $\ell$ range. 
These levels are appreciable compared to the goals of PICO and other next-decade experiments. 

%%%%%%%%%%%%%%%%%%%%%%%%%%%%%%%%%%%
\subsubsection{Foreground Separation Assessment and Methodology}
\label{sec:foreground_separation_methodology}
%%%%%%%%%%%%%%%%%%%%%%%%%%%%%%%%%%%

To investigate the efficacy of PICO in addressing the foreground-separation challenge, we used both an analytic forecast and map-domain simulations. \\ 
\noindent$\bullet$ {\bf Analytic Forecast} \hspace{0.1in} The analytic forecast relies on an established, documented, publicly available, cosmological parameters forecasting code~\citep{errard_and_finney}. The code uses \planck -reported Galactic emissions; it assumes that the foreground spectral indices are constant across patch sizes of $\sim$15\degree\ on a side; it employs a parametric maximum-likelihood approach\footnote{In a parametric approach, foregrounds are assumed to follow emission laws described by a number of free parameters. Parametric models use the frequency dependence of the data along each line of sight to determine the values of the parameters~\citep{eriksen/etal:2008}.} to remove the foregrounds and to forecast $\sigma(r)$; and it uses the cleanest 60\% of the sky. Lensing $B$-modes are included in the input spectra (and are partially removed via delensing, taking into account both noise and foregrounds), but the input for the inflationary signal is $r=0$. \\ 
%Results from the publicly available code have been verified using an independent code that uses similar analytic calculations.  \\
%
%\noindent$\bullet$ {\bf Analytic Forecast} \hspace{0.1in} The analytic forecast relies on an established, documented, publicly available, cosmological parameters forecasting code~\citep{errard_and_finney}. The code uses \planck -reported Galactic emissions; it assumes that the foreground spectral indices are constant across patch sizes of $\sim$15~deg on a side; it employs a parametric maximum-likelihood approach to remove the foregrounds and to forecast $\sigma(r)$: the algorithm propagates the uncertainty due to the noise after component separation, as well as foregrounds residuals and lensing $B$-mode obtained after an iterative, internal delensing. The code uses the cleanest 60\% of the sky and assumes an input inflationary signal with $r=0$. Result from the publicly available code have been verified using a second, independent code that uses similar analytic calculations. 
%
\noindent$\bullet$ {\bf Map-Domain Simulations} \hspace{0.1in} Map-domain simulations have become the `gold standard' in the community. In this approach, we simulate sky maps that are constrained by available data, but otherwise have a mixture of foreground properties. We `observe' these maps just like a realistic experiment would do, and then apply foreground separation techniques -- both parametric and non-parametric\footnote{\label{nonparametric}Non-parametric techniques rely on the fact that CMB emission is uncorrelated with the foregrounds and thus a correlations analysis within a given spatial/frequency data-cube can be used to separate the two sources of emission~\citep{delabrouille2003,Cardoso2008,Delabrouille2009,nilc,gnilc}.} -- to separate the Galactic and CMB emissions.

To test the results we constructed a variety of full-sky models~\citep{gnilc_memo}. All the models were broadly consistent with available data and with uncertainties from \wmap\ and \planck , but they differed in their degrees of Galactic emission complexity. Models included spectral parameters varying spatially and along the line of sight, anomalous microwave emission up to 2\% polarized, dust polarization that rotates slightly as a function of frequency because of projection effects, or dust \ac{SED}s that depart from a simple modified blackbody. All the foreground maps were generated at native resolution of 7\arcmin\ pixels~\citep{gorski/etal:2005}, with widely-used and thoroughly-tested map-generation codes~\citep{thorne2018_pysm,delabrouille/etal:2013}. 

For each of the models, we added CMB signals in both intensity and polarization, matching a $\Lambda$CDM universe. The input inflationary signal was $r=0$, i.e., no signal, and the $BB$-lensing matched the level after 85\% delensing as forecast for PICO. Each of these sky models had 50 realizations of the PICO noise level. 
%, and 50 others had a level of $r=0.003$. \comor{do you want to talk about r=0.003?}
%\vspace{0.1in}
%\noindent{\bf Foreground Separation} \hspace{0.1in} 
The sky models were analyzed with a variety of foreground separation techniques.
%, which were based on the two broad categories described above. 
Because of limited resources for this study not all models were analyzed with all techniques, and not all realizations were used. 

%%%%%%%%%%%%%%%%%%%%%%%%%%%%%%%%%%%
\subsubsection{Results and Discussion}
\label{sec:foregrounds_results}
%%%%%%%%%%%%%%%%%%%%%%%%%%%%%%%%%%%

When using the PICO baseline noise levels with the analytic forecasts we find that $\sigma(r) = 2\times 10^{-5}$, a level that is five times lower than required ($\sigma(r) =1 \times 10^{-4}$, see SO1). We consider this forecast optimistic because it assumes strictly white noise, a specific model for the underlying foregrounds that has only eight parameters\footnote{Six amplitudes for the $Q$ and $U$ Stokes parameters of the CMB, dust, and synchrotron emission, and two spectral indices, for dust and synchrotron.} per $15 \times 15$~deg$^{2}$ pixel, and Gaussian parameter likelihood functions. The foregrounds may be more complex, requiring more  parameters (for example, spatially varying temperature for the dust, or more than a single spectral index per source of emission), and may have stronger spatial variations. Additionally, the parameter likelihoods may not be Gaussian. 
\begin{figure}[h]
\includegraphics[width=3.2in]{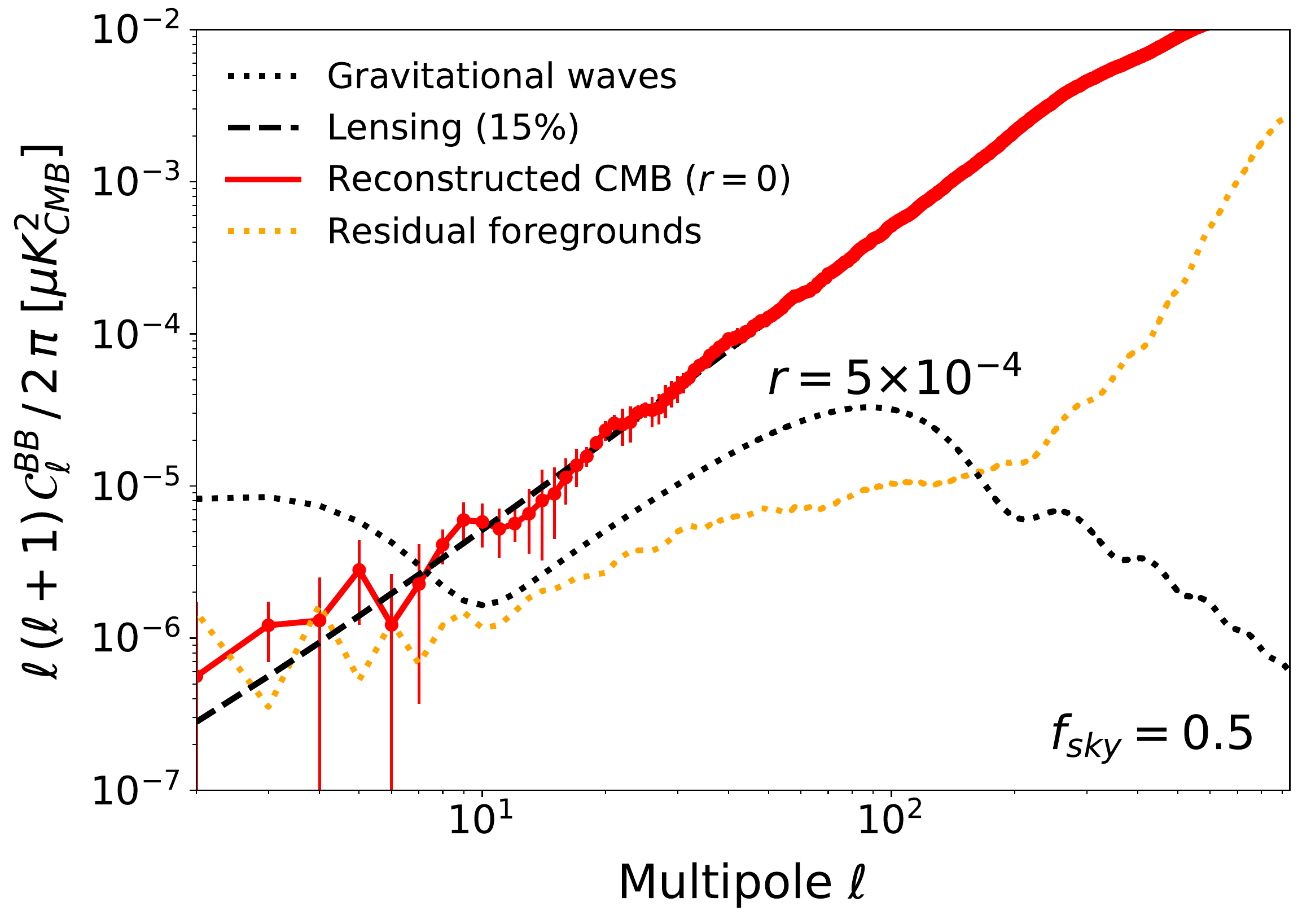}
\hspace{0.1in}
\includegraphics[width=3.2in]{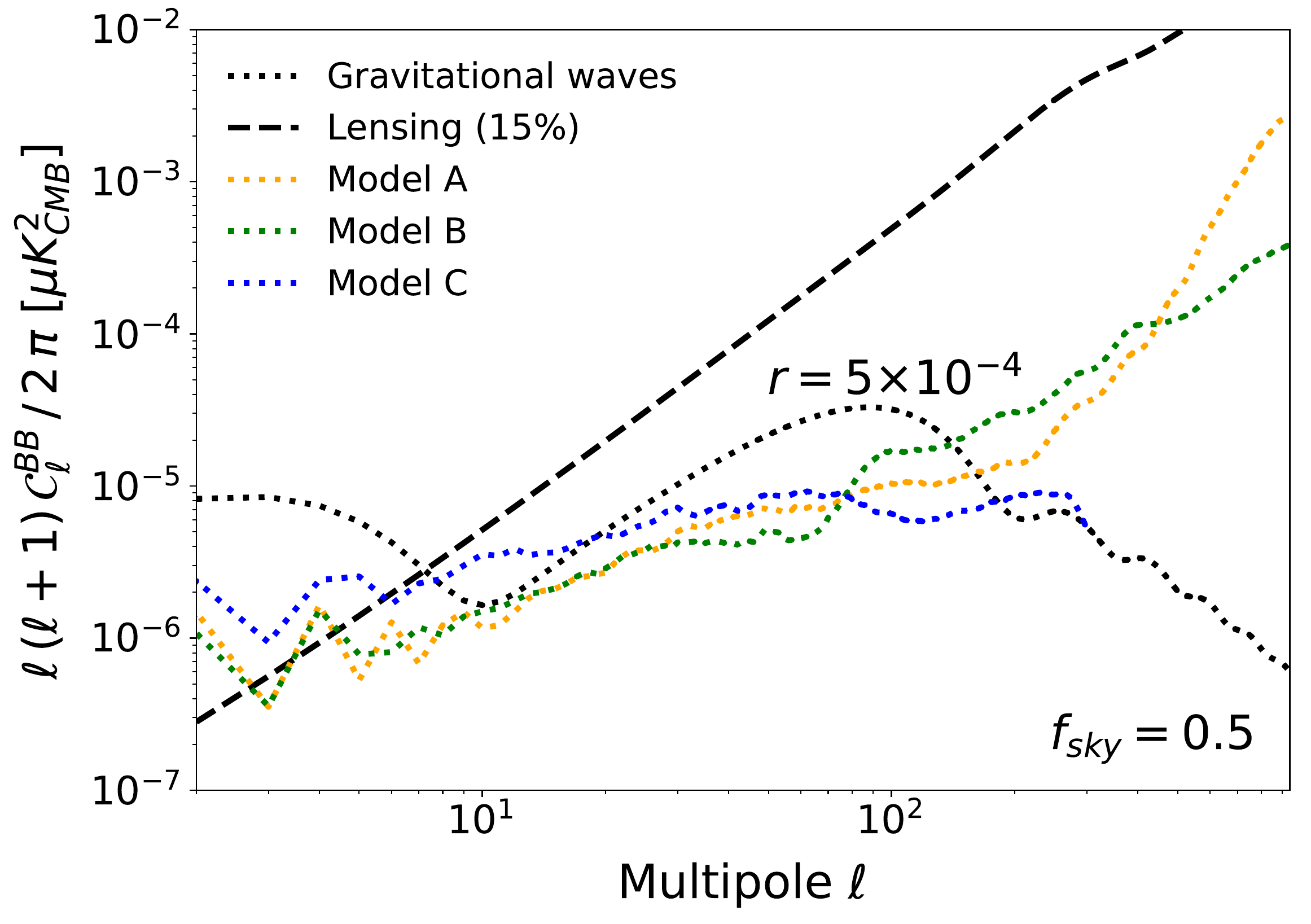}
\vspace{-0.3in}
\caption{\captiontext Angular power spectra of $BB$ due to the CMB and of residual foregrounds after an end-to-end map-based foreground-separation exercise. The PICO low noise levels and breadth in frequency coverage enable separation of model A foregrounds such that the residual foreground spectrum (left, yellow dotted) is a factor of ten (four) below a $BB$ inflationary signal with $r=5\times10^{-4}$ (black dotted) at $\ell=5 (80)$. Within errors, the recovered CMB (red) matches the input CMB, which consists of only lensing $BB$ (dashed black), over all angular scales $\ell \gtrsim 6$.  The results for model B are similar (right, green dots), while model C has somewhat higher residuals at low $\ell$. In this exercise we used 50\% of the sky. Lower foreground residual levels are obtainable with smaller, cleaner patches of $\sim$5\% of sky, which would reduce the residual foregrounds at $\ell \simeq 80$. 
\label{fig:nilc} } 
\vspace{-0.05in}
\end{figure}

The `gold-standard' map-based simulations give initial evidence that the combination of PICO's sensitivity and broad frequency coverage are effective in foreground removal and that PICO will reach the requirement of $r = 5\times 10^{-4} \,(5\sigma)$. Figure~\ref{fig:nilc} shows the results of a foreground-separation exercise over 50\% of the sky, with three representative models of Galactic emissions, labeled A, B, and C~\citep{gnilc_memo}.  This exercise used GNILC, a non-parametric technique\cref{nonparametric}~\citep{gnilc}, tuned to give low foregrounds on the largest angular scales, that is, the lowest $\ell$ modes. The input CMB $BB$ signal, consisting of only lensing $B$-modes, is reconstructed within errors for all $\ell \gtrsim 5$.  With models A and B, the residual foreground $BB$ power spectrum, encoding the levels of remaining foreground emission after foreground separation is a factor of ten below an inflationary $BB$ signal for  $r = 5\times 10^{-4}$ at $\ell \simeq 4$.  These are the angular scales at which the inflationary signal is stronger than the signal from lensing.  Comparing the residual foregrounds for models A and B at this $\ell$ range to the input $BB$ foregrounds at 155~GHz (for example, Fig.~\ref{fig:clbb}) we find a strong suppression (a factor of 1000 in temperature), which is a consequence of PICO's multiplicity of bands and high sensitivity.  The residual in model C is a factor of 2 higher than for A and B at $\ell <30$. Of all models, this model is least constrained to match existing sky measurements~\citep{gnilc_memo}.  

At intermediate angular scales, $\ell \simeq 80$, the residual foreground is a factor of four lower than the inflationary signal. We expect lower residuals when the GNILC analysis is optimized for this $\ell$ range. Furthermore, for reconstructing signals at this $\ell$ range, it is sufficient to analyze data from smaller $\sim$5\% regions of the sky. These will have lower mean foreground levels, making the foreground-separation exercise easier, and pushing residuals to levels lower than demonstrated for 50\% of the sky. With its full-sky coverage, PICO will have access to several independent 5\% sky patches, and will thus make several independent measurements of its $r$ target. 

\begin{figure}
%\hspace{-0.1in}
%\parbox{4.5in}
\centering
\includegraphics[width=3in]{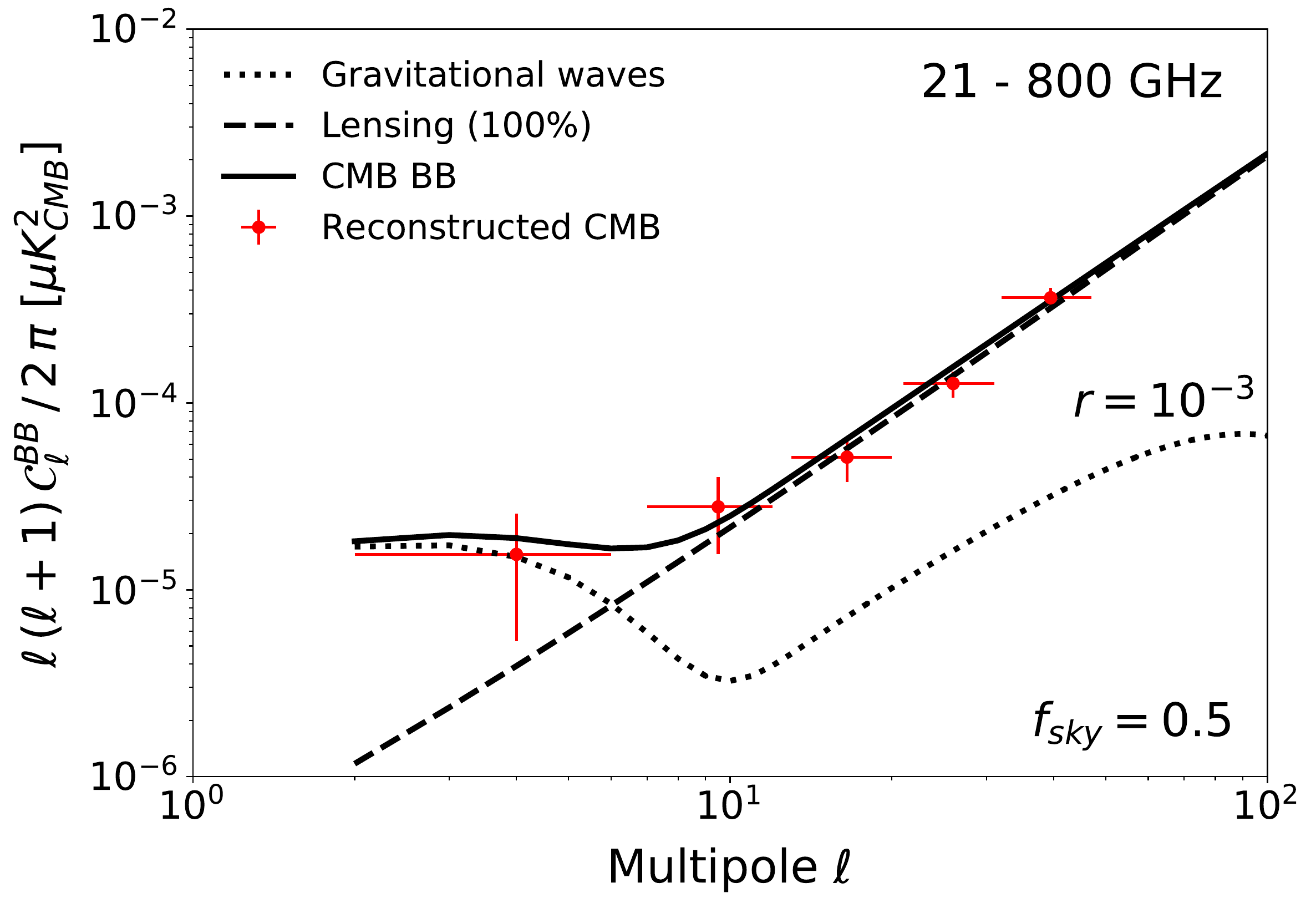}
\hspace{-0.0in}
\includegraphics[width=3in]{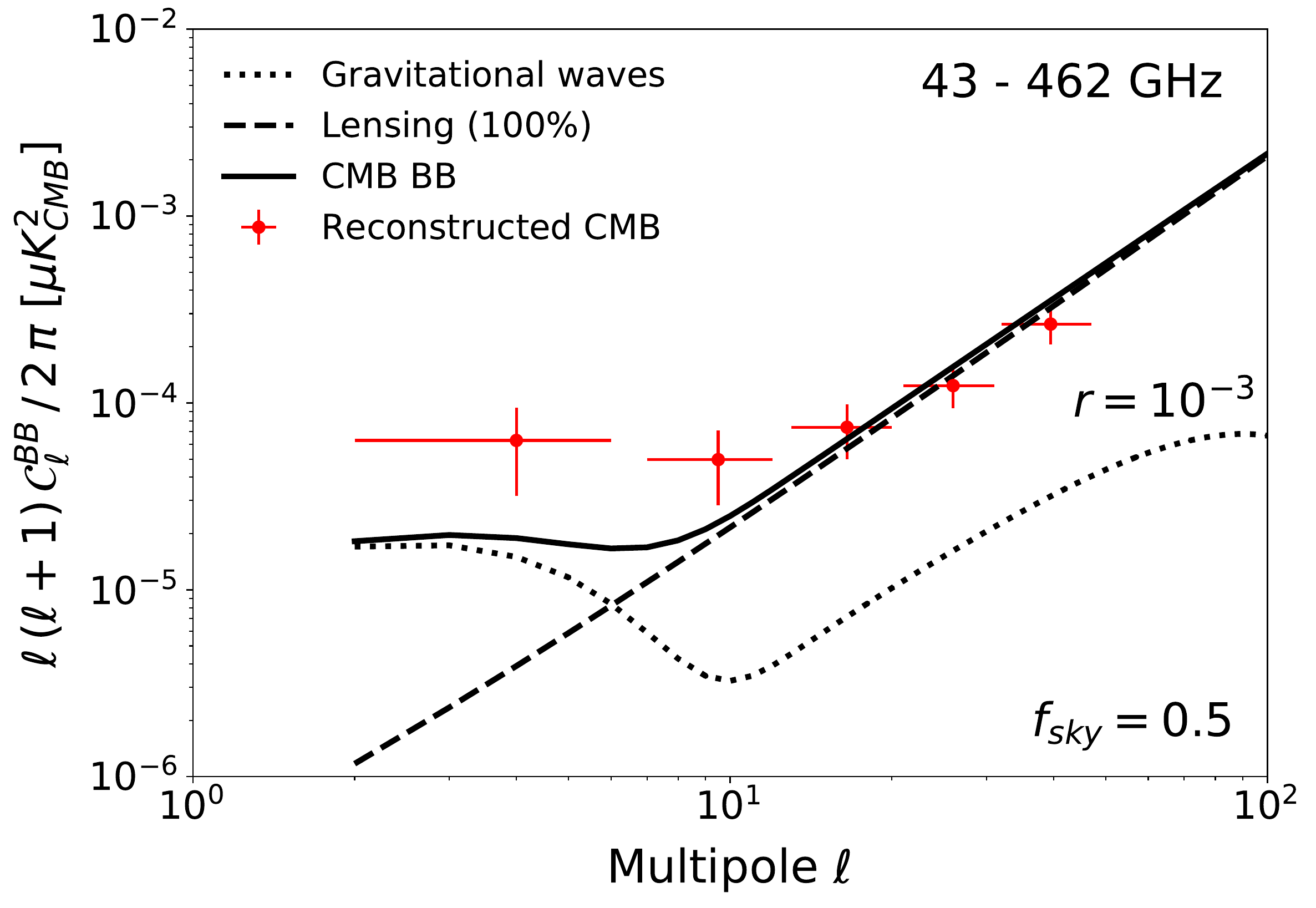}
%\hspace{0.in}
%\parbox{2.0in}{
\vspace{-0.1in}
\caption{\captiontext
{\bf Left:} Foreground separation with all of PICO's 21 frequency bands recovers the input CMB $BB$ power spectrum (solid black) without bias (red). The input CMB spectrum has a contribution from lensing (dashed) and an inflationary signal with $r=0.001$ (dotted). This exercise uses a parametric approach~\citep{eriksen/etal:2008} with foregrounds varying on 4$^\circ$ pixels, and using 50\% sky fraction. {\bf Right:} Running the same foreground separation algorithm on the same sky but using only PICO's bands between 43 and 462~GHz produces an output spectrum (red) that is biased at low multipoles relative to the input. With real data, such a bias would be erroneously interpreted as a higher value of $r$. 
\label{fig:commander}}
\vspace{-0.1in}
\end{figure}

Some of our results validate the need for a broad frequency coverage with a strong lever arm on Galactic emissions outside the primary CMB bands. Figure~\ref{fig:commander} shows that removing several of PICO's frequency bands, particularly those that monitor dust at high frequencies and synchrotron at low frequencies, can significantly bias the extracted $BB$ power spectrum, especially at the lowest multipoles. In this exercise the input CMB contained the lensing signal {\it and} an inflationary signal with $r=0.001$, and a parametric technique was used for foreground separation~\citep{eriksen/etal:2008,gnilc_memo}. 

%While our results are encouraging, as they suggest that PICO's frequency coverage and sensitivity will be adequate for this level  of $r$, more work should be invested to gain complete confidence. 
While these results suggest that PICO's frequency coverage and sensitivity will be adequate for this level of $r$,  more work should be invested to gain complete confidence. For example, some of the other sky models yield a level of residual foregrounds that would result in biased measurements, reflecting larger values of $r$; and some of the foreground-separation techniques appear to give consistently higher foreground residuals than others. To make progress, it is important to continue the simulations and algorithm development program, by: running numerous realizations of different sky models and analyzing them with various approaches; optimizing sky masks; and potentially using a combination of techniques to handle large, intermediate, and small angular scale foregrounds differently. It would also be valuable to continue measurements of Galactic emissions with ground- and balloon-based experiments to further reduce the current level of Galactic emission uncertainties.  

% ------------

\subsection{Systematic Uncertainties}%  (3 pgs)}
\label{sec:systematics}

Having flown the \wmap\ and \planck\ space missions and fielded numerous sub-orbital experiments to measure polarization, the mm/sub-mm wavelength community has gained extensive experience with systematic uncertainties that occur in various experimental configurations. A rich literature investigates the types of systematic errors due to the environment, the instrumentation, observation strategies, and data analysis that could confound polarization measurements by creating a bias or an increased variance \cite{hu03,shimon2008,yadav2010,Griffiths2014,LFI_systematics,Kaplan2002,Miller2009,Pagano2009,SO_sys_optical,SO_sys_detector, bicep_systematics,SPIDER_systematics}.%Teams have used the accumulated experience to incorporate technological solutions during the design phase, and to optimize data analysis techniques to identify and compensate for systematic errors. 
%As a consequence, all of the cosmological results reported to date are noise, not systematic uncertainty limited. In several cases systematic effects have been identified in the data, but then suppressed to levels below the noise. 

Just as requirements on signal separation %(\S~\ref{sec:signal_separation}) 
are determined by the need to reach the faint inflationary signal, so are the requirements on control of systematic uncertainties. Since an inflationary $BB$ power spectrum with $r = 5 \times 10^{-4}$ has a peak signal level of 7~nK, systematic effects need to be controlled to a level of 1~nK. It has long been recognized that exquisite control of systematic uncertainties will be required from any experiment attempting to reach levels of $r \lesssim 1\times 10^{-3}$, and it is widely accepted that the stability provided aboard a space platform makes it best suited to control systematic uncertainties compared to other platforms. This is one of the most compelling reasons to observe from space.  As \wmap\ and \planck~ demonstrated, an L2 orbit offers excellent thermal stability, as well as flexibility in the choice of scan strategy.  

Sources of systematic effects and their ultimate degree of severity are a function of the instrument implementation, the spacecraft scan strategy, and mitigation methods developed during the data analysis phase. Thus, a proper assessment requires end-to-end simulation of the mission. Such a simulation should include realistic non-idealities of the spacecraft, telescope, and instrument, and fold in data post-processing and analysis techniques. Developing such a simulation is a significant undertaking, which took years  for the \planck\ mission, and was beyond the scope of this study. We have instead opted to: (1)~implement design features within PICO that would provide strong data redundancy and enable cross-checks during the data analysis~(\S~\ref{sec:systematics_key}); and (2)~enumerate the sources of possible systematic errors, assess their effects, and investigate three that were deemed the highest priority~(\S~\ref{sec:systematics_list}--\S~\ref{sec:fsl}). 

\subsubsection{Potential Systematic Effects}
\label{sec:systematics_list}

The systematic effects faced by PICO can be grouped into three broad categories: (1)~coupling between signals; (2)~stability; and (3)~stray light. For the first category, the most important are the intensity coupling into polarization (both $E$ and $B$) and $E$ coupling into $B$. This is because $T$ (denoting intensity) is approximately ten times stronger than $E$, which is approximately ten times stronger than $B$. The systematic effects are listed in Table \ref{tbl:SystematicsList2col} and were prioritized for further study using a priority level incorporating a PICO Systematics Working Group's assessment of how mission-limiting the effect is, how well these effects are understood by the community, and whether mitigation techniques exist.  

We used simulations to investigate the following three effects that had the highest priority: error in the absolute calibration of polarization angle; error in the relative calibration between orthogonally oriented detectors; and the effect of the telescope sidelobes. We adapted tools developed for \planck~\cite{plank2015_xii_focalplane} and in the context of a European-led mission concept~\citep{core_systematics}. To understand the severity of the effects, we analyzed each in isolation, and in most cases without complicating effects such as inclusion of foreground-separation steps. More detailed studies of the combination of effects and the inclusion of a foreground-separation step are important but are left to the future. 

\begin{table}[h!]
\scriptsize
%\footnotesize
{\centering
\caption{\captiontext
Enumeration of potential systematic errors anticipated in \pico's measurements, their assessed priority level,
%for affecting measurement of the inflationary signal,
their effects on the measurements, and subsections with further discussion for effects with priority level~5.
%Risk level rises with value. Effects with risk level 5 have been analyzed in more detail.
\label{tbl:SystematicsList2col}
}
\vspace{-2mm}
 \begin{tabular}{@{}p{4.5cm} @{}c p{1.5cm} @{~}c@{} p{0.5cm} @{}p{3.0cm} @{}c p{1.5cm} @{~}c@{}}
 %\hline
 \cline{1-4} \cline{6-9}
 \noalign{ \vskip 1pt}
\multicolumn{1}{c}{\textbf{Name}} & \textbf{Priority$^a$}&\multicolumn{1}{c}{\textbf{Effect$^b$}}& &  &\multicolumn{1}{c}{\textbf{Name}}&\textbf{Priority$^a$}&\multicolumn{1}{c}{\textbf{Effect$^b$}}\\
 %\hline
 \cline{1-4} \cline{6-9}
  \noalign{ \vskip 1pt}
\textbf{Coupling of Signals}& & & & &   \textbf{Stability} & & \\
\quad Polarization angle calibration\dotfill&
5&
$E{\to}B$ &
\S~\ref{sec:angle} & &
% 2nd half table
\quad Gain stability\dotfill&
5&
$T{\to}P$, $E{\to} B$
&
\S~\ref{sec:gain_stability}
\\
\quad Bandpass mismatch\dotfill&
 4&
$T{\to}P$, $E{\to}B$ & & &
% 2nd half table
\quad Pointing jitter\dotfill&
3&
$T{\to}P$, $E{\to}B$
   \\
%\cline{6-9}
 \noalign{ \vskip 1pt}
\quad Beam mismatch\dotfill&
4&
$T{\to}P$, $E{\to}B$ & &
& %Sect.~\ref{sec:angle}
% 2nd half table
\textbf{Straylight}& &
\\
\quad Time response accuracy, stability\dotfill&
4&
$T{\to}P$, $E{\to}B$ & & &
% 2nd half table
\quad Far sidelobes\dotfill&
5&
spurious $P$
&
\S~\ref{sec:fsl}
\\
%\cline{6-9}
\noalign{ \vskip 1pt}
\quad Readout cross-talk\dotfill&
4&
spurious $P$ & & &
% 2nd half table
\textbf{Other}
\\
\quad Chromatic beam shape\dotfill&
4&
spurious $P$ & & &
% 2nd half table
\multirow{2}{3.3cm}{\quad Residual correlated noise \\\quad($1/f$, cosmic ray hits)\dotfill}&
3 &
\multirow{2}{1.4cm}{increased variance}
\\
\quad Gain mismatch\dotfill&
3&
$T{\to}P$ & & &
\\
\quad Cross-polarization\dotfill&
3&
$E{\to}B$
\\
\cline{1-4}
\cline{6-9}
\end{tabular}
\vskip 3pt
} % end centering
 \noindent
 \footnotesize
 {$^{a}$}~Level 5 indicates a highly significant, design-driving effect; it may have limited past measurements, or is not well understood.  Level 4 is an effect that is either known to be large but is understood reasonably well, or is a smaller effect that requires precise modeling.  In Level 3 we expect the effect to be small, but it is not sufficiently well understood and detailed modeling will be done during a Phase A study. Level 2 indicates a well-understood or minimal effect that may not need modeling, and Level 1 is for an effect that is not significant and  does not need modeling.\qquad
 {$^{b}$}~$T \rightarrow P $ denotes coupling of the intensity signal (labeled as $T$ to denote temperature) into polarization, which would generally be both $E$ and $B$. Similar meaning holds for $E \rightarrow B$.\par
%\vspace{-0.1in}
\end{table}

\subsubsection{Absolute Polarization Angle Calibration}
\label{sec:angle}

In PICO, each of the Stokes $Q$ and $U$ parameters along any line of sight is evaluated through having sensitivity to two orthogonal polarization states. The relative designation of $Q$ and $U$ is derived from having sensitivity to pairs of polarization orientations that are 45\degree\ apart (\S\,\ref{sec:polarimetry}). A systematic error in the implementation (or estimation) of these angles by an amount $\alpha$ causes signals in $Q$ and $U$, and thus in $E$ and $B$, to mix. Because the CMB $E$-mode is much larger than $B$, mixing between $E$ and $B$ leads to the generation of a spurious $BB$ angular power spectrum that mirrors the shape of the $EE$ spectrum~(Fig.~\ref{fig:rot_bb_tb_eb}). The level of spurious $BB$ is proportional to $\alpha^{2} \times EE$. At angular multipoles $\ell \lesssim 100$ a systematic error $\alpha\approx 10\arcmin$ will result in a spurious $BB$ level that is approximately equivalent to $r = 1\times10^{-4}$~\citep{shimon2008,Aumont+2018}.  The mixing of $E$ and $B$ also leads to spurious cross-spectra $EB$ and $TB$, which respectively mimic the $EE$ and $TE$ spectra. 

This systematic error is most usefully split into two contributions: an overall `absolute' error in the assumed instrument's sensitivity to polarization orientations relative to fixed sky coordinates; and a `relative' rotation error between various pairs of detectors. For PICO, the relative rotation of the detectors will be measured to $0.1\arcmin$ by comparing the measured polarization signals between many independent detectors and pairs. However, directly measuring the overall rotation in flight -- which is the process of calibrating the polarization angles --  is challenging, since there are no sufficiently well calibrated polarized astronomical sources. For example, \citet{Aumont+2018} showed that for the best characterized source -- the Crab Nebula -- the current uncertainty of $0.33$\degree\ on the polarization orientation limits measurements to $r \sim 0.01$. 

PICO will overcome this potential source of error 
%through using its high \ac{SNR} measurement of the polarization to identify and reduce it below relevant levels 
in data analysis. \citet{yadav2010} showed that because the $T$ and $E$ signals are much stronger than $B$, an experiment that searches for a specific level of cosmological $BB$ will have high \ac{SNR} for detecting spurious $EB$ and $TB$ cross-spectra arising from error in the calibration of the polarization angle. Applying their method to the PICO baseline specifications we find a constraint of $\alpha<0.2\arcmin$ and $0.6\arcmin$ $(3\sigma)$ using the $EB$ and $TB$ spectra, respectively, and thus a suppression of this systematic effect to negligible levels~(Fig.~\ref{fig:rot_bb_tb_eb}). The constraints quoted include a delensing level of 73\%, which is the PICO forecast including foreground separation.

\subsubsection{Differential Gain}
\label{sec:gain_stability}

Photometric calibration is the process of converting the raw output of each detector -- typically given in digital readout units -- to physical units via a calibration factor $C(t)$, which is a function of time. One straightforward way for PICO to derive $Q$ and $U$ is through differencing detectors that are sensitive to two orthogonal polarization states $A$ and $B$. A systematic error in the determination of either $A$ or $B$ calibration factors will translate to a biased $Q$ or $U$. We investigated whether the anticipated error on $C_{A,B} (t)$ is adequate for PICO's requirements on measuring the inflationary signal.  
%However, statistical or systematic errors in the determination of $G_{A}(t)$ and $G_{B}(t)$ translate to increased uncertainties, or biased results, respectively.  We investigated whether the anticipated error on $G(t)$ is adequate for PICO's requirements on measuring the faint inflationary signal.  

\begin{figure}[thb]
\centerline{
\includegraphics[width=0.375\textwidth]{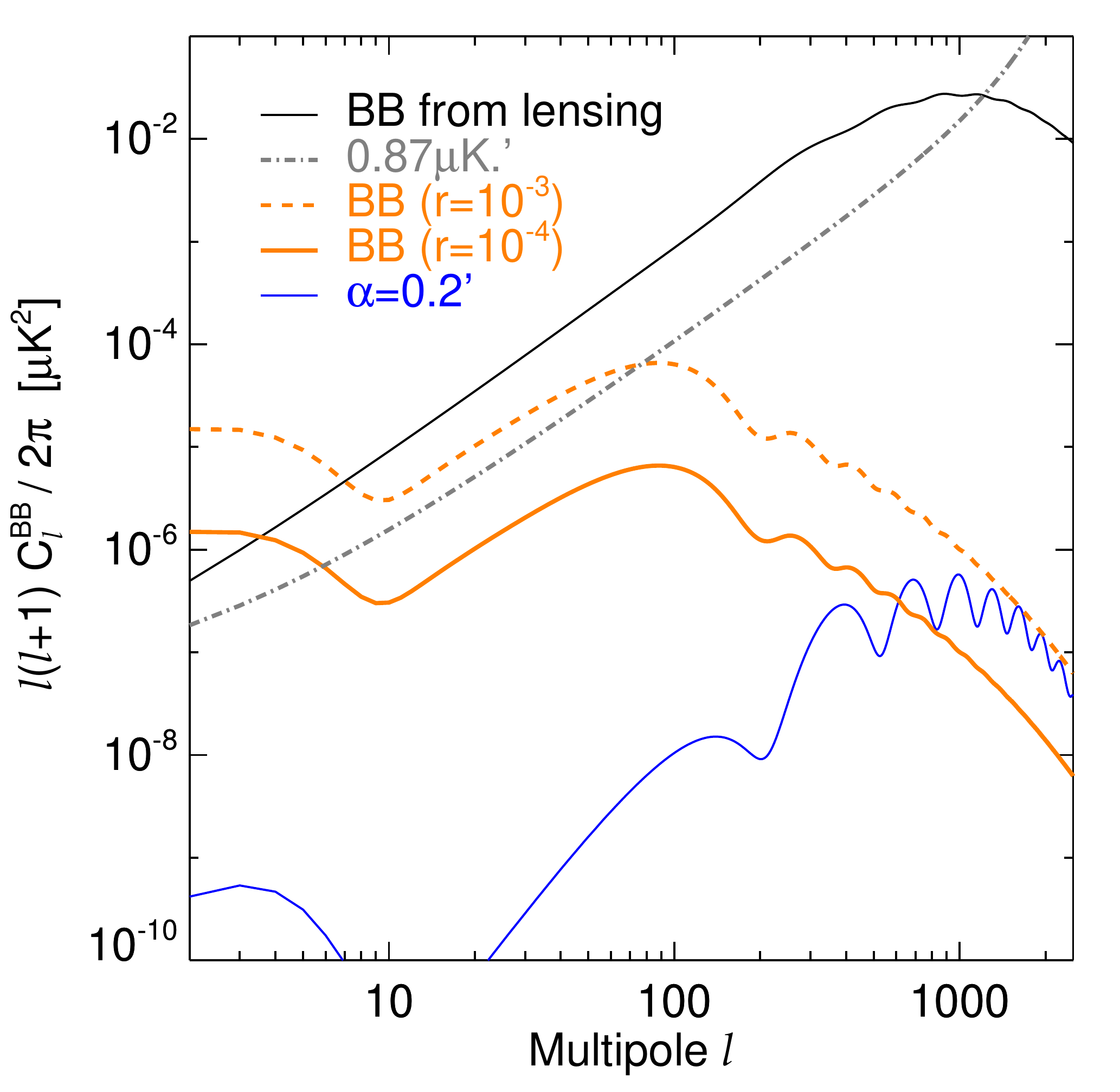} 
\hspace{0.3in}
\includegraphics[width=0.41\textwidth]{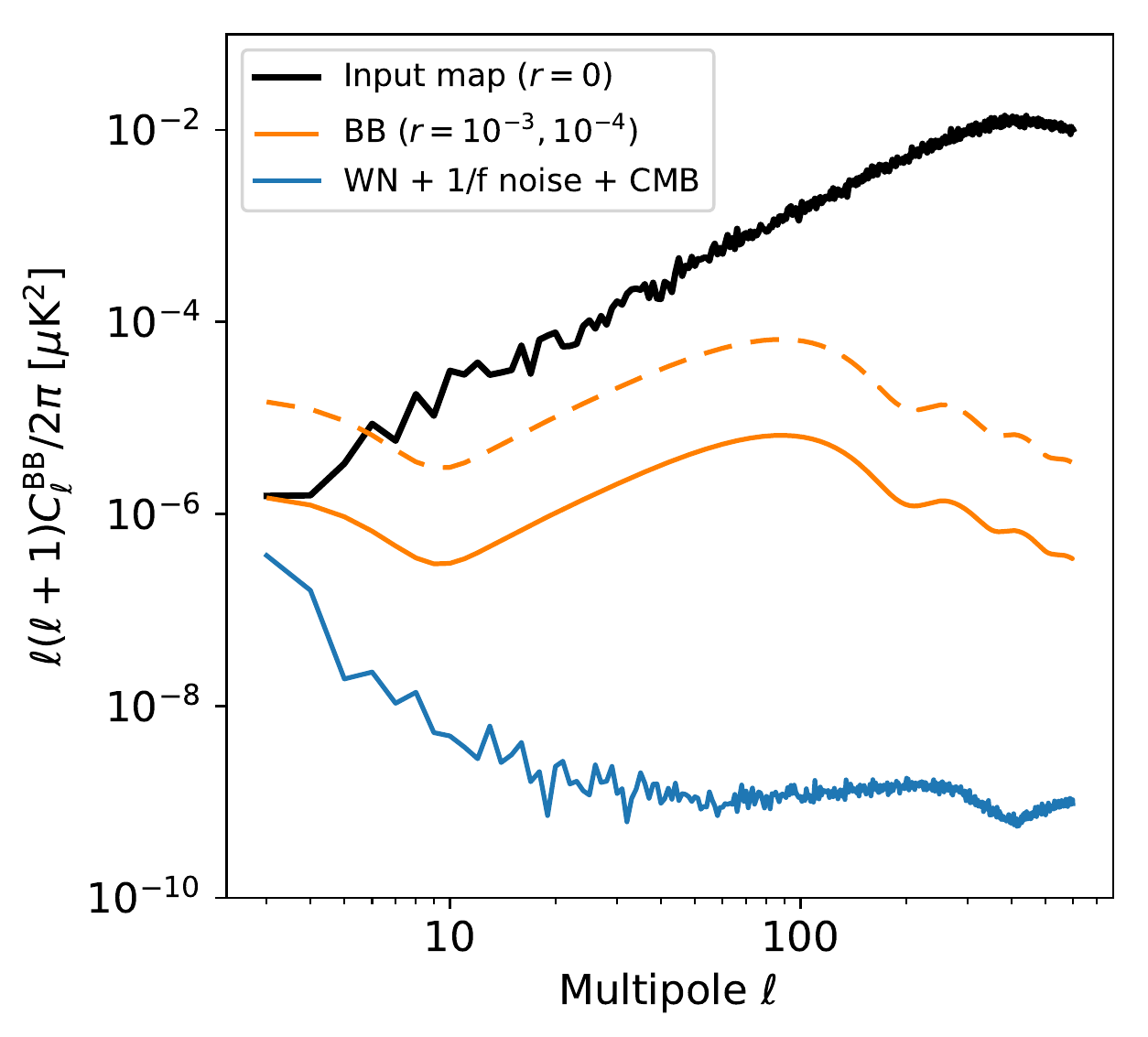} }
\vspace{-0.1in}
\caption{\captiontext
Two of the initially-estimated highest priority systematic effects for PICO can be suppressed to low levels relative to requirements; here we show inflationary signals with  $r = 1\times 10^{-4}\,\, \mbox{and} \,\, 1\times 10^{-3}$ (solid and dash orange, respectively), and $BB$ lensing (black, theory on the left; realization on right). {\bf Left:} The residual spurious $BB$ spectrum due to 0.2\arcmin\ mis-calibration of PICO's angles of polarization sensitivity (solid blue) has the shape of the $EE$ spectrum, and is small compared to the requirement for $\ell<200$ and compared to the baseline statistical noise level (grey dash). {\bf Right:} Simulated residual $BB$ power after accounting for calibration drifts (solid blue). 
\label{fig:rot_bb_tb_eb} }
\vspace{-0.1in}
\end{figure}

We assume that the inflationary signal will be extracted from data in the primary CMB bands between 60 and 300~GHz.  Detectors in these bands will be calibrated using measurements of the CMB dipole, a signal that will be measured once per minute as the telescope scans the sky (\S~\ref{sec:survey_design}).  We evaluated the combined impact of the scan strategy and white- and $1/f$ noise in the estimation of $C(t)$.
%\footnote{We use software tools developed for and validated in the context of analysis of the \planck/LFI data.} 
The simulation included signals from the anisotropy of the CMB, including the dipole and $BB$ lensing. Full details of the simulation pipeline are available in the PICO website~\citep{picoweb_dipole}. Figure~\ref{fig:rot_bb_tb_eb} demonstrates that the power spectrum due to error in $C_{A,B} (T)$ is much lower than the PICO requirement of $\sigma(r) = 1\times 10^{-4}$. 

\subsubsection{Far Sidelobes}
\label{sec:fsl}

Differences between the assumed and actual antenna pattern of the detectors will give rise to systematic errors. Such differences are particularly hard to detect in the `far-sidelobes' where the antenna pattern is below the noise level. Unknown far-sidelobe response can couple to bright Galactic signals when the telescope points tens of degrees away from the Galactic plane, and cause spurious signals. To evaluate PICO's susceptibility to this systematic effect we 
%used the same physical optics software as used by \planck\ to 
computed PICO's $4\pi$~sr antenna response for four 155~GHz detectors located at the center of the focal plane. We simulated the time domain response of the detectors as they scan the sky over a year of PICO observations. We convolved their antenna response with a full-sky Galactic emission model~\citep{thorne2018_pysm}, reconstructed maps of $I$, $Q$, and $U$, and calculated the resulting $BB$ angular power spectrum when using a \planck\ Galactic mask excluding 60\% of the sky~\citep{planck_2013_xv}. 

The largest sidelobe in the antenna response is at a level of $-80$~dB from the main lobe. We find that if that sidelobe is known to a level of $-95$~dB (\ac{SNR}$>$20), or further suppressed to that level, the contamination from the sidelobe is a factor of ten below the requirement of $\sigma(r) = 5 \times 10^{-4}$. This suppression can be achieved by adding baffles and through ground-based measurements. \planck 's ground-based measurements mapped the antenna response to levels between $-90$ and $-100$~dB from the main lobe~\citep{planck_sidelobes_IEEE}. The combination of measurements and modeling will be used to remove sidelobe pickup during data analysis.

\subsubsection{Additional Key Findings}
\label{sec:systematics_key}

Properly modeling, engineering for, and controlling systematic effects are key for the success of any experimental endeavor striving to achieve $\sigma(r) \lesssim 1 \times 10^{-3}$. Based on extensive community experience with both hardware and analysis of data we make the following points. \\
$\bullet$ \hspace{0.1in}  Relative to other platforms, a space-based mission provides the most thermally stable platform, and thus the prerequisite for improved control of systematic effects. PICO's orbit at L2 is among the most thermally stable of possible orbits. \\
$\bullet$ \hspace{0.1in} PICO's sky scan pattern gives strong data redundancy, which enables numerous cross-checks. Each of the 12,996 detectors makes independent maps of the $I,\,Q$, and $U$ Stokes parameters enabling many comparisons within and across frequency bands, within and across sections of the focal plane, and within and across bolometers that have either the same or different polarization sensitivities. Half the sky is scanned every two weeks, and the entire sky is scanned in 6 months. Thus combinations of maps constructed at different times during of the mission will be differenced to search for residual time-dependent systematic effects. \\
$\bullet$ \hspace{0.1in}  The scan pattern gives almost continuous scans of planets and large amplitude ($\geq 4$~mK) CMB dipole signals~\citep{picoweb_dipole}. These features result in continuous, high \ac{SNR} calibration and antenna-pattern characterization. In comparison, \planck\ observed each of the planets with only a 6 month cadence and had nearly 100~days/year during which the dipole calibration signals were below 4~mK, at times dipping below 1~mK. \\
$\bullet$ \hspace{0.1in}  We showed that two of the highest priority systematic effects can be controlled to levels that are small compared to requirements. More analysis and planning is required to address systematic uncertainties arising from the far-sidelobe response of the telescope. 

We strongly recommend that further support be provided for further analysis of systematic effects, their combinations, and their coupling with foreground separation. Specifically, support for suborbital efforts is essential to continue the development of means to identify systematic effects, and to develop new techniques to mitigate them. We also endorse support for the development of a complete end-to-end software simulation facility, which is the most robust way to quantify mission trade-offs under the influence of a combination of systematic effects that are coupled to the task of signal separation.

% ------------

\subsection{Complementarity with Sub-Orbital Measurements}
\label{sec:complementarity}

\begin{table}[tb]
\caption{Relative characteristics of ground, balloon, and space platforms for experiments in the CMB bands.\label{tab:comparison}}
\begingroup
%\openup 5pt
\newdimen\tblskip \tblskip=5pt
\nointerlineskip
\vskip -5mm
\footnotesize %\footnotesize
\setbox\tablebox=\vbox{
    \newdimen\digitwidth
    \setbox0=\hbox{\rm 0}
    \digitwidth=\wd0
    \catcode`*=\active
    \def*{\kern\digitwidth}
    \newdimen\signwidth
    \setbox0=\hbox{+}
    \signwidth=\wd0
    \catcode`!=\active
    \def!{\kern\signwidth}
\halign to \textwidth{
\hbox to 1.8in{#\leaderfil}\tabskip=0.6em plus 0.6em&
#\hfil&
#\hfil&
#\hfil\tabskip=0pt\cr
\noalign{\doubleline}
\omit\hfil{\bf Characteristic}\hfil&\omit\hfil{\bf Ground}\hfil&\omit\hfil{\bf Balloon}\hfil&\omit\hfil{\bf Space}\hfil\cr
\noalign{\vskip 3pt\hrule\vskip 5pt}
Sky coverage&Partial from single site&Partial from single flight&Full\cr
\noalign{\vskip 3pt}
Frequency coverage&70~GHz inaccessible,$^{a}$&70~GHz inaccessible,$^{a}$&Unrestricted\cr
\omit                       &$\nu \ge 300$\,GHz unusable,&otherwise, almost unlimited&\cr
\omit                       &limited atmospheric windows&&\cr
\noalign{\vskip 3pt}
Angular resolution at 150\,GHz\,$^{b}$& $1.\!'5$ with 6\,m telescope&$6'$ with 1.5\,m telescope&$6'$ with 1.5\,m telescope \cr
\noalign{\vskip 3pt}
Detector noise\,$^{c}$&$265\,\mu$K$_{\rm CMB}\sqrt{\rm s}$& $162\,\mu$K$_{\rm CMB}\sqrt{\rm s}$ & $38\,\mu$K$_{\rm CMB}\sqrt{\rm s}$ \cr
\noalign{\vskip 3pt}
Integration time  &  Unlimited, with interruptions & Weeks, continuous  & Several years, continuous \cr
\noalign{\vskip 3pt}
Repairability, Upgradeability & Good & None; multiple flights possible & None \cr
\noalign{\vskip 5pt\hrule\vskip 3pt}
} % close halign
\noindent {$^{a}$}~70\,GHz is the frequency at which large angular scale $B$-mode Galactic emissions have a minimum (Fig.~\ref{fig:pico-channels-and-fg}).
\qquad {$^{b}$}~We give representative telescope apertures. Significantly larger apertures for balloons and in space result in higher mass, volume, and cost.
\qquad {$^{c}$}~The noise-equivalent temperatures given are illustrative of general capabilities. Detailed comparisons depend on detector heat-sink temperatures, bandwidths, and other factors that differ among specific implementations. Ground -- median detector noise at 95\,GHz from BICEP3~\citep{kang20182017}; balloon -- median detector noise at 94\,GHz from SPIDER~\citep{privatecommunication}; space -- 90\,GHz from PICO CBE.    
} % close vbox
\endPlancktable
\endgroup
\end{table}

Since the first \ac{CMB} measurements, more than 50 years ago, important observations have been made from the ground, from balloons, and from space. Each of the CMB satellites flown to date -- \cobe, \wmap, and \planck\ -- has relied on technologies and experience that were the result of sub-orbital efforts. PICO is no different. Examples include: the arrays of micro-fabricated, multi-color pixels and the multiplexed readout that are baselined for the PICO focal plane are a consequence of this decade's technical developments (\S\,\ref{sec:focal_plane}, \S\,\ref{sec:detector_readout}); and the recent results from \planck \ and ground-based experiments that established the need for a multitude of frequency bands to characterize and control foregrounds.  A healthy sub-orbital program is essential for the success of PICO. 

%\comor{original text above this line}

The phenomenal success and the immense science outcomes of past space missions are a direct consequence of their relative advantages (Table~\ref{tab:comparison}). In every respect, with the exception of repairability and upgradeability, space has the advantage. When the entire sky is needed, as for measurements on the largest angular scales, space is by far the most suitable platform.  When broad frequency coverage is needed, space will be required to reach the ultimate limits set by astronomical foregrounds because ground-based observations are limited to a handful of atmospheric windows, mostly below 300~GHz. Balloons can provide useful information at higher frequencies, but their limited observing time limits \ac{SNR}. The stability offered in space can not be matched on any other platform, and it translates to superb control of systematic uncertainties. %There is consensus within the CMB community that for levels of $r \lesssim 0.01$ the challenges in the measurement are the ability to remove Galactic emissions (\S~\ref{sec:signal_separation}) and to control systematic uncertainties (\S~\ref{sec:systematics}). 

The relative advantages of a space mission used to come with higher costs relative to sub-orbital experiments. However, this balance now shifts. To make further advances in CMB science it is now required to mount massive ground-based efforts.  By the early 2020s, S3 experiments plan to implement more than 100,000 detectors in 9 receivers in Chile and the South Pole. The total cost is in the vicinity of \$100M. The cost for a subsequent scale-up, a $\sim$500,000-detector ground-based CMB experiment planned for the next decade, is squarely within the cost window of this Probe. Even at that cost, the PICO goal of reaching $r = 5 \times 10^{-4}\,(5 \sigma) $ is beyond the reach of sub-orbital observations in the foreseeable future.  

%There is consensus within the CMB community that for levels of $r \lesssim 0.01$ the challenges in the measurement are the ability to remove Galactic emissions and to control systematic uncertainties. 
For measuring $r$ and for achieving the other PICO SOs, a space-based platform is either necessary or has strong advantages.  For science requiring higher angular resolution, such as observations of galaxy clusters with 1 arcmin resolution at 150~GHz, the ground has an advantage. An appropriately large aperture on the ground will also provide high-resolution information at lower frequencies, which may be important for separating Galactic emissions at high $\ell$. We therefore recommend to pursue a space mission in the next decade, and to complement it with a ground-based program that will overlap in $\ell$ space, and will add science at the highest angular resolution, beyond the reach of a space mission.

%A recommended plan for the next decade is therefore to pursue a space mission, and complement it with a ground program that will overlap in $\ell$ space, and will add science at the highest angular resolution, beyond the reach of a space mission.

Balloon observations have been exceedingly valuable in the past, and will continue to play an important role through making measurements at frequency bands above 280~GHz. Because balloon observations are largely free from the noise induced by atmospheric turbulence, they are suited for probing the low $\ell$ multipoles. The balloon environment is the best available for elevating the TRL of relevant technologies.

% ------------

\subsection{Measurement Requirements} % (2 pgs)}
\label{sec:requirements}

The set of physical parameters and observables that derive from the PICO \ac{SOs} place 
requirements on the depth of the mission, the fraction of sky the instrument scans, the frequency range the 
instrument probes and the number of frequency bands, the angular resolution provided by the reflectors, and
the specific pattern with which PICO will observe the sky.  \\
%combination of PICO's diverse science goals are achievable using a single instrument that executes one, continuous, simple observing pattern of the entire sky. This pattern integrates noise down to unprecedented levels, and provides for multiple checks of possible systematic errors in the data analysis. \\
%
$\bullet$ {\bf Depth} \hspace{0.1in} We quantify survey depth in terms of the RMS fluctuations that would give
an \ac{SNR} ratio of~1 in 1\arcmin$\times$1\arcmin\ sky pixel.  The science objective driving 
the depth requirement is SO1, the search for the inflationary signal, which 
requires a combined depth of 0.87~\microkamin. This requirement is a combination of the low level of the signal, the need
to separate the various signals detected in each band, and the need to detect and subtract systematic effects 
to anticipated levels.  The map depth requirement flows to instrument sensitivity requirements (Table~\ref{tab:STM}) and to the mission duration requirement (5 years), assuming $95\%$ survey efficiency. \\
% The \ac{CBE} value is 0.61~\microkamin\ coming from a realistic estimate of detector noise, and giving 40\% margin on mission performance. \\
%The required depth assumes full sky coverage, which is d. We expect to only use 50-60\% of the full sky, which do not include the brightest emission from the Milky Way, for the data analysis leading to a constraint on $r$. But as discussed below, other science objectives, require a scan of the Galaxy, leading to the requirement  
%To achieve this combined depth we implement the focal plane of detectors listed inTable~\ref{tab:reqs} The requirement on performance includes the following factors  a factor of 1.5 degradation in noise relative to our baseline expectations. 
%
$\bullet$ {\bf Sky Coverage} \hspace{0.1in} There are several \ac{SOs} driving a full-sky survey for PICO. The term `full-sky' refers to the entire area of sky available after separating astrophysical sources of confusion. In practice this implies an area of 50--70\% of the sky for probing non-Galactic signals, and 100\% of sky for achieving the Galactic science goals. 

(1) Probing the optical depth to the epoch of reionization (SO5) requires full sky coverage as the signal peaks in the $EE$ power spectrum on angular scales of 20$^\circ$ to 90$^\circ$ $(2 \leq \ell \leq 10)$. Measuring this optical depth to limits imposed by cosmic variance\cref{CVL} is key for minimizing the error on the neutrino-mass measurement. 
%\comblue{and for other astrophysical surveys} 

(2) The inflationary $BB$ power spectrum (SO1) has local maxima in the `reionization peak' ($ 2 \leq \ell \leq 10$), and in the `recombination peak' ($ \ell \simeq 80$)~(Fig.~\ref{fig:clbb}). A detection would strongly benefit from confirmation at {\it both} angular scales. Measurements of the reionization peak are currently beyond the capabilities of ground-based instruments. A detection would also strongly benefit from confirmation  {\it in several independent patches of the sky}. This is achievable with PICO through observing the recombination peak in several small (3--5\% sky fraction) patches of the sky. No similar capability is currently planned for any next-decade instrument.  

%(3) The PICO constraint on $N_{\rm eff}$ (SO4) requires all the $\ell$ modes available in the $TT,\, TE$, and $EE$ power spectra, limited by cosmic variance at $\ell=3500\,(2500)$ for $TT\,(EE)$. To achieve this, full sky coverage is required.  
(3) The PICO constraint on $N_{\rm eff}$ (SO4) requires determination of the $TT,\, TE$, and $EE$ power spectra, limited by cosmic variance at $\ell=3500\,(2500)$ for $TT\,(EE)$. To achieve this, full sky coverage is required.  

(4) Achieving the targeted neutrino mass limits (SO3), giving two independent $4\sigma$ constraints on the minimal sum of 58~meV, requires a lensing map and cluster counts from as large a sky fraction as possible. 

(5) PICO's survey of the Galactic plane and regions outside of it is essential to achieving its Galactic structure  and star-formation science goals (SO6, 7). \\
$\bullet$ {\bf Frequency Bands} \hspace{0.1in} The multitude of astrophysical signals that PICO will characterize determine the frequency range and number of bands that the mission requires. The Galactic and cosmological signals are separable using their spectral signatures. The cosmological signals peak in the frequency range between 60 and 300~GHz. Galactic signals, specifically the make-up of Galactic dust (SO6), require spectral characterization at frequencies between 100 and 800~GHz. Simulations indicate that 21 bands, each with 25\% bandwidth, that are spread across the range 20--800~GHz can achieve the separation between Galactic and cosmological signals at the level of fidelity required by PICO~(\S~\ref{sec:signal_separation}). \\
$\bullet$ {\bf Resolution} \hspace{0.1in} 
Several \ac{SOs} require the resolution per frequency listed in Table~\ref{tab:specs}. To reach $\sigma(r) = 1\times10^{-4}$ we will need to `delens' the $B$-mode map, as described in \S~\ref{sec:fundamentalsci} and~\S~\ref{sec:gravitationallensing}. Delensing efficacy is a function of noise and resolution. For PICO, the combination of the two gives between 73 and 85\% delensing, which is adequate for achieving our \ac{SOs}. The process of delensing may be affected by contamination from Galactic dust. It is thus required to map Galactic dust to at least the same resolution as in the main CMB bands.  Higher resolution is mandated by SO6 and 7, which require resolution of 1\arcmin\ at 800~GHz. 

The constraints on the number of light relics (SO4) will be extracted from the $TT,\,TE$ and $EE$ power spectra at $\ell \lesssim 4000$, which requires the resolution specified in Table~\ref{tab:specs}. \\
%We have thus chosen to implement diffracted-limited resolution between 20 and 800~GHz. \\
%
$\bullet$ {\bf Sky Scan Pattern} \hspace{0.1in} 
Control of polarization systematics uncertainties at anticipated levels is enabled by: (1) making $I$, $Q$, and $U$ Stokes-parameter maps of the entire sky from each independent detector; (2) by enabling sub-percent absolute gain calibration of the detectors through observations of the CMB dipole; and (3) by enabling cross-checks on the results through comparing multiple cuts of the data, a process known as `jack-knife test'.   With these requirements we chose a sky scan pattern (\S~\ref{sec:survey_design}) that enables each detector to scan a given pixel of the sky in a multitude of directions, satisfying requirement (1). The scan gives large amplitude CMB dipole signals in spacecraft rotations throughout the lifetime of the mission, satisfying requirement (2). With PICO's sky scan pattern, more than 50\% of the sky is  scanned within two weeks of the start of the survey. The entire sky is surveyed within 6 months, and then this pattern repeats. Thus the PICO scan pattern gives 10 independent maps and multiple ways to perform data jack-knives, satisfying requirement (3).   
% need to be more quantitative. 

%\newpage

% ------------

%\section{Instrument (6 pgs, Hanany \& Transgrud)}

%\subfile{instrument.tex}

% ------------

%\section{Mission (5 pgs, Trangsrud)}

%\subfile{mission.tex}

% ------------

%\section{Technology Maturation (4 pgs, O'Brient \& Trangsrud)}

%\subfile{technology.tex}

% ------------

%\section{Management, Risk, Heritage, and Cost (4 pgs, Trangsrud)}

%\subfile{cost.tex}

% ------------
 \newcommand\pdeg{.\!\!\degree}
\newcommand\parcm{.\!\!'}

\section{Instrument}
\label{sec:instrument} %3

%\begin{wrapfigure}{r}{0.6\textwidth}
%\vskip -8pt
%\hfill
%\includegraphics[width=0.6\textwidth]{images/ArchitectureBlockDiagram.png} 
%\vskip -2pt
%\caption{\captiontext
%PICO instrument block diagram. Active coolers provide cooling to the 100\,mK focal-plane, the surrounding 1\,K box, the 4.5\,K secondary reflector, and the 4.5\,K thermal liner that acts as a cold aperture stop. Data from the focal-plane flows to (redundant, cross-strapped) warm readout electronics on the spun module of the spacecraft bus.
%\label{fig:ArchitectureBlockDiagram} }  
%\end{wrapfigure}

\begin{figure}[h] 
\hspace{-0.1in}
\parbox{5.1in}{\centerline{
\includegraphics[width=5.25in]{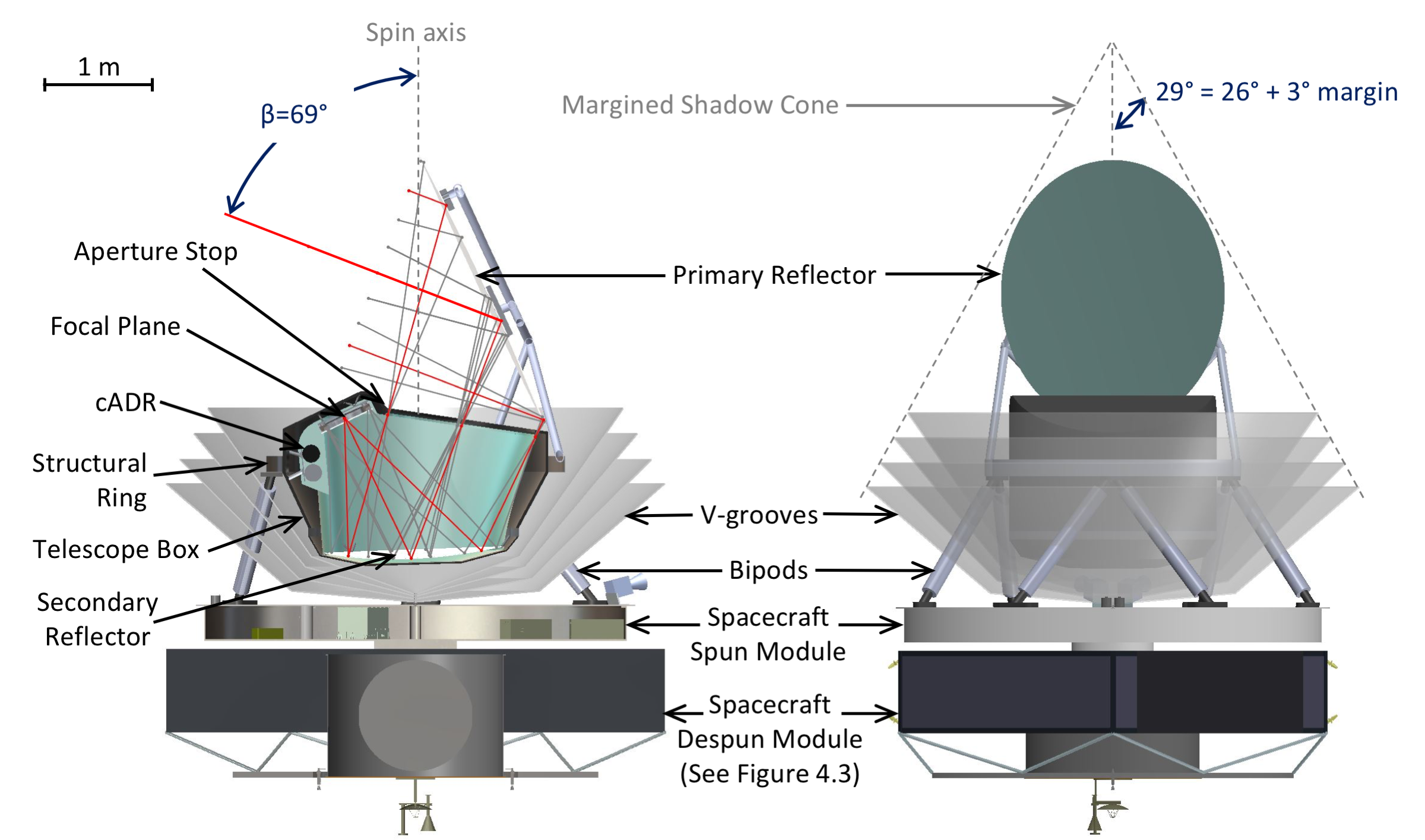} }}
\hspace{0.05in}
\parbox{1.3in}{
\caption{\captiontext
PICO overall configuration in side view and cross section (left), and front view with V-Groove assembly shown semi-transparent (right).  The mission consists of a single science instrument mounted on a structural ring. The ring is supported by bipods on a stage spinning at constant speed relative to a despun module. Figure~\ref{fig:ArchitectureBlockDiagram} shows the functions hosted by each of the modules. 
\label{fig:InstrumentCAD}} }
%\end{center}
\vspace{-0.1in}
\end{figure}

PICO meets all of its science-derived instrument requirements (\S\,\ref{sec:requirements}) with a single instrument: an imaging polarimeter with 21 logarithmically spaced frequency bands centered between 21 and 799\,GHz (Table~\ref{tab:STM}). The instrument has a two-reflector Dragone-style telescope (\S\,\ref{sec:telescope} and Fig.~\ref{fig:InstrumentCAD}). The focal plane is populated by 12,996 \ac{TES} bolometers (\S\,\ref{sec:focal_plane}) and read out using a time-domain multiplexing scheme (\S\,\ref{sec:detector_readout}). PICO employs a single science observing mode: fixed rate imaging while scanning the sky (\S~\ref{sec:survey_design}). 

The instrument is configured inside the shadow of a V-groove assembly that thermally and optically shields it from the Sun (Fig.~\ref{fig:InstrumentCAD} and \S\,\ref{sec:radiative_cooling}). 
%The V-groove assembly consists of four nested radiation shields that provide passive cooling (\S\,\ref{sec:radiative_cooling}). 
The Sun shadow cone depicted in Fig.~\ref{fig:InstrumentCAD} is $29\degree$. The angle to the Sun during the survey, $\alpha = 26\degree$ (\S\,\ref{sec:survey_design} and Fig.~\ref{fig:MissionDesignFigure}), is supplemented with a margin of $3\degree$ to account for the radius of the Sun ($0\pdeg25$), pointing control error, design margin, and alignment tolerances.

The V-groove assembly is attached to the bipod struts that support the instrument structural ring. The ring supports the primary reflector and telescope box. The telescope box contains the actively cooled components (\S\,\ref{sec:cadr}, \S\,\ref{sec:4kcooler}), including the secondary reflector, the focal plane and sub-kelvin refrigerator structures. Just inside the box, a thermal liner serves as a cold optical baffle and aperture stop. Instrument integration and test are described in \S\,\ref{sec:iandt}.

\begin{figure}[h]
\hspace{-0.in}
\parbox{4.1in}{\centering{
\includegraphics[width=4.0in]{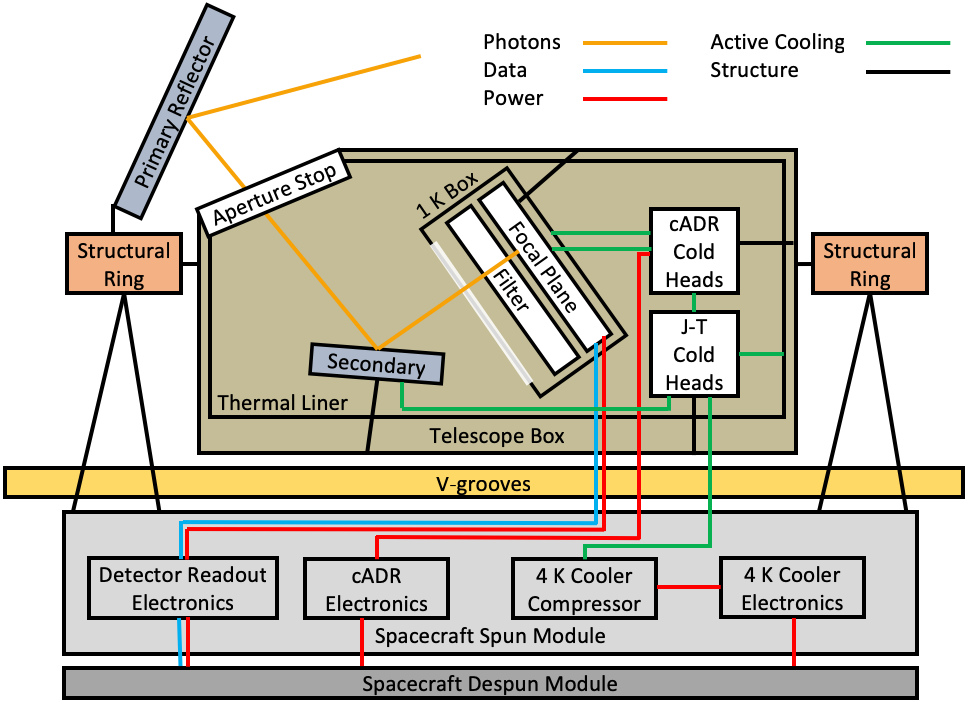} } }
\hspace{0.12in}
\parbox{2.2in}{
\caption{\captiontext
PICO instrument block diagram. Active coolers provide cooling to the 100\,mK focal plane, the surrounding 1\,K box, the 4.5\,K secondary reflector, and the 4.5\,K thermal liner that acts as a cold aperture stop. V-grooves provide passive cooling. The instrument, V-grooves, and spacecraft spun module spin together at a rate of 1\,RPM. The spacecraft spun module hosts the 4\,K cooler compressor and drive electronics, the sub-K cooler drive electronics, and the detector warm readout electronics. Only power and digital data lines cross to the spacecraft despun module, which hosts the spacecraft power, telemetry, attitude control, and communication systems (\S\,\ref{sec:spacecraft}).
\label{fig:ArchitectureBlockDiagram} } } 
\vspace{-0.in}
\end{figure}

During the survey, the instrument is spun at 1~rpm and the spin axis is made to precess about the anti-Sun direction (\S\,\ref{sec:survey_design}). Spacecraft control is simplified by mounting the instrument on a spinning spacecraft module, while a larger non-spinning module houses most spacecraft subsystems (\S\,\ref{sec:spacecraft}). Instrument elements that act as heat sources are accommodated on the spinning module of the spacecraft. Only power and digital data lines cross between the spinning and non-spinning modules.  A functional block diagram of the instrument is shown in Figure~\ref{fig:ArchitectureBlockDiagram}. 

\subsection{Telescope}
\label{sec:telescope} %3.1

The PICO telescope design is driven by a combination of science requirements and physical volume limits. The science requirements are: a large diffraction-limited field of view (DLFOV) sufficient to support approximately $10^4$ detectors; arcminute resolution at 800\,GHz; low spurious polarization; and low sidelobe response. All requirements are met with PICO's 1.4\,m aperture modified open-Dragone design. There are no moving parts in the PICO optical system.

%\begin{figure}%[ht]
%\parbox{4.0in}{
%\includegraphics[width=3.5in]{images/ArchitectureBlockDiagram.png} }
%\hspace{0.1in}
%\parbox{2.3in}{
%\caption{\captiontext
%PICO instrument block diagram. Active coolers provide cooling to the 100\,mK focal-plane, the surrounding 1\,K box, the 4.5\,K secondary reflector, and the 4.5\,K thermal liner that acts as a cold aperture stop. Data from the focal-plane flows to (redundant, cross-strapped) warm readout electronics on the spun module of the spacecraft bus.
%\label{fig:ArchitectureBlockDiagram} }  }
%\end{figure}

The PICO optical design was selected following a trade study examining cross-Dragone, Gregorian Dragone, and open-Dragone designs~\citep{Young2018}.  The open-Dragone and crossed-Dragone systems offer more diffraction-limited focal-plane area than the Gregorian Dragone one~\citep{deBernardis2018} and are able to support enough detectors to provide the required sensitivity. The open-Dragone design does not require the more massive and voluminous baffles that the cross-Dragone does, and hence can satisfy the aperture size requirement within the shadow cone.

PICO's initial open-Dragone design~\citep{dragone78, granet2001} has been modified with the addition of an aperture stop and adding corrections to the primary and secondary reflectors to enlarge the DLFOV. The detailed geometric parameterization of the PICO optical design is described by~\citet{Young2018}. The primary reflector (270\,cm $\times$ 205\,cm) is passively cooled and the secondary reflector (160\,cm $\times$ 158\,cm) is actively cooled. The highest frequency (900\,GHz) sets the surface accuracy requirement of the reflectors at $\lambda/14 =24\,\mu{\rm m}$. The focal ratio is 1.42. The slightly concave focal surface, which has a radius of curvature of 4.55~m, is telecentric to within $0\pdeg12$ across the entire FOV.

%\comblue{what about the reflector material?}
An actively cooled circular aperture stop between the primary and secondary reflectors reduces detector noise and shields the focal plane from stray radiation. Stray-light analysis of the PICO open-Dragone design using GRASP confirms that the focal plane is protected from direct view of the sky, and that spillover past the primary is suppressed by 80~dB relative to the main lobe for both co-pol and cross-pol beams. Detailed baffle design will be performed during mission formulation.

% \begin{figure}
% \begin{center}
% \includegraphics[width=3in]{figures/OpticsDiagram.png}
% \caption{The optical system is compact.\label{fig:OpticsDiagram}}
% \end{center}
% \end{figure}

\subsection{Focal Plane}
\label{sec:focal_plane} %3.2
%
% wrap version of figure 3.3.  Unwrapped version is below.

\begin{wrapfigure}{r}{0.45\textwidth}
\vskip -8pt
\hfill
\includegraphics[width=0.45\textwidth]{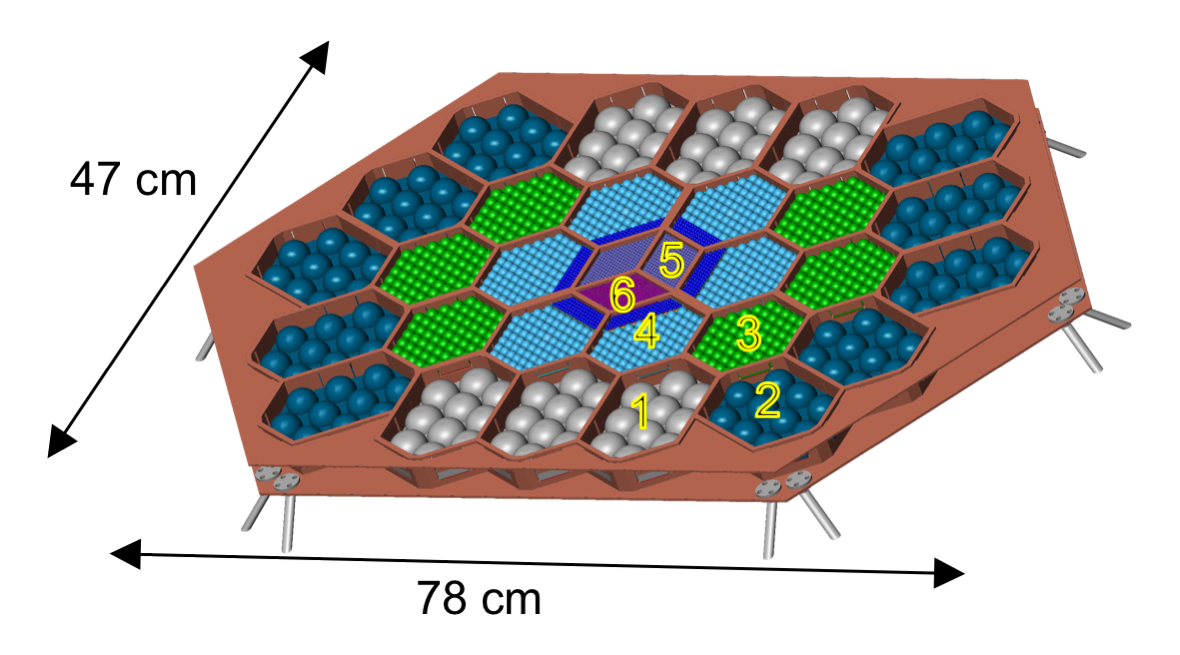}
\vskip -4pt
\caption{\captiontext PICO focal plane. Detectors are fabricated on six types of tiles (shown numbered and colored as in Table~\ref{tab:focal_plane}). The wafers are located on the focal plane such that higher frequency bands, which require better optical performance, are placed nearer to the center. All detectors are within the diffraction-limited performance for their respective frequency bands.  
\label{fig:FocalPlaneMechanical}}
\end{wrapfigure}

PICO's focal plane is populated by an array of \ac{TES} bolometers operating in 21 frequency bands, each with 25\% fractional bandwidth, and band centers ranging from 21 to 799~GHz. 
The layout of the PICO focal plane is shown in Fig.~\ref{fig:FocalPlaneMechanical} and detailed in Table~\ref{tab:focal_plane}. 
%A conceptual layout of the PICO focal-plane is shown in Fig.~\ref{fig:FocalPlaneMechanical} and detailed in Table~\ref{tab:focal_plane}.

Bolometers operating in the mm/sub-mm wave band are photon-noise limited. Therefore, increase in sensitivity is achieved through an increase in detector count. The PICO focal plane has 12,996 detectors, 175~times the number flown aboard \planck , thereby providing a breakthrough increase in sensitivity
with a comparably sized telescope. This breakthrough is enabled by development and demonstration in suborbital projects, which now commonly operate arrays of $10^3$--$10^4$ detectors (\S\,\ref{sec:technology_maturation}). Further technology maturation required for PICO is described in Section~\S\,\ref{sec:technology_maturation}.

\subsubsection{21--462\,GHz Bands}
\label{sec:low_freq_det} % 3.2.1
\vspace{0.02in}

Several optical-coupling technologies have matured over the past ten years to efficiently use focal-plane area: horns with ortho-mode transducers (OMTs) \citep{Duff2016}; lithographed antenna arrays~\citep{BICEP2015}; and sinuous antennas under lenslets~\citep{Edwards2012}. Horn-coupling and sinuous antenna/lenslet-coupling deliver quantum efficiency $>70\,\%$ over more than an octave of bandwidth, which have been partitioned into two or three colors per pixel.  Only single-color pixels have been demonstrated to date with antenna-arrays, but this coupling enables smaller pixels and therefore they can be more densely packed.
%%%%%%%%
% Table 3.1

\begin{wraptable}[20]{r}{0.45\textwidth}
\vskip -20pt
\hfill
\begin{minipage}{0.45\textwidth}
\caption{\captiontext PICO makes efficient use of the focal area with multichroic pixels (three bands per pixel, \S\,\ref{sec:low_freq_det}). The sampling rate is based on the smallest beam (Table~\ref{tab:bands}), with 3 samples per FWHM at a scan speed $(360\degree/{\rm min})\sin(\beta=69\degree) = 336\degree/{\rm min}$. Scaling from suborbital experience, we anticipate that TES bolometers can support these sampling rates with $\sim 4\times$ margin.\label{tab:focal_plane}}
\begingroup
\definecolor{A}{RGB}{209, 209, 209}
\definecolor{B}{RGB}{64, 128, 159}
\definecolor{C}{RGB}{112, 208, 85}
\definecolor{D}{RGB}{109, 196, 232}
\definecolor{E}{RGB}{32, 49, 184}
\definecolor{F}{RGB}{119, 120, 199}
\definecolor{G}{RGB}{235, 92, 243}
%\openup 5pt
\newdimen\tblskip \tblskip=5pt
\nointerlineskip
\vskip -5mm
\footnotesize %\footnotesize
\setbox\tablebox=\vbox{
    \newdimen\digitwidth
    \setbox0=\hbox{\rm 0}
    \digitwidth=\wd0
    \catcode`*=\active
    \def*{\kern\digitwidth}
    \newdimen\signwidth
    \setbox0=\hbox{+}
    \signwidth=\wd0
    \catcode`!=\active
    \def!{\kern\signwidth}
\halign{
\hfil#\hfil\tabskip=0.6em&
\hfil#\hfil&
\hfil#\hfil&
\hfil#\hfil&
\hfil#\hfil&
\hfil#\hfil\tabskip=0pt\cr
\noalign{\doubleline}
\omit \hfil Tile\hfil&&Pixels/&Pixel&Band centers&Sampling\cr
\omit \hfil type\hfil&$N_{\rm tile}$&tile&type&[GHz]&rate [Hz]\cr
\noalign{\vskip 3pt\hrule\vskip 5pt}
1&6&10&\colorbox{A}{A}&21, 30, 43&45\cr
\noalign{\vskip 5pt\hrule height 0.2pt\vskip 3pt}
2&10&10&\colorbox{B}{\textcolor{White}B}&25, 36, 52&55\cr
\noalign{\vskip 5pt\hrule height 0.2pt\vskip 3pt}
3&6&61&\colorbox{C}{C}&62, 90, 129&136\cr
\noalign{\vskip 5pt\hrule height 0.2pt\vskip 3pt}
4&6&85&\colorbox{D}{D}&75, 108, 155&163\cr
\omit&\omit&80&\colorbox{E}{\textcolor{White}E}&186, 268, 385&403\cr
\noalign{\vskip 5pt\hrule height 0.2pt\vskip 3pt}
5&2&450&\colorbox{F}{\textcolor{White}F}&223, 321, 462&480\cr
\noalign{\vskip 5pt\hrule height 0.2pt\vskip 3pt}
6&1&220&\colorbox{G}{G}&555&917\cr
\omit&\omit&200&\colorbox{G}{H}&666\cr
\omit&\omit&180&\colorbox{G}{I}&799\cr
\noalign{\vskip 5pt\hrule\vskip 3pt}
} % close halign
} % close vbox
\endPlancktable
\endgroup
\end{minipage}
\end{wraptable}

%%%%%%%%

The PICO baseline focal plane employs three-color sinuous antenna/lenslet pixels~\citep{Suzuki2014} for the 21--462\,GHz bands. Niobium microstrips mediate the signals between the antenna and detectors, and partition the %feed's 
wide continuous bandwidth into three narrow channels using integrated, on-wafer, micro-machined filter circuits~\citep{OBrient2013}. Six transition edge sensor bolometers per pixel detect the radiation in two orthogonal polarization states. %The technology maturation required for PICO is described in \$\,\ref{sec:bolometers}.

\subsubsection{555--799\,GHz Bands}
\label{sec:high_freq_det} % 3.2.2

PICO's highest three frequency channels are beyond the niobium superconducting band-gap, rendering on-wafer, microstrip filters a poor solution for defining the optical passband. For these bands we use feedhorns to couple the radiation to two single-color polarization-sensitive TES bolometers. %Radiation is coupled through horns directly to an absorber in the throat of a waveguide. TES bolometers detect the incident power.  
The waveguide cut-off defines the lower edge of the band, and quasi-optical metal-mesh filters define the upper edge. Numerous experiments have successfully used similar approaches~\citep{Shirokoff2011,Bleem2012,Turner2001}. 
%The technology maturation required for PICO is described in \S\,\ref{sec:dev_arrays}.

%%%%%
% Table 3.2

\begin{table}[tb]
\caption{\captiontext 
PICO has 21 partially overlapping frequency bands with band centers ($\nu_{\rm c}$) from 21\,GHz to 799\,GHz and each with bandwidth $\Delta\nu/\nu_{\rm c} = 25\,\%$. The beams are single mode, with FWHM sizes of $6\parcm2 \times (155\,{\rm GHz}/\nu_{\rm c})$. The \ac{CBE} per-bolometer sensitivity is photon-noise limited (\S\,\ref{sec:sensitivity}). The total number of bolometers for each band is equal to (number of tiles) $\times$ (pixels per tile) $\times$ (2 polarizations per pixel), from Table~\ref{tab:focal_plane}. Array sensitivity assumes $90\,\%$ detector operability. The map depth assumes 5\,yr of full sky survey at $95\,\%$ survey efficiency, except the 25 and 30\,GHz frequency bands, which are conservatively excluded during 4\,hr/day Ka-band (26\,GHz) telecom periods (\S\,\ref{sec:ground_segment}).\label{tab:bands}}
\begingroup
\def\nunit{[$\mu{\rm K}_{\rm CMB}\,{\rm s}^{1/2}$]}
%\openup 5pt
\newdimen\tblskip \tblskip=5pt
\nointerlineskip
\vskip -5mm
\footnotesize %\footnotesize
\setbox\tablebox=\vbox{
    \newdimen\digitwidth
    \setbox0=\hbox{\rm 0}
    \digitwidth=\wd0
    \catcode`*=\active
    \def*{\kern\digitwidth}
    \newdimen\signwidth
    \setbox0=\hbox{+}
    \signwidth=\wd0
    \catcode`!=\active
    \def!{\kern\signwidth}
\halign to \textwidth{
\hbox to 0.8in{#\leaderfil}\tabskip=0.6em plus 1em&
\hfil#\hfil&
\hfil#\hfil&
\hfil#\hfil&
\hfil#\hfil&
\hfil#\hfil&
\hfil#\hfil&
\hfil#\hfil\tabskip=0pt\cr
\noalign{\doubleline}
\omit \hfil Band\hfil&Beam&CBE&&CBE&Baseline&\multispan2\hfil Baseline polarization\hfil\cr
\omit \hfil center\hfil&FWHM&bolo NET&$N_{\rm bolo}$&array NET&array NET&\multispan2\hfil map depth\hfil\cr
\noalign{\vskip -5pt}
\omit&\omit&\omit&\omit&\omit&\omit&\multispan2{\hrulefill}\cr
\omit \hfil[GHz]\hfil&[arcmin]&\nunit&&\nunit&\nunit&[$\mu{\rm K}_{\rm CMB}\,{\rm arcmin}$]&[Jy\,sr$^{-1}$]\cr
%\omit \hfil Tile\hfil&&Pixels/&Pixel&Bandcenters&Sampling\cr
%\omit \hfil type\hfil&$N_{\rm tile}$&tile&type&[GHz]&rate [Hz]\cr
\noalign{\vskip 3pt\hrule\vskip 5pt}
*21&38.4&*112&*120&12.0&17.0&23.9&*8.3\cr
*25&32.0&*103&*200&*8.4&11.9&18.4&10.9\cr
*30&28.3&*59.4&*120&*5.7&8.0&12.4&11.8\cr
*36&23.6&*54.4&*200&*4.0&5.7&*7.9&12.9\cr
*43&22.2&*41.7&*120&*4.0&5.6&*7.9&19.5\cr
*52&18.4&*38.4&*200&*2.8&4.0&*5.7&23.8\cr
*62&12.8&*69.2&*732&*2.7&3.8&*5.4&45.4\cr
*75&10.7&*65.4&1020&*2.1&3.0&*4.2&58.3\cr
*90&*9.5&*37.7&*732&*1.4&2.0&*2.8&59.3\cr
108&*7.9&*36.2&1020&*1.1&1.6&*2.3&77.3\cr
129&*7.4&*27.8&*732&*1.1&1.5&*2.1&96.0\cr
155&*6.2&*27.5&1020&*0.9&1.3&*1.8&119\cr
186&*4.3&*70.8&*960&*2.0&2.8&*4.0&433\cr
223&*3.6&*84.2&*900&*2.3&3.3&*4.5&604\cr
268&*3.2&*54.8&*960&*1.5&2.2&*3.1&433\cr
321&*2.6&*77.6&*900&*2.1&3.0&*4.2&578\cr
385&*2.5&*69.1&*960&*2.3&3.2&*4.5&429\cr
462&*2.1&*133&*900&*4.5&6.4&*9.1&551\cr
555&*1.5&*658&*440&23.0&32.5&45.8&1580*\cr
666&*1.3&2210&*400&89.0&126&177&2080*\cr
799&*1.1&10400*&*360&526&744&1050&2880*\cr
\noalign{\vskip 3pt\hrule\vskip 5pt}
\multispan3\quad\quad Total\leaderfil&12\,996&0.43&0.61&0.87&\cr
\noalign{\vskip 5pt\hrule\vskip 3pt}
} % close halign
} % close vbox
\endPlancktable
\endgroup
\end{table}

%%%%%

\subsubsection{Polarimetry}
\label{sec:polarimetry} %3.2.3

Polarimetry is achieved by differencing the signals from pairs of two co-pointed bolometers within a pixel that are sensitive to two orthogonal polarization states. Half the pixels in the focal plane are sensitive to the $Q$ and half to the $U$ Stokes parameters of the incident radiation. Two layouts for the distribution of the $Q$ and $U$ pixels on the focal plane have been investigated~\citep{picoweb_QU}; both would satisfy mission requirements. Stokes $I$ is obtained from the sum of the signals of orthogonal detectors.  

\subsubsection{Sensitivity}
\label{sec:sensitivity} %3.2.3

PICO's Current Best Estimate (CBE) sensitivity meets the requirements of the baseline mission with \hbox{$>40\,\%$} margin (Table~\ref{tab:bands}).
%\comblue{this is an odd statement. need to rework}

We developed an end-to-end noise model of the PICO instrument to predict mission sensitivity and provide a metric by which to evaluate mission design trades. The model includes four noise sources per bolometer: photon, phonon, Johnson, and readout (from both cold and warm readout electronics). To validate our calculations, we compared two independent software packages that have been validated with several operating CMB instruments. The calculations agreed within 1\% both for individual noise terms and for overall mission noise. A detailed description of the PICO noise model and its inputs is available in~\citet{Young2018}; small differences between that publication and Table~\ref{tab:bands} are due to refinements of the primary mirror and stop temperatures.

Laboratory experiments have demonstrated that TES bolometers can be made background-limited in the low loading environment they would experience at L2~\citep{Beyer2012}. For PICO, the primary contributor to noise is the optical load. The sources of optical load are the CMB, reflectors, aperture stop, and low-pass filters. The CMB and stop account for at least 50\% of the optical load at all frequencies up to and including 555~GHz. At higher bands emission from the primary mirror dominates.

The sensitivity model assumes white noise at all frequencies. Sub-orbital submillimeter experiments have demonstrated TES detectors that are stable to at least as low as 20\,mHz \citep{Rahlin2014}, meeting the requirements for PICO's scan strategy (\S\,\ref{sec:survey_design}). 

\subsection{Detector Readout}
\label{sec:detector_readout} %3.3

Suborbital experiment teams over the past ten years have chosen to use voltage-biased TESs because their current readout scheme lends itself to superconducting quantum interface device (SQUID)-based multiplexing. Multiplexing reduces the number of wires to the cryogenic stages and thus the total thermal load that the cryocoolers must dissipate. This approach also simplifies the instrument design.

In the multiplexing circuitry, SQUIDs function as low-noise amplifiers and cryogenic switches. The current baseline for PICO is to use a time-domain multiplexer (TDM), which assigns each detector's address in a square matrix of simultaneously read columns, and sequentially cycles through each row of the array~\citep{Henderson2016}. The PICO baseline architecture uses a matrix of 128 rows and 102 columns. The thermal loading on the cold stages from the wire harnesses is subdominant to conductive loading through the mechanical support structures.

Because SQUIDs are sensitive magnetometers, suborbital experiments have developed techniques to shield them from Earth's magnetic field using highly permeable or superconducting materials~\citep{Hui2018}.  Total suppression factors better than $10^7$ have been demonstrated for dynamic magnetic fields~\citep{Runyan2010}. PICO will use these demonstrated techniques to shield SQUID readout chips from the ambient magnetic environment, which is 20,000 times smaller than near Earth, as well as from fields generated by on-board components, including the 0.1\,K cooler (\S\,\ref{sec:cadr}). This cooler is delivered with its own magnetic shielding, which reduces the field at the distance of the SQUIDs to less than 0.1\,G, which is less than Earth's field experienced by SQUIDs aboard suborbital experiments.
SQUIDs are also sensitive to radio-frequency interference (RFI). Several suborbital experiments have demonstrated RFI shielding using aluminized mylar wrapped at cryogenic stages to form a Faraday cage around the SQUIDs~\citep{Kermish2012,EBEX2018,BICEP2014}. Cable shielding extends the Faraday cage to the detector warm readout electronics.

Redundant warm electronics boxes perform detector readout and instrument housekeeping using commercially available radiation-hardened analog-to-digital converters, requiring 75\,W total. The readout electronics compress the data before delivering them to the spacecraft, requiring an additional 15\,W. PICO detectors produce a total of 6.1\,Tbits/day assuming 16\,bits/sample, sampling rates from Table~\ref{tab:focal_plane}, and bolometer counts from Table~\ref{tab:bands}. \planck \ HFI had a typical 4.7$\times$ compression in flight, with information loss increasing noise by only about $10\,\%$~\citep{Pajot2018,PlanckHFI2011}. Suborbital work has demonstrated 6.2$\times$ lossless compression~\citep{EBEX2017}. PICO assumes 4$\times$ lossless compression.

\medskip
\subsection{Thermal}
\label{sec:thermal} %3.4

Like the \planck -HFI instrument, PICO's focal plane is maintained at 0.1\,K to ensure low detector noise while implementing readily available technology~(\S\,\ref{sec:cadr}). To minimize detector noise due to instrument thermal radiation, the aperture stop and reflectors are cooled using both active and radiative cooling (\S\,\ref{sec:4kcooler}, \S\,\ref{sec:radiative_cooling}, Fig.~\ref{fig:ArchitectureBlockDiagram}).  All thermal requirements are met with robust margins (Table~\ref{tab:cooler}).

\definecolor{mygray}{gray}{0.6}
\begin{table}[tb]
\caption{\captiontext Projected cooler heat lift capabilities offer  more than $100\,\%$ heat lift margin, complying with cooler technology best practices~\citep{Donabedian2003}.\label{tab:cooler}}
\begingroup
%\openup 5pt
\newdimen\tblskip \tblskip=5pt
\nointerlineskip
\vskip -5mm
\footnotesize %\footnotesize
\setbox\tablebox=\vbox{
    \newdimen\digitwidth
    \setbox0=\hbox{\rm 0}
    \digitwidth=\wd0
    \catcode`*=\active
    \def*{\kern\digitwidth}
    \newdimen\signwidth
    \setbox0=\hbox{+}
    \signwidth=\wd0
    \catcode`!=\active
    \def!{\kern\signwidth}
\halign to \textwidth{
\hbox to 2.5in{#\leaderfil}\tabskip=1em plus 1em&
\hfil#\hfil\tabskip=1em&
\hfil#\hfil\tabskip=1.5em plus1em&
\hfil#\hfil\tabskip=1em&
\hfil#\hfil&
\hfil#\hfil\tabskip=0pt\cr
\noalign{\doubleline}
\omit&\multispan2 Temperature [K]&\multispan3 \hfil Active heat lift [mW]\hfil\cr
\noalign{\vskip -3pt}
\omit&\multispan2\hrulefill&\multispan3\hrulefill\cr
\omit\hfil Component\hfil&Required&CBE&Required&Capability&Projected\cr
\omit&\omit&\omit&per model$^{a}$&today&capability\cr
\noalign{\vskip 3pt\hrule\vskip 3pt}
Primary reflector & $< 40$& $17$& \multispan3\hfil N/A (radiatively cooled) \hfil\cr
\noalign{\vskip 3pt{\color{mygray}\hrule}\vskip 3pt}
Secondary reflector & $<8$&$4.5$\cr%\noalign{\vskip-1em}
Aperture stop& $4.5$&$4.5$&$42$ at 4.5\,K&$> 55$ at 6.2\,K$^{b}$&$>100$ at 4.5\,K$^{c}$\cr%\noalign{\vskip-1em}
cADR heat rejection$^{d}$ & $4.5$&$4.5$\cr
\noalign{\vskip 3pt{\color{mygray}\hrule}\vskip 3pt}
Focal plane enclosure and filter&$1.0$&$1.0$&$0.36$&$1.0$&  N/A$^{e}$  \cr
\noalign{\vskip 3pt{\color{mygray}\hrule}\vskip 3pt}
Focal plane&$0.1$&$0.1$&$5.7\times10^{-3}$&$32\times10^{-3}$&  N/A$^{e}$  \cr
\noalign{\vskip 5pt\hrule\vskip 3pt}
} % close halign
\noindent{$^{a}$}~The required loads were calculated using Thermal Desktop. Reference \citep{Ross2004} was used to estimate the thermal conductive loads through mechanical supports. In addition to the listed components, the total 4.5\,K heat load includes the intercept on the focal plane mechanical supports.\qquad {$^{b}$}~Reference \cite{Petach2014}.\qquad {$^{c}$}~Both NGAS and Ball project $>100$\,mW lift capability at $4.5\,$K using higher compression-ratio compressors currently in development (\S\,\ref{sec:4kcooler} and Fig.~\ref{fig:CoolerFigure}).\qquad {$^{d}$}~The cADR lift capability at 1\,K and 0.1\,K is from a GSFC quote.\qquad {$^{e}$} Capability today already exceeds requirement.\par
} % close vbox
\endPlancktable
\endgroup
\end{table}

\smallskip
\subsubsection{cADR Sub-Kelvin Cooling}
\label{sec:cadr} %3.4.1

A multi-stage continuous adiabatic demagnetization refrigerator (cADR) maintains the PICO focal plane at 0.1\,K and the surrounding enclosure, filter, and readout components at 1\,K. The cADR employs three refrigerant assemblies operating sequentially to absorb heat from the focal plane at 0.1\,K and reject it to~1\,K. Two additional assemblies, also operating sequentially, absorb this rejected heat at~1\,K, cool other components to 1\,K, and reject heat at~4.5\,K. This configuration provides continuous cooling with small temperature variations at both the 0.1\,K and 1\,K. Heat straps connect the two cADR cold sinks to multiple points on the focal-plane assembly,
%\comor{was 'focal-plane assembly' defined? isn't it at 0.1K?}
which has high thermal conductance paths built in, to provide spatial temperature uniformity and stability during operation. The detector arrays are thermally sunk to the mounting frame.  Heat loads in the range of 30~$\mu$W at 0.1\,K and 1\,mW at 1\,K (time-average) are within the capabilities of current cADRs developed by GSFC (\S~\ref{sec:heritage})~\citep{Shirron2012,Shirron2016}. The PICO sub-kelvin heat loads are estimated at less than half of this capability (Table~\ref{tab:cooler}).
%\comblue{this paragraph doesn't say anything about the maturity of the multiple stage cADR, and about flight heritage. It should, even if to point to later paragraphs.}

\subsubsection{The 4.5 K Cooler}
\label{sec:4kcooler} %3.4.2

\begin{wrapfigure}{r}{0.4\textwidth}
\vspace{-8pt}
\includegraphics[width=2.6in]{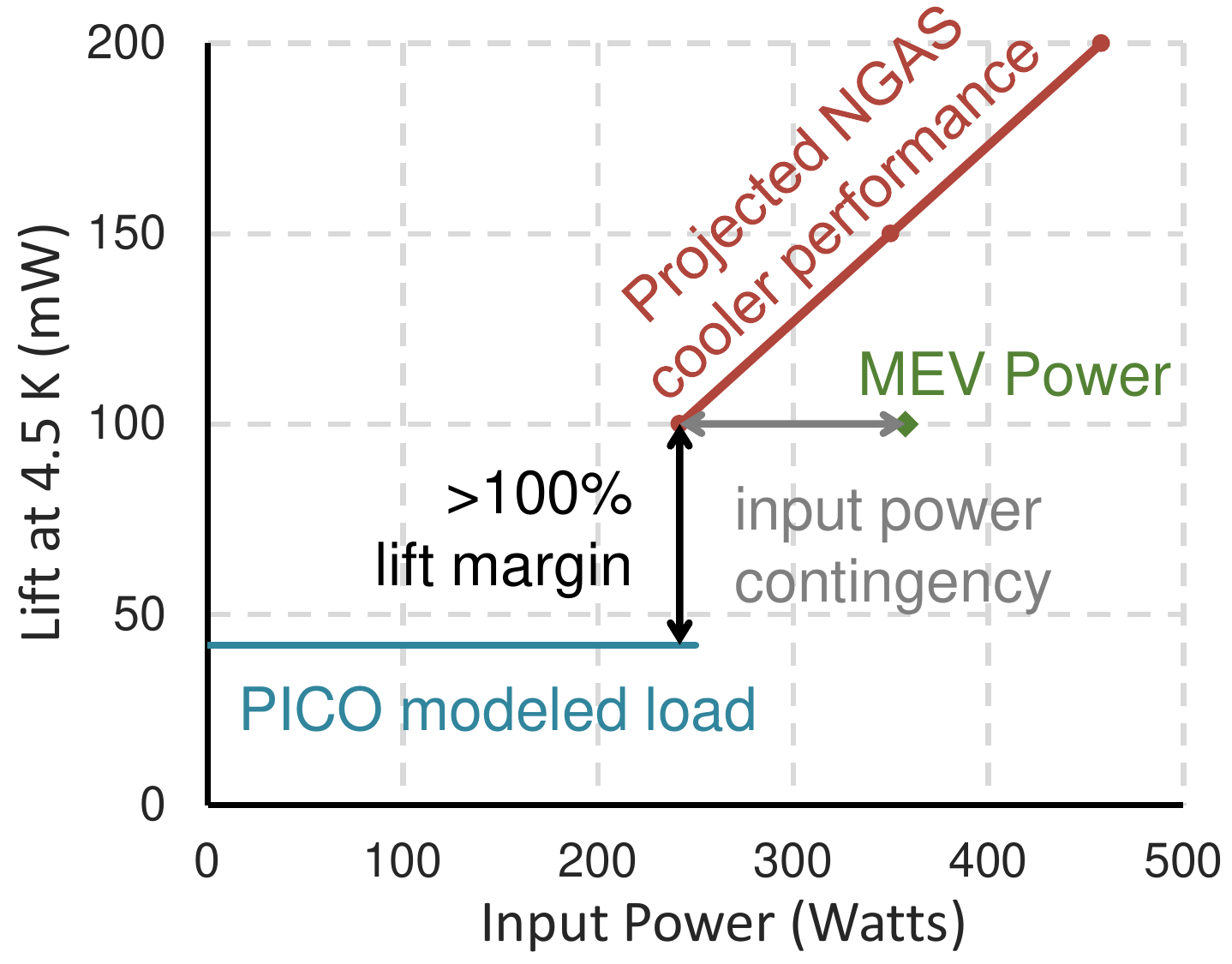} 
\caption{\captiontext
Projected performance of the NGAS cooler using a multi-stage
compressor and $^4$He circulating gas~\citep{Rabb2013} meets PICO's requirements
with $>100\,\%$ margin. PICO requires heat lift of 42\,mW at 4.5\,K (Table~\ref{tab:cooler}). With 250\,W of input power the NGAS cooler is projected to provide 100\,mW of heat lift. We conservatively specify a maximum expected value (MEV) of 350\,W as the compressor's input power, giving 100\,W of additional input power contingency.
\label{fig:CoolerFigure}} 
%\vspace{-0.15in}
\end{wrapfigure}
A cryocooler system similar to that used on JWST to cool the MIRI detectors~\citep{Durand2008,Rabb2013} removes the heat rejected from the cADR and cools the aperture stop and secondary reflector to 4.5\,K. Both NGAS (which provided the MIRI coolers) and Ball Aerospace have developed such coolers under the NASA-sponsored Advanced Cryocooler Technology Development Program~\citep{Glaister2006}. NGAS and Ball use slightly different but functionally-equivalent hardware approaches. A 3-stage precooler provides 16\,K precooling to a separate circulated-gas loop. The circulated-gas loop utilizes Joule--Thomson (J-T) expansion, further cooling the gas to 4.5\,K. The J-T expansion point is located close to the cADR heat rejection point and provides to it the lowest temperature. Subsequently, the gas flow intercepts heat conducted to the focal-plane enclosure, then cools the aperture stop and the secondary reflector before returning to the circulation compressor.  %Model-based projections indicate that the coolers delivered for MIRI could meet the PICO 4.5~K heat lift requirement with more than $100\,\%$ margin with these straightforward modifications: replacement of the $^4$He gas used for MIRI's J-T  with $^3$He; and resizing the $^3$He heat exchangers to take advantage of the different gas properties.

NGAS and Ball are actively working on increasing the flow rate and compression ratio of the J-T compressor,  which should result in higher system efficiency and greater heat-lift relative to the current MIRI cooler.  NGAS uses $^4$He as the circulating gas, as was used for MIRI. Ball uses a somewhat larger compressor and $^3$He as the circulating gas. Both employ re-optimized heat exchangers. The NGAS project has completed PDR-level development, and is expected to reach CDR well before PICO begins Phase-A. The projected performance of this cooler is shown in Fig.~\ref{fig:CoolerFigure}; it gives 100~mW at 250\,W input power, which is more than 100\,\% heat lift margin relative to PICO's requirements (Table~\ref{tab:cooler}). For PICO we have assumed an input power of 350\,W.

The entire precooler assembly and the J-T circulator compressor are located on the warm spacecraft spun module (Fig.~\ref{fig:ArchitectureBlockDiagram}).
%, with relatively short tubing lengths conducting the gas flow from the precooling point to the J-T expansion point.
All waste heat rejected by the cooler compressors and drive electronics is transferred to the spacecraft heat-rejection system. Unlike JWST, the PICO cooler does not require deployment of the remote cold head.

\smallskip
\subsubsection{Radiative Cooling}
\label{sec:radiative_cooling} %3.4.3

%The V-groove assembly consists of four nested radiation shields that provide passive cooling (\S\,\ref{sec:radiative_cooling}). 
An assembly of four nested V-groove radiators, acting as radiation shields, provides passive cooling (Fig.~\ref{fig:InstrumentCAD}). This is standard, 30-years old technology~(\S~\ref{sec:heritage}). The outermost shield shadows the interior ones from the Sun. The V-grooves radiate to space, each reaching successively cooler temperatures. The assembly provides a cold radiative environment to the primary reflector, structural ring, and telescope box. As a consequence radiative loads on those elements are smaller than the conductive loads through the mechanical support structures.

\subsection{Instrument Integration and Test}
\label{sec:iandt} % 3.5

The PICO instrument integration and testing plan benefits from heritage and experience with the \planck\ HFI instrument~\citep{Pajot2010}.

We screen detector wafers prior to selection of flight wafers and focal-plane integration. The cADR and 4\,K cryocooler vendors will qualify them prior to delivery. We will determine the relative alignment of the two reflectors under in-flight thermal conditions using a thermal vacuum (TVAC) chamber and photogrammetry. We integrate the flight focal-plane assembly and flight cADR in a dedicated sub-kelvin cryogenic testbed. We characterize noise, responsivity, and focal-plane temperature stability using a representative optical load for each frequency band (temperature-controlled blackbody), and we perform polarimetric and spectroscopic calibration.

%PICO screens detector wafer performance prior to selection of flight wafers and focal-plane integration. The cADR and 4\,K cryocooler are qualified prior to delivery. The relative alignment of the two reflectors under thermal contraction is photogrammetrically verified in a thermal vacuum (TVAC) chamber.

%PICO integrates the flight focal-plane assembly and flight cADR in a dedicated sub-kelvin cryogenic testbed. Noise, responsivity, and focal-plane temperature stability are characterized using a representative optical load for each frequency band (temperature-controlled blackbody). Polarimetric and spectroscopic calibration are performed.

The focal plane is integrated with the reflectors and structures, and alignment verified with photogrammetry at cold temperatures in a TVAC chamber.  The completely integrated observatory (instrument and spacecraft bus) is tested in TVAC to measure parasitic optical loading from the instrument, noise, microphonics, and RFI. The observatory is 4.5\,m in diameter and 6.1\,m tall. There are no deployables.

\bigskip
\section{Design Reference Mission}
\label{sec:design_reference} %4
%
%% Table 4.1

%\begin{table}
\begin{wraptable}[12]{r}{3.45in}
\vspace{-7pt}
\begin{minipage}{3.5in}
\caption{\captiontext PICO carries margin on key mission parameters. Maximum Expected Value (MEV) includes contingency.}\label{tab:mission_parameters}
\begingroup
%\openup 5pt
\newdimen\tblskip \tblskip=5pt
\nointerlineskip
\vskip -7mm
\footnotesize %\footnotesize
\setbox\tablebox=\vbox{
    \newdimen\digitwidth
    \setbox0=\hbox{\rm 0}
    \digitwidth=\wd0
    \catcode`*=\active
    \def*{\kern\digitwidth}
    \newdimen\signwidth
    \setbox0=\hbox{+}
    \signwidth=\wd0
    \catcode`!=\active
    \def!{\kern\signwidth}
\halign{
\hbox to 1.7in{#\leaderfil}\tabskip=0.6em&
\vtop{\hsize 1.75in\raggedright\hangafter=1\hangindent=1em\noindent\strut#\strut\par}\tabskip=0pt\cr
\noalign{\doubleline}
Orbit type&Sun-Earth L2 Quasi-Halo\cr
Mission class&Class B\cr
Mission duration&5 years\cr
Propellant (hydrazine)&213\,kg (77\,\% tank fill)\cr
Launch mass (MEV)&2147\,kg (3195\,kg capability)\cr
Max power (MEV)&1320\,W (with 125\,\% margin on available solar array area)\cr
Onboard data storage&4.6\,Tb (3 days of compressed data, enabling retransmission)\cr
Survey implementation&Instrument on spin table\cr
Attitude control&Zero-momentum 3-axis stabilized\cr
\noalign{\vskip 5pt\hrule\vskip 3pt}
} % close halign
} % close vbox
\endPlancktable
\endgroup
%\end{table}
\end{minipage}
\end{wraptable}

The PICO design reference mission is summarized in Table~\ref{tab:mission_parameters}.

\subsection{Concept of Operations}
\label{sec:operations} %4.1

The PICO concept of operations is similar to that of the successful \wmap~\citep{Bennett2003} and \planck~\citep{Tauber2010} missions. After launch, PICO cruises to a quasi-halo orbit around the Earth--Sun L2 Lagrange point (\S\,\ref{sec:mission_design}). A two-week decontamination period is followed by instrument cooldown, lasting about two months. After in-orbit checkout is complete, PICO begins its science survey.

PICO has a single science observing mode, surveying the sky continuously for 5 years using a pre-planned repetitive survey pattern (\S\,\ref{sec:survey_design}). Instrument data are compressed and stored on-board, then returned to Earth in daily 4-hr Ka-band science downlink passes (concurrent with science observations). Because PICO is observing relatively static Galactic, extragalactic, and cosmological targets, there are no requirements for time-critical observations or data latency. Presently, there are no plans for targets of opportunity or guest observer programs during the prime mission. The PICO instrument does not require cryogenic consumables (as the \textit{Planck} mission did), permitting consideration of significant mission extension beyond the prime mission.

\begin{figure}[!b]
  \begin{minipage}[b]{0.29\textwidth}
    \begin{center}
    \includegraphics[width=1.5in]{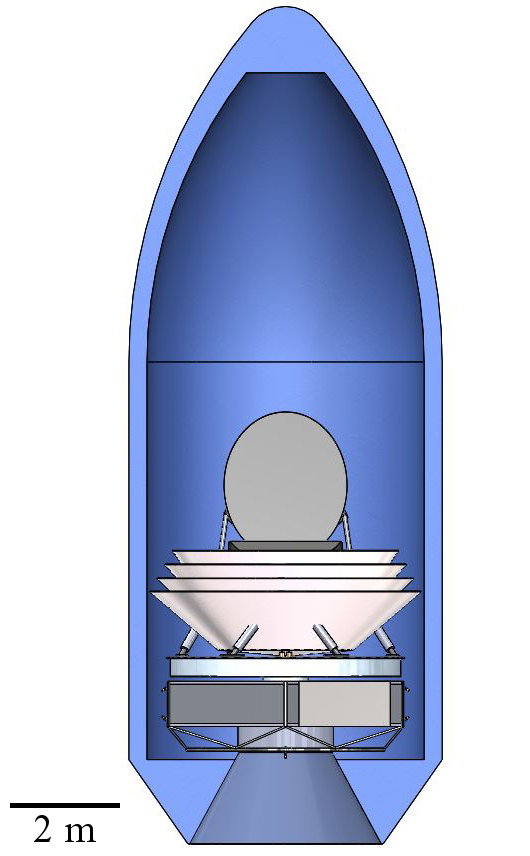}
\caption{\captiontext PICO is compatible with the Falcon~9.\label{fig:InFairing}}
    \end{center}
  \end{minipage}
\hfill
\begin{minipage}[b]{0.67\textwidth}
    \begin{center}
    \includegraphics[width=4in]{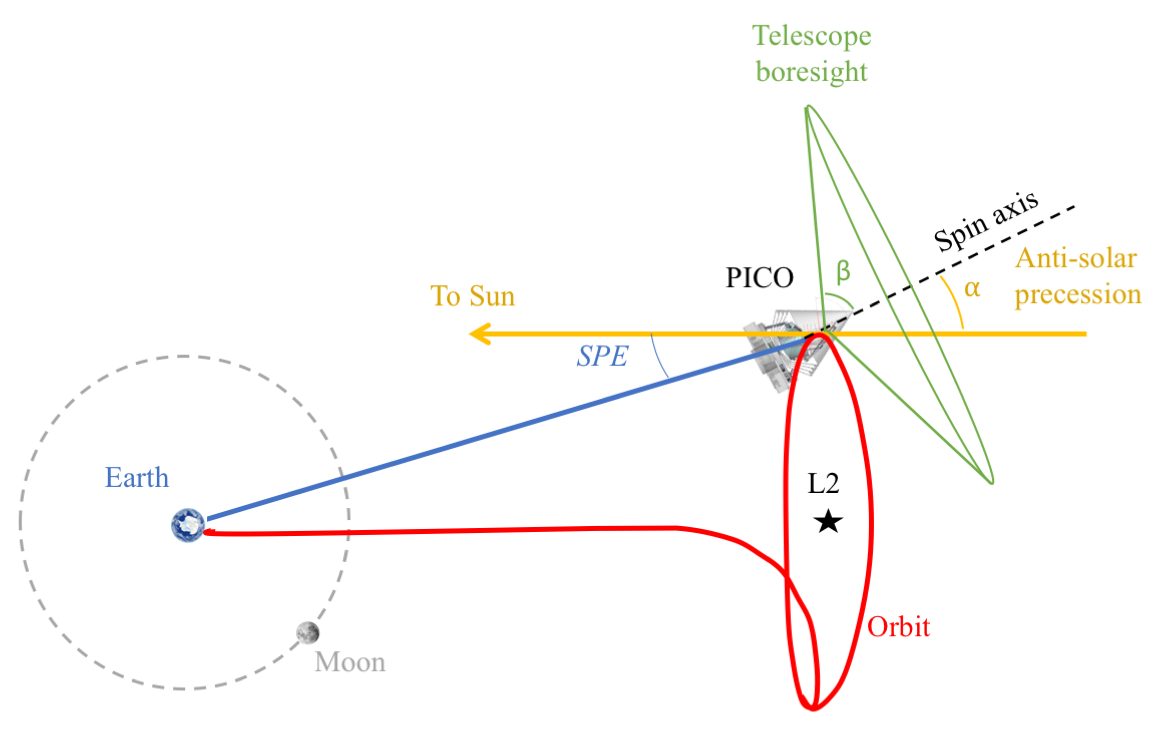}
\caption{\captiontext
  PICO surveys by continuously spinning the instrument about a
  precessing axis.\label{fig:MissionDesignFigure}}
   \end{center}
  \end{minipage}
\end{figure}

\subsubsection{Mission Design and Launch}
\label{sec:mission_design} %4.1.1

The science survey is conducted from a quasi-halo orbit around the Earth--Sun L2 Lagrange point. \planck \ and \wmap\ also operated in L2 orbits. L2 orbits provide favorable survey geometry relative to Earth orbits by mitigating viewing restrictions imposed by terrestrial and lunar stray light. The PICO orbit around L2 is small enough to ensure that the Sun--Probe--Earth (SPE) angle is less than $15\degree$. This maintains the telescope boresight $>70\degree$ away from the Earth (Fig.~\ref{fig:MissionDesignFigure}, $70\degree = 180\degree -\alpha - \beta - \rm{SPE}$).

High data-rate downlink to the Deep Space Network (DSN) is available from L2 using near-Earth Ka bands. L2 provides a stable thermal environment, simplifying thermal control. The PICO orbit exhibits no post-launch eclipses.
 
NASA requires that Probes be compatible with an Evolved Expendable Launch Vehicle (EELV). For the purpose of this study, the Falcon~9~\citep{SpaceX2015} is used as the reference vehicle. Figure~\ref{fig:InFairing} shows PICO configured for launch in a Falcon~9 fairing. The Falcon~9 launch capability for ocean recovery exceeds PICO's 2147\,kg total launch mass (including contingency) by a $50\,\%$ margin.

Insertion to the halo manifold and associated trajectory correction maneuvers %(TCMs) 
require 150\,m\,s$^{-1}$ of total $\Delta V$ by the spacecraft. Orbit maintenance requires minimal propellant (statistical $\Delta V\sim 2$\,m\,s$^{-1}$\,year$^{-1}$). The orbital period is $\sim6$\,months. There are no disposal requirements for L2 orbits, but spacecraft are customarily decommissioned to heliocentric orbit.

\subsubsection{Survey Design}
\label{sec:survey_design} %4.1.2
 
PICO employs a highly repetitive scan strategy to map the full
sky. During the survey, PICO spins with a period
$T_{\rm spin} = 1$\,min about a spin axis oriented $\alpha=26\degree$
from the anti-solar direction (Fig.~\ref{fig:MissionDesignFigure}). This spin axis
is forced to precess about the anti-solar direction with a period
$T_{\rm prec}= 10$\,hr. The telescope boresight is oriented at an
angle $\beta=69\degree$ away from the spin axis (Fig.~\ref{fig:InstrumentCAD}). This $\beta$ angle is
chosen such that $\alpha + \beta > 90\degree$, enabling mapping of all
ecliptic latitudes. The precession axis tracks along with the Earth in its
yearly orbit around the Sun, so this scan strategy maps the full sky
(all ecliptic longitudes) within 6 months.

PICO's $\alpha=26\degree$ value is chosen to be substantially larger than
the \textit{Planck} mission's $\alpha$ angle ($7.5\degree$) to
mitigate systematic effects by scanning across each sky pixel with a
greater diversity of orientations \citep{Hu2003}. Increasing $\alpha$
further would decrease the Sun-shadowed volume available for the
optics and consequently reduce the telescope aperture size. A
deployable Sunshade was considered, but found not to be required, and
was thus excluded in favor of a more conservative and less costly
approach.

The instrument spin rate, selected through a trade study, matches that
of the \textit{Planck} mission. The study balanced low-frequency
($1/f$) noise subtraction (improves with spin rate) against
implementation cost and heritage, pointing reconstruction ability
(anti-correlated with spin rate), and data volume (linearly correlated
with spin rate).  The CMB dipole appears in the PICO data timestream
at the spin frequency (1\,rpm = 16.7\,mHz). Higher multipole signals
appear at harmonics of the spin frequency, starting at 33\,mHz, above
the knee in the detector low-frequency noise (\S\,\ref{sec:sensitivity}). A destriping mapmaker applied in data
post-processing effectively operates as a high-pass filter, as
demonstrated by \planck~\citep{Kurki-Suonio2009}. PICO's spin-axis precession frequency is 
more than 400 times faster than that of \planck , greatly reducing the effects of any residual $1/f$
noise by spreading the effects more isotropically across pixels.

\subsection{Ground Segment}
\label{sec:ground_segment} %4.2

The PICO Mission Operations System (MOS) and Ground Data System (GDS)
can be built with extensive reuse of standard tools. The PICO concept
of operations is described in \S\,\ref{sec:operations}.
% There are
% no time critical events, and no driving data latency
% requirements. Routine orbit maintenance activities are required
% roughly every three months (\S\,\ref{sec:mission_design}). The payload
% consists of a single instrument with a single science observing mode
% (a repetitive survey pattern, \S\,\ref{sec:survey_design}).
All space-ground communications, ranging, and tracking are performed
by the DSN 34\,m Beam Wave Guide (BWG). X-band is
used to transmit spacecraft commanding, return engineering data, and
provide navigation information (S-band is a viable alternative, and
could be considered in a future trade). Ka-band is used for high-rate
return of science data.  The baseline 150\,Mb/s transfer rate
(130\,Mb/s information rate after CCSDS encoding) is an existing DSN
catalog service~\citep{DSN2015}.  The instrument produces 6.1\,Tb/day,
which is compressed to 1.5\,Tb/day
(\S\,\ref{sec:detector_readout}). Daily 4\,hr DSN passes return PICO
data in 3.1\,hr, with the remaining 0.9\,hr available as needed for
retransmission or missed-pass recovery.

\begin{figure}
\hspace{-0.15in}
\parbox{5.1in}{\centering{
\includegraphics[width=5.25in]{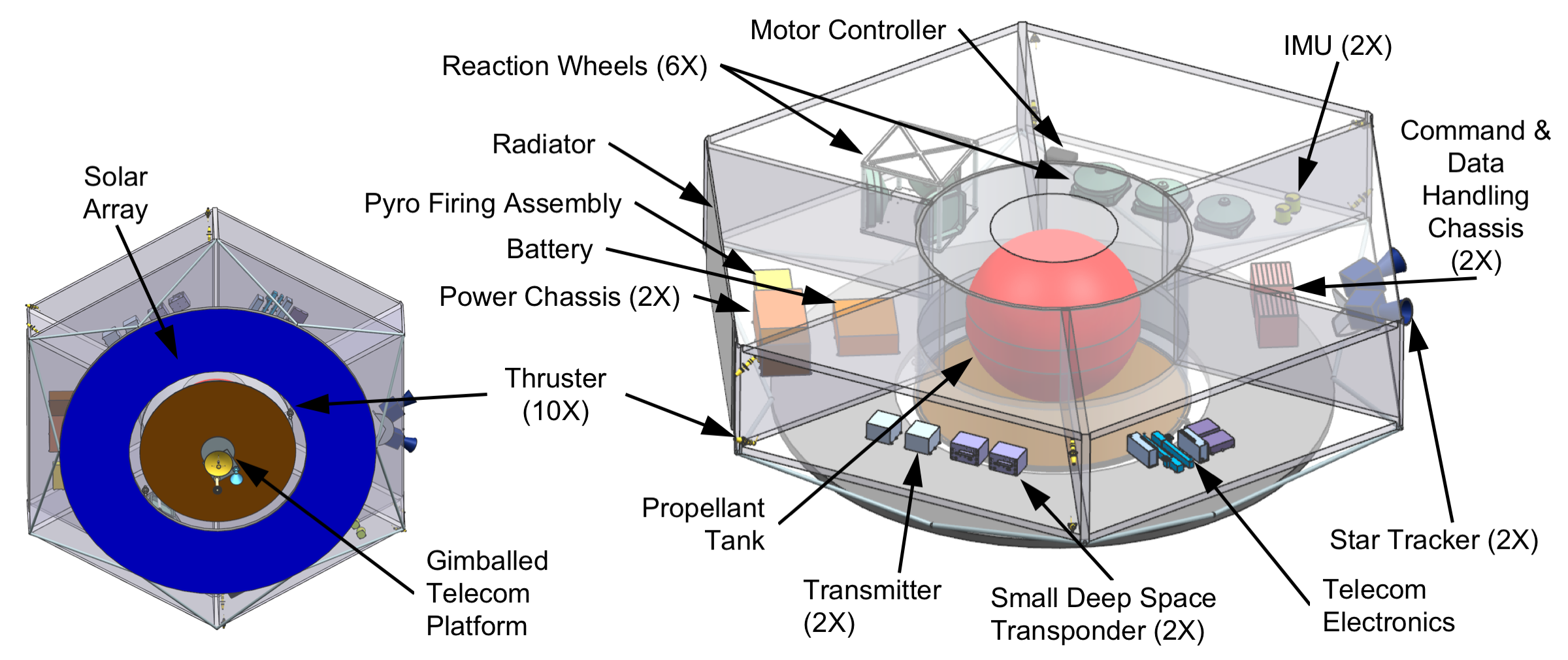} } }
\hspace{0.12in}
\parbox{1.3in}{
\caption{\captiontext
Modular equipment bays provide easy access to all components in the spacecraft de-spun module and enable parallel integration of spacecraft subsystems.\label{fig:Spacecraft}} }
\vspace{-0.25in}
\end{figure}

\subsection{Spacecraft}
\label{sec:spacecraft} %4.3

The PICO spacecraft bus is Class~B and designed for a minimum lifetime of 5\,years in the L2 environment. Mission-critical elements are redundant. Flight spares, engineering models, and prototypes appropriate to Class~B are budgeted.

The aft end of the spacecraft (the ``de-spun module'') is comprised of
six equipment bays that house standard components
(Fig.~\ref{fig:Spacecraft}).  The instrument and V-grooves are mounted on
bipods from the spacecraft ``spun module,'' which contains hosted
instrument elements (Fig.~\ref{fig:InstrumentCAD}). A motor drives the
spun module at 1\,rpm to support the science survey requirements
(\S\,\ref{sec:survey_design}). Reaction wheels on the despun module
cancel the angular momentum of the spun module and provide three-axis
control (\S\,\ref{sec:attitude_determination}).

The bipods that mechanically support the instrument are thermally
insulating. The passively radiating V-groove assembly thermally
isolates the instrument from solar radiation and from the bus
(\S\,\ref{sec:radiative_cooling}). Like \textit{Planck} \citep{Tauber2010}, the V-grooves are
manufactured using honeycomb material. Additional radiators on the
spun and despun spacecraft modules ($\sim1$\,m$^2$ each) reject heat
dissipated by spacecraft subsystems and hosted instrument elements.

PICO's avionics are dual-string with standard interfaces. Solid-state recorders provide three days of science data storage (4.6 Tbit, \S\,\ref{sec:heritage}), enabling retransmission of missed data (\S\,\ref{sec:ground_segment}).

PICO employs a fully redundant Ka- and X-band telecommunications
architecture. The Ka-band system uses a 0.3\,m high-gain antenna to
support a science data downlink information rate of 130 Mb/s to a
34\,m BWG DSN ground station with a link margin of 4.8\,dB. The X-band
system provides command and engineering telemetry communication
through all mission phases using medium- and low-gain
antennas. Amplifiers, switches, and all three antennas are on a
gimballed platform, enabling Ka and X-band downlink concurrent with
science observations.

 The heritage power electronics are dual-string.
A 74\,A-hr Li-ion battery is sized for a 3\,hr launch phase with 44\,\% depth of discharge.
After the launch phase, the driving
mode is telecom concurrent with science survey (1320\,W including 43\,\% contingency).
Solar cells on the aft side of the bus (5.8\,m$^2$ array, $\alpha=26\degree$ off-Sun) support this mode with positive power,  and unused area in the solar array plane (7.4\,m$^2$ more area by growing to 4.5\,m diameter) affords 125\,\% margin
(Fig.~\ref{fig:Spacecraft}).

The propulsion design is a simple mono-propellant blow-down hydrazine
system with standard redundancy. Two aft-pointed 22\,N thrusters
provide $\Delta V$ and attitude control for orbit insertion and
maintenance (\S\,\ref{sec:mission_design}), requiring 140 kg of
propellant.  Eight 4\,N thrusters provide reaction-wheel momentum
management and backup attitude-control authority (60\,kg of
propellant). Accounting for ullage (14\,kg), the baseline propellant
tank fill fraction is 77\,\%.

\subsubsection{Attitude Determination and Control}
\label{sec:attitude_determination} %4.3.1

PICO uses a zero net angular momentum control architecture with heritage from the SMAP mission (\S\,\ref{sec:heritage}). PICO's instrument spin rate (1\,rpm) matches that of the \planck\ mission, but the precession of the spin axis is faster (10\,hr vs 6\, months), and the precession angle larger ($26\degree$ vs $7.5\degree$). These differences make the spin-stabilized \planck \ control architecture impractical. 
% because of the amount of torque that would be required to drive precession.

The PICO instrument spin rate is achieved and maintained using a spin motor. The spin motor drive electronics provide the coarse spin rate knowledge used for controlling the spin rate to meet the $\pm0.1$\,rpm requirement. Data and power are passed across the interface using slip rings.

PICO requires $220$\,N\,m\,s to cancel the angular momentum of the instrument and spacecraft spun module at 1\,RPM. This value includes mass contingency and is based on the CAD model. Three Honeywell HR-16 reaction wheel assemblies (RWAs), each capable of 150\,N\,m\,s, are mounted on the despun module parallel to the instrument spin axis, and spin opposite to the instrument to achieve zero net angular momentum. The despun module is three-axis stabilized. The spin axis is precessed using three RWAs mounted normal to the spin axis in a triangle configuration. Each set of three RWAs is sized such that two could perform the required function with margin, providing single fault tolerance.

Spin-axis pointing and spin-rate knowledge are achieved and maintained using star tracker and inertial measurement unit (IMU) data. The attitude determination system is single-fault tolerant, with two IMUs each on the spun and despun modules, and two star trackers each on the spun and despun modules. Two Sun sensors on the despun module are used for safe-mode contingencies and instrument Sun avoidance. All attitude control and reconstruction requirements are met, including spin axis control $< 60$\,arcmin with $< 1$\,arcmin/min stability, and reconstructed pointing knowledge $< 10$\,arcsec (each axis, $3\sigma$).

Additional pointing reconstruction is performed in post-processing using the science data.  The PICO instrument will observe planets (compact, bright sources) nearly every day.  By fitting the telescope pointing to the known planetary ephemerides, the knowledge of the telescope boresight pointing and the relative pointing of each detector will improve to better than 1\,arcsec (each axis, $3\sigma$). \planck , with fewer detectors, making lower \ac{SNR} measurements of the planets, and observing with a scan strategy that acquired measurements of each planet only once every 6\,months, demonstrated 0.8\,arcsec ($1\sigma$) pointing reconstruction uncertainty in-scan and 1.9\,arcsec ($1\sigma$) cross-scan~\citep{2016A&A...594A...1P}.
%(Planck, Planck 2015 results. I. Overview of products and scientific results 2016)

%\bigskip
\newpage
\section{Technology Maturation}
\label{sec:technology_maturation} %5
\begin{wrapfigure}[10]{r}{3.80in}  % r is right aligned, l is left. Capital letters allow figure to float on page.
%\vspace{-15pt} % move figure up to align with section headings.
\parbox{2.35in}{
\includegraphics[width=2.35in]{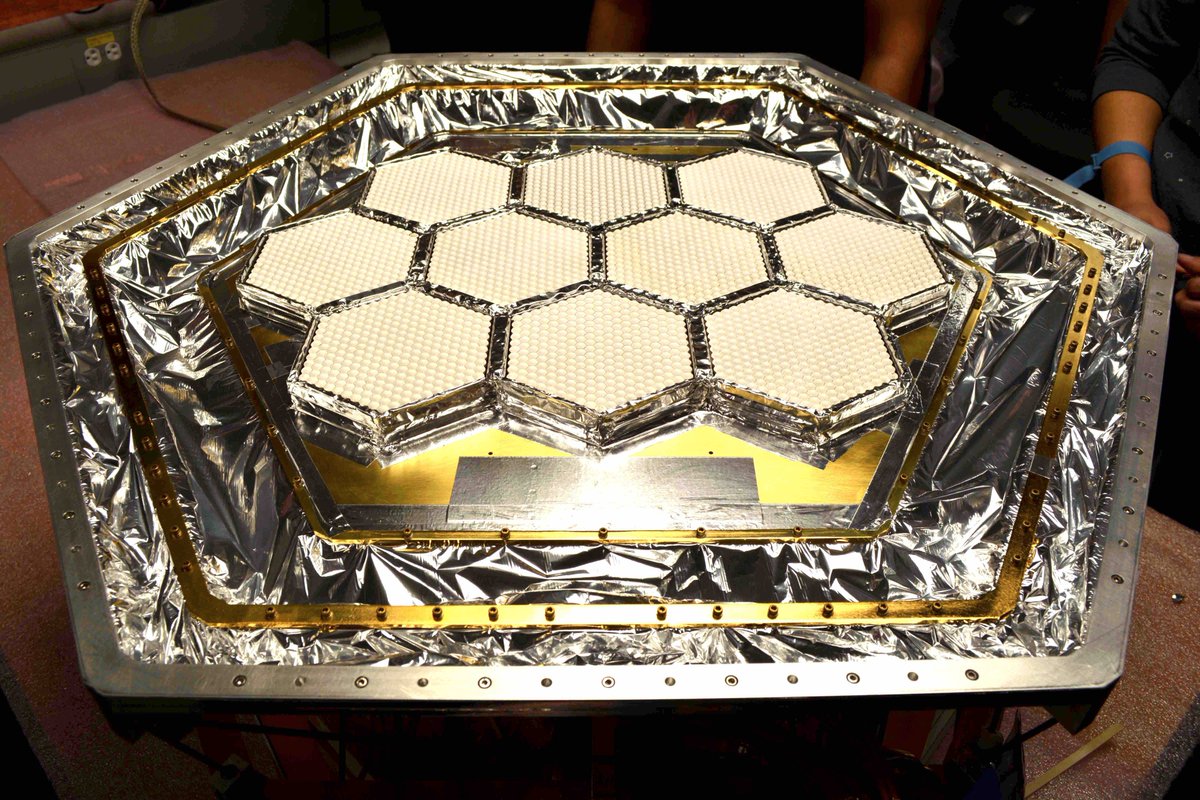} }  %% image was 3 in wide in past version.
\parbox{1.40in}{
\caption{\captiontext SPT-3G operates a focal plane with sinuous antenna-coupled, three-band pixels with
16,000 bolometers~\citep{Dutcher2018}. Each pixel couples radiation to bands at 95, 150, and
220~GHz.
\label{fig:spt_fp} }
}
\end{wrapfigure}
PICO builds off of the heritage of \planck-HFI and \textit{Herschel}.  Since the time of \planck\ and \textit{Herschel}, suborbital experiments have used monolithically fabricated TES bolometers and multiplexing schemes to field instruments with thousands of \ac{TES} bolometers per camera (Fig.~\ref{fig:spt_fp}). By the time PICO enters Phase~A, S3 experiments plan to be operating nearly 100,000 \ac{TES} bolometers in several independent cameras~\citep{Simons2018,biceparray,spt3g}.

 The remaining technology developments required to enable the PICO baseline design are:
\begin{enumerate}
\item extension of three-color antenna-coupled bolometers down to 21\,GHz and up to 462\,GHz (\S\,\ref{sec:bolometers});
\item construction of high-frequency direct absorbing arrays and laboratory testing (\S\,\ref{sec:dev_arrays});
\item beam line and 100\,mK testing to simulate the cosmic ray environment at L2 (\S\,\ref{sec:env_testing});
\item expansion of time-division multiplexing to support 128 switched rows per readout column (\S\,\ref{sec:multiplexing}).
%\item Simulation software (\S\,\ref{sec:simulation}). {\color{red}Shaul to fill this in.}
\end{enumerate}
All of these developments are straightforward extensions of technologies already available today.  We recommend APRA and SAT support to complete development of these technologies through the milestones described in Table~\ref{tab:technologies}.

\medskip
\subsection{21--462\,GHz Bands}
\label{sec:bolometers} %5.1

Suborbital teams have successfully demonstrated a variety of optical-coupling schemes, including horns with ortho-mode transducers (OMTs), lithographed antenna arrays, and sinuous antennas under lenslets (Table~\ref{tab:suborbital1}). All have achieved background-limited performance with sufficient margin on design parameters to achieve this performance in the lower background environment at L2. All have been packaged into modules and focal-plane units in working cameras representative of the PICO integration. Experiments have already used a number of PICO's observing bands between 27\,GHz and 270\,GHz (Table~\ref{tab:suborbital1}).  To date, statistical map depths of 3\,$\mu$K$_{\rm CMB}$\,arcmin have been achieved over small sky areas, which is within a factor of five of PICO's CBE over the entire sky (Table~\ref{tab:bands}). %and have demonstrated systematic control better than this level through full-pipeline simulations and null-test analysis (jackknife tests).

% Other experiments have
% successfully deployed two-color pixels. All of these detector arrays
% have been packaged into modules and focal-plane units in working
% cameras representative of the PICO integration.

%% Table 5.1

\begin{table}
\caption{\captiontext 
  PICO technologies can be developed to TRL~5 prior to a 2023 Phase~A start using the APRA and SAT programs, requiring a total of about \$\,13M. Per NASA guidance, these costs are outside the mission cost (\S\,\ref{sec:mission_cost}).\label{tab:technologies}}
\begingroup
%\openup 5pt
\newdimen\tblskip \tblskip=5pt
\nointerlineskip
\vskip -5mm
\footnotesize %\footnotesize
\setbox\tablebox=\vbox{
    \newdimen\digitwidth
    \setbox0=\hbox{\rm 0}
    \digitwidth=\wd0
    \catcode`*=\active
    \def*{\kern\digitwidth}
    \newdimen\signwidth
    \setbox0=\hbox{+}
    \signwidth=\wd0
    \catcode`!=\active
    \def!{\kern\signwidth}
\halign{
\vtop{\hsize 1.3in\raggedright\hangafter=1\hangindent=2em\noindent\strut#\strut\par}\leaderfil\tabskip=0.6em&
\vtop{\hsize 0.7in\raggedright\hangafter=1\hangindent=0em\noindent\strut#\strut\par}&
\vtop{\hsize 0.9in\raggedright\hangafter=1\hangindent=0em\noindent\strut#\strut\par}&
\vtop{\hsize 0.7in\raggedright\hangafter=1\hangindent=0em\noindent\strut#\strut\par}&
\vtop{\hsize 0.6in\raggedright\hangafter=1\hangindent=0em\noindent\strut#\strut\par}&
\vtop{\hsize 0.5in\raggedright\hangafter=1\hangindent=0em\noindent\strut#\strut\par}&
\vtop{\hsize 0.7in\raggedright\hangafter=1\hangindent=0em\noindent\strut#\strut\par}&
\hfil#\hfil \tabskip=0pt\cr
\noalign{\doubleline}
\omit\hfil Task\hfil&\omit\hfil Current\hfil&\omit\hfil Milestone A\hfil&\omit\hfil Milestone B\hfil&\omit\hfil Milestone C\hfil&\omit\hfil Current\hfil&\omit\hfil Required\hfil&\omit\hfil Date TRL5\hfil\cr
\omit&\omit\hfil status\hfil&&&&\omit\hfil funding\hfil&\omit\hfil funding\hfil&\omit\hfil achieved\hfil\cr
\noalign{\vskip 3pt\hrule\vskip 5pt}
1a. Three-color arrays $\nu<90$\,GHz&2-color lab demos $\nu > 30$\,GHz&Field demo of 30--40\,GHz (2020)&Lab demos 20--90 GHz (2022)& \hspace{0.12in} ----- &APRA \& SAT&\$2.5M over 4\,yr (1 APRA + 1 SAT) &2022\cr
1b. Three-color arrays $\nu > 220$\,GHz&2-color lab demos $\nu < 300$\,GHz&Field demo of 150--270\,GHz (2021)&Lab demos 150-460 GHz (2022) &\hspace{0.12in} ----- &APRA \& SAT&\$3.5M over 4\,yr (2 SATs) & 2022\cr
2. Direct absorbing arrays $\nu > 50$\,GHz& 0.1--5\,THz unpolarized&Design \& prototype of arrays (2021)&Lab demo of 555\,GHz (2022)& Lab demo of 799\,GHz (2023) & None & \$2M over 5\,yr (1 SAT) & 2023 \cr
3. Cosmic ray studies& 250\,mK w/ sources&100\,mK tests with sources (2021)&Beamline tests (2023)& \hspace{0.12in} ----- & APRA \& SAT & \$0.5--1M over 5\,yr (part of 1 SAT) & \hspace{0.0in} ----- \cr
4a. Fast readout electronics& MUX66 demo&Engineering and Fab of electronics (2020)&Lab demo (2021)&Field demo (2023)& No direct funds &\$4M over 5\,yr (1 SAT)&2023\cr
4b. System engin\-eering; 128$\times$ MUX demo&MUX66 demo&Design of cables (2020)&Lab demo (2021)& Field demo (2023) & No direct funds & \hspace{0.12in} ----- & \hspace{0.0in} ----- \cr
\noalign{\vskip 5pt\hrule\vskip 3pt}
} % close halign
} % close vbox
\endPlancktable
\endgroup
\end{table}

The baseline PICO instrument requires three-color dual-polarized antenna-coupled bolometers covering bands from 21 to 462\,GHz (\S\,\ref{sec:low_freq_det}).  The sinuous antenna has the bandwidth to service three bands per pixel, whereas horns and antenna arrays have only been used for two. Our baseline is to use a three-band sinuous antenna, although we have designs that use two- or one-band per pixel and have the same or similar baseline noise as PICO~(\S~\ref{sec:technology_descopes}). SPT-3G has used the PICO-baselined three-color pixel design to deploy 16,000 detectors covering 90/150/220\,GHz~\citep{Dutcher2018}.

The extension to lower frequencies requires larger antennas and therefore control of film properties and lithography over larger areas. Scaling to higher frequencies requires tighter fabrication tolerances and electromagnetic wave transmission losses tend to increase due to material properties. Current anti-reflection technologies for the lenslets need to be extended with thicker and thinner layers to cover the lowest and highest frequency channels. These developments will require control of cleanliness and understanding of process parameters. Changes to elements in the light path will require characterization of beam properties.

%% Table 5.2

\begin{table}
\caption{\captiontext Multiple active suborbital efforts are advancing technologies relevant to PICO.\label{tab:suborbital1}}
\begingroup
%\openup 5pt
\newdimen\tblskip \tblskip=5pt
\nointerlineskip
\vskip -5mm
\footnotesize %\footnotesize
\setbox\tablebox=\vbox{
    \newdimen\digitwidth
    \setbox0=\hbox{\rm 0}
    \digitwidth=\wd0
    \catcode`*=\active
    \def*{\kern\digitwidth}
    \newdimen\signwidth
    \setbox0=\hbox{+}
    \signwidth=\wd0
    \catcode`!=\active
    \def!{\kern\signwidth}
\halign to \textwidth{
\hbox to 1.2in{#\leaderfil}\tabskip=0.6em plus 0.6em&
\hfil#\hfil&
\hfil#\hfil&
\hfil#\hfil&
\hfil#\hfil&
\hfil#\hfil&
\hfil#\hfil&
\hfil#\hfil\tabskip=0pt\cr
\noalign{\doubleline}
\omit\hfil Project\hfil&Type&Optical Coupling&$\nu_c$&Colors&$N_{\rm bolo}$&Significance&Reference\cr
 \omit  &       &               &[GHz]&per pixel&\cr
\noalign{\vskip 3pt\hrule\vskip 5pt}
PICO baseline&Flight&&21 -- 462&Three&11,796&&\S\,\ref{sec:low_freq_det}\cr
SPT-3G&Ground&Sinuous&90 -- 220&Three&16,260&Trichroic&\cite{Dutcher2018}\cr
Advanced ACT-pol&Ground&Horns&27 -- 230&Two&3,072&Dichroic&\cite{Li2018}\cr
BICEP/Keck&Ground&Antenna arrays&90 -- 270&One&5,120&50\,nK-deg&\cite{bicep_keck2018}\cr
Berkeley, Caltech, NIST&Lab&Various&30 -- 270&Various&--&Band coverage&\cite{Westbrook2016,Hui2018,Simon2018}\cr
SPIDER&Balloon&Antenna arrays&90 -- 150&One&2,400&Stable to 10\,mHz&\cite{Rahlin2014}\cr
\noalign{\vskip 5pt\hrule\vskip 3pt}
} % close halign
} % close vbox
\endPlancktable
%\tablenote {{\rm a}} 143--343\,GHz only.\par
\endgroup
\end{table}

The direction of polarization sensitivity of the sinuous antenna varies with frequency, thus presenting a potential source of systematic error. Over 25\% bandwidth, the variation is approximately $\pm 5$~deg~\citep{obrient2008b}. There are solutions to this in the focal-plane design, measurements, data analysis, and free parameters of the sinuous antenna geometry.  A recent study found that pre-flight characterization of the effect through measurements can readily mitigate it as a source of systematic uncertainty~\citep{picoweb_wobble}. Studies with current field demonstrations, such as with the data of SPT-3G, will be particularly important. The PICO concept is robust to any challenges in developing three-color pixels; \S\,\ref{sec:technology_descopes} describes options to descope to two- and one-color pixels, technologies for which the polarization sensitivity is constant as a function of frequency.

\medskip
\subsection{555--799\,GHz bands}
\label{sec:dev_arrays}

The baseline PICO instrument requires single-color, horn-coupled, dual-polarization, direct-absorbing bolometers from 555 to 799\,GHz (\S\,\ref{sec:high_freq_det}).  \planck\ and \textit{Herschel} demonstrated the architecture of horns coupled to direct absorbing bolometers. 
%(Fig.~\ref{fig:DirectAbsorbing})    
Ground experiments with similar designs have deployed focal planes with hundreds of horn-coupled spiderweb bolometers, replacing the \textit{Planck} and \textit{Herschel} NTD-Ge thermistors with TESs, and adjusting time constants as necessary (Table~\ref{tab:suborbital2}). \planck -HFI, SPT-pol, and BICEP demonstrated dual-polarized detectors. \textit{Herschel} and SPT-SZ demonstrated monolithic unpolarized detectors. PICO will require detectors that merge these two designs in monolithic dual-polarized arrays. Since all the components of the technology already exist, the remaining necessary development is the packaging. Filled arrays of detectors such as Backshort Under Ground (BUG) bolometers are also an option~\citep{Staguhn2006}.

%% Table 5.3

\begin{table}
\caption{\captiontext PICO high-frequency detectors leverage development and demonstration by \textit{Planck}, \textit{Herschel}, and SPT.\label{tab:suborbital2}}
\begingroup
%\openup 5pt
\newdimen\tblskip \tblskip=5pt
\nointerlineskip
\vskip -5mm
\footnotesize %\footnotesize
\setbox\tablebox=\vbox{
    \newdimen\digitwidth
    \setbox0=\hbox{\rm 0}
    \digitwidth=\wd0
    \catcode`*=\active
    \def*{\kern\digitwidth}
    \newdimen\signwidth
    \setbox0=\hbox{+}
    \signwidth=\wd0
    \catcode`!=\active
    \def!{\kern\signwidth}
\halign to \textwidth{
\hbox to 1.2in{#\leaderfil}\tabskip=0.6em plus 0.6em&
\hfil#\hfil&
\hfil#\hfil&
\hfil#\hfil&
\hfil#\hfil&
\hfil#\hfil&
\hfil#\hfil&
\hfil#\hfil&
\hfil#\hfil\tabskip=0pt\cr
\noalign{\doubleline}
\omit\hfil Project\hfil&Type&Polarized&Mono-&$\nu_c$&Colors&$N_{\rm bolo}$&Significance&Reference\cr
 \omit  &       &              &lithic  &[GHz]&per pixel&\cr
\noalign{\vskip 3pt\hrule\vskip 5pt}
PICO baseline&Flight&Yes&Yes&555 -- 799&One&1,200&&\S\,\ref{sec:high_freq_det}\cr
\textit{Planck} HFI&Flight&143--343\,GHz&No&143 -- 857&One&48&TRL 9 polarized&\cite{Turner2001}\cr
\textit{Herschel}&Flight&No&Yes&570 -- 1200&One&270&TRL 9 monolothic&\cite{Ferlet2008}\cr
SPT-SZ&Ground&No&Yes&90 -- 220&One&840&Monolithic array TESs&\cite{Shirokoff2011}\cr
SPT-pol-90&Ground&Yes&No&90&One&180&Dual pol absorbing TESs&\cite{Sayre2012}\cr
\noalign{\vskip 5pt\hrule\vskip 3pt}
} % close halign
} % close vbox
\endPlancktable
%\tablenote {{\rm a}} 143--343\,GHz only.\par
\endgroup
\end{table}

%The greatest remaining challenge is the low risk development of a packaging design.

\subsection{Environmental Testing}
\label{sec:env_testing}

%\begin{wrapfigure}[27]{r}{0.35\textwidth}  % r is right aligned, l is left. Capital letters allow figure to float on page.
%\centering
%\vspace{-5pt} % move figure up to align with section headings.
%\includegraphics[width=0.35\textwidth]{figures/DirectAbsorbing.png}  %% figure was 2.5 in, .4 /textwidth is 2.6in.
%\vspace{-0.25in}
%\caption{\captiontext Top: \planck\ used horns to couple the electromagnetic radiation to its detectors. Horn coupling has been used in other experiments, and is the baseline for PICO's coupling between 555 and 799~GHz. Bottom: The photograph shows a dual-polarization, direct-absorbing bolometer from BICEP. The technology was also used with SPT-Pol and \planck-HFI for 143--343\,GHz bands.
%\label{fig:DirectAbsorbing} }
%\end{wrapfigure}

Laboratory tests and in-flight data from balloons suggest that TES
bolometer arrays may be more naturally robust against cosmic rays than
the individual NTD-Ge bolometers used in \textit{Planck}. PICO will leverage lessons
learned from \textit{Planck} and ensure robust thermal sinking of
detector array substrates. Cosmic-ray
glitches have fast recovery times and low coincidence rates
\citep{SPIDER2018,Filippini_inprep}. Residual risk can be retired with 100\,mK
testing where the array heat sinking may be weaker, and beam-line
tests to simulate the expected flight environment.

\subsection{Multiplexing}
\label{sec:multiplexing}

More than ten experiments have used time-domain multiplexer (TDM) readout. SCUBA2 on JCMT has 10,000 pixels, nearly as many detectors as planned for PICO~\citep{Holland2013}. Most of these experiments have used 32-row multiplexing. Recently ACT has expanded this to 64-row multiplexing~\citep{Henderson2016}.

PICO's sensitivity requirements dictate the use of 13,000 transition-edge-sensor bolometers and a multiplexed system.  Our baseline design is to use TDM readout with 128 switched rows per readout column (TDM-128$\times$). The leap to TDM-128$\times$ requires: \\
$\bullet$ development of fast-switched room temperature electronics; and \\
$\bullet$ system engineering of room temperature to cryogenic row-select cabling to ensure sufficiently fast row-switch settling times.

The historical row revisit rate for bolometric instruments using 32$\times$ TDM has been 25\,kHz \cite[e.g.,][]{BICEP2015}. However, X-ray instruments using TDM routinely switch between rows at 6.25~MHz~\citep{Doriese2016}. The PICO baseline assumes a 6.25~MHz switch rate and TDM-128$\times$, which dictates a row-revisit rate %(effective sampling rate) 
of 48.8\,kHz. To limit aliased noise, PICO implements low-pass filters in each readout channel with a bandwidth of 6\,kHz, dictated by detector stability considerations and the required $\sim1$\,kHz signal bandwidth.  With these parameters and using the same TDM multiplexer SQUID design, the increased total noise due to aliasing is less than 15\,\% and is included in our detector noise budget.  The system engineering study will culminate in a demonstration of TDM-128$\times$ SQUID aliased noise below PICO detector sensitivity requirements.

\subsection{Technology Descopes}
\label{sec:technology_descopes} %5.3

A descope from three-color sinuous antenna/lenslet-coupled pixels to two-color horn-coupled, or to single color antenna-array pixels remains a viable alternative should the three-color technology not mature as planned. In both alternative options, bands above 555~GHz are the same as the baseline. For the lower frequencies, the two-color horn-coupled pixel option contains 8,840 detectors and has 19 colors. Because horns have a $2.3:1$ bandwidth, each of the two bands in a pixel has 35\,\% bandwidth (compared to the baseline 25\,\%), which compensates for pixel count, resulting in 0.61\,$\mu$K$_{\rm CMB}$\,arcmin aggregate CBE map depth. This is the same as the three-color CBE map depth, and affords the same $40\,\%$ margin relative to the 0.87\,$\mu$K$_{\rm CMB}$\,arcmin baseline requirement (Table~\ref{tab:bands}). Detailed analysis would be performed to assess the impact of the coarser spectral resolution on signal component separation. Single color antenna-array pixels can have higher packing density than the other two architectures. This option has 6,540 detectors, 21 colors, each with 30\,\% bandwidth, and a noise level of 0.74\,$\mu$K$_{\rm CMB}$\,arcmin, leaving only 17\% noise margin relative to the requirement. 

%A descope from three-color sinuous antenna/lenslet-coupled pixels to two-color horn-coupled pixels remain a viable alternative should the three-color technology not mature as planned. We have a design for a PICO-size focal-plane with two-color horn-coupled pixels at the lower frequencies and the baseline one-color pixels at the higher frequencies. It contains 8,840 detectors (compared to the baseline with 12,966) and has 19 colors (baseline has 21 colors). Because horns have a $2.3:1$ bandwidth, each of the two bands in a pixel has 35\,\% bandwidth (compared to the baseline 25\,\%), which compensates for pixel count, resulting in 0.61\,$\mu$K$_{\rm CMB}$\,arcmin aggregate CBE map depth, which is the same as the three-color CBE map depth, and affords the same $40\,\%$ margin relative to the 0.87\,$\mu$K$_{\rm CMB}$\,arcmin baseline requirement (Table~\ref{tab:bands}). Detailed analysis would be performed to assess the impact of the coarser spectral resolution on signal component separation (\S\,\ref{sec:signal_separation}).

\subsection{Enhancing Technologies}
\label{sec:enhancing_technologies} %5.4

The following technologies are neither required nor assumed by the PICO baseline concept. However, they represent opportunities to extend scientific capabilities or simplify engineering.

PICO baselines TDM readout because of its relative maturity and demonstrated sensitivity and stability in relevant science missions. Lab tests of frequency-domain multiplexing (FDM) give comparable performance with higher multiplexing factors and lower thermal loads on cryogenic stages relative to TDM, but with higher ambient temperature power consumption. Suborbital experiments such as SPT-3G are using FDM to read out focal planes comparable in size to PICO.

Microwave frequency SQUID multiplexing can increase the multiplexing density and reduce the number of wires between the 4\,K and ambient temperature stages~\citep{Dober2017,Irwin2004}. Kinetic inductance detectors and Thermal KIDs can further reduce the wire count, obviate the need for SQUID-based amplifiers, and simplify integration by integrating the multiplexing function on the same substrate as the detectors~\citep{McCarrick2018,Steinbach2018,Johnson2018}. The cost to develop these technologies is \$3--4M/year, with a high chance of reaching TRL-5 before Phase~A.
%\costfootnote

%\newpage
\bigskip
\section{Project Management, Heritage, Risk, and Cost}
\label{sec:project_management} %6

\subsection{PICO Study Participants}
\label{sec:study_participants} %6.1

The PICO study was open to the entire mm/sub-mm science community. Seven working groups were led by members of PICO's Executive Committee, which had a telephone conference weekly under the leadership of PI Shaul Hanany. Several of the working groups also had weekly telecon conferences. More than 60 people participated in-person in each of two community workshops (November 2017 and May 2018). This report has contributions from 82 authors, and it has been endorsed by additional 131 members of the community.  The full list of authors and endorsers is on page~\pageref{authorlist}.

The PICO engineering concept definition package was generated by Team~X.\footnote{\label{teamx} Team~X is JPL's concurrent design facility.} The Team~X study was supported by inputs from a JPL engineering team and Lockheed Martin.

\subsection{Project Management Plan}
\label{sec:management_plan} %6.2

PICO benefits from the experience of predecessor missions such as \planck\ and \wmap, as well as many years of investment in technology development and a multitude of suborbital experiments. In addition to demonstrated science and engineering capabilities, this heritage has developed a community of people with the expertise required to field a successful mission.

\begin{figure}[hb]
\begin{center}
\includegraphics[width=\textwidth]{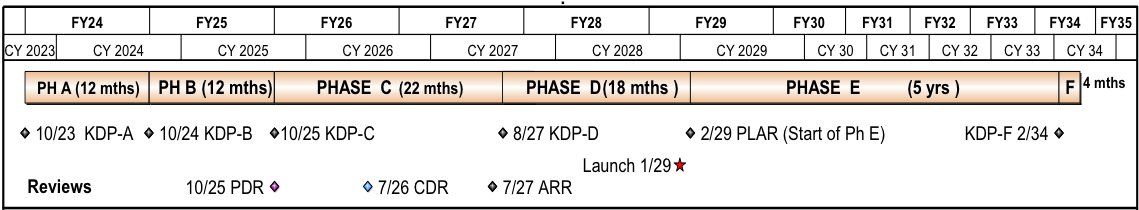}
\end{center}
\vspace{-0.25in}
\caption{\captiontext The PICO baseline schedule is based on historical actuals from similarly-sized missions such as Juno and SMAP. Per NASA direction, Probe studies assume a Phase~A start in October 2023.\label{fig:Schedule}}
\vspace{-0.05in}
\end{figure}

This study assumes mission management by JPL with a Principal Investigator leading a single science team. A Project Manager provides project oversight for schedule, budget, and deliverables. A Project Systems Engineer leads systems engineering activities and serves as the Engineering Technical Authority. A Mission Assurance Manager serves as the Independent Technical Authority. The PICO mission development schedule is shown in Fig.~\ref{fig:Schedule}.

Probes are medium-class missions, similar in cost scope to NASA's
New Frontiers missions, which are Category~1 and Risk Classification A
or B, with Phase A--D costs capped at $\sim$ \$850M (not including the
launch vehicle). JPL is well-prepared to manage Probe missions, having
managed the Juno New Frontiers mission (launched 2011) and also the
development of the medium-class \textit{Spitzer} Space Telescope (launched
2003). JPL delivered the bolometric detectors for the \textit{Planck}
HFI instrument (launched 2009). Presently, JPL is managing NEOCam, a
Discovery class infrared space telescope.

The PICO spacecraft provider will be selected during mission formulation. Multiple organizations are capable of providing a spacecraft bus to meet PICO's requirements. Lockheed Martin contributed to the PICO concept study, leveraging their experience with New Frontiers missions Juno and OSIRIS-REx.
 
\subsection{Heritage}
\label{sec:heritage} %6.3

%The successful \textit{Planck} mission provides science heritage for PICO. 
The technical heritage for PICO traces to multiple missions. Because PICO observes in the mm/sub-mm regime, the surface accuracy requirement for the reflectors is relatively easy to meet. PICO's reflectors are similar to \planck 's, but somewhat larger ($270 \,\, {\rm cm} \times 205\,{\rm cm}$ primary versus $189\,{\rm cm} \times 155\,{\rm cm}$)~\citep{Gloesener2006}. \textit{Herschel} observed at shorter wavelengths that required higher surface accuracy and had a larger reflector ($350\, \,{\rm cm}$ diameter primary)~\citep{Toulemont2004}.

The heritage of the PICO detectors and readout electronics (which are described in \S\,\ref{sec:focal_plane}, \S\,\ref{sec:detector_readout}) is described in \S\,\ref{sec:technology_maturation}.

PICO's detectors are cooled by a cADR (\S\,\ref{sec:cadr}) with requirements that are within the capabilities of current ADRs developed by Goddard Space Flight Center. These systems have been applied to several JAXA missions, including \textit{Hitomi}~\citep{Shirron2016}. PICO's 4\,K cryocooler (\S\,\ref{sec:4kcooler}) is a direct extension of the JWST MIRI design~\citep{Durand2008,Rabb2013}. PICO benefits from a simpler and more reliable implementation of the J-T system than was required for MIRI, in that no deployment of cooling lines is required, and all flow valving is performed on the warm spacecraft. Cooling multiple independent points with a J-T loop has been demonstrated on \planck\ with the JPL-supplied 18\,K cooler~\citep{Planck2011}. Structures similar to PICO's V-groove radiator assembly (\S\,\ref{sec:radiative_cooling}) are a standard approach for passive cooling, and were first described more than thirty years ago~\citep{Bard1987}. We baselined a simple honeycomb material construction like that successfully flown by \planck~\citep{ESA2009,Planck2011}.

%PICO's 4.5\,m diameter V-groove assembly fits inside the launch vehicle fairing.

Most requirements on the PICO spacecraft are well within typical ranges and can be met with standard high heritage systems (\S\,\ref{sec:spacecraft}). PICO's spin architecture and data volume requirements are less typical, and are discussed below.

The spin system is less demanding than the successful SMAP spin system. The PICO spin rate is 1\,rpm, and the mission requires $\sim220$\,N\,m\,s of spin angular momentum cancellation (\S\,\ref{sec:attitude_determination}).
Only data and power lines pass across the spin interface between the spinning and non-spinning modules (Fig.~\ref{fig:ArchitectureBlockDiagram}). With SMAP, a 6-m instrument antenna is spun at 14.6\,rpm, and it requires  359\,N\,m\,s of spin angular momentum cancellation~\citep{Brown2016}. Data and power successfully pass across the spin interface. 

Though PICO's data volume is notable by current standards, it is already surpassed by missions in development. The mission produces 6.1\,Tb/day of raw data which is compressed to 1.5\,Tb/day (\S\,\ref{sec:detector_readout}). Data downlinks occur daily, but we baseline storage of 3\,days of (compressed) data to mitigate missed telecom passes. This requires 4.5\,Tb of onboard storage, in family with the 3.14\,Tb solid-state recorder currently in use by Landsat~8 and much smaller than the 12\,Tb flash memory planned for NISAR~\citep{Jasper2017}. The baseline 150\,Mb/s Ka-band data downlink is an existing DSN catalog service~\citep{DSN2015}. The baseline mission generates 2,200\,Tb of raw (uncompressed) data per year, less than the 6,800\,Tb/year currently returned by Landsat~8 and 9,300\,Tb/yr planned by NISAR~\citep{Jasper2017}.

\medskip
\subsection{Risk Assessment}
\label{sec:risk_assessment} %6.4

\subsubsection{Pre-Mission Risks}
\label{sec:premission_risks} %6.4.1

Technology development (\S\,\ref{sec:technology_maturation}) is performed prior to the beginning of mission development, and is outside of the mission cost (per NASA direction), so associated risks do not represent threats to the cost of mission development. Rather, these technology development risks affect the availability of components for the baseline mission. A technology-related mission descope is described in \S\,\ref{sec:technology_descopes}.

\subsubsection{Development Risks}
\label{sec:development_risks} %6.4.2

PICO's healthy contingencies, margins, and reserves provide
flexibility to address risks realized during mission development. PICO
carries $>40\,\%$ instrument sensitivity margin (Table~\ref{tab:bands}),
$>100\,\%$ heat lift margin (Table~\ref{tab:cooler}), $43\,\%$ system
power contingency, $31\,\%$ payload mass contingency, and $25\,\%$
spacecraft mass contingency. The Falcon~9 launch capability (assuming ocean
recovery) exceeds PICO's total launch mass (including contingency) by
a $50\,\%$ margin. The PICO budget includes $30\,\%$ cost
reserves for Phases A--D (\S\,\ref{sec:mission_cost}).

%During mission development the Project Systems Engineer continually assesses risks, tracks progress toward retiring them, and updates mitigations. 
Mitigations for %a few top
the risks identified during this study are described below. \\
$\bullet$ \hspace{0.05in} Thermal risk can be mitigated through extensive thermal modeling and
review in Phase A, and design for early test verification. \\
$\bullet$ \hspace{0.05in} Risks associated with the instrument spin architecture can be mitigated by engaging JPL engineers who were involved in the SMAP mission. \\
$\bullet$ \hspace{0.05in} Detector delivery schedule risk can be mitigated by beginning fabrication early in the project life cycle and fabricating a generous number of detector wafers to ensure adequate yield. Multiple institutions (including, for example, JPL, GSFC, NIST, and ANL) would be capable of producing the PICO detectors. \\
%Suborbital programs generally achieve $>66\,\%$ detector wafer yield. \\
$\bullet$ \hspace{0.05in} Risks associated with the integration and test of a cryogenic instrument can be mitigated through advanced planning and allocation of appropriate schedule and schedule margin.

\subsubsection{Operations Risks}
\label{sec:operations_risks} %6.4.3
%% Table 6.1
\definecolor{mygray}{gray}{0.6}
\begin{wraptable}[26]{r}{0.49\textwidth}
\vspace{-3pt}
\begin{minipage}{0.48\textwidth}
\caption{\captiontext 
  Detailed breakdown of Team X and PICO Team cost estimates (in FY18\$). Costs are based on the schedule in Fig.~\ref{fig:Schedule}, which includes 5~years of operations.}\label{tab:cost}
\begingroup
%\openup 5pt
\newdimen\tblskip \tblskip=5pt
\nointerlineskip
\vskip -7mm
\footnotesize %\footnotesize
\setbox\tablebox=\vbox{
    \newdimen\digitwidth
    \setbox0=\hbox{\rm 0}
    \digitwidth=\wd0
    \catcode`*=\active
    \def*{\kern\digitwidth}
    \newdimen\signwidth
    \setbox0=\hbox{+}
    \signwidth=\wd0
    \catcode`!=\active
    \def!{\kern\signwidth}
\halign{
\hbox {\vtop{\hsize1.95in\hangafter=1\hangindent=2em\noindent\raggedright\strut#\strut\par}}\tabskip=0.6em&
\hfil#\hfil&
\hfil#\hfil\tabskip=0pt\cr
\noalign{\doubleline}
\omit\hfil Work Breakdown Structure\hfil&&\cr
\omit\hfil (WBS) elements\hfil&Team X&PICO\cr
\noalign{\vskip 3pt\hrule\vskip 5pt}
Development Cost (Phases A--D)&\$\,724M&\$\,634--677M\cr
%\noalign{\vskip 3pt\hrule\vskip 5pt}
\quad 1.0, 2.0, 3.0 Management, Systems Engineering, and Mission Assurance&\$\,*54M&\$\,*47--*50M\cr
\quad 4.0 Science&\multispan2 \hfil\$\,*19M\hfil\cr
\quad 5.0 Payload System&\multispan2 \hfil\$\,168M\hfil\cr
\quad 6.0 Flight System&\$\,248M&\$\,210--240M\cr
\quad 10.0 Assembly, Test, and Launch Operations (ATLO)&\$\,*24M&\cr	
\quad 7.0 Mission Operations Preparation&\multispan2 \hfil\$\,*16M\hfil\cr
\quad 9.0 Ground Data Systems&\multispan2 \hfil\$\,*21M\hfil\cr
\quad 12.0 Mission and Navigation Design&\multispan2 \hfil\$\,**7M\hfil\cr
\quad Development Reserves (30\%)&\$167M&\$\,146--156M\cr
\noalign{\vskip 3pt{\color{mygray}\hrule}\vskip 5pt}
Operations Cost (Phase E)&\multispan2 \hfil \$\,*84M\hfil\cr
%\noalign{\vskip 3pt\hrule\vskip 5pt}
\quad 1.0 Management&\multispan2 \hfil \$\,**6M \hfil\cr
\quad 4.0 Science&\multispan2 \hfil\$\,*20M\hfil\cr
\quad 7.0 Mission Operations&\multispan2 \hfil\$\,*34M\hfil\cr
\quad 9.0 Ground Data Systems&\multispan2 \hfil \$\,*14M\hfil\cr
\quad Operations Reserves (13\%)&\multispan2 \hfil \$\,*10M\hfil\cr
\noalign{\vskip 3pt{\color{mygray}\hrule}\vskip 5pt}
Launch Vehicle Cost&\multispan2 \hfil\$\,150M\hfil\cr
\noalign{\vskip 3pt\hrule\vskip 5pt}
Total Cost&\$\,958M&\$\,868--911M\cr
\noalign{\vskip 5pt\hrule\vskip 3pt}
} % close halign
} % close vbox
\endPlancktable
\endgroup
\end{minipage}
\end{wraptable}

The PICO design meets the requirements associated with the NASA Class~B risk classification. For Class~B missions, essential spacecraft and instrument functions are typically fully redundant. This increases mission cost, but significantly reduces the risk of mission failure.

The PICO mission utilizes a single instrument with a single observing mode mapping the sky using a repetitive survey pattern. The mission does not require any time-critical activities. The observatory fits into 
the launch vehicle fairing in its operational configuration, therefore no hardware deployments are required. 
Because PICO observes at long wavelengths, the telescope does not require a dust cover (nor the %associated
 mission-critical cover release).

The spacecraft incorporates a fault protection system for anomaly
detection and resolution. The Sun-pointed, command receptive,
thermally stable safe-mode attitude allows ground intervention for
fault resolution without time constraints. PICO's high degree of
hardware redundancy and onboard fault protection ensure spacecraft
safety in the event of unforeseen failures and faults.

As described in \S\,\ref{sec:signal_separation} and
\S\,\ref{sec:systematics}, pre-Phase A simulation software maturation
is recommended to mitigate the challenges associated with foreground
separation and systematics control.

%\newpage  % temporay newpage to make wrapfig work and allow better length estimate of document.  to be removed!!

\medskip
\subsection{Mission Cost}
\label{sec:mission_cost} %6.5
%\costfootnote
%
%\input tables/table6.1-2_vert_wrap.tex

We estimate PICO's total Phase A--E lifecycle cost between \$870M and \$960M, including the \$150M allocation for the Launch Vehicle (per NASA direction). These cost estimates include 30\,\% reserves for development (Phases A--D) and 13\,\% reserves for operations (Phase~E). Pre-Phase-A technology maturation 
%(\S\,\ref{sec:technology_maturation}) 
will be accomplished through the normal APRA and SAT processes, and is not included in the mission cost. 
%(per NASA direction).

Table~\ref{tab:cost} shows the mission cost breakdown, including the JPL Team~X\cref{teamx} cost estimate, as well as the PICO team cost estimate. Team~X estimates are generally model-based, and were generated after a series of instrument and mission-level studies. Their accuracy is commensurate with the level of understanding typical to Pre-Phase-A concept development. They do not constitute an implementation or cost commitment on the part of JPL or Caltech.

The PICO team has adopted the Team~X estimates, but also obtained a parametrically estimated cost range for the Flight System (WBS 6) and Assembly, Test, and Launch Operations (ATLO, WBS~7) from Lockheed Martin Corporation to represent the cost benefits that might be realized by working with an industry partner. After adding estimated JPL overhead and Team~X estimated V-groove assembly costs (not included in the Lockheed estimate), the PICO team cost is in-family with but lower than the Team~X cost.

Management, Systems Engineering, and Mission Assurance (WBS 1--3)
development costs scale linearly with the WBS 4--12 development costs
in the Team~X model, and are adjusted accordingly in the PICO team
estimate. Science team (WBS~4) costs are assessed by Team~X based on PICO
science team estimates of the numbers and types of contributors and
meetings required for each year of PICO mission development and
operations. These workforce estimates are informed by recent
experience with the \textit{Planck} mission.

Payload system (WBS~5) costs are discussed in detail in
\S\,\ref{sec:instrument_cost}.  PICO's spacecraft (WBS~6) cost
reflects a robust Class~B architecture
(\S\,\ref{sec:spacecraft}). Mission-critical elements are
redundant. Appropriate flight spares, engineering models and
prototypes are budgeted. The V-groove assembly (\S\,\ref{sec:radiative_cooling})
is costed in WBS~6.  Mission operations (WBS~7), Ground Data Systems
(WBS~9), and Mission Navigation and Design (WBS~12) costs reflect a
relatively simple concept of operations (\S\,\ref{sec:operations}). PICO has a single
instrument and a single science observing mode. It surveys the sky
continuously using a pre-planned repetitive survey pattern. Orbit
maintenance activities are simple and infrequent.

\subsubsection{Payload Cost}
\label{sec:instrument_cost} %6.5.1

%% Table 6.2

\begin{wraptable}[10]{R}{2.6in}
\vspace{-8pt}
 \begin{minipage}{2.6in}
\caption{\captiontext  Detailed breakdown of PICO instrument costs.}\label{tab:instrument_cost}
\begingroup
%\openup 5pt
\newdimen\tblskip \tblskip=5pt
\nointerlineskip
\vskip -7mm
\footnotesize %\footnotesize
\setbox\tablebox=\vbox{
    \newdimen\digitwidth
    \setbox0=\hbox{\rm 0}
    \digitwidth=\wd0
    \catcode`*=\active
    \def*{\kern\digitwidth}
    \newdimen\signwidth
    \setbox0=\hbox{+}
    \signwidth=\wd0
    \catcode`!=\active
    \def!{\kern\signwidth}
\halign{
\hbox to 2.1in{#\leaderfil}\tabskip=0.6em&
\hfil#\hfil\tabskip=0pt\cr
\noalign{\doubleline}
\omit\hfil Instrument Elements\hfil&Cost\cr
\noalign{\vskip 3pt\hrule\vskip 5pt}
Management, Systems Eng., Assurance&\$\,*18M\cr
4\,K Cooler and 0.1\,K cADR&\$\,*71M\cr
Focal plane and electronics&\$\,*27M\cr
Mechanical, Thermal, Software&\$\,*17M\cr
Telescope&\$\,**6M\cr
Instrument integration and test&\$\,*29M\cr
\noalign{\vskip 3pt\hrule\vskip 5pt}
Total Instrument Cost&\$\,168M\cr
\noalign{\vskip 5pt\hrule\vskip 3pt}
} % close halign
} % close vbox
\endPlancktable
\endgroup
\end{minipage}
\end{wraptable}

The PICO payload consists of a single instrument: an imaging
polarimeter. Payload costs are tabulated in
Table~\ref{tab:instrument_cost}.

The superconducting detectors require sub-kelvin cooling to
operate. The active cooling system (the 0.1\,K cADR and 4\,K
cryocooler, \S\,\ref{sec:cadr} and \S\,\ref{sec:4kcooler}) comprises nearly half of the payload
cost. The cADR cost for this study is an estimate from Goddard
Space Flight Center, and assumes the provision of both a flight
model and an engineering model. GSFC has produced ADRs for multiple
spaceflight missions. The 4\,K cryocooler cost for this study is based
on the NASA Instrument Cost Model (NICM) VIII CER Cryocooler model
\cite{Mrozinski2017}, assuming a commercial build. PICO benefits
greatly from recent and ongoing investment by commercial suppliers of
4\,K coolers (as described in \S\,\ref{sec:4kcooler}).  Team~X used NICM~VIII to model
the cost of the focal plane and dual string readout electronics (\S\,\ref{sec:focal_plane},
\S\,\ref{sec:detector_readout}).  Team~X estimated the telescope cost using the Stahl model
\cite{Stahl2016}. The telescope is not a major cost driver, primarily
because the reflectors only need to be diffraction limited at 330\,$\mu$m
(900\,GHz) (\S\,\ref{sec:telescope}).

Based on JPL experience, 18\,\% of the instrument cost is allocated for integration and testing (I\&T). This includes I\&T of the flight focal-plane assembly with the flight cADR and then I\&T of the complete instrument including the focal-plane assembly, reflectors, structures, and coolers (\S\,\ref{sec:iandt}). I\&T of the instrument with the spacecraft is costed in WBS~10 (ATLO).
%\costfootnote

\newpage

% [Amy] NASA recently dictated that all Probes add a cost table using
% their standard template. This does not count against the 50 page
% limit. We should paste this in as a graphic rather than converting
% it to a LaTeX table – the whole idea is to standardize, so
% reformatting is not appropriate. The existing cost section (6.5)
% remains (including Table 6.1) – this is an add-on, and should get
% its own page.

\section*{NASA Standard Template Cost Table}

\begin{centering}
\includegraphics{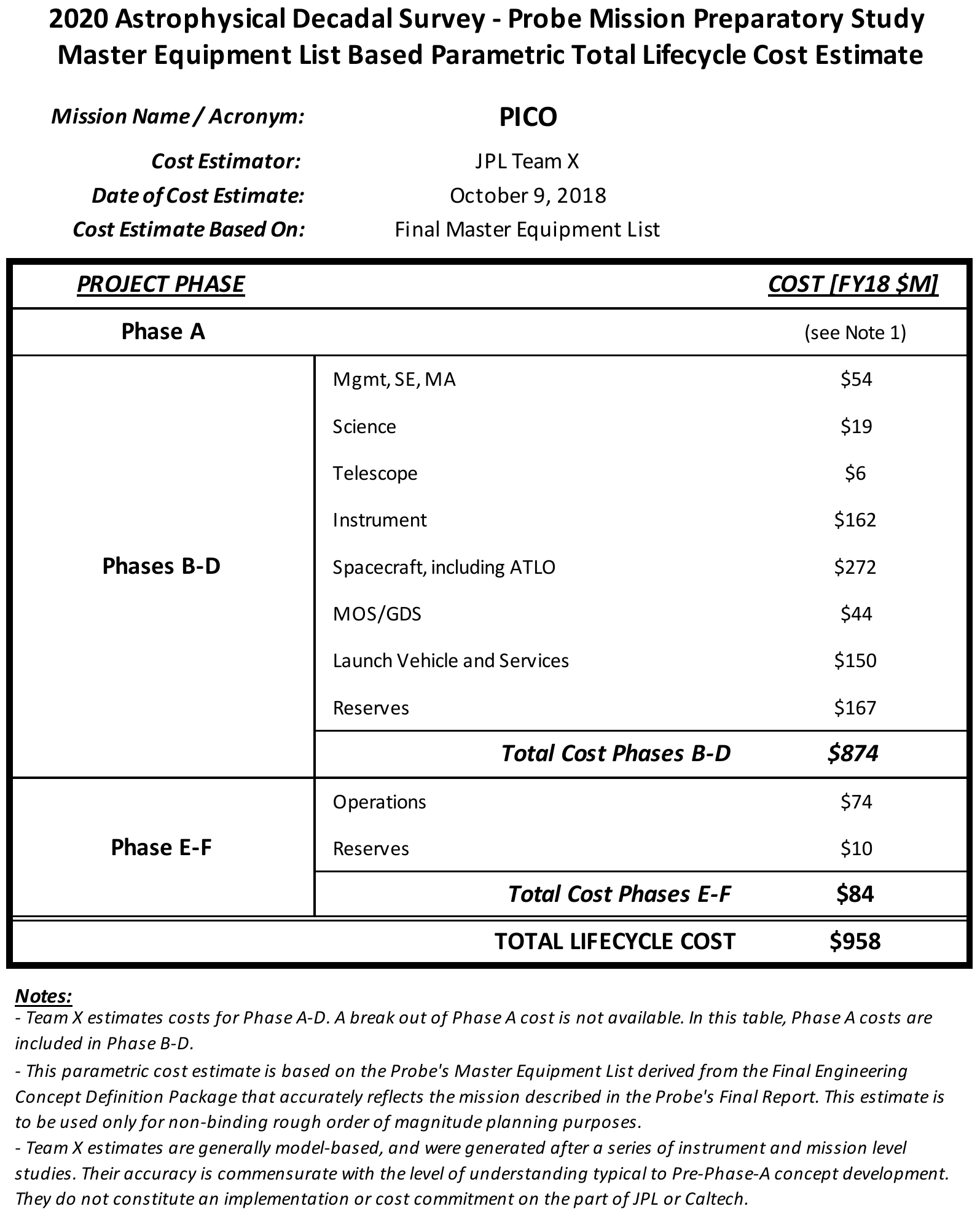}
\end{centering}

\newpage
\def\bibfont{\footnotesize}
\setlength{\bibsep}{1pt}
\bibliographystyle{IEEEtranN}
\bibliography{mybib,pico-report}

\begin{acronym}
    %A
    \acro{ACS}{attitude control system}
    \acro{ADC}{analog-to-digital converters}
    \acro{ADS}{attitude determination software}
    \acro{AHWP}{achromatic half-wave plate}
    \acro{AMC}{Advanced Motion Controls}
    \acro{AME}{anomalous microwave emission}
    \acro{ARC}{anti-reflection coatings}
    \acro{ATA}{advanced technology attachment}
    %B
    \acro{BAO}{baryon acoustic oscillations}
    \acro{BRC}{bolometer readout crates}
    \acro{BLAST}{Balloon-borne Large-Aperture Submillimeter Telescope}
    %C
    \acro{CANbus}{controller area network bus}
    \acro{CBE}{current best estimate}
    \acro{CIB}{cosmic infrared background}
    \acro{CMB}{cosmic microwave background}
    \acro{CMM}{coordinate measurement machine}
    \acro{CSBF}{Columbia Scientific Balloon Facility}
    \acro{CCD}{charge coupled device}
    %D
    \acro{DAC}{digital-to-analog converters}
    \acro{DASI}{Degree~Angular~Scale~Interferometer}
    \acro{dGPS}{differential global positioning system}
    \acro{DfMUX}{digital~frequency~domain~multiplexer}
    \acro{DLFOV}{diffraction limited field of view}
    \acro{DSP}{digital signal processing}
    %E
    \acro{EBEX}{E~and~B~Experiment}
    \acro{EBEX2013}{EBEX2013}
    \acro{ELIS}{EBEX low inductance striplines}
    \acro{ETC}{EBEX test cryostat}
    %F
    \acro{FDM}{frequency domain multiplexing}
    \acro{FPGA}{field programmable gate array}
    \acro{FCP}{flight control program}
    \acro{FOV}{field of view}
    \acro{FWHM}{full width half maximum}
    %G
    \acro{GPS}{global positioning system}
    %H
    \acro{HDPE}{high density polyethylene}
    \acro{HIM}{high index materials}
    \acro{HWP}{half-wave plate} 
    %I
    \acro{IA}{integrated attitude}
    \acro{ICM}{intercluster medium}
    \acro{IGM}{intergalactic medium}
    \acro{IGW}{inflationary gravity wave} 
    \acro{ILC}{independent linear combination}
    \acro{IP}{instrumental polarization} 
    \acro{ISM}{interstellar medium}
    %J
    \acro{JSON}{JavaScript Object Notation}
    %L
    \acro{LDB}{long duration balloon}
    \acro{LED}{light emitting diode}
    \acro{LC}{inductor and capacitor}
    \acro{LCS}{liquid cooling system}
    \acro{LZH}{Lazer Zentrum Hannover}
%M
    \acro{MCP}{multi-color pixel}
    \acro{MSM}{millimeter and sub-millimeter}    
    \acro{MLR}{multilayer reflective}
    \acro{MAXIMA}{Millimeter~Anisotropy~eXperiment~IMaging~Array}
    %N
    \acro{NASA}{National Aeronautics and Space Administration}
    \acro{NDF}{neutral density filter}
    %P
    \acro{PCB}{printed circuit board}
    \acro{PE}{polyethylene}
%    \acro{PTFE}{polytetrafluoroethylene}
    \acro{PME}{polarization modulation efficiency}
    \acro{PSF}{point spread function}
    \acro{PV}{pressure vessel}
    \acro{PWM}{pulse width modulation}
    %R
    \acro{RMS}{root mean square}
%S
    \acro{SED}{spectral energy distribution}
    \acro{SLR}{single layer reflective}
    \acro{SMB}{superconducting magnetic bearing}
    \acro{SNR}{signal-to-noise ratio}
    \acro{SOs}{science objectives}
    \acro{SO}{science objective}
    \acro{SQUID}{superconducting quantum interference device}
    \acro{SQL}{structured query language}
    \acro{STARS}{star tracking attitude reconstruction software}
    \acro{SZ}{Sunyaev--Zeldovich}
    \acro{SWS}{sub-wavelength structures}
%T
    \acro{tSZ}{thermal Sunyaev--Zeldovich}
    \acro{TES}{transition-edge-sensor}
    \acro{TDRSS}{tracking and data relay satellites}
   \acro{TM}{transformation matrix}
   \acro{TRL}{Technology Readiness Level}
% U
    \acro{UHMWPE}{ultra high molecular weight polyethylene}   
    \acro{UMN}{University of Minnesota}
    
\end{acronym}

\end{document}